\documentclass[apjl,ip,twocolumn]{aastex61}
\usepackage{comment}
\usepackage{ifthen}

\newcommand{\forloop}[5][1]%
{%
\setcounter{#2}{#3}%
\ifthenelse{#4}%
	{%
	#5%
	\addtocounter{#2}{#1}%
	\forloop[#1]{#2}{\value{#2}}{#4}{#5}%
	}%
	{%
	}%
}%

\newcommand{\ctbd}[1]{}

\newcommand{\Lc}{Light curve}

\newcommand{\masy}{\ensuremath{\rm mas\,yr^{-1}}}
\newcommand{\kms}{\ensuremath{\rm km\,s^{-1}}}
\newcommand{\ms}{\ensuremath{\rm m\,s^{-1}}}

\newcommand{\gcmc}{\ensuremath{\rm g\,cm^{-3}}}

\newcommand{\C}{\ensuremath{^{\circ}C\;}}

\newcommand{\vsini}{\ensuremath{v \sin{i}}}
\newcommand{\feh}{\ensuremath{\rm [Fe/H]}}

\newcommand{\rsun}{\ensuremath{R_\sun}}
\newcommand{\msun}{\ensuremath{M_\sun}}
\newcommand{\lsun}{\ensuremath{L_\sun}}

\newcommand{\rstar}{\ensuremath{R_\star}}
\newcommand{\mstar}{\ensuremath{M_\star}}
\newcommand{\lstar}{\ensuremath{L_\star}}

\newcommand{\teffstar}{\ensuremath{T_{\rm eff\star}}}
\newcommand{\rhostar}{\ensuremath{\rho_\star}}
\newcommand{\loggstar}{\ensuremath{\log{g_{\star}}}}

\newcommand{\rpl}{\ensuremath{R_{p}}}
\newcommand{\mpl}{\ensuremath{M_{p}}}

\newcommand{\rhopl}{\ensuremath{\rho_{p}}}

\newcommand{\arstar}{\ensuremath{a/\rstar}}
\newcommand{\zrstar}{\ensuremath{\zeta/\rstar}}

\newcommand{\rjup}{\ensuremath{R_{\rm J}}}
\newcommand{\mjup}{\ensuremath{M_{\rm J}}}

\newcommand{\reftabl}[1]{Table~\ref{tab:#1}}

\newcommand{\loopand}{\ifnum\value{planetcounter}=2 and \else\fi}
\newcommand{\loopcomma}{\ifnum\value{planetcounter}<2 ,\else. \fi}
\newcommand{\loopcommanoperiod}{\ifnum\value{planetcounter}<2 ,\else \space\fi}
\newcommand{\loopcommanospace}{\ifnum\value{planetcounter}<2 ,\else \fi}

\newcommand{\hatcurhtrxxxxxA}{HATS563-025}                      
\newcommand{\hatcurfieldxxxxxA}{\ensuremath{string}}            
\newcommand{\hatcurCCraxxxxxA}{\ensuremath{11^{\mathrm h}24^{\mathrm m}03.5929{\mathrm s}}}                   
\newcommand{\hatcurCCdecxxxxxA}{\ensuremath{-19{\arcdeg}33{\arcmin}25.6653{\arcsec}}}                 
\newcommand{\hatcurCCmagxxxxxA}{NULL}                           
\newcommand{\hatcurCCtwomassxxxxxA}{2MASS~11240360-1933257}     
\newcommand{\hatcurCCgscxxxxxA}{GSC~NULL}                       
\newcommand{\hatcurCCgaiaxxxxxA}{GAIA~3545653561940141696}      
\newcommand{\hatcurCCgaiadrtwoxxxxxA}{GAIA~DR2~3545653561942122368} 
\newcommand{\hatcurCCgaiadrtwoshortxxxxxA}{3545653561942122368} 
\newcommand{\hatcurCCtassmvxxxxxA}{\ensuremath{nff\pmnff}}      
\newcommand{\hatcurCCtassmvshortxxxxxA}{\ensuremath{0.0}}       
\newcommand{\hatcurCCtassmBxxxxxA}{\ensuremath{nff\pmnff}}      
\newcommand{\hatcurCCtassmBshortxxxxxA}{\ensuremath{0.0}}       
\newcommand{\hatcurCCtassmIxxxxxA}{\ensuremath{nff\pmnff}}      
\newcommand{\hatcurCCtassmIshortxxxxxA}{\ensuremath{0.0}}       
\newcommand{\hatcurCCtassmgxxxxxA}{\ensuremath{nff\pmnff}}      
\newcommand{\hatcurCCtassmgshortxxxxxA}{\ensuremath{0.0}}       
\newcommand{\hatcurCCtassmrxxxxxA}{\ensuremath{nff\pmnff}}      
\newcommand{\hatcurCCtassmrshortxxxxxA}{\ensuremath{0.0}}       
\newcommand{\hatcurCCtassmixxxxxA}{\ensuremath{nff\pmnff}}      
\newcommand{\hatcurCCtassmishortxxxxxA}{\ensuremath{0.0}}       
\newcommand{\hatcurCCparallaxxxxxxA}{\ensuremath{3.425\pm0.042}} 
\newcommand{\hatcurCCgaiamGxxxxxA}{\ensuremath{15.9706\pm0.0029}} 
\newcommand{\hatcurCCgaiamBPxxxxxA}{\ensuremath{0\pm0}}         
\newcommand{\hatcurCCgaiamRPxxxxxA}{\ensuremath{0\pm0}}         
\newcommand{\hatcurCCtwomassJmagxxxxxA}{\ensuremath{13.341\pm0.023}} 
\newcommand{\hatcurCCtwomassHmagxxxxxA}{\ensuremath{12.687\pm0.029}} 
\newcommand{\hatcurCCtwomassKmagxxxxxA}{\ensuremath{12.452\pm0.031}} 
\newcommand{\hatcurCCcitJmagxxxxxA}{\ensuremath{13.239\pm0.025}} 
\newcommand{\hatcurCCcitHmagxxxxxA}{\ensuremath{12.592\pm0.030}} 
\newcommand{\hatcurCCcitKmagxxxxxA}{\ensuremath{12.385\pm0.032}} 
\newcommand{\hatcurCCbbJmagxxxxxA}{\ensuremath{13.331\pm0.027}} 
\newcommand{\hatcurCCbbHmagxxxxxA}{\ensuremath{12.619\pm0.031}} 
\newcommand{\hatcurCCbbKmagxxxxxA}{\ensuremath{12.405\pm0.032}} 
\newcommand{\hatcurCCesoJmagxxxxxA}{\ensuremath{13.340\pm0.032}} 
\newcommand{\hatcurCCesoHmagxxxxxA}{\ensuremath{12.619\pm0.046}} 
\newcommand{\hatcurCCesoKmagxxxxxA}{\ensuremath{12.401\pm0.033}} 
\newcommand{\hatcurCCesoJHmagxxxxxA}{\ensuremath{0.720\pm0.052}} 
\newcommand{\hatcurCCesoJKmagxxxxxA}{\ensuremath{0.939\pm0.044}} 
\newcommand{\hatcurCCesoHKmagxxxxxA}{\ensuremath{0.218\pm0.056}} 
\newcommand{\hatcurCCWonemagxxxxxA}{\ensuremath{12.342\pm0.023}} 
\newcommand{\hatcurCCWtwomagxxxxxA}{\ensuremath{12.326\pm0.023}} 
\newcommand{\hatcurCCWthreemagxxxxxA}{\ensuremath{nff\pmnff}}   
\newcommand{\hatcurCCWfourmagxxxxxA}{\ensuremath{nff\pmnff}}    
\newcommand{\hatcurLCdipxxxxxA}{\ensuremath{45.4}}              
\newcommand{\hatcurLCrprstarxxxxxA}{\ensuremath{0.1844\pm0.0030}} 
\newcommand{\hatcurLCbsqxxxxxA}{\ensuremath{0.165_{-0.033}^{+0.029}}} 
\newcommand{\hatcurLCimpxxxxxA}{\ensuremath{0.406_{-0.043}^{+0.034}}} 
\newcommand{\hatcurLCzetaxxxxxA}{\ensuremath{35.31_{-0.42}^{+0.56}}} 
\newcommand{\hatcurLCdurxxxxxA}{\ensuremath{0.06889\pm0.00064}} 
\newcommand{\hatcurLCdurshortxxxxxA}{\ensuremath{0.0689}}       
\newcommand{\hatcurLCdurhrxxxxxA}{\ensuremath{1.653\pm0.015}}   
\newcommand{\hatcurLCdurhrshortxxxxxA}{\ensuremath{1.653}}      
\newcommand{\hatcurLCqxxxxxA}{\ensuremath{0.03980\pm0.00037}}   
\newcommand{\hatcurLCqshortxxxxxA}{\ensuremath{0.040}}          
\newcommand{\hatcurLCingdurxxxxxA}{\ensuremath{0.01256\pm0.00049}} 
\newcommand{\hatcurLCPxxxxxA}{\ensuremath{1.73185606\pm0.00000055}} 
\newcommand{\hatcurLCPprecxxxxxA}{\ensuremath{1.7318561}}       
\newcommand{\hatcurLCPshortxxxxxA}{\ensuremath{1.7319}}         
\newcommand{\hatcurLCTxxxxxA}{\ensuremath{2458392.02654\pm0.00024}} 
\newcommand{\hatcurLCTAxxxxxA}{\ensuremath{2455217.5344\pm0.0011}} 
\newcommand{\hatcurLCTBxxxxxA}{\ensuremath{2458566.94400\pm0.00023}} 
\newcommand{\hatcurLChatnetmAxxxxxA}{\ensuremath{15.17166\pm0.00040}} 
\newcommand{\hatcurLCiblendAxxxxxA}{\ensuremath{0.954\pm0.046}} 
\newcommand{\hatcurLChatnetmBxxxxxA}{\ensuremath{19.26561\pm0.00022}} 
\newcommand{\hatcurLCiblendBxxxxxA}{\ensuremath{0.963\pm0.033}} 
\newcommand{\hatcurLCrhoxxxxxA}{\ensuremath{4.43_{-0.14}^{+0.18}}} 
\newcommand{\hatcurSMEiteffxxxxxA}{\ensuremath{3775\pm54}}      
\newcommand{\hatcurSMEizfehxxxxxA}{\ensuremath{0.294\pm0.088}}  
\newcommand{\hatcurSMEizfehshortxxxxxA}{\ensuremath{0.29}}      
\newcommand{\hatcurSMEiloggxxxxxA}{\ensuremath{4.50\pm0.50}}    
\newcommand{\hatcurSMEivsinxxxxxA}{\ensuremath{0\pm50}}         
\newcommand{\hatcurSMEivmacxxxxxA}{\ensuremath{nff\pmnff}}      
\newcommand{\hatcurSMEivmicxxxxxA}{\ensuremath{nff\pmnff}}      
\newcommand{\hatcurextraerrMJxxxxxA}{\ensuremath{0\pm0}}        
\newcommand{\hatcurextraerrMJtwosiglimxxxxxA}{\ensuremath{<0.0200}} 
\newcommand{\hatcurextraerrMHxxxxxA}{\ensuremath{0\pm0}}        
\newcommand{\hatcurextraerrMHtwosiglimxxxxxA}{\ensuremath{<0.0200}} 
\newcommand{\hatcurextraerrMKsxxxxxA}{\ensuremath{0\pm0}}       
\newcommand{\hatcurextraerrMKstwosiglimxxxxxA}{\ensuremath{<0.0200}} 
\newcommand{\hatcurextraerrMGxxxxxA}{\ensuremath{0\pm0}}        
\newcommand{\hatcurextraerrMGtwosiglimxxxxxA}{\ensuremath{<0.0200}} 
\newcommand{\hatcurextraerrMWonexxxxxA}{\ensuremath{0\pm0}}     
\newcommand{\hatcurextraerrMWonetwosiglimxxxxxA}{\ensuremath{<0.0200}} 
\newcommand{\hatcurextraerrMWtwoxxxxxA}{\ensuremath{0\pm0}}     
\newcommand{\hatcurextraerrMWtwotwosiglimxxxxxA}{\ensuremath{<0.0200}} 
\newcommand{\hatcurLBiBxxxxxA}{\ensuremath{0.5710}}             
\newcommand{\hatcurLBiiBxxxxxA}{\ensuremath{0.2510}}            
\newcommand{\hatcurLBiVxxxxxA}{\ensuremath{0.4895}}             
\newcommand{\hatcurLBiiVxxxxxA}{\ensuremath{0.3055}}            
\newcommand{\hatcurLBiRxxxxxA}{\ensuremath{0.30\pm0.14}}        
\newcommand{\hatcurLBiiRxxxxxA}{\ensuremath{0.29\pm0.17}}       
\newcommand{\hatcurLBiIxxxxxA}{\ensuremath{0.2558}}             
\newcommand{\hatcurLBiiIxxxxxA}{\ensuremath{0.3251}}            
\newcommand{\hatcurLBiuxxxxxA}{\ensuremath{0.5943}}             
\newcommand{\hatcurLBiiuxxxxxA}{\ensuremath{0.2214}}            
\newcommand{\hatcurLBigxxxxxA}{\ensuremath{0.39\pm0.17}}        
\newcommand{\hatcurLBiigxxxxxA}{\ensuremath{0.29\pm0.20}}       
\newcommand{\hatcurLBirxxxxxA}{\ensuremath{0.40\pm0.15}}        
\newcommand{\hatcurLBiirxxxxxA}{\ensuremath{0.23\pm0.18}}       
\newcommand{\hatcurLBiixxxxxA}{\ensuremath{0.45\pm0.14}}        
\newcommand{\hatcurLBiiixxxxxA}{\ensuremath{0.27\pm0.18}}       
\newcommand{\hatcurLBizxxxxxA}{\ensuremath{0.32\pm0.12}}        
\newcommand{\hatcurLBiizxxxxxA}{\ensuremath{0.17\pm0.16}}       
\newcommand{\hatcurLBiJxxxxxA}{\ensuremath{0.1499}}             
\newcommand{\hatcurLBiiJxxxxxA}{\ensuremath{0.2488}}            
\newcommand{\hatcurLBiHxxxxxA}{\ensuremath{0.1199}}             
\newcommand{\hatcurLBiiHxxxxxA}{\ensuremath{0.2641}}            
\newcommand{\hatcurLBiKxxxxxA}{\ensuremath{0.0910}}             
\newcommand{\hatcurLBiiKxxxxxA}{\ensuremath{0.2450}}            
\newcommand{\hatcurLBiTxxxxxA}{\ensuremath{0.19_{-0.10}^{+0.14}}} 
\newcommand{\hatcurLBiiTxxxxxA}{\ensuremath{0.17\pm0.17}}       
\newcommand{\hatcurLBikepxxxxxA}{\ensuremath{0.3716}}           
\newcommand{\hatcurLBiikepxxxxxA}{\ensuremath{0.3464}}          
\newcommand{\hatcurLBiCxxxxxA}{\ensuremath{0.3335}}             
\newcommand{\hatcurLBiiCxxxxxA}{\ensuremath{0.3503}}            
\newcommand{\hatcurLBiMxxxxxA}{\ensuremath{0.4561}}             
\newcommand{\hatcurLBiiMxxxxxA}{\ensuremath{0.3301}}            
\newcommand{\hatcurLBiSonexxxxxA}{\ensuremath{0.0689}}            
\newcommand{\hatcurLBiiSonexxxxxA}{\ensuremath{0.1890}}           
\newcommand{\hatcurLBiStwoxxxxxA}{\ensuremath{0.0582}}            
\newcommand{\hatcurLBiiStwoxxxxxA}{\ensuremath{0.1489}}           
\newcommand{\hatcurLBiSthreexxxxxA}{\ensuremath{0.0560}}            
\newcommand{\hatcurLBiiSthreexxxxxA}{\ensuremath{0.1294}}           
\newcommand{\hatcurLBiSfourxxxxxA}{\ensuremath{0.0640}}            
\newcommand{\hatcurLBiiSfourxxxxxA}{\ensuremath{0.1031}}           
\newcommand{\hatcurISOmxxxxxA}{\ensuremath{0.6010\pm0.0080}}    
\newcommand{\hatcurISOmshortxxxxxA}{\ensuremath{0.60}}          
\newcommand{\hatcurISOmlongxxxxxA}{\ensuremath{0.6010\pm0.0080}} 
\newcommand{\hatcurISOrxxxxxA}{\ensuremath{0.5758\pm0.0055}}    
\newcommand{\hatcurISOrshortxxxxxA}{\ensuremath{0.58}}          
\newcommand{\hatcurISOrlongxxxxxA}{\ensuremath{0.5758\pm0.0055}} 
\newcommand{\hatcurISOrhoxxxxxA}{\ensuremath{4.43_{-0.14}^{+0.18}}} 
\newcommand{\hatcurISOrholongxxxxxA}{\ensuremath{4.43_{-0.14}^{+0.18}}} 
\newcommand{\hatcurISOloggxxxxxA}{\ensuremath{4.696\pm0.011}}   
\newcommand{\hatcurISOlumxxxxxA}{\ensuremath{0.0608\pm0.0015}}  
\newcommand{\hatcurISOlumshortxxxxxA}{\ensuremath{0.06}}        
\newcommand{\hatcurISOteffxxxxxA}{\ensuremath{3776.9\pm9.5}}    
\newcommand{\hatcurISOzfehxxxxxA}{\ensuremath{0.514_{-0.021}^{+0.033}}} 
\newcommand{\hatcurISOagexxxxxA}{\ensuremath{11.0\pm5.1}}       
\newcommand{\hatcurISOspecxxxxxA}{M}                            
\newcommand{\hatcurRVKxxxxxA}{\ensuremath{346\pm34}}            
\newcommand{\hatcurRVKtwosiglimxxxxxA}{\ensuremath{<362.2}}     
\newcommand{\hatcurRVrkxxxxxA}{\ensuremath{0\pm0}}              
\newcommand{\hatcurRVrhxxxxxA}{\ensuremath{0\pm0}}              
\newcommand{\hatcurRVkxxxxxA}{\ensuremath{0\pm0}}               
\newcommand{\hatcurRVhxxxxxA}{\ensuremath{0\pm0}}               
\newcommand{\hatcurRVtronexxxxxA}{\ensuremath{0\pm0}}           
\newcommand{\hatcurRVtrtwoxxxxxA}{\ensuremath{0\pm0}}           
\newcommand{\hatcurRVgammaxxxxxA}{\ensuremath{15850\pm12}}      
\newcommand{\hatcurRVjitterxxxxxA}{\ensuremath{0\pm42}}         
\newcommand{\hatcurRVjittertwosiglimxxxxxA}{\ensuremath{<32.1}} 
\newcommand{\hatcurRVfitrmsxxxxxA}{\ensuremath{.1fym}}          %
\newcommand{\hatcurRVeccenxxxxxA}{\ensuremath{0\pm0}}           
\newcommand{\hatcurRVeccentwosiglimxxxxxA}{\ensuremath{<0.000}} 
\newcommand{\hatcurRVomegaxxxxxA}{\ensuremath{0\pm0}}           
\newcommand{\hatcurPPixxxxxA}{\ensuremath{87.39\pm0.29}}        
\newcommand{\hatcurPPgxxxxxA}{\ensuremath{33.8\pm3.7}}          
\newcommand{\hatcurPPloggxxxxxA}{\ensuremath{3.529\pm0.063}}    
\newcommand{\hatcurPParxxxxxA}{\ensuremath{8.90\pm0.10}}        
\newcommand{\hatcurPParelxxxxxA}{\ensuremath{0.02384\pm0.00011}} 
\newcommand{\hatcurPPrhoxxxxxA}{\ensuremath{1.64\pm0.19}}       
\newcommand{\hatcurPPmxxxxxA}{\ensuremath{1.46\pm0.14}}         
\newcommand{\hatcurPPmtwosiglimxxxxxA}{\ensuremath{<1.54}}      
\newcommand{\hatcurPPmshortxxxxxA}{\ensuremath{1.46}}           
\newcommand{\hatcurPPmlongxxxxxA}{\ensuremath{1.46\pm0.14}}     
\newcommand{\hatcurPPmexxxxxA}{\ensuremath{464\pm46}}           
\newcommand{\hatcurPPmeshortxxxxxA}{\ensuremath{464.1}}         
\newcommand{\hatcurPPmelongxxxxxA}{\ensuremath{464\pm46}}       
\newcommand{\hatcurPPrxxxxxA}{\ensuremath{1.032\pm0.021}}       
\newcommand{\hatcurPPrshortxxxxxA}{\ensuremath{1.03}}           
\newcommand{\hatcurPPrlongxxxxxA}{\ensuremath{1.032\pm0.021}}   
\newcommand{\hatcurPPrexxxxxA}{\ensuremath{11.57\pm0.24}}       
\newcommand{\hatcurPPreshortxxxxxA}{\ensuremath{11.6}}          
\newcommand{\hatcurPPrelongxxxxxA}{\ensuremath{11.57\pm0.24}}   
\newcommand{\hatcurPPmrcorrxxxxxA}{\ensuremath{-0.01}}          
\newcommand{\hatcurPPteffxxxxxA}{\ensuremath{895.1\pm5.7}}      
\newcommand{\hatcurPPthetaxxxxxA}{\ensuremath{0.112\pm0.011}}   
\newcommand{\hatcurPPfluxperixxxxxA}{\ensuremath{1.456\pm0.037}} 
\newcommand{\hatcurPPfluxperidimxxxxxA}{\ensuremath{8}}         
\newcommand{\hatcurPPfluxapxxxxxA}{\ensuremath{1.456\pm0.037}}  
\newcommand{\hatcurPPfluxapdimxxxxxA}{\ensuremath{8}}           
\newcommand{\hatcurPPfluxavgxxxxxA}{\ensuremath{1.456\pm0.037}} 
\newcommand{\hatcurPPfluxavgdimxxxxxA}{\ensuremath{8}}          
\newcommand{\hatcurPPfluxavglogxxxxxA}{\ensuremath{8.163\pm0.011}} 
\newcommand{\hatcurXsecphasexxxxxA}{\ensuremath{0\pm0}}         
\newcommand{\hatcurXsecondaryxxxxxA}{\ensuremath{2458392.89247\pm0.00024}} 
\newcommand{\hatcurXsecdurxxxxxA}{\ensuremath{0.06889\pm0.00064}} 
\newcommand{\hatcurXsecingdurxxxxxA}{\ensuremath{0.01256\pm0.00049}} 
\newcommand{\hatcurPPphiconjxxxxxA}{\ensuremath{0\pm0}}         
\newcommand{\hatcurPPperixxxxxA}{\ensuremath{2458391.59358\pm0.00024}} 
\newcommand{\hatcurPPaequivxxxxxA}{\ensuremath{0.0967\pm0.0012}} 
\newcommand{\hatcurPPtcircxxxxxA}{\ensuremath{29.1\pm4.4}}      
\newcommand{\hatcurPPtinfallxxxxxA}{\ensuremath{378_{-23}^{+30}}} 
\newcommand{\hatcurXdistxxxxxA}{\ensuremath{286.6\pm3.0}}       
\newcommand{\hatcurXAvxxxxxA}{\ensuremath{0.1960_{-0.0150}^{+0.0100}}} 
\newcommand{\hatcurXdistredxxxxxA}{\ensuremath{286.6\pm3.0}}    
\newcommand{\hatcurXEBVxxxxxA}{\ensuremath{0.0630_{-0.0050}^{+0.0030}}} 
\newcommand{\hatcurCCpmraxxxxxA}{\ensuremath{-38.86\pm0.12}}    
\newcommand{\hatcurCCpmdecxxxxxA}{\ensuremath{39.679\pm0.077}}  
\newcommand{\hatcurCCpmxxxxxA}{\ensuremath{55.54\pm0.14}}       

\newcommand{\hatcurhtrxxxxxB}{HATS548-007}                      
\newcommand{\hatcurfieldxxxxxB}{\ensuremath{string}}            
\newcommand{\hatcurCCraxxxxxB}{\ensuremath{04^{\mathrm h}03^{\mathrm m}47.8440{\mathrm s}}}                   
\newcommand{\hatcurCCdecxxxxxB}{\ensuremath{-25{\arcdeg}24{\arcmin}32.1170{\arcsec}}}                 
\newcommand{\hatcurCCmagxxxxxB}{15.759}                         
\newcommand{\hatcurCCtwomassxxxxxB}{2MASS~04034783-2524320}     
\newcommand{\hatcurCCgscxxxxxB}{GSC~}                           
\newcommand{\hatcurCCgaiaxxxxxB}{GAIA~5082914333903138048}      
\newcommand{\hatcurCCgaiadrtwoxxxxxB}{GAIA~DR2~5082914338199586560} 
\newcommand{\hatcurCCgaiadrtwoshortxxxxxB}{5082914338199586560} 
\newcommand{\hatcurCCtassmvxxxxxB}{\ensuremath{15.759\pm0.036}} 
\newcommand{\hatcurCCtassmvshortxxxxxB}{\ensuremath{15.8}}      
\newcommand{\hatcurCCtassmBxxxxxB}{\ensuremath{17.120\pm0.019}} 
\newcommand{\hatcurCCtassmBshortxxxxxB}{\ensuremath{17.1}}      
\newcommand{\hatcurCCtassmIxxxxxB}{\ensuremath{nff\pmnff}}      
\newcommand{\hatcurCCtassmIshortxxxxxB}{\ensuremath{0.0}}       
\newcommand{\hatcurCCtassmgxxxxxB}{\ensuremath{16.503\pm0.070}} 
\newcommand{\hatcurCCtassmgshortxxxxxB}{\ensuremath{16.5}}      
\newcommand{\hatcurCCtassmrxxxxxB}{\ensuremath{15.156\pm0.090}} 
\newcommand{\hatcurCCtassmrshortxxxxxB}{\ensuremath{15.2}}      
\newcommand{\hatcurCCtassmixxxxxB}{\ensuremath{14.230\pm0.070}} 
\newcommand{\hatcurCCtassmishortxxxxxB}{\ensuremath{14.2}}      
\newcommand{\hatcurCCparallaxxxxxxB}{\ensuremath{5.100\pm0.029}} 
\newcommand{\hatcurCCgaiamGxxxxxB}{\ensuremath{14.90330\pm0.00040}} 
\newcommand{\hatcurCCgaiamBPxxxxxB}{\ensuremath{16.0189\pm0.0026}} 
\newcommand{\hatcurCCgaiamRPxxxxxB}{\ensuremath{13.8535\pm0.0013}} 
\newcommand{\hatcurCCtwomassJmagxxxxxB}{\ensuremath{12.481\pm0.023}} 
\newcommand{\hatcurCCtwomassHmagxxxxxB}{\ensuremath{11.756\pm0.026}} 
\newcommand{\hatcurCCtwomassKmagxxxxxB}{\ensuremath{11.584\pm0.021}} 
\newcommand{\hatcurCCcitJmagxxxxxB}{\ensuremath{12.470\pm0.025}} 
\newcommand{\hatcurCCcitHmagxxxxxB}{\ensuremath{11.748\pm0.027}} 
\newcommand{\hatcurCCcitKmagxxxxxB}{\ensuremath{11.608\pm0.022}} 
\newcommand{\hatcurCCbbJmagxxxxxB}{\ensuremath{12.562\pm0.027}} 
\newcommand{\hatcurCCbbHmagxxxxxB}{\ensuremath{11.772\pm0.028}} 
\newcommand{\hatcurCCbbKmagxxxxxB}{\ensuremath{11.628\pm0.022}} 
\newcommand{\hatcurCCesoJmagxxxxxB}{\ensuremath{12.571\pm0.032}} 
\newcommand{\hatcurCCesoHmagxxxxxB}{\ensuremath{11.770\pm0.037}} 
\newcommand{\hatcurCCesoKmagxxxxxB}{\ensuremath{11.624\pm0.024}} 
\newcommand{\hatcurCCesoJHmagxxxxxB}{\ensuremath{0.801\pm0.045}} 
\newcommand{\hatcurCCesoJKmagxxxxxB}{\ensuremath{0.947\pm0.037}} 
\newcommand{\hatcurCCesoHKmagxxxxxB}{\ensuremath{0.145\pm0.042}} 
\newcommand{\hatcurCCWonemagxxxxxB}{\ensuremath{11.486\pm0.024}} 
\newcommand{\hatcurCCWtwomagxxxxxB}{\ensuremath{11.502\pm0.022}} 
\newcommand{\hatcurCCWthreemagxxxxxB}{\ensuremath{11.61\pm0.19}} 
\newcommand{\hatcurCCWfourmagxxxxxB}{\ensuremath{nff\pmnff}}    
\newcommand{\hatcurLCdipxxxxxB}{\ensuremath{0.0}}               
\newcommand{\hatcurLCrprstarxxxxxB}{\ensuremath{0.1555\pm0.0021}} 
\newcommand{\hatcurLCbsqxxxxxB}{\ensuremath{0.165_{-0.029}^{+0.024}}} 
\newcommand{\hatcurLCimpxxxxxB}{\ensuremath{0.406_{-0.038}^{+0.029}}} 
\newcommand{\hatcurLCzetaxxxxxB}{\ensuremath{29.71\pm0.38}}     
\newcommand{\hatcurLCdurxxxxxB}{\ensuremath{0.07969\pm0.00065}} 
\newcommand{\hatcurLCdurshortxxxxxB}{\ensuremath{0.0797}}       
\newcommand{\hatcurLCdurhrxxxxxB}{\ensuremath{1.913\pm0.016}}   
\newcommand{\hatcurLCdurhrshortxxxxxB}{\ensuremath{1.913}}      
\newcommand{\hatcurLCqxxxxxB}{\ensuremath{0.02860\pm0.00023}}   
\newcommand{\hatcurLCqshortxxxxxB}{\ensuremath{0.029}}          
\newcommand{\hatcurLCingdurxxxxxB}{\ensuremath{0.01256\pm0.00036}} 
\newcommand{\hatcurLCPxxxxxB}{\ensuremath{2.7886556\pm0.0000011}} 
\newcommand{\hatcurLCPprecxxxxxB}{\ensuremath{2.7886556}}       
\newcommand{\hatcurLCPshortxxxxxB}{\ensuremath{2.7887}}         
\newcommand{\hatcurLCTxxxxxB}{\ensuremath{2458611.05487\pm0.00027}} 
\newcommand{\hatcurLCTAxxxxxB}{\ensuremath{2456840.25858\pm0.00068}} 
\newcommand{\hatcurLCTBxxxxxB}{\ensuremath{2459110.22422\pm0.00036}} 
\newcommand{\hatcurLChatnetmAxxxxxB}{\ensuremath{14.49372\pm0.00046}} 
\newcommand{\hatcurLCiblendAxxxxxB}{\ensuremath{0.756\pm0.061}} 
\newcommand{\hatcurLChatnetmBxxxxxB}{\ensuremath{14.502420\pm0.000086}} 
\newcommand{\hatcurLCiblendBxxxxxB}{\ensuremath{0.956\pm0.030}} 
\newcommand{\hatcurLChatnetmCxxxxxB}{\ensuremath{-0.000670\pm0.000083}} 
\newcommand{\hatcurLCiblendCxxxxxB}{\ensuremath{0.937\pm0.032}} 
\newcommand{\hatcurLChatnetmDxxxxxB}{\ensuremath{-0.000520\pm0.000090}} 
\newcommand{\hatcurLCiblendDxxxxxB}{\ensuremath{0.949\pm0.028}} 
\newcommand{\hatcurLCrhoxxxxxB}{\ensuremath{4.239_{-0.065}^{+0.091}}} 
\newcommand{\hatcurSMEiteffxxxxxB}{\ensuremath{3812\pm79}}      
\newcommand{\hatcurSMEizfehxxxxxB}{\ensuremath{0.28\pm0.12}}    
\newcommand{\hatcurSMEizfehshortxxxxxB}{\ensuremath{0.28}}      
\newcommand{\hatcurSMEiloggxxxxxB}{\ensuremath{4.50\pm0.50}}    
\newcommand{\hatcurSMEivsinxxxxxB}{\ensuremath{0\pm50}}         
\newcommand{\hatcurSMEivmacxxxxxB}{\ensuremath{nff\pmnff}}      
\newcommand{\hatcurSMEivmicxxxxxB}{\ensuremath{nff\pmnff}}      
\newcommand{\hatcurextraerrMJxxxxxB}{\ensuremath{0\pm0}}        
\newcommand{\hatcurextraerrMJtwosiglimxxxxxB}{\ensuremath{<0.0200}} 
\newcommand{\hatcurextraerrMHxxxxxB}{\ensuremath{0\pm0}}        
\newcommand{\hatcurextraerrMHtwosiglimxxxxxB}{\ensuremath{<0.0200}} 
\newcommand{\hatcurextraerrMKsxxxxxB}{\ensuremath{0\pm0}}       
\newcommand{\hatcurextraerrMKstwosiglimxxxxxB}{\ensuremath{<0.0200}} 
\newcommand{\hatcurextraerrMGxxxxxB}{\ensuremath{0\pm0}}        
\newcommand{\hatcurextraerrMGtwosiglimxxxxxB}{\ensuremath{<0.0200}} 
\newcommand{\hatcurextraerrMBPtwoxxxxxB}{\ensuremath{0\pm0}}    
\newcommand{\hatcurextraerrMBPtwotwosiglimxxxxxB}{\ensuremath{<0.0200}} 
\newcommand{\hatcurextraerrMRPxxxxxB}{\ensuremath{0\pm0}}       
\newcommand{\hatcurextraerrMRPtwosiglimxxxxxB}{\ensuremath{<0.0200}} 
\newcommand{\hatcurextraerrMgxxxxxB}{\ensuremath{0\pm0}}        
\newcommand{\hatcurextraerrMgtwosiglimxxxxxB}{\ensuremath{<0.0200}} 
\newcommand{\hatcurextraerrMrxxxxxB}{\ensuremath{0\pm0}}        
\newcommand{\hatcurextraerrMrtwosiglimxxxxxB}{\ensuremath{<0.0200}} 
\newcommand{\hatcurextraerrMixxxxxB}{\ensuremath{0\pm0}}        
\newcommand{\hatcurextraerrMitwosiglimxxxxxB}{\ensuremath{<0.0200}} 
\newcommand{\hatcurextraerrMWonexxxxxB}{\ensuremath{0\pm0}}     
\newcommand{\hatcurextraerrMWonetwosiglimxxxxxB}{\ensuremath{<0.0200}} 
\newcommand{\hatcurextraerrMWtwoxxxxxB}{\ensuremath{0\pm0}}     
\newcommand{\hatcurextraerrMWtwotwosiglimxxxxxB}{\ensuremath{<0.0200}} 
\newcommand{\hatcurLBiBxxxxxB}{\ensuremath{0.5844}}             
\newcommand{\hatcurLBiiBxxxxxB}{\ensuremath{0.2352}}            
\newcommand{\hatcurLBiVxxxxxB}{\ensuremath{0.4945}}             
\newcommand{\hatcurLBiiVxxxxxB}{\ensuremath{0.2944}}            
\newcommand{\hatcurLBiRxxxxxB}{\ensuremath{0.4370}}             
\newcommand{\hatcurLBiiRxxxxxB}{\ensuremath{0.2876}}            
\newcommand{\hatcurLBiIxxxxxB}{\ensuremath{0.2655}}             
\newcommand{\hatcurLBiiIxxxxxB}{\ensuremath{0.3147}}            
\newcommand{\hatcurLBiuxxxxxB}{\ensuremath{0.6157}}             
\newcommand{\hatcurLBiiuxxxxxB}{\ensuremath{0.2014}}            
\newcommand{\hatcurLBigxxxxxB}{\ensuremath{0.36\pm0.13}}        
\newcommand{\hatcurLBiigxxxxxB}{\ensuremath{0.34\pm0.17}}       
\newcommand{\hatcurLBirxxxxxB}{\ensuremath{0.29\pm0.12}}        
\newcommand{\hatcurLBiirxxxxxB}{\ensuremath{0.21\pm0.17}}       
\newcommand{\hatcurLBiixxxxxB}{\ensuremath{0.3087}}             
\newcommand{\hatcurLBiiixxxxxB}{\ensuremath{0.3075}}            
\newcommand{\hatcurLBizxxxxxB}{\ensuremath{0.170_{-0.090}^{+0.120}}} 
\newcommand{\hatcurLBiizxxxxxB}{\ensuremath{0.10\pm0.16}}       
\newcommand{\hatcurLBiJxxxxxB}{\ensuremath{0.1590}}             
\newcommand{\hatcurLBiiJxxxxxB}{\ensuremath{0.2457}}            
\newcommand{\hatcurLBiHxxxxxB}{\ensuremath{0.1275}}             
\newcommand{\hatcurLBiiHxxxxxB}{\ensuremath{0.2607}}            
\newcommand{\hatcurLBiKxxxxxB}{\ensuremath{0.0959}}             
\newcommand{\hatcurLBiiKxxxxxB}{\ensuremath{0.2414}}            
\newcommand{\hatcurLBiTxxxxxB}{\ensuremath{0.28\pm0.12}}        
\newcommand{\hatcurLBiiTxxxxxB}{\ensuremath{0.34\pm0.16}}       
\newcommand{\hatcurLBikepxxxxxB}{\ensuremath{0.3797}}           
\newcommand{\hatcurLBiikepxxxxxB}{\ensuremath{0.3390}}          
\newcommand{\hatcurLBiCxxxxxB}{\ensuremath{0.3436}}             
\newcommand{\hatcurLBiiCxxxxxB}{\ensuremath{0.3422}}            
\newcommand{\hatcurLBiMxxxxxB}{\ensuremath{0.4623}}             
\newcommand{\hatcurLBiiMxxxxxB}{\ensuremath{0.3228}}            
\newcommand{\hatcurLBiSonexxxxxB}{\ensuremath{0.0708}}            
\newcommand{\hatcurLBiiSonexxxxxB}{\ensuremath{0.1866}}           
\newcommand{\hatcurLBiStwoxxxxxB}{\ensuremath{0.0602}}            
\newcommand{\hatcurLBiiStwoxxxxxB}{\ensuremath{0.1477}}           
\newcommand{\hatcurLBiSthreexxxxxB}{\ensuremath{0.0569}}            
\newcommand{\hatcurLBiiSthreexxxxxB}{\ensuremath{0.1283}}           
\newcommand{\hatcurLBiSfourxxxxxB}{\ensuremath{0.0638}}            
\newcommand{\hatcurLBiiSfourxxxxxB}{\ensuremath{0.1041}}           
\newcommand{\hatcurISOmxxxxxB}{\ensuremath{0.6017_{-0.0055}^{+0.0074}}} 
\newcommand{\hatcurISOmshortxxxxxB}{\ensuremath{0.60}}          
\newcommand{\hatcurISOmlongxxxxxB}{\ensuremath{0.6017_{-0.0055}^{+0.0074}}} 
\newcommand{\hatcurISOrxxxxxB}{\ensuremath{0.5848\pm0.0026}}    
\newcommand{\hatcurISOrshortxxxxxB}{\ensuremath{0.58}}          
\newcommand{\hatcurISOrlongxxxxxB}{\ensuremath{0.5848\pm0.0026}} 
\newcommand{\hatcurISOrhoxxxxxB}{\ensuremath{4.239_{-0.065}^{+0.091}}} 
\newcommand{\hatcurISOrholongxxxxxB}{\ensuremath{4.239_{-0.065}^{+0.091}}} 
\newcommand{\hatcurISOloggxxxxxB}{\ensuremath{4.6831\pm0.0069}} 
\newcommand{\hatcurISOlumxxxxxB}{\ensuremath{0.06359\pm0.00069}} 
\newcommand{\hatcurISOlumshortxxxxxB}{\ensuremath{0.06}}        
\newcommand{\hatcurISOteffxxxxxB}{\ensuremath{3790.4\pm5.7}}    
\newcommand{\hatcurISOzfehxxxxxB}{\ensuremath{0.522_{-0.028}^{+0.051}}} 
\newcommand{\hatcurISOagexxxxxB}{\ensuremath{14.9_{-4.3}^{+3.3}}} 
\newcommand{\hatcurISOspecxxxxxB}{***TODO***}                   
\newcommand{\hatcurRVKxxxxxB}{\ensuremath{99.2\pm7.9}}          
\newcommand{\hatcurRVKtwosiglimxxxxxB}{\ensuremath{<112.1}}     
\newcommand{\hatcurRVrkxxxxxB}{\ensuremath{0\pm0}}              
\newcommand{\hatcurRVrhxxxxxB}{\ensuremath{0\pm0}}              
\newcommand{\hatcurRVkxxxxxB}{\ensuremath{0\pm0}}               
\newcommand{\hatcurRVhxxxxxB}{\ensuremath{0\pm0}}               
\newcommand{\hatcurRVtronexxxxxB}{\ensuremath{0\pm0}}           
\newcommand{\hatcurRVtrtwoxxxxxB}{\ensuremath{0\pm0}}           
\newcommand{\hatcurRVgammaxxxxxB}{\ensuremath{39997.7\pm5.4}}   
\newcommand{\hatcurRVjitterxxxxxB}{\ensuremath{0.0\pm1.0}}      
\newcommand{\hatcurRVjittertwosiglimxxxxxB}{\ensuremath{<2.8}}  
\newcommand{\hatcurRVfitrmsxxxxxB}{\ensuremath{.1fym}}          %
\newcommand{\hatcurRVeccenxxxxxB}{\ensuremath{0\pm0}}           
\newcommand{\hatcurRVeccentwosiglimxxxxxB}{\ensuremath{<0.000}} 
\newcommand{\hatcurRVomegaxxxxxB}{\ensuremath{0\pm0}}           
\newcommand{\hatcurPPixxxxxB}{\ensuremath{88.07\pm0.15}}        
\newcommand{\hatcurPPgxxxxxB}{\ensuremath{15.5\pm1.4}}          
\newcommand{\hatcurPPloggxxxxxB}{\ensuremath{3.192\pm0.038}}    
\newcommand{\hatcurPParxxxxxB}{\ensuremath{12.037_{-0.062}^{+0.086}}} 
\newcommand{\hatcurPParelxxxxxB}{\ensuremath{0.032742_{-0.000099}^{+0.000134}}} 
\newcommand{\hatcurPPrhoxxxxxB}{\ensuremath{0.878\pm0.084}}     
\newcommand{\hatcurPPmxxxxxB}{\ensuremath{0.491\pm0.039}}       
\newcommand{\hatcurPPmtwosiglimxxxxxB}{\ensuremath{<0.55}}      
\newcommand{\hatcurPPmshortxxxxxB}{\ensuremath{0.49}}           
\newcommand{\hatcurPPmlongxxxxxB}{\ensuremath{0.491\pm0.039}}   
\newcommand{\hatcurPPmexxxxxB}{\ensuremath{156\pm12}}           
\newcommand{\hatcurPPmeshortxxxxxB}{\ensuremath{156.0}}         
\newcommand{\hatcurPPmelongxxxxxB}{\ensuremath{156\pm12}}       
\newcommand{\hatcurPPrxxxxxB}{\ensuremath{0.884\pm0.013}}       
\newcommand{\hatcurPPrshortxxxxxB}{\ensuremath{0.88}}           
\newcommand{\hatcurPPrlongxxxxxB}{\ensuremath{0.884\pm0.013}}   
\newcommand{\hatcurPPrexxxxxB}{\ensuremath{9.91\pm0.14}}        
\newcommand{\hatcurPPreshortxxxxxB}{\ensuremath{9.9}}           
\newcommand{\hatcurPPrelongxxxxxB}{\ensuremath{9.91\pm0.14}}    
\newcommand{\hatcurPPmrcorrxxxxxB}{\ensuremath{-0.11}}          
\newcommand{\hatcurPPteffxxxxxB}{\ensuremath{772.3\pm2.3}}      
\newcommand{\hatcurPPthetaxxxxxB}{\ensuremath{0.0602\pm0.0049}} 
\newcommand{\hatcurPPfluxperixxxxxB}{\ensuremath{8.065\pm0.094}} 
\newcommand{\hatcurPPfluxperidimxxxxxB}{\ensuremath{7}}         
\newcommand{\hatcurPPfluxapxxxxxB}{\ensuremath{8.065\pm0.094}}  
\newcommand{\hatcurPPfluxapdimxxxxxB}{\ensuremath{7}}           
\newcommand{\hatcurPPfluxavgxxxxxB}{\ensuremath{8.065\pm0.094}} 
\newcommand{\hatcurPPfluxavgdimxxxxxB}{\ensuremath{7}}          
\newcommand{\hatcurPPfluxavglogxxxxxB}{\ensuremath{7.9066\pm0.0051}} 
\newcommand{\hatcurXsecphasexxxxxB}{\ensuremath{0\pm0}}         
\newcommand{\hatcurXsecondaryxxxxxB}{\ensuremath{2458612.44920\pm0.00027}} 
\newcommand{\hatcurXsecdurxxxxxB}{\ensuremath{0.07969\pm0.00065}} 
\newcommand{\hatcurXsecingdurxxxxxB}{\ensuremath{0.01256\pm0.00036}} 
\newcommand{\hatcurPPphiconjxxxxxB}{\ensuremath{0\pm0}}         
\newcommand{\hatcurPPperixxxxxB}{\ensuremath{2458610.35770\pm0.00026}} 
\newcommand{\hatcurPPaequivxxxxxB}{\ensuremath{0.12990\pm0.00076}} 
\newcommand{\hatcurPPtcircxxxxxB}{\ensuremath{168\pm20}}        
\newcommand{\hatcurPPtinfallxxxxxB}{\ensuremath{8270\pm720}}    
\newcommand{\hatcurXdistxxxxxB}{\ensuremath{195.3\pm1.0}}       
\newcommand{\hatcurXAvxxxxxB}{\ensuremath{0.0560\pm0.0095}}     
\newcommand{\hatcurXdistredxxxxxB}{\ensuremath{195.3\pm1.0}}    
\newcommand{\hatcurXEBVxxxxxB}{\ensuremath{0.0180\pm0.0031}}    
\newcommand{\hatcurCCpmraxxxxxB}{\ensuremath{12.872\pm0.038}}   
\newcommand{\hatcurCCpmdecxxxxxB}{\ensuremath{-1.753\pm0.048}}  
\newcommand{\hatcurCCpmxxxxxB}{\ensuremath{12.991\pm0.061}}     

\newcommand{\hatcurhtrxxxxxC}{HATS597-003}                      
\newcommand{\hatcurfieldxxxxxC}{\ensuremath{string}}            
\newcommand{\hatcurCCraxxxxxC}{\ensuremath{04^{\mathrm h}41^{\mathrm m}21.5520{\mathrm s}}}                   
\newcommand{\hatcurCCdecxxxxxC}{\ensuremath{-32{\arcdeg}19{\arcmin}13.5029{\arcsec}}}                 
\newcommand{\hatcurCCmagxxxxxC}{NULL}                           
\newcommand{\hatcurCCtwomassxxxxxC}{2MASS~04412154-3219128}     
\newcommand{\hatcurCCgscxxxxxC}{GSC~}                           
\newcommand{\hatcurCCgaiaxxxxxC}{GAIA~4877426571427909248}      
\newcommand{\hatcurCCgaiadrtwoxxxxxC}{GAIA~DR2~4877426575724467456} 
\newcommand{\hatcurCCgaiadrtwoshortxxxxxC}{4877426575724467456} 
\newcommand{\hatcurCCtassmvxxxxxC}{\ensuremath{nff\pmnff}}      
\newcommand{\hatcurCCtassmvshortxxxxxC}{\ensuremath{0.0}}       
\newcommand{\hatcurCCtassmBxxxxxC}{\ensuremath{nff\pmnff}}      
\newcommand{\hatcurCCtassmBshortxxxxxC}{\ensuremath{0.0}}       
\newcommand{\hatcurCCtassmIxxxxxC}{\ensuremath{nff\pmnff}}      
\newcommand{\hatcurCCtassmIshortxxxxxC}{\ensuremath{0.0}}       
\newcommand{\hatcurCCtassmgxxxxxC}{\ensuremath{nff\pmnff}}      
\newcommand{\hatcurCCtassmgshortxxxxxC}{\ensuremath{0.0}}       
\newcommand{\hatcurCCtassmrxxxxxC}{\ensuremath{nff\pmnff}}      
\newcommand{\hatcurCCtassmrshortxxxxxC}{\ensuremath{0.0}}       
\newcommand{\hatcurCCtassmixxxxxC}{\ensuremath{nff\pmnff}}      
\newcommand{\hatcurCCtassmishortxxxxxC}{\ensuremath{0.0}}       
\newcommand{\hatcurCCparallaxxxxxxC}{\ensuremath{2.564\pm0.038}} 
\newcommand{\hatcurCCgaiamGxxxxxC}{\ensuremath{15.79420\pm0.00060}} 
\newcommand{\hatcurCCgaiamBPxxxxxC}{\ensuremath{16.6962\pm0.0057}} 
\newcommand{\hatcurCCgaiamRPxxxxxC}{\ensuremath{14.8610\pm0.0020}} 
\newcommand{\hatcurCCtwomassJmagxxxxxC}{\ensuremath{13.690\pm0.029}} 
\newcommand{\hatcurCCtwomassHmagxxxxxC}{\ensuremath{12.984\pm0.024}} 
\newcommand{\hatcurCCtwomassKmagxxxxxC}{\ensuremath{12.812\pm0.033}} 
\newcommand{\hatcurCCcitJmagxxxxxC}{\ensuremath{13.680\pm0.030}} 
\newcommand{\hatcurCCcitHmagxxxxxC}{\ensuremath{12.976\pm0.025}} 
\newcommand{\hatcurCCcitKmagxxxxxC}{\ensuremath{12.836\pm0.034}} 
\newcommand{\hatcurCCbbJmagxxxxxC}{\ensuremath{13.771\pm0.032}} 
\newcommand{\hatcurCCbbHmagxxxxxC}{\ensuremath{13.000\pm0.026}} 
\newcommand{\hatcurCCbbKmagxxxxxC}{\ensuremath{12.856\pm0.034}} 
\newcommand{\hatcurCCesoJmagxxxxxC}{\ensuremath{13.779\pm0.037}} 
\newcommand{\hatcurCCesoHmagxxxxxC}{\ensuremath{12.998\pm0.036}} 
\newcommand{\hatcurCCesoKmagxxxxxC}{\ensuremath{12.852\pm0.035}} 
\newcommand{\hatcurCCesoJHmagxxxxxC}{\ensuremath{0.781\pm0.047}} 
\newcommand{\hatcurCCesoJKmagxxxxxC}{\ensuremath{0.928\pm0.049}} 
\newcommand{\hatcurCCesoHKmagxxxxxC}{\ensuremath{0.146\pm0.049}} 
\newcommand{\hatcurCCWonemagxxxxxC}{\ensuremath{12.758\pm0.024}} 
\newcommand{\hatcurCCWtwomagxxxxxC}{\ensuremath{12.784\pm0.026}} 
\newcommand{\hatcurCCWthreemagxxxxxC}{\ensuremath{nff\pmnff}}   
\newcommand{\hatcurCCWfourmagxxxxxC}{\ensuremath{nff\pmnff}}    
\newcommand{\hatcurLCdipxxxxxC}{\ensuremath{0.0}}               
\newcommand{\hatcurLCrprstarxxxxxC}{\ensuremath{0.1772\pm0.0045}} 
\newcommand{\hatcurLCbsqxxxxxC}{\ensuremath{0.079_{-0.048}^{+0.051}}} 
\newcommand{\hatcurLCimpxxxxxC}{\ensuremath{0.281_{-0.105}^{+0.080}}} 
\newcommand{\hatcurLCzetaxxxxxC}{\ensuremath{30.76\pm0.67}}     
\newcommand{\hatcurLCdurxxxxxC}{\ensuremath{0.0774\pm0.0011}}   
\newcommand{\hatcurLCdurshortxxxxxC}{\ensuremath{0.0774}}       
\newcommand{\hatcurLCdurhrxxxxxC}{\ensuremath{1.858\pm0.027}}   
\newcommand{\hatcurLCdurhrshortxxxxxC}{\ensuremath{1.858}}      
\newcommand{\hatcurLCqxxxxxC}{\ensuremath{0.03990\pm0.00058}}   
\newcommand{\hatcurLCqshortxxxxxC}{\ensuremath{0.040}}          
\newcommand{\hatcurLCingdurxxxxxC}{\ensuremath{0.01254\pm0.00067}} 
\newcommand{\hatcurLCPxxxxxC}{\ensuremath{1.9416423\pm0.0000014}} 
\newcommand{\hatcurLCPprecxxxxxC}{\ensuremath{1.9416423}}       
\newcommand{\hatcurLCPshortxxxxxC}{\ensuremath{1.9416}}         
\newcommand{\hatcurLCTxxxxxC}{\ensuremath{2458424.55556\pm0.00053}} 
\newcommand{\hatcurLCTAxxxxxC}{\ensuremath{2456543.1042\pm0.0015}} 
\newcommand{\hatcurLCTBxxxxxC}{\ensuremath{2458746.86821\pm0.00057}} 
\newcommand{\hatcurLChatnetmAxxxxxC}{\ensuremath{15.64062\pm0.00036}} 
\newcommand{\hatcurLCiblendAxxxxxC}{\ensuremath{0.892\pm0.056}} 
\newcommand{\hatcurLChatnetmBxxxxxC}{\ensuremath{-0.00304\pm0.00015}} 
\newcommand{\hatcurLCiblendBxxxxxC}{\ensuremath{0.966\pm0.033}} 
\newcommand{\hatcurLCrhoxxxxxC}{\ensuremath{3.80\pm0.17}}       
\newcommand{\hatcurSMEiteffxxxxxC}{\ensuremath{3990\pm120}}     
\newcommand{\hatcurSMEizfehxxxxxC}{\ensuremath{0.29\pm0.13}}    
\newcommand{\hatcurSMEizfehshortxxxxxC}{\ensuremath{0.29}}      
\newcommand{\hatcurSMEiloggxxxxxC}{\ensuremath{4.50\pm0.50}}    
\newcommand{\hatcurSMEivsinxxxxxC}{\ensuremath{0\pm50}}         
\newcommand{\hatcurSMEivmacxxxxxC}{\ensuremath{nff\pmnff}}      
\newcommand{\hatcurSMEivmicxxxxxC}{\ensuremath{nff\pmnff}}      
\newcommand{\hatcurextraerrMJxxxxxC}{\ensuremath{0\pm0}}        
\newcommand{\hatcurextraerrMJtwosiglimxxxxxC}{\ensuremath{<0.0200}} 
\newcommand{\hatcurextraerrMHxxxxxC}{\ensuremath{0\pm0}}        
\newcommand{\hatcurextraerrMHtwosiglimxxxxxC}{\ensuremath{<0.0200}} 
\newcommand{\hatcurextraerrMKsxxxxxC}{\ensuremath{0\pm0}}       
\newcommand{\hatcurextraerrMKstwosiglimxxxxxC}{\ensuremath{<0.0200}} 
\newcommand{\hatcurextraerrMGxxxxxC}{\ensuremath{0\pm0}}        
\newcommand{\hatcurextraerrMGtwosiglimxxxxxC}{\ensuremath{<0.0200}} 
\newcommand{\hatcurextraerrMBPtwoxxxxxC}{\ensuremath{0\pm0}}    
\newcommand{\hatcurextraerrMBPtwotwosiglimxxxxxC}{\ensuremath{<0.0200}} 
\newcommand{\hatcurextraerrMRPxxxxxC}{\ensuremath{0\pm0}}       
\newcommand{\hatcurextraerrMRPtwosiglimxxxxxC}{\ensuremath{<0.0200}} 
\newcommand{\hatcurextraerrMWonexxxxxC}{\ensuremath{0\pm0}}     
\newcommand{\hatcurextraerrMWonetwosiglimxxxxxC}{\ensuremath{<0.0200}} 
\newcommand{\hatcurextraerrMWtwoxxxxxC}{\ensuremath{0\pm0}}     
\newcommand{\hatcurextraerrMWtwotwosiglimxxxxxC}{\ensuremath{<0.0200}} 
\newcommand{\hatcurLBiBxxxxxC}{\ensuremath{0.7014}}             
\newcommand{\hatcurLBiiBxxxxxC}{\ensuremath{0.1182}}            
\newcommand{\hatcurLBiVxxxxxC}{\ensuremath{0.5707}}             
\newcommand{\hatcurLBiiVxxxxxC}{\ensuremath{0.2046}}            
\newcommand{\hatcurLBiRxxxxxC}{\ensuremath{0.4830}}             
\newcommand{\hatcurLBiiRxxxxxC}{\ensuremath{0.2327}}            
\newcommand{\hatcurLBiIxxxxxC}{\ensuremath{0.34\pm0.15}}        
\newcommand{\hatcurLBiiIxxxxxC}{\ensuremath{0.27\pm0.17}}       
\newcommand{\hatcurLBiuxxxxxC}{\ensuremath{0.7492}}             
\newcommand{\hatcurLBiiuxxxxxC}{\ensuremath{0.0747}}            
\newcommand{\hatcurLBigxxxxxC}{\ensuremath{0.47\pm0.14}}        
\newcommand{\hatcurLBiigxxxxxC}{\ensuremath{0.34\pm0.15}}       
\newcommand{\hatcurLBirxxxxxC}{\ensuremath{0.34\pm0.16}}        
\newcommand{\hatcurLBiirxxxxxC}{\ensuremath{0.31\pm0.17}}       
\newcommand{\hatcurLBiixxxxxC}{\ensuremath{0.3712}}             
\newcommand{\hatcurLBiiixxxxxC}{\ensuremath{0.2527}}            
\newcommand{\hatcurLBizxxxxxC}{\ensuremath{0.2882}}             
\newcommand{\hatcurLBiizxxxxxC}{\ensuremath{0.2510}}            
\newcommand{\hatcurLBiJxxxxxC}{\ensuremath{0.2146}}             
\newcommand{\hatcurLBiiJxxxxxC}{\ensuremath{0.2279}}            
\newcommand{\hatcurLBiHxxxxxC}{\ensuremath{0.1632}}             
\newcommand{\hatcurLBiiHxxxxxC}{\ensuremath{0.2566}}            
\newcommand{\hatcurLBiKxxxxxC}{\ensuremath{0.1218}}             
\newcommand{\hatcurLBiiKxxxxxC}{\ensuremath{0.2332}}            
\newcommand{\hatcurLBiTxxxxxC}{\ensuremath{0.34\pm0.14}}        
\newcommand{\hatcurLBiiTxxxxxC}{\ensuremath{0.36\pm0.17}}       
\newcommand{\hatcurLBikepxxxxxC}{\ensuremath{0.4512}}           
\newcommand{\hatcurLBiikepxxxxxC}{\ensuremath{0.2754}}          
\newcommand{\hatcurLBiCxxxxxC}{\ensuremath{0.4199}}             
\newcommand{\hatcurLBiiCxxxxxC}{\ensuremath{0.2799}}            
\newcommand{\hatcurLBiMxxxxxC}{\ensuremath{0.5363}}             
\newcommand{\hatcurLBiiMxxxxxC}{\ensuremath{0.2478}}            
\newcommand{\hatcurLBiSonexxxxxC}{\ensuremath{0.0864}}            
\newcommand{\hatcurLBiiSonexxxxxC}{\ensuremath{0.1752}}           
\newcommand{\hatcurLBiStwoxxxxxC}{\ensuremath{0.0736}}            
\newcommand{\hatcurLBiiStwoxxxxxC}{\ensuremath{0.1430}}           
\newcommand{\hatcurLBiSthreexxxxxC}{\ensuremath{0.0640}}            
\newcommand{\hatcurLBiiSthreexxxxxC}{\ensuremath{0.1238}}           
\newcommand{\hatcurLBiSfourxxxxxC}{\ensuremath{0.0634}}            
\newcommand{\hatcurLBiiSfourxxxxxC}{\ensuremath{0.1063}}           
\newcommand{\hatcurISOmxxxxxC}{\ensuremath{0.662_{-0.021}^{+0.016}}} 
\newcommand{\hatcurISOmshortxxxxxC}{\ensuremath{0.66}}          
\newcommand{\hatcurISOmlongxxxxxC}{\ensuremath{0.662_{-0.021}^{+0.016}}} 
\newcommand{\hatcurISOrxxxxxC}{\ensuremath{0.6259\pm0.0079}}    
\newcommand{\hatcurISOrshortxxxxxC}{\ensuremath{0.63}}          
\newcommand{\hatcurISOrlongxxxxxC}{\ensuremath{0.6259\pm0.0079}} 
\newcommand{\hatcurISOrhoxxxxxC}{\ensuremath{3.80\pm0.17}}      
\newcommand{\hatcurISOrholongxxxxxC}{\ensuremath{3.80\pm0.17}}  
\newcommand{\hatcurISOloggxxxxxC}{\ensuremath{4.665\pm0.016}}   
\newcommand{\hatcurISOlumxxxxxC}{\ensuremath{0.0916\pm0.0029}}  
\newcommand{\hatcurISOlumshortxxxxxC}{\ensuremath{0.09}}        
\newcommand{\hatcurISOteffxxxxxC}{\ensuremath{4016\pm17}}       
\newcommand{\hatcurISOzfehxxxxxC}{\ensuremath{0.322_{-0.049}^{+0.065}}} 
\newcommand{\hatcurISOagexxxxxC}{\ensuremath{4.6_{-4.0}^{+8.7}}} 
\newcommand{\hatcurISOspecxxxxxC}{***TODO***}                   
\newcommand{\hatcurRVKxxxxxC}{\ensuremath{562\pm15}}            
\newcommand{\hatcurRVKtwosiglimxxxxxC}{\ensuremath{<587.5}}     
\newcommand{\hatcurRVrkxxxxxC}{\ensuremath{0\pm0}}              
\newcommand{\hatcurRVrhxxxxxC}{\ensuremath{0\pm0}}              
\newcommand{\hatcurRVkxxxxxC}{\ensuremath{0\pm0}}               
\newcommand{\hatcurRVhxxxxxC}{\ensuremath{0\pm0}}               
\newcommand{\hatcurRVtronexxxxxC}{\ensuremath{0\pm0}}           
\newcommand{\hatcurRVtrtwoxxxxxC}{\ensuremath{0\pm0}}           
\newcommand{\hatcurRVgammaxxxxxC}{\ensuremath{8597\pm11}}       
\newcommand{\hatcurRVjitterxxxxxC}{\ensuremath{5.6\pm6.6}}      
\newcommand{\hatcurRVjittertwosiglimxxxxxC}{\ensuremath{<19.0}} 
\newcommand{\hatcurRVfitrmsxxxxxC}{\ensuremath{.1fym}}          %
\newcommand{\hatcurRVeccenxxxxxC}{\ensuremath{0\pm0}}           
\newcommand{\hatcurRVeccentwosiglimxxxxxC}{\ensuremath{<0.000}} 
\newcommand{\hatcurRVomegaxxxxxC}{\ensuremath{0\pm0}}           
\newcommand{\hatcurPPixxxxxC}{\ensuremath{88.24\pm0.59}}        
\newcommand{\hatcurPPgxxxxxC}{\ensuremath{55.8\pm3.6}}          
\newcommand{\hatcurPPloggxxxxxC}{\ensuremath{3.747\pm0.029}}    
\newcommand{\hatcurPParxxxxxC}{\ensuremath{9.12\pm0.14}}        
\newcommand{\hatcurPParelxxxxxC}{\ensuremath{0.02658_{-0.00029}^{+0.00021}}} 
\newcommand{\hatcurPPrhoxxxxxC}{\ensuremath{2.58\pm0.23}}       
\newcommand{\hatcurPPmxxxxxC}{\ensuremath{2.629\pm0.089}}       
\newcommand{\hatcurPPmtwosiglimxxxxxC}{\ensuremath{<2.77}}      
\newcommand{\hatcurPPmshortxxxxxC}{\ensuremath{2.63}}           
\newcommand{\hatcurPPmlongxxxxxC}{\ensuremath{2.629\pm0.089}}   
\newcommand{\hatcurPPmexxxxxC}{\ensuremath{836\pm28}}           
\newcommand{\hatcurPPmeshortxxxxxC}{\ensuremath{835.5}}         
\newcommand{\hatcurPPmelongxxxxxC}{\ensuremath{836\pm28}}       
\newcommand{\hatcurPPrxxxxxC}{\ensuremath{1.079\pm0.031}}       
\newcommand{\hatcurPPrshortxxxxxC}{\ensuremath{1.08}}           
\newcommand{\hatcurPPrlongxxxxxC}{\ensuremath{1.079\pm0.031}}   
\newcommand{\hatcurPPrexxxxxC}{\ensuremath{12.10\pm0.35}}       
\newcommand{\hatcurPPreshortxxxxxC}{\ensuremath{12.1}}          
\newcommand{\hatcurPPrelongxxxxxC}{\ensuremath{12.10\pm0.35}}   
\newcommand{\hatcurPPmrcorrxxxxxC}{\ensuremath{0.01}}           
\newcommand{\hatcurPPteffxxxxxC}{\ensuremath{939.8\pm6.7}}      
\newcommand{\hatcurPPthetaxxxxxC}{\ensuremath{0.1947\pm0.0075}} 
\newcommand{\hatcurPPfluxperixxxxxC}{\ensuremath{1.767\pm0.051}} 
\newcommand{\hatcurPPfluxperidimxxxxxC}{\ensuremath{8}}         
\newcommand{\hatcurPPfluxapxxxxxC}{\ensuremath{1.767\pm0.051}}  
\newcommand{\hatcurPPfluxapdimxxxxxC}{\ensuremath{8}}           
\newcommand{\hatcurPPfluxavgxxxxxC}{\ensuremath{1.767\pm0.051}} 
\newcommand{\hatcurPPfluxavgdimxxxxxC}{\ensuremath{8}}          
\newcommand{\hatcurPPfluxavglogxxxxxC}{\ensuremath{8.247\pm0.012}} 
\newcommand{\hatcurXsecphasexxxxxC}{\ensuremath{0\pm0}}         
\newcommand{\hatcurXsecondaryxxxxxC}{\ensuremath{2458425.52638\pm0.00053}} 
\newcommand{\hatcurXsecdurxxxxxC}{\ensuremath{0.0774\pm0.0011}} 
\newcommand{\hatcurXsecingdurxxxxxC}{\ensuremath{0.01254\pm0.00067}} 
\newcommand{\hatcurPPphiconjxxxxxC}{\ensuremath{0\pm0}}         
\newcommand{\hatcurPPperixxxxxC}{\ensuremath{2458424.07015\pm0.00053}} 
\newcommand{\hatcurPPaequivxxxxxC}{\ensuremath{0.0878\pm0.0013}} 
\newcommand{\hatcurPPtcircxxxxxC}{\ensuremath{73\pm11}}         
\newcommand{\hatcurPPtinfallxxxxxC}{\ensuremath{294\pm25}}      
\newcommand{\hatcurXdistxxxxxC}{\ensuremath{389.9\pm5.6}}       
\newcommand{\hatcurXAvxxxxxC}{\ensuremath{0.062\pm0.012}}       
\newcommand{\hatcurXdistredxxxxxC}{\ensuremath{389.9\pm5.6}}    
\newcommand{\hatcurXEBVxxxxxC}{\ensuremath{0.0200\pm0.0039}}    
\newcommand{\hatcurCCpmraxxxxxC}{\ensuremath{-5.429\pm0.057}}   
\newcommand{\hatcurCCpmdecxxxxxC}{\ensuremath{-38.609\pm0.090}} 
\newcommand{\hatcurCCpmxxxxxC}{\ensuremath{38.99\pm0.11}}       

\newcommand{\hatcurhtrxxxxxD}{HATS607-010}                      
\newcommand{\hatcurfieldxxxxxD}{\ensuremath{string}}            
\newcommand{\hatcurCCraxxxxxD}{\ensuremath{09^{\mathrm h}59^{\mathrm m}17.6640{\mathrm s}}}                   
\newcommand{\hatcurCCdecxxxxxD}{\ensuremath{-27{\arcdeg}23{\arcmin}34.1427{\arcsec}}}                 
\newcommand{\hatcurCCmagxxxxxD}{16.354}                         
\newcommand{\hatcurCCtwomassxxxxxD}{2MASS~09591770-2723339}     
\newcommand{\hatcurCCgscxxxxxD}{GSC~}                           
\newcommand{\hatcurCCgaiaxxxxxD}{GAIA~5466556137225499136}      
\newcommand{\hatcurCCgaiadrtwoxxxxxD}{GAIA~DR2~5466556141521710592} 
\newcommand{\hatcurCCgaiadrtwoshortxxxxxD}{5466556141521710592} 
\newcommand{\hatcurCCtassmvxxxxxD}{\ensuremath{16.354\pm0.010}} 
\newcommand{\hatcurCCtassmvshortxxxxxD}{\ensuremath{16.4}}      
\newcommand{\hatcurCCtassmBxxxxxD}{\ensuremath{17.716\pm0.010}} 
\newcommand{\hatcurCCtassmBshortxxxxxD}{\ensuremath{17.7}}      
\newcommand{\hatcurCCtassmIxxxxxD}{\ensuremath{nff\pmnff}}      
\newcommand{\hatcurCCtassmIshortxxxxxD}{\ensuremath{0.0}}       
\newcommand{\hatcurCCtassmgxxxxxD}{\ensuremath{17.10\pm0.25}}   
\newcommand{\hatcurCCtassmgshortxxxxxD}{\ensuremath{17.1}}      
\newcommand{\hatcurCCtassmrxxxxxD}{\ensuremath{15.746\pm0.060}} 
\newcommand{\hatcurCCtassmrshortxxxxxD}{\ensuremath{15.7}}      
\newcommand{\hatcurCCtassmixxxxxD}{\ensuremath{15.212\pm0.010}} 
\newcommand{\hatcurCCtassmishortxxxxxD}{\ensuremath{15.2}}      
\newcommand{\hatcurCCparallaxxxxxxD}{\ensuremath{2.265\pm0.056}} 
\newcommand{\hatcurCCgaiamGxxxxxD}{\ensuremath{15.7364\pm0.0011}} 
\newcommand{\hatcurCCgaiamBPxxxxxD}{\ensuremath{16.5518\pm0.0054}} 
\newcommand{\hatcurCCgaiamRPxxxxxD}{\ensuremath{14.8547\pm0.0038}} 
\newcommand{\hatcurCCtwomassJmagxxxxxD}{\ensuremath{13.779\pm0.029}} 
\newcommand{\hatcurCCtwomassHmagxxxxxD}{\ensuremath{13.047\pm0.029}} 
\newcommand{\hatcurCCtwomassKmagxxxxxD}{\ensuremath{12.934\pm0.032}} 
\newcommand{\hatcurCCcitJmagxxxxxD}{\ensuremath{13.770\pm0.029}} 
\newcommand{\hatcurCCcitHmagxxxxxD}{\ensuremath{13.041\pm0.030}} 
\newcommand{\hatcurCCcitKmagxxxxxD}{\ensuremath{12.958\pm0.032}} 
\newcommand{\hatcurCCbbJmagxxxxxD}{\ensuremath{13.859\pm0.032}} 
\newcommand{\hatcurCCbbHmagxxxxxD}{\ensuremath{13.063\pm0.031}} 
\newcommand{\hatcurCCbbKmagxxxxxD}{\ensuremath{12.978\pm0.032}} 
\newcommand{\hatcurCCesoJmagxxxxxD}{\ensuremath{13.866\pm0.036}} 
\newcommand{\hatcurCCesoHmagxxxxxD}{\ensuremath{13.057\pm0.036}} 
\newcommand{\hatcurCCesoKmagxxxxxD}{\ensuremath{12.975\pm0.033}} 
\newcommand{\hatcurCCesoJHmagxxxxxD}{\ensuremath{0.809\pm0.021}} 
\newcommand{\hatcurCCesoJKmagxxxxxD}{\ensuremath{0.892\pm0.048}} 
\newcommand{\hatcurCCesoHKmagxxxxxD}{\ensuremath{0.083\pm0.048}} 
\newcommand{\hatcurCCWonemagxxxxxD}{\ensuremath{12.847\pm0.024}} 
\newcommand{\hatcurCCWtwomagxxxxxD}{\ensuremath{12.870\pm0.027}} 
\newcommand{\hatcurCCWthreemagxxxxxD}{\ensuremath{nff\pmnff}}   
\newcommand{\hatcurCCWfourmagxxxxxD}{\ensuremath{nff\pmnff}}    
\newcommand{\hatcurLCdipxxxxxD}{\ensuremath{41.4}}              
\newcommand{\hatcurLCrprstarxxxxxD}{\ensuremath{0.1865\pm0.0024}} 
\newcommand{\hatcurLCbsqxxxxxD}{\ensuremath{0.107_{-0.033}^{+0.028}}} 
\newcommand{\hatcurLCimpxxxxxD}{\ensuremath{0.328_{-0.054}^{+0.041}}} 
\newcommand{\hatcurLCzetaxxxxxD}{\ensuremath{26.02\pm0.28}}     
\newcommand{\hatcurLCdurxxxxxD}{\ensuremath{0.09282\pm0.00070}} 
\newcommand{\hatcurLCdurshortxxxxxD}{\ensuremath{0.0928}}       
\newcommand{\hatcurLCdurhrxxxxxD}{\ensuremath{2.228\pm0.017}}   
\newcommand{\hatcurLCdurhrshortxxxxxD}{\ensuremath{2.228}}      
\newcommand{\hatcurLCqxxxxxD}{\ensuremath{0.03010\pm0.00023}}   
\newcommand{\hatcurLCqshortxxxxxD}{\ensuremath{0.030}}          
\newcommand{\hatcurLCingdurxxxxxD}{\ensuremath{0.01610\pm0.00061}} 
\newcommand{\hatcurLCPxxxxxD}{\ensuremath{3.0876262\pm0.0000016}} 
\newcommand{\hatcurLCPprecxxxxxD}{\ensuremath{3.0876262}}       
\newcommand{\hatcurLCPshortxxxxxD}{\ensuremath{3.0876}}         
\newcommand{\hatcurLCTxxxxxD}{\ensuremath{2459136.69378\pm0.00020}} 
\newcommand{\hatcurLCTAxxxxxD}{\ensuremath{2455573.5732\pm0.0018}} 
\newcommand{\hatcurLCTBxxxxxD}{\ensuremath{2459220.05968\pm0.00021}} 
\newcommand{\hatcurLChatnetmAxxxxxD}{\ensuremath{15.71266\pm0.00035}} 
\newcommand{\hatcurLCiblendAxxxxxD}{\ensuremath{0.906\pm0.054}} 
\newcommand{\hatcurLChatnetmBxxxxxD}{\ensuremath{-0.00309\pm0.00055}} 
\newcommand{\hatcurLCiblendBxxxxxD}{\ensuremath{0.9927\pm0.0095}} 
\newcommand{\hatcurLCrhoxxxxxD}{\ensuremath{3.48_{-0.12}^{+0.17}}} 
\newcommand{\hatcurSMEiteffxxxxxD}{\ensuremath{4082\pm69}}      
\newcommand{\hatcurSMEizfehxxxxxD}{\ensuremath{-0.12\pm0.10}}   
\newcommand{\hatcurSMEizfehshortxxxxxD}{\ensuremath{-0.12}}     
\newcommand{\hatcurSMEiloggxxxxxD}{\ensuremath{4.50\pm0.50}}    
\newcommand{\hatcurSMEivsinxxxxxD}{\ensuremath{0\pm50}}         
\newcommand{\hatcurSMEivmacxxxxxD}{\ensuremath{nff\pmnff}}      
\newcommand{\hatcurSMEivmicxxxxxD}{\ensuremath{nff\pmnff}}      
\newcommand{\hatcurextraerrMJxxxxxD}{\ensuremath{0\pm0}}        
\newcommand{\hatcurextraerrMJtwosiglimxxxxxD}{\ensuremath{<0.0200}} 
\newcommand{\hatcurextraerrMHxxxxxD}{\ensuremath{0\pm0}}        
\newcommand{\hatcurextraerrMHtwosiglimxxxxxD}{\ensuremath{<0.0200}} 
\newcommand{\hatcurextraerrMKsxxxxxD}{\ensuremath{0\pm0}}       
\newcommand{\hatcurextraerrMKstwosiglimxxxxxD}{\ensuremath{<0.0200}} 
\newcommand{\hatcurextraerrMGxxxxxD}{\ensuremath{0\pm0}}        
\newcommand{\hatcurextraerrMGtwosiglimxxxxxD}{\ensuremath{<0.0200}} 
\newcommand{\hatcurextraerrMBPtwoxxxxxD}{\ensuremath{0\pm0}}    
\newcommand{\hatcurextraerrMBPtwotwosiglimxxxxxD}{\ensuremath{<0.0200}} 
\newcommand{\hatcurextraerrMRPxxxxxD}{\ensuremath{0\pm0}}       
\newcommand{\hatcurextraerrMRPtwosiglimxxxxxD}{\ensuremath{<0.0200}} 
\newcommand{\hatcurextraerrMgxxxxxD}{\ensuremath{0\pm0}}        
\newcommand{\hatcurextraerrMgtwosiglimxxxxxD}{\ensuremath{<0.0200}} 
\newcommand{\hatcurextraerrMrxxxxxD}{\ensuremath{0\pm0}}        
\newcommand{\hatcurextraerrMrtwosiglimxxxxxD}{\ensuremath{<0.0200}} 
\newcommand{\hatcurextraerrMixxxxxD}{\ensuremath{0\pm0}}        
\newcommand{\hatcurextraerrMitwosiglimxxxxxD}{\ensuremath{<0.0200}} 
\newcommand{\hatcurextraerrMWonexxxxxD}{\ensuremath{0\pm0}}     
\newcommand{\hatcurextraerrMWonetwosiglimxxxxxD}{\ensuremath{<0.0200}} 
\newcommand{\hatcurextraerrMWtwoxxxxxD}{\ensuremath{0\pm0}}     
\newcommand{\hatcurextraerrMWtwotwosiglimxxxxxD}{\ensuremath{<0.0200}} 
\newcommand{\hatcurLBiBxxxxxD}{\ensuremath{0.7848}}             
\newcommand{\hatcurLBiiBxxxxxD}{\ensuremath{0.0472}}            
\newcommand{\hatcurLBiVxxxxxD}{\ensuremath{0.6317}}             
\newcommand{\hatcurLBiiVxxxxxD}{\ensuremath{0.1535}}            
\newcommand{\hatcurLBiRxxxxxD}{\ensuremath{0.5264}}             
\newcommand{\hatcurLBiiRxxxxxD}{\ensuremath{0.1979}}            
\newcommand{\hatcurLBiIxxxxxD}{\ensuremath{0.3783}}             
\newcommand{\hatcurLBiiIxxxxxD}{\ensuremath{0.2304}}            
\newcommand{\hatcurLBiuxxxxxD}{\ensuremath{0.8436}}             
\newcommand{\hatcurLBiiuxxxxxD}{\ensuremath{-0.0118}}           
\newcommand{\hatcurLBigxxxxxD}{\ensuremath{0.47\pm0.12}}        
\newcommand{\hatcurLBiigxxxxxD}{\ensuremath{0.38\pm0.14}}       
\newcommand{\hatcurLBirxxxxxD}{\ensuremath{0.365_{-0.087}^{+0.117}}} 
\newcommand{\hatcurLBiirxxxxxD}{\ensuremath{0.34\pm0.15}}       
\newcommand{\hatcurLBiixxxxxD}{\ensuremath{0.34\pm0.10}}        
\newcommand{\hatcurLBiiixxxxxD}{\ensuremath{0.21_{-0.14}^{+0.18}}} 
\newcommand{\hatcurLBizxxxxxD}{\ensuremath{0.23\pm0.12}}        
\newcommand{\hatcurLBiizxxxxxD}{\ensuremath{0.25\pm0.17}}       
\newcommand{\hatcurLBiJxxxxxD}{\ensuremath{0.2428}}             
\newcommand{\hatcurLBiiJxxxxxD}{\ensuremath{0.2237}}            
\newcommand{\hatcurLBiHxxxxxD}{\ensuremath{0.1724}}             
\newcommand{\hatcurLBiiHxxxxxD}{\ensuremath{0.2648}}            
\newcommand{\hatcurLBiKxxxxxD}{\ensuremath{0.1300}}             
\newcommand{\hatcurLBiiKxxxxxD}{\ensuremath{0.2371}}            
\newcommand{\hatcurLBiTxxxxxD}{\ensuremath{0.55\pm0.15}}        
\newcommand{\hatcurLBiiTxxxxxD}{\ensuremath{0.35\pm0.15}}       
\newcommand{\hatcurLBikepxxxxxD}{\ensuremath{0.5067}}           
\newcommand{\hatcurLBiikepxxxxxD}{\ensuremath{0.2326}}          
\newcommand{\hatcurLBiCxxxxxD}{\ensuremath{0.4758}}             
\newcommand{\hatcurLBiiCxxxxxD}{\ensuremath{0.2392}}            
\newcommand{\hatcurLBiMxxxxxD}{\ensuremath{0.5991}}             
\newcommand{\hatcurLBiiMxxxxxD}{\ensuremath{0.1948}}            
\newcommand{\hatcurLBiSonexxxxxD}{\ensuremath{0.0945}}            
\newcommand{\hatcurLBiiSonexxxxxD}{\ensuremath{0.1746}}           
\newcommand{\hatcurLBiStwoxxxxxD}{\ensuremath{0.0810}}            
\newcommand{\hatcurLBiiStwoxxxxxD}{\ensuremath{0.1431}}           
\newcommand{\hatcurLBiSthreexxxxxD}{\ensuremath{0.0688}}            
\newcommand{\hatcurLBiiSthreexxxxxD}{\ensuremath{0.1236}}           
\newcommand{\hatcurLBiSfourxxxxxD}{\ensuremath{0.0659}}            
\newcommand{\hatcurLBiiSfourxxxxxD}{\ensuremath{0.1065}}           
\newcommand{\hatcurISOmxxxxxD}{\ensuremath{0.655\pm0.014}}      
\newcommand{\hatcurISOmshortxxxxxD}{\ensuremath{0.66}}          
\newcommand{\hatcurISOmlongxxxxxD}{\ensuremath{0.655\pm0.014}}  
\newcommand{\hatcurISOrxxxxxD}{\ensuremath{0.6428\pm0.0066}}    
\newcommand{\hatcurISOrshortxxxxxD}{\ensuremath{0.64}}          
\newcommand{\hatcurISOrlongxxxxxD}{\ensuremath{0.6428\pm0.0066}} 
\newcommand{\hatcurISOrhoxxxxxD}{\ensuremath{3.48_{-0.12}^{+0.17}}} 
\newcommand{\hatcurISOrholongxxxxxD}{\ensuremath{3.48_{-0.12}^{+0.17}}} 
\newcommand{\hatcurISOloggxxxxxD}{\ensuremath{4.638\pm0.014}}   
\newcommand{\hatcurISOlumxxxxxD}{\ensuremath{0.1019\pm0.0028}}  
\newcommand{\hatcurISOlumshortxxxxxD}{\ensuremath{0.10}}        
\newcommand{\hatcurISOteffxxxxxD}{\ensuremath{4071\pm13}}       
\newcommand{\hatcurISOzfehxxxxxD}{\ensuremath{0.253\pm0.039}}   
\newcommand{\hatcurISOagexxxxxD}{\ensuremath{12.1\pm5.0}}       
\newcommand{\hatcurISOspecxxxxxD}{K}                            
\newcommand{\hatcurRVKxxxxxD}{\ensuremath{253\pm63}}            
\newcommand{\hatcurRVKtwosiglimxxxxxD}{\ensuremath{<293.1}}     
\newcommand{\hatcurRVrkxxxxxD}{\ensuremath{0\pm0}}              
\newcommand{\hatcurRVrhxxxxxD}{\ensuremath{0\pm0}}              
\newcommand{\hatcurRVkxxxxxD}{\ensuremath{0\pm0}}               
\newcommand{\hatcurRVhxxxxxD}{\ensuremath{0\pm0}}               
\newcommand{\hatcurRVtronexxxxxD}{\ensuremath{0\pm0}}           
\newcommand{\hatcurRVtrtwoxxxxxD}{\ensuremath{0\pm0}}           
\newcommand{\hatcurRVgammaxxxxxD}{\ensuremath{-7758\pm43}}      
\newcommand{\hatcurRVjitterxxxxxD}{\ensuremath{0\pm94}}         
\newcommand{\hatcurRVjittertwosiglimxxxxxD}{\ensuremath{<87.3}} 
\newcommand{\hatcurRVfitrmsxxxxxD}{\ensuremath{.1fym}}          %
\newcommand{\hatcurRVeccenxxxxxD}{\ensuremath{0\pm0}}           
\newcommand{\hatcurRVeccentwosiglimxxxxxD}{\ensuremath{<0.000}} 
\newcommand{\hatcurRVomegaxxxxxD}{\ensuremath{0\pm0}}           
\newcommand{\hatcurPPixxxxxD}{\ensuremath{88.44\pm0.27}}        
\newcommand{\hatcurPPgxxxxxD}{\ensuremath{25.0_{-1.6}^{+2.3}}}  
\newcommand{\hatcurPPloggxxxxxD}{\ensuremath{3.398_{-0.028}^{+0.038}}} 
\newcommand{\hatcurPParxxxxxD}{\ensuremath{12.06_{-0.14}^{+0.19}}} 
\newcommand{\hatcurPParelxxxxxD}{\ensuremath{0.03607\pm0.00025}} 
\newcommand{\hatcurPPrhoxxxxxD}{\ensuremath{1.077_{-0.081}^{+0.112}}} 
\newcommand{\hatcurPPmxxxxxD}{\ensuremath{1.374_{-0.074}^{+0.100}}} 
\newcommand{\hatcurPPmtwosiglimxxxxxD}{\ensuremath{<1.60}}      
\newcommand{\hatcurPPmshortxxxxxD}{\ensuremath{1.37}}           
\newcommand{\hatcurPPmlongxxxxxD}{\ensuremath{1.374_{-0.074}^{+0.100}}} 
\newcommand{\hatcurPPmexxxxxD}{\ensuremath{437_{-24}^{+32}}}    
\newcommand{\hatcurPPmeshortxxxxxD}{\ensuremath{436.6}}         
\newcommand{\hatcurPPmelongxxxxxD}{\ensuremath{437_{-24}^{+32}}} 
\newcommand{\hatcurPPrxxxxxD}{\ensuremath{1.165\pm0.021}}       
\newcommand{\hatcurPPrshortxxxxxD}{\ensuremath{1.17}}           
\newcommand{\hatcurPPrlongxxxxxD}{\ensuremath{1.165\pm0.021}}   
\newcommand{\hatcurPPrexxxxxD}{\ensuremath{13.06\pm0.23}}       
\newcommand{\hatcurPPreshortxxxxxD}{\ensuremath{13.1}}          
\newcommand{\hatcurPPrelongxxxxxD}{\ensuremath{13.06\pm0.23}}   
\newcommand{\hatcurPPmrcorrxxxxxD}{\ensuremath{-0.01}}          
\newcommand{\hatcurPPteffxxxxxD}{\ensuremath{828.3\pm5.9}}      
\newcommand{\hatcurPPthetaxxxxxD}{\ensuremath{0.129\pm0.032}}   
\newcommand{\hatcurPPfluxperixxxxxD}{\ensuremath{1.067\pm0.030}} 
\newcommand{\hatcurPPfluxperidimxxxxxD}{\ensuremath{8}}         
\newcommand{\hatcurPPfluxapxxxxxD}{\ensuremath{1.067\pm0.030}}  
\newcommand{\hatcurPPfluxapdimxxxxxD}{\ensuremath{8}}           
\newcommand{\hatcurPPfluxavgxxxxxD}{\ensuremath{1.067\pm0.030}} 
\newcommand{\hatcurPPfluxavgdimxxxxxD}{\ensuremath{8}}          
\newcommand{\hatcurPPfluxavglogxxxxxD}{\ensuremath{8.028\pm0.012}} 
\newcommand{\hatcurXsecphasexxxxxD}{\ensuremath{0\pm0}}         
\newcommand{\hatcurXsecondaryxxxxxD}{\ensuremath{2459138.23759\pm0.00020}} 
\newcommand{\hatcurXsecdurxxxxxD}{\ensuremath{0.09282\pm0.00070}} 
\newcommand{\hatcurXsecingdurxxxxxD}{\ensuremath{0.01610\pm0.00061}} 
\newcommand{\hatcurPPphiconjxxxxxD}{\ensuremath{0\pm0}}         
\newcommand{\hatcurPPperixxxxxD}{\ensuremath{2459135.92187\pm0.00020}} 
\newcommand{\hatcurPPaequivxxxxxD}{\ensuremath{0.1130\pm0.0016}} 
\newcommand{\hatcurPPtcircxxxxxD}{\ensuremath{195_{-20}^{+29}}} 
\newcommand{\hatcurPPtinfallxxxxxD}{\ensuremath{3620\pm560}}    
\newcommand{\hatcurXdistxxxxxD}{\ensuremath{413.9\pm5.9}}       
\newcommand{\hatcurXAvxxxxxD}{\ensuremath{0.111_{-0.019}^{+0.013}}} 
\newcommand{\hatcurXdistredxxxxxD}{\ensuremath{413.9\pm5.9}}    
\newcommand{\hatcurXEBVxxxxxD}{\ensuremath{0.0360_{-0.0060}^{+0.0040}}} 
\newcommand{\hatcurCCpmraxxxxxD}{\ensuremath{-24.64\pm0.10}}    
\newcommand{\hatcurCCpmdecxxxxxD}{\ensuremath{-8.286\pm0.098}}  
\newcommand{\hatcurCCpmxxxxxD}{\ensuremath{25.99\pm0.14}}       

\newcommand{\hatcurCCbbHmag}[1]{\ifnum#1=74 %
\hatcurCCbbHmagxxxxxA
\else
\ifnum#1=75 %
\hatcurCCbbHmagxxxxxB
\else
\ifnum#1=76 %
\hatcurCCbbHmagxxxxxC
\else
\ifnum#1=77 %
\hatcurCCbbHmagxxxxxD
\else
??????\fi
\fi
\fi
\fi
}
\newcommand{\hatcurCCbbJmag}[1]{\ifnum#1=74 %
\hatcurCCbbJmagxxxxxA
\else
\ifnum#1=75 %
\hatcurCCbbJmagxxxxxB
\else
\ifnum#1=76 %
\hatcurCCbbJmagxxxxxC
\else
\ifnum#1=77 %
\hatcurCCbbJmagxxxxxD
\else
??????\fi
\fi
\fi
\fi
}
\newcommand{\hatcurCCbbKmag}[1]{\ifnum#1=74 %
\hatcurCCbbKmagxxxxxA
\else
\ifnum#1=75 %
\hatcurCCbbKmagxxxxxB
\else
\ifnum#1=76 %
\hatcurCCbbKmagxxxxxC
\else
\ifnum#1=77 %
\hatcurCCbbKmagxxxxxD
\else
??????\fi
\fi
\fi
\fi
}
\newcommand{\hatcurCCcitHmag}[1]{\ifnum#1=74 %
\hatcurCCcitHmagxxxxxA
\else
\ifnum#1=75 %
\hatcurCCcitHmagxxxxxB
\else
\ifnum#1=76 %
\hatcurCCcitHmagxxxxxC
\else
\ifnum#1=77 %
\hatcurCCcitHmagxxxxxD
\else
??????\fi
\fi
\fi
\fi
}
\newcommand{\hatcurCCcitJmag}[1]{\ifnum#1=74 %
\hatcurCCcitJmagxxxxxA
\else
\ifnum#1=75 %
\hatcurCCcitJmagxxxxxB
\else
\ifnum#1=76 %
\hatcurCCcitJmagxxxxxC
\else
\ifnum#1=77 %
\hatcurCCcitJmagxxxxxD
\else
??????\fi
\fi
\fi
\fi
}
\newcommand{\hatcurCCcitKmag}[1]{\ifnum#1=74 %
\hatcurCCcitKmagxxxxxA
\else
\ifnum#1=75 %
\hatcurCCcitKmagxxxxxB
\else
\ifnum#1=76 %
\hatcurCCcitKmagxxxxxC
\else
\ifnum#1=77 %
\hatcurCCcitKmagxxxxxD
\else
??????\fi
\fi
\fi
\fi
}
\newcommand{\hatcurCCdec}[1]{\ifnum#1=74 %
\hatcurCCdecxxxxxA
\else
\ifnum#1=75 %
\hatcurCCdecxxxxxB
\else
\ifnum#1=76 %
\hatcurCCdecxxxxxC
\else
\ifnum#1=77 %
\hatcurCCdecxxxxxD
\else
??????\fi
\fi
\fi
\fi
}
\newcommand{\hatcurCCesoHKmag}[1]{\ifnum#1=74 %
\hatcurCCesoHKmagxxxxxA
\else
\ifnum#1=75 %
\hatcurCCesoHKmagxxxxxB
\else
\ifnum#1=76 %
\hatcurCCesoHKmagxxxxxC
\else
\ifnum#1=77 %
\hatcurCCesoHKmagxxxxxD
\else
??????\fi
\fi
\fi
\fi
}
\newcommand{\hatcurCCesoHmag}[1]{\ifnum#1=74 %
\hatcurCCesoHmagxxxxxA
\else
\ifnum#1=75 %
\hatcurCCesoHmagxxxxxB
\else
\ifnum#1=76 %
\hatcurCCesoHmagxxxxxC
\else
\ifnum#1=77 %
\hatcurCCesoHmagxxxxxD
\else
??????\fi
\fi
\fi
\fi
}
\newcommand{\hatcurCCesoJHmag}[1]{\ifnum#1=74 %
\hatcurCCesoJHmagxxxxxA
\else
\ifnum#1=75 %
\hatcurCCesoJHmagxxxxxB
\else
\ifnum#1=76 %
\hatcurCCesoJHmagxxxxxC
\else
\ifnum#1=77 %
\hatcurCCesoJHmagxxxxxD
\else
??????\fi
\fi
\fi
\fi
}
\newcommand{\hatcurCCesoJKmag}[1]{\ifnum#1=74 %
\hatcurCCesoJKmagxxxxxA
\else
\ifnum#1=75 %
\hatcurCCesoJKmagxxxxxB
\else
\ifnum#1=76 %
\hatcurCCesoJKmagxxxxxC
\else
\ifnum#1=77 %
\hatcurCCesoJKmagxxxxxD
\else
??????\fi
\fi
\fi
\fi
}
\newcommand{\hatcurCCesoJmag}[1]{\ifnum#1=74 %
\hatcurCCesoJmagxxxxxA
\else
\ifnum#1=75 %
\hatcurCCesoJmagxxxxxB
\else
\ifnum#1=76 %
\hatcurCCesoJmagxxxxxC
\else
\ifnum#1=77 %
\hatcurCCesoJmagxxxxxD
\else
??????\fi
\fi
\fi
\fi
}
\newcommand{\hatcurCCesoKmag}[1]{\ifnum#1=74 %
\hatcurCCesoKmagxxxxxA
\else
\ifnum#1=75 %
\hatcurCCesoKmagxxxxxB
\else
\ifnum#1=76 %
\hatcurCCesoKmagxxxxxC
\else
\ifnum#1=77 %
\hatcurCCesoKmagxxxxxD
\else
??????\fi
\fi
\fi
\fi
}
\newcommand{\hatcurCCgaia}[1]{\ifnum#1=74 %
\hatcurCCgaiaxxxxxA
\else
\ifnum#1=75 %
\hatcurCCgaiaxxxxxB
\else
\ifnum#1=76 %
\hatcurCCgaiaxxxxxC
\else
\ifnum#1=77 %
\hatcurCCgaiaxxxxxD
\else
??????\fi
\fi
\fi
\fi
}
\newcommand{\hatcurCCgaiadrtwo}[1]{\ifnum#1=74 %
\hatcurCCgaiadrtwoxxxxxA
\else
\ifnum#1=75 %
\hatcurCCgaiadrtwoxxxxxB
\else
\ifnum#1=76 %
\hatcurCCgaiadrtwoxxxxxC
\else
\ifnum#1=77 %
\hatcurCCgaiadrtwoxxxxxD
\else
??????\fi
\fi
\fi
\fi
}
\newcommand{\hatcurCCgaiadrtwoshort}[1]{\ifnum#1=74 %
\hatcurCCgaiadrtwoshortxxxxxA
\else
\ifnum#1=75 %
\hatcurCCgaiadrtwoshortxxxxxB
\else
\ifnum#1=76 %
\hatcurCCgaiadrtwoshortxxxxxC
\else
\ifnum#1=77 %
\hatcurCCgaiadrtwoshortxxxxxD
\else
??????\fi
\fi
\fi
\fi
}
\newcommand{\hatcurCCgaiamBP}[1]{\ifnum#1=74 %
\hatcurCCgaiamBPxxxxxA
\else
\ifnum#1=75 %
\hatcurCCgaiamBPxxxxxB
\else
\ifnum#1=76 %
\hatcurCCgaiamBPxxxxxC
\else
\ifnum#1=77 %
\hatcurCCgaiamBPxxxxxD
\else
??????\fi
\fi
\fi
\fi
}
\newcommand{\hatcurCCgaiamG}[1]{\ifnum#1=74 %
\hatcurCCgaiamGxxxxxA
\else
\ifnum#1=75 %
\hatcurCCgaiamGxxxxxB
\else
\ifnum#1=76 %
\hatcurCCgaiamGxxxxxC
\else
\ifnum#1=77 %
\hatcurCCgaiamGxxxxxD
\else
??????\fi
\fi
\fi
\fi
}
\newcommand{\hatcurCCgaiamRP}[1]{\ifnum#1=74 %
\hatcurCCgaiamRPxxxxxA
\else
\ifnum#1=75 %
\hatcurCCgaiamRPxxxxxB
\else
\ifnum#1=76 %
\hatcurCCgaiamRPxxxxxC
\else
\ifnum#1=77 %
\hatcurCCgaiamRPxxxxxD
\else
??????\fi
\fi
\fi
\fi
}
\newcommand{\hatcurCCgsc}[1]{\ifnum#1=74 %
\hatcurCCgscxxxxxA
\else
\ifnum#1=75 %
\hatcurCCgscxxxxxB
\else
\ifnum#1=76 %
\hatcurCCgscxxxxxC
\else
\ifnum#1=77 %
\hatcurCCgscxxxxxD
\else
??????\fi
\fi
\fi
\fi
}
\newcommand{\hatcurCCmag}[1]{\ifnum#1=74 %
\hatcurCCmagxxxxxA
\else
\ifnum#1=75 %
\hatcurCCmagxxxxxB
\else
\ifnum#1=76 %
\hatcurCCmagxxxxxC
\else
\ifnum#1=77 %
\hatcurCCmagxxxxxD
\else
??????\fi
\fi
\fi
\fi
}
\newcommand{\hatcurCCparallax}[1]{\ifnum#1=74 %
\hatcurCCparallaxxxxxxA
\else
\ifnum#1=75 %
\hatcurCCparallaxxxxxxB
\else
\ifnum#1=76 %
\hatcurCCparallaxxxxxxC
\else
\ifnum#1=77 %
\hatcurCCparallaxxxxxxD
\else
??????\fi
\fi
\fi
\fi
}
\newcommand{\hatcurCCpm}[1]{\ifnum#1=74 %
\hatcurCCpmxxxxxA
\else
\ifnum#1=75 %
\hatcurCCpmxxxxxB
\else
\ifnum#1=76 %
\hatcurCCpmxxxxxC
\else
\ifnum#1=77 %
\hatcurCCpmxxxxxD
\else
??????\fi
\fi
\fi
\fi
}
\newcommand{\hatcurCCpmdec}[1]{\ifnum#1=74 %
\hatcurCCpmdecxxxxxA
\else
\ifnum#1=75 %
\hatcurCCpmdecxxxxxB
\else
\ifnum#1=76 %
\hatcurCCpmdecxxxxxC
\else
\ifnum#1=77 %
\hatcurCCpmdecxxxxxD
\else
??????\fi
\fi
\fi
\fi
}
\newcommand{\hatcurCCpmra}[1]{\ifnum#1=74 %
\hatcurCCpmraxxxxxA
\else
\ifnum#1=75 %
\hatcurCCpmraxxxxxB
\else
\ifnum#1=76 %
\hatcurCCpmraxxxxxC
\else
\ifnum#1=77 %
\hatcurCCpmraxxxxxD
\else
??????\fi
\fi
\fi
\fi
}
\newcommand{\hatcurCCra}[1]{\ifnum#1=74 %
\hatcurCCraxxxxxA
\else
\ifnum#1=75 %
\hatcurCCraxxxxxB
\else
\ifnum#1=76 %
\hatcurCCraxxxxxC
\else
\ifnum#1=77 %
\hatcurCCraxxxxxD
\else
??????\fi
\fi
\fi
\fi
}
\newcommand{\hatcurCCtassmB}[1]{\ifnum#1=74 %
\hatcurCCtassmBxxxxxA
\else
\ifnum#1=75 %
\hatcurCCtassmBxxxxxB
\else
\ifnum#1=76 %
\hatcurCCtassmBxxxxxC
\else
\ifnum#1=77 %
\hatcurCCtassmBxxxxxD
\else
??????\fi
\fi
\fi
\fi
}
\newcommand{\hatcurCCtassmBshort}[1]{\ifnum#1=74 %
\hatcurCCtassmBshortxxxxxA
\else
\ifnum#1=75 %
\hatcurCCtassmBshortxxxxxB
\else
\ifnum#1=76 %
\hatcurCCtassmBshortxxxxxC
\else
\ifnum#1=77 %
\hatcurCCtassmBshortxxxxxD
\else
??????\fi
\fi
\fi
\fi
}
\newcommand{\hatcurCCtassmg}[1]{\ifnum#1=74 %
\hatcurCCtassmgxxxxxA
\else
\ifnum#1=75 %
\hatcurCCtassmgxxxxxB
\else
\ifnum#1=76 %
\hatcurCCtassmgxxxxxC
\else
\ifnum#1=77 %
\hatcurCCtassmgxxxxxD
\else
??????\fi
\fi
\fi
\fi
}
\newcommand{\hatcurCCtassmgshort}[1]{\ifnum#1=74 %
\hatcurCCtassmgshortxxxxxA
\else
\ifnum#1=75 %
\hatcurCCtassmgshortxxxxxB
\else
\ifnum#1=76 %
\hatcurCCtassmgshortxxxxxC
\else
\ifnum#1=77 %
\hatcurCCtassmgshortxxxxxD
\else
??????\fi
\fi
\fi
\fi
}
\newcommand{\hatcurCCtassmi}[1]{\ifnum#1=74 %
\hatcurCCtassmixxxxxA
\else
\ifnum#1=75 %
\hatcurCCtassmixxxxxB
\else
\ifnum#1=76 %
\hatcurCCtassmixxxxxC
\else
\ifnum#1=77 %
\hatcurCCtassmixxxxxD
\else
??????\fi
\fi
\fi
\fi
}
\newcommand{\hatcurCCtassmI}[1]{\ifnum#1=74 %
\hatcurCCtassmIxxxxxA
\else
\ifnum#1=75 %
\hatcurCCtassmIxxxxxB
\else
\ifnum#1=76 %
\hatcurCCtassmIxxxxxC
\else
\ifnum#1=77 %
\hatcurCCtassmIxxxxxD
\else
??????\fi
\fi
\fi
\fi
}
\newcommand{\hatcurCCtassmishort}[1]{\ifnum#1=74 %
\hatcurCCtassmishortxxxxxA
\else
\ifnum#1=75 %
\hatcurCCtassmishortxxxxxB
\else
\ifnum#1=76 %
\hatcurCCtassmishortxxxxxC
\else
\ifnum#1=77 %
\hatcurCCtassmishortxxxxxD
\else
??????\fi
\fi
\fi
\fi
}
\newcommand{\hatcurCCtassmIshort}[1]{\ifnum#1=74 %
\hatcurCCtassmIshortxxxxxA
\else
\ifnum#1=75 %
\hatcurCCtassmIshortxxxxxB
\else
\ifnum#1=76 %
\hatcurCCtassmIshortxxxxxC
\else
\ifnum#1=77 %
\hatcurCCtassmIshortxxxxxD
\else
??????\fi
\fi
\fi
\fi
}
\newcommand{\hatcurCCtassmr}[1]{\ifnum#1=74 %
\hatcurCCtassmrxxxxxA
\else
\ifnum#1=75 %
\hatcurCCtassmrxxxxxB
\else
\ifnum#1=76 %
\hatcurCCtassmrxxxxxC
\else
\ifnum#1=77 %
\hatcurCCtassmrxxxxxD
\else
??????\fi
\fi
\fi
\fi
}
\newcommand{\hatcurCCtassmrshort}[1]{\ifnum#1=74 %
\hatcurCCtassmrshortxxxxxA
\else
\ifnum#1=75 %
\hatcurCCtassmrshortxxxxxB
\else
\ifnum#1=76 %
\hatcurCCtassmrshortxxxxxC
\else
\ifnum#1=77 %
\hatcurCCtassmrshortxxxxxD
\else
??????\fi
\fi
\fi
\fi
}
\newcommand{\hatcurCCtassmv}[1]{\ifnum#1=74 %
\hatcurCCtassmvxxxxxA
\else
\ifnum#1=75 %
\hatcurCCtassmvxxxxxB
\else
\ifnum#1=76 %
\hatcurCCtassmvxxxxxC
\else
\ifnum#1=77 %
\hatcurCCtassmvxxxxxD
\else
??????\fi
\fi
\fi
\fi
}
\newcommand{\hatcurCCtassmvshort}[1]{\ifnum#1=74 %
\hatcurCCtassmvshortxxxxxA
\else
\ifnum#1=75 %
\hatcurCCtassmvshortxxxxxB
\else
\ifnum#1=76 %
\hatcurCCtassmvshortxxxxxC
\else
\ifnum#1=77 %
\hatcurCCtassmvshortxxxxxD
\else
??????\fi
\fi
\fi
\fi
}
\newcommand{\hatcurCCtwomass}[1]{\ifnum#1=74 %
\hatcurCCtwomassxxxxxA
\else
\ifnum#1=75 %
\hatcurCCtwomassxxxxxB
\else
\ifnum#1=76 %
\hatcurCCtwomassxxxxxC
\else
\ifnum#1=77 %
\hatcurCCtwomassxxxxxD
\else
??????\fi
\fi
\fi
\fi
}
\newcommand{\hatcurCCtwomassHmag}[1]{\ifnum#1=74 %
\hatcurCCtwomassHmagxxxxxA
\else
\ifnum#1=75 %
\hatcurCCtwomassHmagxxxxxB
\else
\ifnum#1=76 %
\hatcurCCtwomassHmagxxxxxC
\else
\ifnum#1=77 %
\hatcurCCtwomassHmagxxxxxD
\else
??????\fi
\fi
\fi
\fi
}
\newcommand{\hatcurCCtwomassJmag}[1]{\ifnum#1=74 %
\hatcurCCtwomassJmagxxxxxA
\else
\ifnum#1=75 %
\hatcurCCtwomassJmagxxxxxB
\else
\ifnum#1=76 %
\hatcurCCtwomassJmagxxxxxC
\else
\ifnum#1=77 %
\hatcurCCtwomassJmagxxxxxD
\else
??????\fi
\fi
\fi
\fi
}
\newcommand{\hatcurCCtwomassKmag}[1]{\ifnum#1=74 %
\hatcurCCtwomassKmagxxxxxA
\else
\ifnum#1=75 %
\hatcurCCtwomassKmagxxxxxB
\else
\ifnum#1=76 %
\hatcurCCtwomassKmagxxxxxC
\else
\ifnum#1=77 %
\hatcurCCtwomassKmagxxxxxD
\else
??????\fi
\fi
\fi
\fi
}
\newcommand{\hatcurCCWfourmag}[1]{\ifnum#1=74 %
\hatcurCCWfourmagxxxxxA
\else
\ifnum#1=75 %
\hatcurCCWfourmagxxxxxB
\else
\ifnum#1=76 %
\hatcurCCWfourmagxxxxxC
\else
\ifnum#1=77 %
\hatcurCCWfourmagxxxxxD
\else
??????\fi
\fi
\fi
\fi
}
\newcommand{\hatcurCCWonemag}[1]{\ifnum#1=74 %
\hatcurCCWonemagxxxxxA
\else
\ifnum#1=75 %
\hatcurCCWonemagxxxxxB
\else
\ifnum#1=76 %
\hatcurCCWonemagxxxxxC
\else
\ifnum#1=77 %
\hatcurCCWonemagxxxxxD
\else
??????\fi
\fi
\fi
\fi
}
\newcommand{\hatcurCCWthreemag}[1]{\ifnum#1=74 %
\hatcurCCWthreemagxxxxxA
\else
\ifnum#1=75 %
\hatcurCCWthreemagxxxxxB
\else
\ifnum#1=76 %
\hatcurCCWthreemagxxxxxC
\else
\ifnum#1=77 %
\hatcurCCWthreemagxxxxxD
\else
??????\fi
\fi
\fi
\fi
}
\newcommand{\hatcurCCWtwomag}[1]{\ifnum#1=74 %
\hatcurCCWtwomagxxxxxA
\else
\ifnum#1=75 %
\hatcurCCWtwomagxxxxxB
\else
\ifnum#1=76 %
\hatcurCCWtwomagxxxxxC
\else
\ifnum#1=77 %
\hatcurCCWtwomagxxxxxD
\else
??????\fi
\fi
\fi
\fi
}
\newcommand{\hatcurextraerrMBPtwo}[1]{\ifnum#1=75 %
\hatcurextraerrMBPtwoxxxxxB
\else
\ifnum#1=76 %
\hatcurextraerrMBPtwoxxxxxC
\else
\ifnum#1=77 %
\hatcurextraerrMBPtwoxxxxxD
\else
??????\fi
\fi
\fi
}
\newcommand{\hatcurextraerrMBPtwotwosiglim}[1]{\ifnum#1=75 %
\hatcurextraerrMBPtwotwosiglimxxxxxB
\else
\ifnum#1=76 %
\hatcurextraerrMBPtwotwosiglimxxxxxC
\else
\ifnum#1=77 %
\hatcurextraerrMBPtwotwosiglimxxxxxD
\else
??????\fi
\fi
\fi
}
\newcommand{\hatcurextraerrMg}[1]{\ifnum#1=75 %
\hatcurextraerrMgxxxxxB
\else
\ifnum#1=77 %
\hatcurextraerrMgxxxxxD
\else
??????\fi
\fi
}
\newcommand{\hatcurextraerrMG}[1]{\ifnum#1=74 %
\hatcurextraerrMGxxxxxA
\else
\ifnum#1=75 %
\hatcurextraerrMGxxxxxB
\else
\ifnum#1=76 %
\hatcurextraerrMGxxxxxC
\else
\ifnum#1=77 %
\hatcurextraerrMGxxxxxD
\else
??????\fi
\fi
\fi
\fi
}
\newcommand{\hatcurextraerrMgtwosiglim}[1]{\ifnum#1=75 %
\hatcurextraerrMgtwosiglimxxxxxB
\else
\ifnum#1=77 %
\hatcurextraerrMgtwosiglimxxxxxD
\else
??????\fi
\fi
}
\newcommand{\hatcurextraerrMGtwosiglim}[1]{\ifnum#1=74 %
\hatcurextraerrMGtwosiglimxxxxxA
\else
\ifnum#1=75 %
\hatcurextraerrMGtwosiglimxxxxxB
\else
\ifnum#1=76 %
\hatcurextraerrMGtwosiglimxxxxxC
\else
\ifnum#1=77 %
\hatcurextraerrMGtwosiglimxxxxxD
\else
??????\fi
\fi
\fi
\fi
}
\newcommand{\hatcurextraerrMH}[1]{\ifnum#1=74 %
\hatcurextraerrMHxxxxxA
\else
\ifnum#1=75 %
\hatcurextraerrMHxxxxxB
\else
\ifnum#1=76 %
\hatcurextraerrMHxxxxxC
\else
\ifnum#1=77 %
\hatcurextraerrMHxxxxxD
\else
??????\fi
\fi
\fi
\fi
}
\newcommand{\hatcurextraerrMHtwosiglim}[1]{\ifnum#1=74 %
\hatcurextraerrMHtwosiglimxxxxxA
\else
\ifnum#1=75 %
\hatcurextraerrMHtwosiglimxxxxxB
\else
\ifnum#1=76 %
\hatcurextraerrMHtwosiglimxxxxxC
\else
\ifnum#1=77 %
\hatcurextraerrMHtwosiglimxxxxxD
\else
??????\fi
\fi
\fi
\fi
}
\newcommand{\hatcurextraerrMi}[1]{\ifnum#1=75 %
\hatcurextraerrMixxxxxB
\else
\ifnum#1=77 %
\hatcurextraerrMixxxxxD
\else
??????\fi
\fi
}
\newcommand{\hatcurextraerrMitwosiglim}[1]{\ifnum#1=75 %
\hatcurextraerrMitwosiglimxxxxxB
\else
\ifnum#1=77 %
\hatcurextraerrMitwosiglimxxxxxD
\else
??????\fi
\fi
}
\newcommand{\hatcurextraerrMJ}[1]{\ifnum#1=74 %
\hatcurextraerrMJxxxxxA
\else
\ifnum#1=75 %
\hatcurextraerrMJxxxxxB
\else
\ifnum#1=76 %
\hatcurextraerrMJxxxxxC
\else
\ifnum#1=77 %
\hatcurextraerrMJxxxxxD
\else
??????\fi
\fi
\fi
\fi
}
\newcommand{\hatcurextraerrMJtwosiglim}[1]{\ifnum#1=74 %
\hatcurextraerrMJtwosiglimxxxxxA
\else
\ifnum#1=75 %
\hatcurextraerrMJtwosiglimxxxxxB
\else
\ifnum#1=76 %
\hatcurextraerrMJtwosiglimxxxxxC
\else
\ifnum#1=77 %
\hatcurextraerrMJtwosiglimxxxxxD
\else
??????\fi
\fi
\fi
\fi
}
\newcommand{\hatcurextraerrMKs}[1]{\ifnum#1=74 %
\hatcurextraerrMKsxxxxxA
\else
\ifnum#1=75 %
\hatcurextraerrMKsxxxxxB
\else
\ifnum#1=76 %
\hatcurextraerrMKsxxxxxC
\else
\ifnum#1=77 %
\hatcurextraerrMKsxxxxxD
\else
??????\fi
\fi
\fi
\fi
}
\newcommand{\hatcurextraerrMKstwosiglim}[1]{\ifnum#1=74 %
\hatcurextraerrMKstwosiglimxxxxxA
\else
\ifnum#1=75 %
\hatcurextraerrMKstwosiglimxxxxxB
\else
\ifnum#1=76 %
\hatcurextraerrMKstwosiglimxxxxxC
\else
\ifnum#1=77 %
\hatcurextraerrMKstwosiglimxxxxxD
\else
??????\fi
\fi
\fi
\fi
}
\newcommand{\hatcurextraerrMr}[1]{\ifnum#1=75 %
\hatcurextraerrMrxxxxxB
\else
\ifnum#1=77 %
\hatcurextraerrMrxxxxxD
\else
??????\fi
\fi
}
\newcommand{\hatcurextraerrMRP}[1]{\ifnum#1=75 %
\hatcurextraerrMRPxxxxxB
\else
\ifnum#1=76 %
\hatcurextraerrMRPxxxxxC
\else
\ifnum#1=77 %
\hatcurextraerrMRPxxxxxD
\else
??????\fi
\fi
\fi
}
\newcommand{\hatcurextraerrMRPtwosiglim}[1]{\ifnum#1=75 %
\hatcurextraerrMRPtwosiglimxxxxxB
\else
\ifnum#1=76 %
\hatcurextraerrMRPtwosiglimxxxxxC
\else
\ifnum#1=77 %
\hatcurextraerrMRPtwosiglimxxxxxD
\else
??????\fi
\fi
\fi
}
\newcommand{\hatcurextraerrMrtwosiglim}[1]{\ifnum#1=75 %
\hatcurextraerrMrtwosiglimxxxxxB
\else
\ifnum#1=77 %
\hatcurextraerrMrtwosiglimxxxxxD
\else
??????\fi
\fi
}
\newcommand{\hatcurextraerrMWone}[1]{\ifnum#1=74 %
\hatcurextraerrMWonexxxxxA
\else
\ifnum#1=75 %
\hatcurextraerrMWonexxxxxB
\else
\ifnum#1=76 %
\hatcurextraerrMWonexxxxxC
\else
\ifnum#1=77 %
\hatcurextraerrMWonexxxxxD
\else
??????\fi
\fi
\fi
\fi
}
\newcommand{\hatcurextraerrMWonetwosiglim}[1]{\ifnum#1=74 %
\hatcurextraerrMWonetwosiglimxxxxxA
\else
\ifnum#1=75 %
\hatcurextraerrMWonetwosiglimxxxxxB
\else
\ifnum#1=76 %
\hatcurextraerrMWonetwosiglimxxxxxC
\else
\ifnum#1=77 %
\hatcurextraerrMWonetwosiglimxxxxxD
\else
??????\fi
\fi
\fi
\fi
}
\newcommand{\hatcurextraerrMWtwo}[1]{\ifnum#1=74 %
\hatcurextraerrMWtwoxxxxxA
\else
\ifnum#1=75 %
\hatcurextraerrMWtwoxxxxxB
\else
\ifnum#1=76 %
\hatcurextraerrMWtwoxxxxxC
\else
\ifnum#1=77 %
\hatcurextraerrMWtwoxxxxxD
\else
??????\fi
\fi
\fi
\fi
}
\newcommand{\hatcurextraerrMWtwotwosiglim}[1]{\ifnum#1=74 %
\hatcurextraerrMWtwotwosiglimxxxxxA
\else
\ifnum#1=75 %
\hatcurextraerrMWtwotwosiglimxxxxxB
\else
\ifnum#1=76 %
\hatcurextraerrMWtwotwosiglimxxxxxC
\else
\ifnum#1=77 %
\hatcurextraerrMWtwotwosiglimxxxxxD
\else
??????\fi
\fi
\fi
\fi
}
\newcommand{\hatcurfield}[1]{\ifnum#1=74 %
\hatcurfieldxxxxxA
\else
\ifnum#1=75 %
\hatcurfieldxxxxxB
\else
\ifnum#1=76 %
\hatcurfieldxxxxxC
\else
\ifnum#1=77 %
\hatcurfieldxxxxxD
\else
??????\fi
\fi
\fi
\fi
}
\newcommand{\hatcurhtr}[1]{\ifnum#1=74 %
\hatcurhtrxxxxxA
\else
\ifnum#1=75 %
\hatcurhtrxxxxxB
\else
\ifnum#1=76 %
\hatcurhtrxxxxxC
\else
\ifnum#1=77 %
\hatcurhtrxxxxxD
\else
??????\fi
\fi
\fi
\fi
}
\newcommand{\hatcurISOage}[1]{\ifnum#1=74 %
\hatcurISOagexxxxxA
\else
\ifnum#1=75 %
\hatcurISOagexxxxxB
\else
\ifnum#1=76 %
\hatcurISOagexxxxxC
\else
\ifnum#1=77 %
\hatcurISOagexxxxxD
\else
??????\fi
\fi
\fi
\fi
}
\newcommand{\hatcurISOlogg}[1]{\ifnum#1=74 %
\hatcurISOloggxxxxxA
\else
\ifnum#1=75 %
\hatcurISOloggxxxxxB
\else
\ifnum#1=76 %
\hatcurISOloggxxxxxC
\else
\ifnum#1=77 %
\hatcurISOloggxxxxxD
\else
??????\fi
\fi
\fi
\fi
}
\newcommand{\hatcurISOlum}[1]{\ifnum#1=74 %
\hatcurISOlumxxxxxA
\else
\ifnum#1=75 %
\hatcurISOlumxxxxxB
\else
\ifnum#1=76 %
\hatcurISOlumxxxxxC
\else
\ifnum#1=77 %
\hatcurISOlumxxxxxD
\else
??????\fi
\fi
\fi
\fi
}
\newcommand{\hatcurISOlumshort}[1]{\ifnum#1=74 %
\hatcurISOlumshortxxxxxA
\else
\ifnum#1=75 %
\hatcurISOlumshortxxxxxB
\else
\ifnum#1=76 %
\hatcurISOlumshortxxxxxC
\else
\ifnum#1=77 %
\hatcurISOlumshortxxxxxD
\else
??????\fi
\fi
\fi
\fi
}
\newcommand{\hatcurISOm}[1]{\ifnum#1=74 %
\hatcurISOmxxxxxA
\else
\ifnum#1=75 %
\hatcurISOmxxxxxB
\else
\ifnum#1=76 %
\hatcurISOmxxxxxC
\else
\ifnum#1=77 %
\hatcurISOmxxxxxD
\else
??????\fi
\fi
\fi
\fi
}
\newcommand{\hatcurISOmlong}[1]{\ifnum#1=74 %
\hatcurISOmlongxxxxxA
\else
\ifnum#1=75 %
\hatcurISOmlongxxxxxB
\else
\ifnum#1=76 %
\hatcurISOmlongxxxxxC
\else
\ifnum#1=77 %
\hatcurISOmlongxxxxxD
\else
??????\fi
\fi
\fi
\fi
}
\newcommand{\hatcurISOmshort}[1]{\ifnum#1=74 %
\hatcurISOmshortxxxxxA
\else
\ifnum#1=75 %
\hatcurISOmshortxxxxxB
\else
\ifnum#1=76 %
\hatcurISOmshortxxxxxC
\else
\ifnum#1=77 %
\hatcurISOmshortxxxxxD
\else
??????\fi
\fi
\fi
\fi
}
\newcommand{\hatcurISOr}[1]{\ifnum#1=74 %
\hatcurISOrxxxxxA
\else
\ifnum#1=75 %
\hatcurISOrxxxxxB
\else
\ifnum#1=76 %
\hatcurISOrxxxxxC
\else
\ifnum#1=77 %
\hatcurISOrxxxxxD
\else
??????\fi
\fi
\fi
\fi
}
\newcommand{\hatcurISOrho}[1]{\ifnum#1=74 %
\hatcurISOrhoxxxxxA
\else
\ifnum#1=75 %
\hatcurISOrhoxxxxxB
\else
\ifnum#1=76 %
\hatcurISOrhoxxxxxC
\else
\ifnum#1=77 %
\hatcurISOrhoxxxxxD
\else
??????\fi
\fi
\fi
\fi
}
\newcommand{\hatcurISOrholong}[1]{\ifnum#1=74 %
\hatcurISOrholongxxxxxA
\else
\ifnum#1=75 %
\hatcurISOrholongxxxxxB
\else
\ifnum#1=76 %
\hatcurISOrholongxxxxxC
\else
\ifnum#1=77 %
\hatcurISOrholongxxxxxD
\else
??????\fi
\fi
\fi
\fi
}
\newcommand{\hatcurISOrlong}[1]{\ifnum#1=74 %
\hatcurISOrlongxxxxxA
\else
\ifnum#1=75 %
\hatcurISOrlongxxxxxB
\else
\ifnum#1=76 %
\hatcurISOrlongxxxxxC
\else
\ifnum#1=77 %
\hatcurISOrlongxxxxxD
\else
??????\fi
\fi
\fi
\fi
}
\newcommand{\hatcurISOrshort}[1]{\ifnum#1=74 %
\hatcurISOrshortxxxxxA
\else
\ifnum#1=75 %
\hatcurISOrshortxxxxxB
\else
\ifnum#1=76 %
\hatcurISOrshortxxxxxC
\else
\ifnum#1=77 %
\hatcurISOrshortxxxxxD
\else
??????\fi
\fi
\fi
\fi
}
\newcommand{\hatcurISOspec}[1]{\ifnum#1=74 %
\hatcurISOspecxxxxxA
\else
\ifnum#1=75 %
\hatcurISOspecxxxxxB
\else
\ifnum#1=76 %
\hatcurISOspecxxxxxC
\else
\ifnum#1=77 %
\hatcurISOspecxxxxxD
\else
??????\fi
\fi
\fi
\fi
}
\newcommand{\hatcurISOteff}[1]{\ifnum#1=74 %
\hatcurISOteffxxxxxA
\else
\ifnum#1=75 %
\hatcurISOteffxxxxxB
\else
\ifnum#1=76 %
\hatcurISOteffxxxxxC
\else
\ifnum#1=77 %
\hatcurISOteffxxxxxD
\else
??????\fi
\fi
\fi
\fi
}
\newcommand{\hatcurISOzfeh}[1]{\ifnum#1=74 %
\hatcurISOzfehxxxxxA
\else
\ifnum#1=75 %
\hatcurISOzfehxxxxxB
\else
\ifnum#1=76 %
\hatcurISOzfehxxxxxC
\else
\ifnum#1=77 %
\hatcurISOzfehxxxxxD
\else
??????\fi
\fi
\fi
\fi
}
\newcommand{\hatcurLBiB}[1]{\ifnum#1=74 %
\hatcurLBiBxxxxxA
\else
\ifnum#1=75 %
\hatcurLBiBxxxxxB
\else
\ifnum#1=76 %
\hatcurLBiBxxxxxC
\else
\ifnum#1=77 %
\hatcurLBiBxxxxxD
\else
??????\fi
\fi
\fi
\fi
}
\newcommand{\hatcurLBiC}[1]{\ifnum#1=74 %
\hatcurLBiCxxxxxA
\else
\ifnum#1=75 %
\hatcurLBiCxxxxxB
\else
\ifnum#1=76 %
\hatcurLBiCxxxxxC
\else
\ifnum#1=77 %
\hatcurLBiCxxxxxD
\else
??????\fi
\fi
\fi
\fi
}
\newcommand{\hatcurLBig}[1]{\ifnum#1=74 %
\hatcurLBigxxxxxA
\else
\ifnum#1=75 %
\hatcurLBigxxxxxB
\else
\ifnum#1=76 %
\hatcurLBigxxxxxC
\else
\ifnum#1=77 %
\hatcurLBigxxxxxD
\else
??????\fi
\fi
\fi
\fi
}
\newcommand{\hatcurLBiH}[1]{\ifnum#1=74 %
\hatcurLBiHxxxxxA
\else
\ifnum#1=75 %
\hatcurLBiHxxxxxB
\else
\ifnum#1=76 %
\hatcurLBiHxxxxxC
\else
\ifnum#1=77 %
\hatcurLBiHxxxxxD
\else
??????\fi
\fi
\fi
\fi
}
\newcommand{\hatcurLBii}[1]{\ifnum#1=74 %
\hatcurLBiixxxxxA
\else
\ifnum#1=75 %
\hatcurLBiixxxxxB
\else
\ifnum#1=76 %
\hatcurLBiixxxxxC
\else
\ifnum#1=77 %
\hatcurLBiixxxxxD
\else
??????\fi
\fi
\fi
\fi
}
\newcommand{\hatcurLBiI}[1]{\ifnum#1=74 %
\hatcurLBiIxxxxxA
\else
\ifnum#1=75 %
\hatcurLBiIxxxxxB
\else
\ifnum#1=76 %
\hatcurLBiIxxxxxC
\else
\ifnum#1=77 %
\hatcurLBiIxxxxxD
\else
??????\fi
\fi
\fi
\fi
}
\newcommand{\hatcurLBiiB}[1]{\ifnum#1=74 %
\hatcurLBiiBxxxxxA
\else
\ifnum#1=75 %
\hatcurLBiiBxxxxxB
\else
\ifnum#1=76 %
\hatcurLBiiBxxxxxC
\else
\ifnum#1=77 %
\hatcurLBiiBxxxxxD
\else
??????\fi
\fi
\fi
\fi
}
\newcommand{\hatcurLBiiC}[1]{\ifnum#1=74 %
\hatcurLBiiCxxxxxA
\else
\ifnum#1=75 %
\hatcurLBiiCxxxxxB
\else
\ifnum#1=76 %
\hatcurLBiiCxxxxxC
\else
\ifnum#1=77 %
\hatcurLBiiCxxxxxD
\else
??????\fi
\fi
\fi
\fi
}
\newcommand{\hatcurLBiig}[1]{\ifnum#1=74 %
\hatcurLBiigxxxxxA
\else
\ifnum#1=75 %
\hatcurLBiigxxxxxB
\else
\ifnum#1=76 %
\hatcurLBiigxxxxxC
\else
\ifnum#1=77 %
\hatcurLBiigxxxxxD
\else
??????\fi
\fi
\fi
\fi
}
\newcommand{\hatcurLBiiH}[1]{\ifnum#1=74 %
\hatcurLBiiHxxxxxA
\else
\ifnum#1=75 %
\hatcurLBiiHxxxxxB
\else
\ifnum#1=76 %
\hatcurLBiiHxxxxxC
\else
\ifnum#1=77 %
\hatcurLBiiHxxxxxD
\else
??????\fi
\fi
\fi
\fi
}
\newcommand{\hatcurLBiii}[1]{\ifnum#1=74 %
\hatcurLBiiixxxxxA
\else
\ifnum#1=75 %
\hatcurLBiiixxxxxB
\else
\ifnum#1=76 %
\hatcurLBiiixxxxxC
\else
\ifnum#1=77 %
\hatcurLBiiixxxxxD
\else
??????\fi
\fi
\fi
\fi
}
\newcommand{\hatcurLBiiI}[1]{\ifnum#1=74 %
\hatcurLBiiIxxxxxA
\else
\ifnum#1=75 %
\hatcurLBiiIxxxxxB
\else
\ifnum#1=76 %
\hatcurLBiiIxxxxxC
\else
\ifnum#1=77 %
\hatcurLBiiIxxxxxD
\else
??????\fi
\fi
\fi
\fi
}
\newcommand{\hatcurLBiiJ}[1]{\ifnum#1=74 %
\hatcurLBiiJxxxxxA
\else
\ifnum#1=75 %
\hatcurLBiiJxxxxxB
\else
\ifnum#1=76 %
\hatcurLBiiJxxxxxC
\else
\ifnum#1=77 %
\hatcurLBiiJxxxxxD
\else
??????\fi
\fi
\fi
\fi
}
\newcommand{\hatcurLBiiK}[1]{\ifnum#1=74 %
\hatcurLBiiKxxxxxA
\else
\ifnum#1=75 %
\hatcurLBiiKxxxxxB
\else
\ifnum#1=76 %
\hatcurLBiiKxxxxxC
\else
\ifnum#1=77 %
\hatcurLBiiKxxxxxD
\else
??????\fi
\fi
\fi
\fi
}
\newcommand{\hatcurLBiikep}[1]{\ifnum#1=74 %
\hatcurLBiikepxxxxxA
\else
\ifnum#1=75 %
\hatcurLBiikepxxxxxB
\else
\ifnum#1=76 %
\hatcurLBiikepxxxxxC
\else
\ifnum#1=77 %
\hatcurLBiikepxxxxxD
\else
??????\fi
\fi
\fi
\fi
}
\newcommand{\hatcurLBiiM}[1]{\ifnum#1=74 %
\hatcurLBiiMxxxxxA
\else
\ifnum#1=75 %
\hatcurLBiiMxxxxxB
\else
\ifnum#1=76 %
\hatcurLBiiMxxxxxC
\else
\ifnum#1=77 %
\hatcurLBiiMxxxxxD
\else
??????\fi
\fi
\fi
\fi
}
\newcommand{\hatcurLBiir}[1]{\ifnum#1=74 %
\hatcurLBiirxxxxxA
\else
\ifnum#1=75 %
\hatcurLBiirxxxxxB
\else
\ifnum#1=76 %
\hatcurLBiirxxxxxC
\else
\ifnum#1=77 %
\hatcurLBiirxxxxxD
\else
??????\fi
\fi
\fi
\fi
}
\newcommand{\hatcurLBiiR}[1]{\ifnum#1=74 %
\hatcurLBiiRxxxxxA
\else
\ifnum#1=75 %
\hatcurLBiiRxxxxxB
\else
\ifnum#1=76 %
\hatcurLBiiRxxxxxC
\else
\ifnum#1=77 %
\hatcurLBiiRxxxxxD
\else
??????\fi
\fi
\fi
\fi
}
\newcommand{\hatcurLBiiSfour}[1]{\ifnum#1=74 %
\hatcurLBiiSfourxxxxxA
\else
\ifnum#1=75 %
\hatcurLBiiSfourxxxxxB
\else
\ifnum#1=76 %
\hatcurLBiiSfourxxxxxC
\else
\ifnum#1=77 %
\hatcurLBiiSfourxxxxxD
\else
??????\fi
\fi
\fi
\fi
}
\newcommand{\hatcurLBiiSone}[1]{\ifnum#1=74 %
\hatcurLBiiSonexxxxxA
\else
\ifnum#1=75 %
\hatcurLBiiSonexxxxxB
\else
\ifnum#1=76 %
\hatcurLBiiSonexxxxxC
\else
\ifnum#1=77 %
\hatcurLBiiSonexxxxxD
\else
??????\fi
\fi
\fi
\fi
}
\newcommand{\hatcurLBiiSthree}[1]{\ifnum#1=74 %
\hatcurLBiiSthreexxxxxA
\else
\ifnum#1=75 %
\hatcurLBiiSthreexxxxxB
\else
\ifnum#1=76 %
\hatcurLBiiSthreexxxxxC
\else
\ifnum#1=77 %
\hatcurLBiiSthreexxxxxD
\else
??????\fi
\fi
\fi
\fi
}
\newcommand{\hatcurLBiiStwo}[1]{\ifnum#1=74 %
\hatcurLBiiStwoxxxxxA
\else
\ifnum#1=75 %
\hatcurLBiiStwoxxxxxB
\else
\ifnum#1=76 %
\hatcurLBiiStwoxxxxxC
\else
\ifnum#1=77 %
\hatcurLBiiStwoxxxxxD
\else
??????\fi
\fi
\fi
\fi
}
\newcommand{\hatcurLBiiT}[1]{\ifnum#1=74 %
\hatcurLBiiTxxxxxA
\else
\ifnum#1=75 %
\hatcurLBiiTxxxxxB
\else
\ifnum#1=76 %
\hatcurLBiiTxxxxxC
\else
\ifnum#1=77 %
\hatcurLBiiTxxxxxD
\else
??????\fi
\fi
\fi
\fi
}
\newcommand{\hatcurLBiiu}[1]{\ifnum#1=74 %
\hatcurLBiiuxxxxxA
\else
\ifnum#1=75 %
\hatcurLBiiuxxxxxB
\else
\ifnum#1=76 %
\hatcurLBiiuxxxxxC
\else
\ifnum#1=77 %
\hatcurLBiiuxxxxxD
\else
??????\fi
\fi
\fi
\fi
}
\newcommand{\hatcurLBiiV}[1]{\ifnum#1=74 %
\hatcurLBiiVxxxxxA
\else
\ifnum#1=75 %
\hatcurLBiiVxxxxxB
\else
\ifnum#1=76 %
\hatcurLBiiVxxxxxC
\else
\ifnum#1=77 %
\hatcurLBiiVxxxxxD
\else
??????\fi
\fi
\fi
\fi
}
\newcommand{\hatcurLBiiz}[1]{\ifnum#1=74 %
\hatcurLBiizxxxxxA
\else
\ifnum#1=75 %
\hatcurLBiizxxxxxB
\else
\ifnum#1=76 %
\hatcurLBiizxxxxxC
\else
\ifnum#1=77 %
\hatcurLBiizxxxxxD
\else
??????\fi
\fi
\fi
\fi
}
\newcommand{\hatcurLBiJ}[1]{\ifnum#1=74 %
\hatcurLBiJxxxxxA
\else
\ifnum#1=75 %
\hatcurLBiJxxxxxB
\else
\ifnum#1=76 %
\hatcurLBiJxxxxxC
\else
\ifnum#1=77 %
\hatcurLBiJxxxxxD
\else
??????\fi
\fi
\fi
\fi
}
\newcommand{\hatcurLBiK}[1]{\ifnum#1=74 %
\hatcurLBiKxxxxxA
\else
\ifnum#1=75 %
\hatcurLBiKxxxxxB
\else
\ifnum#1=76 %
\hatcurLBiKxxxxxC
\else
\ifnum#1=77 %
\hatcurLBiKxxxxxD
\else
??????\fi
\fi
\fi
\fi
}
\newcommand{\hatcurLBikep}[1]{\ifnum#1=74 %
\hatcurLBikepxxxxxA
\else
\ifnum#1=75 %
\hatcurLBikepxxxxxB
\else
\ifnum#1=76 %
\hatcurLBikepxxxxxC
\else
\ifnum#1=77 %
\hatcurLBikepxxxxxD
\else
??????\fi
\fi
\fi
\fi
}
\newcommand{\hatcurLBiM}[1]{\ifnum#1=74 %
\hatcurLBiMxxxxxA
\else
\ifnum#1=75 %
\hatcurLBiMxxxxxB
\else
\ifnum#1=76 %
\hatcurLBiMxxxxxC
\else
\ifnum#1=77 %
\hatcurLBiMxxxxxD
\else
??????\fi
\fi
\fi
\fi
}
\newcommand{\hatcurLBir}[1]{\ifnum#1=74 %
\hatcurLBirxxxxxA
\else
\ifnum#1=75 %
\hatcurLBirxxxxxB
\else
\ifnum#1=76 %
\hatcurLBirxxxxxC
\else
\ifnum#1=77 %
\hatcurLBirxxxxxD
\else
??????\fi
\fi
\fi
\fi
}
\newcommand{\hatcurLBiR}[1]{\ifnum#1=74 %
\hatcurLBiRxxxxxA
\else
\ifnum#1=75 %
\hatcurLBiRxxxxxB
\else
\ifnum#1=76 %
\hatcurLBiRxxxxxC
\else
\ifnum#1=77 %
\hatcurLBiRxxxxxD
\else
??????\fi
\fi
\fi
\fi
}
\newcommand{\hatcurLBiSfour}[1]{\ifnum#1=74 %
\hatcurLBiSfourxxxxxA
\else
\ifnum#1=75 %
\hatcurLBiSfourxxxxxB
\else
\ifnum#1=76 %
\hatcurLBiSfourxxxxxC
\else
\ifnum#1=77 %
\hatcurLBiSfourxxxxxD
\else
??????\fi
\fi
\fi
\fi
}
\newcommand{\hatcurLBiSone}[1]{\ifnum#1=74 %
\hatcurLBiSonexxxxxA
\else
\ifnum#1=75 %
\hatcurLBiSonexxxxxB
\else
\ifnum#1=76 %
\hatcurLBiSonexxxxxC
\else
\ifnum#1=77 %
\hatcurLBiSonexxxxxD
\else
??????\fi
\fi
\fi
\fi
}
\newcommand{\hatcurLBiSthree}[1]{\ifnum#1=74 %
\hatcurLBiSthreexxxxxA
\else
\ifnum#1=75 %
\hatcurLBiSthreexxxxxB
\else
\ifnum#1=76 %
\hatcurLBiSthreexxxxxC
\else
\ifnum#1=77 %
\hatcurLBiSthreexxxxxD
\else
??????\fi
\fi
\fi
\fi
}
\newcommand{\hatcurLBiStwo}[1]{\ifnum#1=74 %
\hatcurLBiStwoxxxxxA
\else
\ifnum#1=75 %
\hatcurLBiStwoxxxxxB
\else
\ifnum#1=76 %
\hatcurLBiStwoxxxxxC
\else
\ifnum#1=77 %
\hatcurLBiStwoxxxxxD
\else
??????\fi
\fi
\fi
\fi
}
\newcommand{\hatcurLBiT}[1]{\ifnum#1=74 %
\hatcurLBiTxxxxxA
\else
\ifnum#1=75 %
\hatcurLBiTxxxxxB
\else
\ifnum#1=76 %
\hatcurLBiTxxxxxC
\else
\ifnum#1=77 %
\hatcurLBiTxxxxxD
\else
??????\fi
\fi
\fi
\fi
}
\newcommand{\hatcurLBiu}[1]{\ifnum#1=74 %
\hatcurLBiuxxxxxA
\else
\ifnum#1=75 %
\hatcurLBiuxxxxxB
\else
\ifnum#1=76 %
\hatcurLBiuxxxxxC
\else
\ifnum#1=77 %
\hatcurLBiuxxxxxD
\else
??????\fi
\fi
\fi
\fi
}
\newcommand{\hatcurLBiV}[1]{\ifnum#1=74 %
\hatcurLBiVxxxxxA
\else
\ifnum#1=75 %
\hatcurLBiVxxxxxB
\else
\ifnum#1=76 %
\hatcurLBiVxxxxxC
\else
\ifnum#1=77 %
\hatcurLBiVxxxxxD
\else
??????\fi
\fi
\fi
\fi
}
\newcommand{\hatcurLBiz}[1]{\ifnum#1=74 %
\hatcurLBizxxxxxA
\else
\ifnum#1=75 %
\hatcurLBizxxxxxB
\else
\ifnum#1=76 %
\hatcurLBizxxxxxC
\else
\ifnum#1=77 %
\hatcurLBizxxxxxD
\else
??????\fi
\fi
\fi
\fi
}
\newcommand{\hatcurLCbsq}[1]{\ifnum#1=74 %
\hatcurLCbsqxxxxxA
\else
\ifnum#1=75 %
\hatcurLCbsqxxxxxB
\else
\ifnum#1=76 %
\hatcurLCbsqxxxxxC
\else
\ifnum#1=77 %
\hatcurLCbsqxxxxxD
\else
??????\fi
\fi
\fi
\fi
}
\newcommand{\hatcurLCdip}[1]{\ifnum#1=74 %
\hatcurLCdipxxxxxA
\else
\ifnum#1=75 %
\hatcurLCdipxxxxxB
\else
\ifnum#1=76 %
\hatcurLCdipxxxxxC
\else
\ifnum#1=77 %
\hatcurLCdipxxxxxD
\else
??????\fi
\fi
\fi
\fi
}
\newcommand{\hatcurLCdur}[1]{\ifnum#1=74 %
\hatcurLCdurxxxxxA
\else
\ifnum#1=75 %
\hatcurLCdurxxxxxB
\else
\ifnum#1=76 %
\hatcurLCdurxxxxxC
\else
\ifnum#1=77 %
\hatcurLCdurxxxxxD
\else
??????\fi
\fi
\fi
\fi
}
\newcommand{\hatcurLCdurhr}[1]{\ifnum#1=74 %
\hatcurLCdurhrxxxxxA
\else
\ifnum#1=75 %
\hatcurLCdurhrxxxxxB
\else
\ifnum#1=76 %
\hatcurLCdurhrxxxxxC
\else
\ifnum#1=77 %
\hatcurLCdurhrxxxxxD
\else
??????\fi
\fi
\fi
\fi
}
\newcommand{\hatcurLCdurhrshort}[1]{\ifnum#1=74 %
\hatcurLCdurhrshortxxxxxA
\else
\ifnum#1=75 %
\hatcurLCdurhrshortxxxxxB
\else
\ifnum#1=76 %
\hatcurLCdurhrshortxxxxxC
\else
\ifnum#1=77 %
\hatcurLCdurhrshortxxxxxD
\else
??????\fi
\fi
\fi
\fi
}
\newcommand{\hatcurLCdurshort}[1]{\ifnum#1=74 %
\hatcurLCdurshortxxxxxA
\else
\ifnum#1=75 %
\hatcurLCdurshortxxxxxB
\else
\ifnum#1=76 %
\hatcurLCdurshortxxxxxC
\else
\ifnum#1=77 %
\hatcurLCdurshortxxxxxD
\else
??????\fi
\fi
\fi
\fi
}
\newcommand{\hatcurLChatnetmA}[1]{\ifnum#1=74 %
\hatcurLChatnetmAxxxxxA
\else
\ifnum#1=75 %
\hatcurLChatnetmAxxxxxB
\else
\ifnum#1=76 %
\hatcurLChatnetmAxxxxxC
\else
\ifnum#1=77 %
\hatcurLChatnetmAxxxxxD
\else
??????\fi
\fi
\fi
\fi
}
\newcommand{\hatcurLChatnetmB}[1]{\ifnum#1=74 %
\hatcurLChatnetmBxxxxxA
\else
\ifnum#1=75 %
\hatcurLChatnetmBxxxxxB
\else
\ifnum#1=76 %
\hatcurLChatnetmBxxxxxC
\else
\ifnum#1=77 %
\hatcurLChatnetmBxxxxxD
\else
??????\fi
\fi
\fi
\fi
}
\newcommand{\hatcurLChatnetmC}[1]{\ifnum#1=75 %
\hatcurLChatnetmCxxxxxB
\else
??????\fi
}
\newcommand{\hatcurLChatnetmD}[1]{\ifnum#1=75 %
\hatcurLChatnetmDxxxxxB
\else
??????\fi
}
\newcommand{\hatcurLCiblendA}[1]{\ifnum#1=74 %
\hatcurLCiblendAxxxxxA
\else
\ifnum#1=75 %
\hatcurLCiblendAxxxxxB
\else
\ifnum#1=76 %
\hatcurLCiblendAxxxxxC
\else
\ifnum#1=77 %
\hatcurLCiblendAxxxxxD
\else
??????\fi
\fi
\fi
\fi
}
\newcommand{\hatcurLCiblendB}[1]{\ifnum#1=74 %
\hatcurLCiblendBxxxxxA
\else
\ifnum#1=75 %
\hatcurLCiblendBxxxxxB
\else
\ifnum#1=76 %
\hatcurLCiblendBxxxxxC
\else
\ifnum#1=77 %
\hatcurLCiblendBxxxxxD
\else
??????\fi
\fi
\fi
\fi
}
\newcommand{\hatcurLCiblendC}[1]{\ifnum#1=75 %
\hatcurLCiblendCxxxxxB
\else
??????\fi
}
\newcommand{\hatcurLCiblendD}[1]{\ifnum#1=75 %
\hatcurLCiblendDxxxxxB
\else
??????\fi
}
\newcommand{\hatcurLCimp}[1]{\ifnum#1=74 %
\hatcurLCimpxxxxxA
\else
\ifnum#1=75 %
\hatcurLCimpxxxxxB
\else
\ifnum#1=76 %
\hatcurLCimpxxxxxC
\else
\ifnum#1=77 %
\hatcurLCimpxxxxxD
\else
??????\fi
\fi
\fi
\fi
}
\newcommand{\hatcurLCingdur}[1]{\ifnum#1=74 %
\hatcurLCingdurxxxxxA
\else
\ifnum#1=75 %
\hatcurLCingdurxxxxxB
\else
\ifnum#1=76 %
\hatcurLCingdurxxxxxC
\else
\ifnum#1=77 %
\hatcurLCingdurxxxxxD
\else
??????\fi
\fi
\fi
\fi
}
\newcommand{\hatcurLCP}[1]{\ifnum#1=74 %
\hatcurLCPxxxxxA
\else
\ifnum#1=75 %
\hatcurLCPxxxxxB
\else
\ifnum#1=76 %
\hatcurLCPxxxxxC
\else
\ifnum#1=77 %
\hatcurLCPxxxxxD
\else
??????\fi
\fi
\fi
\fi
}
\newcommand{\hatcurLCPprec}[1]{\ifnum#1=74 %
\hatcurLCPprecxxxxxA
\else
\ifnum#1=75 %
\hatcurLCPprecxxxxxB
\else
\ifnum#1=76 %
\hatcurLCPprecxxxxxC
\else
\ifnum#1=77 %
\hatcurLCPprecxxxxxD
\else
??????\fi
\fi
\fi
\fi
}
\newcommand{\hatcurLCPshort}[1]{\ifnum#1=74 %
\hatcurLCPshortxxxxxA
\else
\ifnum#1=75 %
\hatcurLCPshortxxxxxB
\else
\ifnum#1=76 %
\hatcurLCPshortxxxxxC
\else
\ifnum#1=77 %
\hatcurLCPshortxxxxxD
\else
??????\fi
\fi
\fi
\fi
}
\newcommand{\hatcurLCq}[1]{\ifnum#1=74 %
\hatcurLCqxxxxxA
\else
\ifnum#1=75 %
\hatcurLCqxxxxxB
\else
\ifnum#1=76 %
\hatcurLCqxxxxxC
\else
\ifnum#1=77 %
\hatcurLCqxxxxxD
\else
??????\fi
\fi
\fi
\fi
}
\newcommand{\hatcurLCqshort}[1]{\ifnum#1=74 %
\hatcurLCqshortxxxxxA
\else
\ifnum#1=75 %
\hatcurLCqshortxxxxxB
\else
\ifnum#1=76 %
\hatcurLCqshortxxxxxC
\else
\ifnum#1=77 %
\hatcurLCqshortxxxxxD
\else
??????\fi
\fi
\fi
\fi
}
\newcommand{\hatcurLCrho}[1]{\ifnum#1=74 %
\hatcurLCrhoxxxxxA
\else
\ifnum#1=75 %
\hatcurLCrhoxxxxxB
\else
\ifnum#1=76 %
\hatcurLCrhoxxxxxC
\else
\ifnum#1=77 %
\hatcurLCrhoxxxxxD
\else
??????\fi
\fi
\fi
\fi
}
\newcommand{\hatcurLCrprstar}[1]{\ifnum#1=74 %
\hatcurLCrprstarxxxxxA
\else
\ifnum#1=75 %
\hatcurLCrprstarxxxxxB
\else
\ifnum#1=76 %
\hatcurLCrprstarxxxxxC
\else
\ifnum#1=77 %
\hatcurLCrprstarxxxxxD
\else
??????\fi
\fi
\fi
\fi
}
\newcommand{\hatcurLCT}[1]{\ifnum#1=74 %
\hatcurLCTxxxxxA
\else
\ifnum#1=75 %
\hatcurLCTxxxxxB
\else
\ifnum#1=76 %
\hatcurLCTxxxxxC
\else
\ifnum#1=77 %
\hatcurLCTxxxxxD
\else
??????\fi
\fi
\fi
\fi
}
\newcommand{\hatcurLCTA}[1]{\ifnum#1=74 %
\hatcurLCTAxxxxxA
\else
\ifnum#1=75 %
\hatcurLCTAxxxxxB
\else
\ifnum#1=76 %
\hatcurLCTAxxxxxC
\else
\ifnum#1=77 %
\hatcurLCTAxxxxxD
\else
??????\fi
\fi
\fi
\fi
}
\newcommand{\hatcurLCTB}[1]{\ifnum#1=74 %
\hatcurLCTBxxxxxA
\else
\ifnum#1=75 %
\hatcurLCTBxxxxxB
\else
\ifnum#1=76 %
\hatcurLCTBxxxxxC
\else
\ifnum#1=77 %
\hatcurLCTBxxxxxD
\else
??????\fi
\fi
\fi
\fi
}
\newcommand{\hatcurLCzeta}[1]{\ifnum#1=74 %
\hatcurLCzetaxxxxxA
\else
\ifnum#1=75 %
\hatcurLCzetaxxxxxB
\else
\ifnum#1=76 %
\hatcurLCzetaxxxxxC
\else
\ifnum#1=77 %
\hatcurLCzetaxxxxxD
\else
??????\fi
\fi
\fi
\fi
}
\newcommand{\hatcurPPaequiv}[1]{\ifnum#1=74 %
\hatcurPPaequivxxxxxA
\else
\ifnum#1=75 %
\hatcurPPaequivxxxxxB
\else
\ifnum#1=76 %
\hatcurPPaequivxxxxxC
\else
\ifnum#1=77 %
\hatcurPPaequivxxxxxD
\else
??????\fi
\fi
\fi
\fi
}
\newcommand{\hatcurPPar}[1]{\ifnum#1=74 %
\hatcurPParxxxxxA
\else
\ifnum#1=75 %
\hatcurPParxxxxxB
\else
\ifnum#1=76 %
\hatcurPParxxxxxC
\else
\ifnum#1=77 %
\hatcurPParxxxxxD
\else
??????\fi
\fi
\fi
\fi
}
\newcommand{\hatcurPParel}[1]{\ifnum#1=74 %
\hatcurPParelxxxxxA
\else
\ifnum#1=75 %
\hatcurPParelxxxxxB
\else
\ifnum#1=76 %
\hatcurPParelxxxxxC
\else
\ifnum#1=77 %
\hatcurPParelxxxxxD
\else
??????\fi
\fi
\fi
\fi
}
\newcommand{\hatcurPPfluxap}[1]{\ifnum#1=74 %
\hatcurPPfluxapxxxxxA
\else
\ifnum#1=75 %
\hatcurPPfluxapxxxxxB
\else
\ifnum#1=76 %
\hatcurPPfluxapxxxxxC
\else
\ifnum#1=77 %
\hatcurPPfluxapxxxxxD
\else
??????\fi
\fi
\fi
\fi
}
\newcommand{\hatcurPPfluxapdim}[1]{\ifnum#1=74 %
\hatcurPPfluxapdimxxxxxA
\else
\ifnum#1=75 %
\hatcurPPfluxapdimxxxxxB
\else
\ifnum#1=76 %
\hatcurPPfluxapdimxxxxxC
\else
\ifnum#1=77 %
\hatcurPPfluxapdimxxxxxD
\else
??????\fi
\fi
\fi
\fi
}
\newcommand{\hatcurPPfluxavg}[1]{\ifnum#1=74 %
\hatcurPPfluxavgxxxxxA
\else
\ifnum#1=75 %
\hatcurPPfluxavgxxxxxB
\else
\ifnum#1=76 %
\hatcurPPfluxavgxxxxxC
\else
\ifnum#1=77 %
\hatcurPPfluxavgxxxxxD
\else
??????\fi
\fi
\fi
\fi
}
\newcommand{\hatcurPPfluxavgdim}[1]{\ifnum#1=74 %
\hatcurPPfluxavgdimxxxxxA
\else
\ifnum#1=75 %
\hatcurPPfluxavgdimxxxxxB
\else
\ifnum#1=76 %
\hatcurPPfluxavgdimxxxxxC
\else
\ifnum#1=77 %
\hatcurPPfluxavgdimxxxxxD
\else
??????\fi
\fi
\fi
\fi
}
\newcommand{\hatcurPPfluxavglog}[1]{\ifnum#1=74 %
\hatcurPPfluxavglogxxxxxA
\else
\ifnum#1=75 %
\hatcurPPfluxavglogxxxxxB
\else
\ifnum#1=76 %
\hatcurPPfluxavglogxxxxxC
\else
\ifnum#1=77 %
\hatcurPPfluxavglogxxxxxD
\else
??????\fi
\fi
\fi
\fi
}
\newcommand{\hatcurPPfluxperi}[1]{\ifnum#1=74 %
\hatcurPPfluxperixxxxxA
\else
\ifnum#1=75 %
\hatcurPPfluxperixxxxxB
\else
\ifnum#1=76 %
\hatcurPPfluxperixxxxxC
\else
\ifnum#1=77 %
\hatcurPPfluxperixxxxxD
\else
??????\fi
\fi
\fi
\fi
}
\newcommand{\hatcurPPfluxperidim}[1]{\ifnum#1=74 %
\hatcurPPfluxperidimxxxxxA
\else
\ifnum#1=75 %
\hatcurPPfluxperidimxxxxxB
\else
\ifnum#1=76 %
\hatcurPPfluxperidimxxxxxC
\else
\ifnum#1=77 %
\hatcurPPfluxperidimxxxxxD
\else
??????\fi
\fi
\fi
\fi
}
\newcommand{\hatcurPPg}[1]{\ifnum#1=74 %
\hatcurPPgxxxxxA
\else
\ifnum#1=75 %
\hatcurPPgxxxxxB
\else
\ifnum#1=76 %
\hatcurPPgxxxxxC
\else
\ifnum#1=77 %
\hatcurPPgxxxxxD
\else
??????\fi
\fi
\fi
\fi
}
\newcommand{\hatcurPPi}[1]{\ifnum#1=74 %
\hatcurPPixxxxxA
\else
\ifnum#1=75 %
\hatcurPPixxxxxB
\else
\ifnum#1=76 %
\hatcurPPixxxxxC
\else
\ifnum#1=77 %
\hatcurPPixxxxxD
\else
??????\fi
\fi
\fi
\fi
}
\newcommand{\hatcurPPlogg}[1]{\ifnum#1=74 %
\hatcurPPloggxxxxxA
\else
\ifnum#1=75 %
\hatcurPPloggxxxxxB
\else
\ifnum#1=76 %
\hatcurPPloggxxxxxC
\else
\ifnum#1=77 %
\hatcurPPloggxxxxxD
\else
??????\fi
\fi
\fi
\fi
}
\newcommand{\hatcurPPm}[1]{\ifnum#1=74 %
\hatcurPPmxxxxxA
\else
\ifnum#1=75 %
\hatcurPPmxxxxxB
\else
\ifnum#1=76 %
\hatcurPPmxxxxxC
\else
\ifnum#1=77 %
\hatcurPPmxxxxxD
\else
??????\fi
\fi
\fi
\fi
}
\newcommand{\hatcurPPme}[1]{\ifnum#1=74 %
\hatcurPPmexxxxxA
\else
\ifnum#1=75 %
\hatcurPPmexxxxxB
\else
\ifnum#1=76 %
\hatcurPPmexxxxxC
\else
\ifnum#1=77 %
\hatcurPPmexxxxxD
\else
??????\fi
\fi
\fi
\fi
}
\newcommand{\hatcurPPmelong}[1]{\ifnum#1=74 %
\hatcurPPmelongxxxxxA
\else
\ifnum#1=75 %
\hatcurPPmelongxxxxxB
\else
\ifnum#1=76 %
\hatcurPPmelongxxxxxC
\else
\ifnum#1=77 %
\hatcurPPmelongxxxxxD
\else
??????\fi
\fi
\fi
\fi
}
\newcommand{\hatcurPPmeshort}[1]{\ifnum#1=74 %
\hatcurPPmeshortxxxxxA
\else
\ifnum#1=75 %
\hatcurPPmeshortxxxxxB
\else
\ifnum#1=76 %
\hatcurPPmeshortxxxxxC
\else
\ifnum#1=77 %
\hatcurPPmeshortxxxxxD
\else
??????\fi
\fi
\fi
\fi
}
\newcommand{\hatcurPPmlong}[1]{\ifnum#1=74 %
\hatcurPPmlongxxxxxA
\else
\ifnum#1=75 %
\hatcurPPmlongxxxxxB
\else
\ifnum#1=76 %
\hatcurPPmlongxxxxxC
\else
\ifnum#1=77 %
\hatcurPPmlongxxxxxD
\else
??????\fi
\fi
\fi
\fi
}
\newcommand{\hatcurPPmrcorr}[1]{\ifnum#1=74 %
\hatcurPPmrcorrxxxxxA
\else
\ifnum#1=75 %
\hatcurPPmrcorrxxxxxB
\else
\ifnum#1=76 %
\hatcurPPmrcorrxxxxxC
\else
\ifnum#1=77 %
\hatcurPPmrcorrxxxxxD
\else
??????\fi
\fi
\fi
\fi
}
\newcommand{\hatcurPPmshort}[1]{\ifnum#1=74 %
\hatcurPPmshortxxxxxA
\else
\ifnum#1=75 %
\hatcurPPmshortxxxxxB
\else
\ifnum#1=76 %
\hatcurPPmshortxxxxxC
\else
\ifnum#1=77 %
\hatcurPPmshortxxxxxD
\else
??????\fi
\fi
\fi
\fi
}
\newcommand{\hatcurPPmtwosiglim}[1]{\ifnum#1=74 %
\hatcurPPmtwosiglimxxxxxA
\else
\ifnum#1=75 %
\hatcurPPmtwosiglimxxxxxB
\else
\ifnum#1=76 %
\hatcurPPmtwosiglimxxxxxC
\else
\ifnum#1=77 %
\hatcurPPmtwosiglimxxxxxD
\else
??????\fi
\fi
\fi
\fi
}
\newcommand{\hatcurPPperi}[1]{\ifnum#1=74 %
\hatcurPPperixxxxxA
\else
\ifnum#1=75 %
\hatcurPPperixxxxxB
\else
\ifnum#1=76 %
\hatcurPPperixxxxxC
\else
\ifnum#1=77 %
\hatcurPPperixxxxxD
\else
??????\fi
\fi
\fi
\fi
}
\newcommand{\hatcurPPphiconj}[1]{\ifnum#1=74 %
\hatcurPPphiconjxxxxxA
\else
\ifnum#1=75 %
\hatcurPPphiconjxxxxxB
\else
\ifnum#1=76 %
\hatcurPPphiconjxxxxxC
\else
\ifnum#1=77 %
\hatcurPPphiconjxxxxxD
\else
??????\fi
\fi
\fi
\fi
}
\newcommand{\hatcurPPr}[1]{\ifnum#1=74 %
\hatcurPPrxxxxxA
\else
\ifnum#1=75 %
\hatcurPPrxxxxxB
\else
\ifnum#1=76 %
\hatcurPPrxxxxxC
\else
\ifnum#1=77 %
\hatcurPPrxxxxxD
\else
??????\fi
\fi
\fi
\fi
}
\newcommand{\hatcurPPre}[1]{\ifnum#1=74 %
\hatcurPPrexxxxxA
\else
\ifnum#1=75 %
\hatcurPPrexxxxxB
\else
\ifnum#1=76 %
\hatcurPPrexxxxxC
\else
\ifnum#1=77 %
\hatcurPPrexxxxxD
\else
??????\fi
\fi
\fi
\fi
}
\newcommand{\hatcurPPrelong}[1]{\ifnum#1=74 %
\hatcurPPrelongxxxxxA
\else
\ifnum#1=75 %
\hatcurPPrelongxxxxxB
\else
\ifnum#1=76 %
\hatcurPPrelongxxxxxC
\else
\ifnum#1=77 %
\hatcurPPrelongxxxxxD
\else
??????\fi
\fi
\fi
\fi
}
\newcommand{\hatcurPPreshort}[1]{\ifnum#1=74 %
\hatcurPPreshortxxxxxA
\else
\ifnum#1=75 %
\hatcurPPreshortxxxxxB
\else
\ifnum#1=76 %
\hatcurPPreshortxxxxxC
\else
\ifnum#1=77 %
\hatcurPPreshortxxxxxD
\else
??????\fi
\fi
\fi
\fi
}
\newcommand{\hatcurPPrho}[1]{\ifnum#1=74 %
\hatcurPPrhoxxxxxA
\else
\ifnum#1=75 %
\hatcurPPrhoxxxxxB
\else
\ifnum#1=76 %
\hatcurPPrhoxxxxxC
\else
\ifnum#1=77 %
\hatcurPPrhoxxxxxD
\else
??????\fi
\fi
\fi
\fi
}
\newcommand{\hatcurPPrlong}[1]{\ifnum#1=74 %
\hatcurPPrlongxxxxxA
\else
\ifnum#1=75 %
\hatcurPPrlongxxxxxB
\else
\ifnum#1=76 %
\hatcurPPrlongxxxxxC
\else
\ifnum#1=77 %
\hatcurPPrlongxxxxxD
\else
??????\fi
\fi
\fi
\fi
}
\newcommand{\hatcurPPrshort}[1]{\ifnum#1=74 %
\hatcurPPrshortxxxxxA
\else
\ifnum#1=75 %
\hatcurPPrshortxxxxxB
\else
\ifnum#1=76 %
\hatcurPPrshortxxxxxC
\else
\ifnum#1=77 %
\hatcurPPrshortxxxxxD
\else
??????\fi
\fi
\fi
\fi
}
\newcommand{\hatcurPPtcirc}[1]{\ifnum#1=74 %
\hatcurPPtcircxxxxxA
\else
\ifnum#1=75 %
\hatcurPPtcircxxxxxB
\else
\ifnum#1=76 %
\hatcurPPtcircxxxxxC
\else
\ifnum#1=77 %
\hatcurPPtcircxxxxxD
\else
??????\fi
\fi
\fi
\fi
}
\newcommand{\hatcurPPteff}[1]{\ifnum#1=74 %
\hatcurPPteffxxxxxA
\else
\ifnum#1=75 %
\hatcurPPteffxxxxxB
\else
\ifnum#1=76 %
\hatcurPPteffxxxxxC
\else
\ifnum#1=77 %
\hatcurPPteffxxxxxD
\else
??????\fi
\fi
\fi
\fi
}
\newcommand{\hatcurPPtheta}[1]{\ifnum#1=74 %
\hatcurPPthetaxxxxxA
\else
\ifnum#1=75 %
\hatcurPPthetaxxxxxB
\else
\ifnum#1=76 %
\hatcurPPthetaxxxxxC
\else
\ifnum#1=77 %
\hatcurPPthetaxxxxxD
\else
??????\fi
\fi
\fi
\fi
}
\newcommand{\hatcurPPtinfall}[1]{\ifnum#1=74 %
\hatcurPPtinfallxxxxxA
\else
\ifnum#1=75 %
\hatcurPPtinfallxxxxxB
\else
\ifnum#1=76 %
\hatcurPPtinfallxxxxxC
\else
\ifnum#1=77 %
\hatcurPPtinfallxxxxxD
\else
??????\fi
\fi
\fi
\fi
}
\newcommand{\hatcurRVeccen}[1]{\ifnum#1=74 %
\hatcurRVeccenxxxxxA
\else
\ifnum#1=75 %
\hatcurRVeccenxxxxxB
\else
\ifnum#1=76 %
\hatcurRVeccenxxxxxC
\else
\ifnum#1=77 %
\hatcurRVeccenxxxxxD
\else
??????\fi
\fi
\fi
\fi
}
\newcommand{\hatcurRVeccentwosiglim}[1]{\ifnum#1=74 %
\hatcurRVeccentwosiglimxxxxxA
\else
\ifnum#1=75 %
\hatcurRVeccentwosiglimxxxxxB
\else
\ifnum#1=76 %
\hatcurRVeccentwosiglimxxxxxC
\else
\ifnum#1=77 %
\hatcurRVeccentwosiglimxxxxxD
\else
??????\fi
\fi
\fi
\fi
}
\newcommand{\hatcurRVfitrms}[1]{\ifnum#1=74 %
\hatcurRVfitrmsxxxxxA
\else
\ifnum#1=75 %
\hatcurRVfitrmsxxxxxB
\else
\ifnum#1=76 %
\hatcurRVfitrmsxxxxxC
\else
\ifnum#1=77 %
\hatcurRVfitrmsxxxxxD
\else
??????\fi
\fi
\fi
\fi
}
\newcommand{\hatcurRVgamma}[1]{\ifnum#1=74 %
\hatcurRVgammaxxxxxA
\else
\ifnum#1=75 %
\hatcurRVgammaxxxxxB
\else
\ifnum#1=76 %
\hatcurRVgammaxxxxxC
\else
\ifnum#1=77 %
\hatcurRVgammaxxxxxD
\else
??????\fi
\fi
\fi
\fi
}
\newcommand{\hatcurRVh}[1]{\ifnum#1=74 %
\hatcurRVhxxxxxA
\else
\ifnum#1=75 %
\hatcurRVhxxxxxB
\else
\ifnum#1=76 %
\hatcurRVhxxxxxC
\else
\ifnum#1=77 %
\hatcurRVhxxxxxD
\else
??????\fi
\fi
\fi
\fi
}
\newcommand{\hatcurRVjitter}[1]{\ifnum#1=74 %
\hatcurRVjitterxxxxxA
\else
\ifnum#1=75 %
\hatcurRVjitterxxxxxB
\else
\ifnum#1=76 %
\hatcurRVjitterxxxxxC
\else
\ifnum#1=77 %
\hatcurRVjitterxxxxxD
\else
??????\fi
\fi
\fi
\fi
}
\newcommand{\hatcurRVjittertwosiglim}[1]{\ifnum#1=74 %
\hatcurRVjittertwosiglimxxxxxA
\else
\ifnum#1=75 %
\hatcurRVjittertwosiglimxxxxxB
\else
\ifnum#1=76 %
\hatcurRVjittertwosiglimxxxxxC
\else
\ifnum#1=77 %
\hatcurRVjittertwosiglimxxxxxD
\else
??????\fi
\fi
\fi
\fi
}
\newcommand{\hatcurRVk}[1]{\ifnum#1=74 %
\hatcurRVkxxxxxA
\else
\ifnum#1=75 %
\hatcurRVkxxxxxB
\else
\ifnum#1=76 %
\hatcurRVkxxxxxC
\else
\ifnum#1=77 %
\hatcurRVkxxxxxD
\else
??????\fi
\fi
\fi
\fi
}
\newcommand{\hatcurRVK}[1]{\ifnum#1=74 %
\hatcurRVKxxxxxA
\else
\ifnum#1=75 %
\hatcurRVKxxxxxB
\else
\ifnum#1=76 %
\hatcurRVKxxxxxC
\else
\ifnum#1=77 %
\hatcurRVKxxxxxD
\else
??????\fi
\fi
\fi
\fi
}
\newcommand{\hatcurRVKtwosiglim}[1]{\ifnum#1=74 %
\hatcurRVKtwosiglimxxxxxA
\else
\ifnum#1=75 %
\hatcurRVKtwosiglimxxxxxB
\else
\ifnum#1=76 %
\hatcurRVKtwosiglimxxxxxC
\else
\ifnum#1=77 %
\hatcurRVKtwosiglimxxxxxD
\else
??????\fi
\fi
\fi
\fi
}
\newcommand{\hatcurRVomega}[1]{\ifnum#1=74 %
\hatcurRVomegaxxxxxA
\else
\ifnum#1=75 %
\hatcurRVomegaxxxxxB
\else
\ifnum#1=76 %
\hatcurRVomegaxxxxxC
\else
\ifnum#1=77 %
\hatcurRVomegaxxxxxD
\else
??????\fi
\fi
\fi
\fi
}
\newcommand{\hatcurRVrh}[1]{\ifnum#1=74 %
\hatcurRVrhxxxxxA
\else
\ifnum#1=75 %
\hatcurRVrhxxxxxB
\else
\ifnum#1=76 %
\hatcurRVrhxxxxxC
\else
\ifnum#1=77 %
\hatcurRVrhxxxxxD
\else
??????\fi
\fi
\fi
\fi
}
\newcommand{\hatcurRVrk}[1]{\ifnum#1=74 %
\hatcurRVrkxxxxxA
\else
\ifnum#1=75 %
\hatcurRVrkxxxxxB
\else
\ifnum#1=76 %
\hatcurRVrkxxxxxC
\else
\ifnum#1=77 %
\hatcurRVrkxxxxxD
\else
??????\fi
\fi
\fi
\fi
}
\newcommand{\hatcurRVtrone}[1]{\ifnum#1=74 %
\hatcurRVtronexxxxxA
\else
\ifnum#1=75 %
\hatcurRVtronexxxxxB
\else
\ifnum#1=76 %
\hatcurRVtronexxxxxC
\else
\ifnum#1=77 %
\hatcurRVtronexxxxxD
\else
??????\fi
\fi
\fi
\fi
}
\newcommand{\hatcurRVtrtwo}[1]{\ifnum#1=74 %
\hatcurRVtrtwoxxxxxA
\else
\ifnum#1=75 %
\hatcurRVtrtwoxxxxxB
\else
\ifnum#1=76 %
\hatcurRVtrtwoxxxxxC
\else
\ifnum#1=77 %
\hatcurRVtrtwoxxxxxD
\else
??????\fi
\fi
\fi
\fi
}
\newcommand{\hatcurSMEilogg}[1]{\ifnum#1=74 %
\hatcurSMEiloggxxxxxA
\else
\ifnum#1=75 %
\hatcurSMEiloggxxxxxB
\else
\ifnum#1=76 %
\hatcurSMEiloggxxxxxC
\else
\ifnum#1=77 %
\hatcurSMEiloggxxxxxD
\else
??????\fi
\fi
\fi
\fi
}
\newcommand{\hatcurSMEiteff}[1]{\ifnum#1=74 %
\hatcurSMEiteffxxxxxA
\else
\ifnum#1=75 %
\hatcurSMEiteffxxxxxB
\else
\ifnum#1=76 %
\hatcurSMEiteffxxxxxC
\else
\ifnum#1=77 %
\hatcurSMEiteffxxxxxD
\else
??????\fi
\fi
\fi
\fi
}
\newcommand{\hatcurSMEivmac}[1]{\ifnum#1=74 %
\hatcurSMEivmacxxxxxA
\else
\ifnum#1=75 %
\hatcurSMEivmacxxxxxB
\else
\ifnum#1=76 %
\hatcurSMEivmacxxxxxC
\else
\ifnum#1=77 %
\hatcurSMEivmacxxxxxD
\else
??????\fi
\fi
\fi
\fi
}
\newcommand{\hatcurSMEivmic}[1]{\ifnum#1=74 %
\hatcurSMEivmicxxxxxA
\else
\ifnum#1=75 %
\hatcurSMEivmicxxxxxB
\else
\ifnum#1=76 %
\hatcurSMEivmicxxxxxC
\else
\ifnum#1=77 %
\hatcurSMEivmicxxxxxD
\else
??????\fi
\fi
\fi
\fi
}
\newcommand{\hatcurSMEivsin}[1]{\ifnum#1=74 %
\hatcurSMEivsinxxxxxA
\else
\ifnum#1=75 %
\hatcurSMEivsinxxxxxB
\else
\ifnum#1=76 %
\hatcurSMEivsinxxxxxC
\else
\ifnum#1=77 %
\hatcurSMEivsinxxxxxD
\else
??????\fi
\fi
\fi
\fi
}
\newcommand{\hatcurSMEizfeh}[1]{\ifnum#1=74 %
\hatcurSMEizfehxxxxxA
\else
\ifnum#1=75 %
\hatcurSMEizfehxxxxxB
\else
\ifnum#1=76 %
\hatcurSMEizfehxxxxxC
\else
\ifnum#1=77 %
\hatcurSMEizfehxxxxxD
\else
??????\fi
\fi
\fi
\fi
}
\newcommand{\hatcurSMEizfehshort}[1]{\ifnum#1=74 %
\hatcurSMEizfehshortxxxxxA
\else
\ifnum#1=75 %
\hatcurSMEizfehshortxxxxxB
\else
\ifnum#1=76 %
\hatcurSMEizfehshortxxxxxC
\else
\ifnum#1=77 %
\hatcurSMEizfehshortxxxxxD
\else
??????\fi
\fi
\fi
\fi
}
\newcommand{\hatcurXAv}[1]{\ifnum#1=74 %
\hatcurXAvxxxxxA
\else
\ifnum#1=75 %
\hatcurXAvxxxxxB
\else
\ifnum#1=76 %
\hatcurXAvxxxxxC
\else
\ifnum#1=77 %
\hatcurXAvxxxxxD
\else
??????\fi
\fi
\fi
\fi
}
\newcommand{\hatcurXdist}[1]{\ifnum#1=74 %
\hatcurXdistxxxxxA
\else
\ifnum#1=75 %
\hatcurXdistxxxxxB
\else
\ifnum#1=76 %
\hatcurXdistxxxxxC
\else
\ifnum#1=77 %
\hatcurXdistxxxxxD
\else
??????\fi
\fi
\fi
\fi
}
\newcommand{\hatcurXdistred}[1]{\ifnum#1=74 %
\hatcurXdistredxxxxxA
\else
\ifnum#1=75 %
\hatcurXdistredxxxxxB
\else
\ifnum#1=76 %
\hatcurXdistredxxxxxC
\else
\ifnum#1=77 %
\hatcurXdistredxxxxxD
\else
??????\fi
\fi
\fi
\fi
}
\newcommand{\hatcurXEBV}[1]{\ifnum#1=74 %
\hatcurXEBVxxxxxA
\else
\ifnum#1=75 %
\hatcurXEBVxxxxxB
\else
\ifnum#1=76 %
\hatcurXEBVxxxxxC
\else
\ifnum#1=77 %
\hatcurXEBVxxxxxD
\else
??????\fi
\fi
\fi
\fi
}
\newcommand{\hatcurXsecdur}[1]{\ifnum#1=74 %
\hatcurXsecdurxxxxxA
\else
\ifnum#1=75 %
\hatcurXsecdurxxxxxB
\else
\ifnum#1=76 %
\hatcurXsecdurxxxxxC
\else
\ifnum#1=77 %
\hatcurXsecdurxxxxxD
\else
??????\fi
\fi
\fi
\fi
}
\newcommand{\hatcurXsecingdur}[1]{\ifnum#1=74 %
\hatcurXsecingdurxxxxxA
\else
\ifnum#1=75 %
\hatcurXsecingdurxxxxxB
\else
\ifnum#1=76 %
\hatcurXsecingdurxxxxxC
\else
\ifnum#1=77 %
\hatcurXsecingdurxxxxxD
\else
??????\fi
\fi
\fi
\fi
}
\newcommand{\hatcurXsecondary}[1]{\ifnum#1=74 %
\hatcurXsecondaryxxxxxA
\else
\ifnum#1=75 %
\hatcurXsecondaryxxxxxB
\else
\ifnum#1=76 %
\hatcurXsecondaryxxxxxC
\else
\ifnum#1=77 %
\hatcurXsecondaryxxxxxD
\else
??????\fi
\fi
\fi
\fi
}
\newcommand{\hatcurXsecphase}[1]{\ifnum#1=74 %
\hatcurXsecphasexxxxxA
\else
\ifnum#1=75 %
\hatcurXsecphasexxxxxB
\else
\ifnum#1=76 %
\hatcurXsecphasexxxxxC
\else
\ifnum#1=77 %
\hatcurXsecphasexxxxxD
\else
??????\fi
\fi
\fi
\fi
}

\newcommand{\hatcurhtreccenxxxxxA}{HATS563-025}                      
\newcommand{\hatcurfieldeccenxxxxxA}{\ensuremath{string}}            
\newcommand{\hatcurCCraeccenxxxxxA}{\ensuremath{11^{\mathrm h}24^{\mathrm m}03.5929{\mathrm s}}}                   
\newcommand{\hatcurCCdececcenxxxxxA}{\ensuremath{-19{\arcdeg}33{\arcmin}25.6653{\arcsec}}}                 
\newcommand{\hatcurCCmageccenxxxxxA}{NULL}                           
\newcommand{\hatcurCCtwomasseccenxxxxxA}{2MASS~11240360-1933257}     
\newcommand{\hatcurCCgsceccenxxxxxA}{GSC~NULL}                       
\newcommand{\hatcurCCgaiaeccenxxxxxA}{GAIA~3545653561940141696}      
\newcommand{\hatcurCCgaiadrtwoeccenxxxxxA}{GAIA~DR2~3545653561942122368} 
\newcommand{\hatcurCCtassmveccenxxxxxA}{\ensuremath{nff\pmnff}}      
\newcommand{\hatcurCCtassmvshorteccenxxxxxA}{\ensuremath{0.0}}       
\newcommand{\hatcurCCtassmBeccenxxxxxA}{\ensuremath{nff\pmnff}}      
\newcommand{\hatcurCCtassmBshorteccenxxxxxA}{\ensuremath{0.0}}       
\newcommand{\hatcurCCtassmIeccenxxxxxA}{\ensuremath{nff\pmnff}}      
\newcommand{\hatcurCCtassmIshorteccenxxxxxA}{\ensuremath{0.0}}       
\newcommand{\hatcurCCtassmgeccenxxxxxA}{\ensuremath{nff\pmnff}}      
\newcommand{\hatcurCCtassmgshorteccenxxxxxA}{\ensuremath{0.0}}       
\newcommand{\hatcurCCtassmreccenxxxxxA}{\ensuremath{nff\pmnff}}      
\newcommand{\hatcurCCtassmrshorteccenxxxxxA}{\ensuremath{0.0}}       
\newcommand{\hatcurCCtassmieccenxxxxxA}{\ensuremath{nff\pmnff}}      
\newcommand{\hatcurCCtassmishorteccenxxxxxA}{\ensuremath{0.0}}       
\newcommand{\hatcurCCparallaxeccenxxxxxA}{\ensuremath{3.425\pm0.042}} 
\newcommand{\hatcurCCgaiamGeccenxxxxxA}{\ensuremath{15.9706\pm0.0029}} 
\newcommand{\hatcurCCgaiamBPeccenxxxxxA}{\ensuremath{0\pm0}}         
\newcommand{\hatcurCCgaiamRPeccenxxxxxA}{\ensuremath{0\pm0}}         
\newcommand{\hatcurCCtwomassJmageccenxxxxxA}{\ensuremath{13.341\pm0.023}} 
\newcommand{\hatcurCCtwomassHmageccenxxxxxA}{\ensuremath{12.687\pm0.029}} 
\newcommand{\hatcurCCtwomassKmageccenxxxxxA}{\ensuremath{12.452\pm0.031}} 
\newcommand{\hatcurCCcitJmageccenxxxxxA}{\ensuremath{13.239\pm0.025}} 
\newcommand{\hatcurCCcitHmageccenxxxxxA}{\ensuremath{12.592\pm0.030}} 
\newcommand{\hatcurCCcitKmageccenxxxxxA}{\ensuremath{12.385\pm0.032}} 
\newcommand{\hatcurCCbbJmageccenxxxxxA}{\ensuremath{13.331\pm0.027}} 
\newcommand{\hatcurCCbbHmageccenxxxxxA}{\ensuremath{12.619\pm0.031}} 
\newcommand{\hatcurCCbbKmageccenxxxxxA}{\ensuremath{12.405\pm0.032}} 
\newcommand{\hatcurCCesoJmageccenxxxxxA}{\ensuremath{13.340\pm0.032}} 
\newcommand{\hatcurCCesoHmageccenxxxxxA}{\ensuremath{12.619\pm0.046}} 
\newcommand{\hatcurCCesoKmageccenxxxxxA}{\ensuremath{12.401\pm0.033}} 
\newcommand{\hatcurCCesoJHmageccenxxxxxA}{\ensuremath{0.720\pm0.052}} 
\newcommand{\hatcurCCesoJKmageccenxxxxxA}{\ensuremath{0.939\pm0.044}} 
\newcommand{\hatcurCCesoHKmageccenxxxxxA}{\ensuremath{0.218\pm0.056}} 
\newcommand{\hatcurCCWonemageccenxxxxxA}{\ensuremath{12.342\pm0.023}} 
\newcommand{\hatcurCCWtwomageccenxxxxxA}{\ensuremath{12.326\pm0.023}} 
\newcommand{\hatcurCCWthreemageccenxxxxxA}{\ensuremath{nff\pmnff}}   
\newcommand{\hatcurCCWfourmageccenxxxxxA}{\ensuremath{nff\pmnff}}    
\newcommand{\hatcurLCdipeccenxxxxxA}{\ensuremath{45.4}}              
\newcommand{\hatcurLCrprstareccenxxxxxA}{\ensuremath{0.1843\pm0.0030}} 
\newcommand{\hatcurLCbsqeccenxxxxxA}{\ensuremath{0.147_{-0.041}^{+0.040}}} 
\newcommand{\hatcurLCimpeccenxxxxxA}{\ensuremath{0.384_{-0.058}^{+0.050}}} 
\newcommand{\hatcurLCzetaeccenxxxxxA}{\ensuremath{35.28\pm0.51}}     
\newcommand{\hatcurLCdureccenxxxxxA}{\ensuremath{0.06879\pm0.00073}} 
\newcommand{\hatcurLCdurshorteccenxxxxxA}{\ensuremath{0.0688}}       
\newcommand{\hatcurLCdurhreccenxxxxxA}{\ensuremath{1.651\pm0.018}}   
\newcommand{\hatcurLCdurhrshorteccenxxxxxA}{\ensuremath{1.651}}      
\newcommand{\hatcurLCqeccenxxxxxA}{\ensuremath{0.03970\pm0.00042}}   
\newcommand{\hatcurLCqshorteccenxxxxxA}{\ensuremath{0.040}}          
\newcommand{\hatcurLCingdureccenxxxxxA}{\ensuremath{0.01233\pm0.00060}} 
\newcommand{\hatcurLCPeccenxxxxxA}{\ensuremath{1.73185597\pm0.00000047}} 
\newcommand{\hatcurLCPprececcenxxxxxA}{\ensuremath{1.7318560}}       
\newcommand{\hatcurLCPshorteccenxxxxxA}{\ensuremath{1.7319}}         
\newcommand{\hatcurLCTeccenxxxxxA}{\ensuremath{2458376.43989\pm0.00025}} 
\newcommand{\hatcurLCTAeccenxxxxxA}{\ensuremath{2455217.53456\pm0.00096}} 
\newcommand{\hatcurLCTBeccenxxxxxA}{\ensuremath{2458566.94404\pm0.00024}} 
\newcommand{\hatcurLChatnetmAeccenxxxxxA}{\ensuremath{15.17174\pm0.00042}} 
\newcommand{\hatcurLCiblendAeccenxxxxxA}{\ensuremath{0.947\pm0.047}} 
\newcommand{\hatcurLChatnetmBeccenxxxxxA}{\ensuremath{19.26565\pm0.00020}} 
\newcommand{\hatcurLCiblendBeccenxxxxxA}{\ensuremath{0.962\pm0.032}} 
\newcommand{\hatcurLCrhoeccenxxxxxA}{\ensuremath{4.48\pm0.13}}       
\newcommand{\hatcurSMEiteffeccenxxxxxA}{\ensuremath{3775\pm54}}      
\newcommand{\hatcurSMEizfeheccenxxxxxA}{\ensuremath{0.294\pm0.088}}  
\newcommand{\hatcurSMEizfehshorteccenxxxxxA}{\ensuremath{0.29}}      
\newcommand{\hatcurSMEiloggeccenxxxxxA}{\ensuremath{4.50\pm0.50}}    
\newcommand{\hatcurSMEivsineccenxxxxxA}{\ensuremath{0\pm50}}         
\newcommand{\hatcurSMEivmaceccenxxxxxA}{\ensuremath{nff\pmnff}}      
\newcommand{\hatcurSMEivmiceccenxxxxxA}{\ensuremath{nff\pmnff}}      
\newcommand{\hatcurextraerrMJeccenxxxxxA}{\ensuremath{0\pm0}}        
\newcommand{\hatcurextraerrMJtwosiglimeccenxxxxxA}{\ensuremath{<0.0200}} 
\newcommand{\hatcurextraerrMHeccenxxxxxA}{\ensuremath{0\pm0}}        
\newcommand{\hatcurextraerrMHtwosiglimeccenxxxxxA}{\ensuremath{<0.0200}} 
\newcommand{\hatcurextraerrMKseccenxxxxxA}{\ensuremath{0\pm0}}       
\newcommand{\hatcurextraerrMKstwosiglimeccenxxxxxA}{\ensuremath{<0.0200}} 
\newcommand{\hatcurextraerrMGeccenxxxxxA}{\ensuremath{0\pm0}}        
\newcommand{\hatcurextraerrMGtwosiglimeccenxxxxxA}{\ensuremath{<0.0200}} 
\newcommand{\hatcurextraerrMWoneeccenxxxxxA}{\ensuremath{0\pm0}}     
\newcommand{\hatcurextraerrMWonetwosiglimeccenxxxxxA}{\ensuremath{<0.0200}} 
\newcommand{\hatcurextraerrMWtwoeccenxxxxxA}{\ensuremath{0\pm0}}     
\newcommand{\hatcurextraerrMWtwotwosiglimeccenxxxxxA}{\ensuremath{<0.0200}} 
\newcommand{\hatcurLBiBeccenxxxxxA}{\ensuremath{0.5710}}             
\newcommand{\hatcurLBiiBeccenxxxxxA}{\ensuremath{0.2510}}            
\newcommand{\hatcurLBiVeccenxxxxxA}{\ensuremath{0.4895}}             
\newcommand{\hatcurLBiiVeccenxxxxxA}{\ensuremath{0.3055}}            
\newcommand{\hatcurLBiReccenxxxxxA}{\ensuremath{0.31\pm0.14}}        
\newcommand{\hatcurLBiiReccenxxxxxA}{\ensuremath{0.26\pm0.17}}       
\newcommand{\hatcurLBiIeccenxxxxxA}{\ensuremath{0.2558}}             
\newcommand{\hatcurLBiiIeccenxxxxxA}{\ensuremath{0.3251}}            
\newcommand{\hatcurLBiueccenxxxxxA}{\ensuremath{0.5943}}             
\newcommand{\hatcurLBiiueccenxxxxxA}{\ensuremath{0.2214}}            
\newcommand{\hatcurLBigeccenxxxxxA}{\ensuremath{0.40\pm0.16}}        
\newcommand{\hatcurLBiigeccenxxxxxA}{\ensuremath{0.28\pm0.18}}       
\newcommand{\hatcurLBireccenxxxxxA}{\ensuremath{0.38\pm0.14}}        
\newcommand{\hatcurLBiireccenxxxxxA}{\ensuremath{0.24\pm0.18}}       
\newcommand{\hatcurLBiieccenxxxxxA}{\ensuremath{0.43\pm0.13}}        
\newcommand{\hatcurLBiiieccenxxxxxA}{\ensuremath{0.30\pm0.17}}       
\newcommand{\hatcurLBizeccenxxxxxA}{\ensuremath{0.34\pm0.14}}        
\newcommand{\hatcurLBiizeccenxxxxxA}{\ensuremath{0.12\pm0.18}}       
\newcommand{\hatcurLBiJeccenxxxxxA}{\ensuremath{0.1499}}             
\newcommand{\hatcurLBiiJeccenxxxxxA}{\ensuremath{0.2488}}            
\newcommand{\hatcurLBiHeccenxxxxxA}{\ensuremath{0.1199}}             
\newcommand{\hatcurLBiiHeccenxxxxxA}{\ensuremath{0.2641}}            
\newcommand{\hatcurLBiKeccenxxxxxA}{\ensuremath{0.0910}}             
\newcommand{\hatcurLBiiKeccenxxxxxA}{\ensuremath{0.2450}}            
\newcommand{\hatcurLBiTeccenxxxxxA}{\ensuremath{0.20\pm0.13}}        
\newcommand{\hatcurLBiiTeccenxxxxxA}{\ensuremath{0.18\pm0.17}}       
\newcommand{\hatcurLBikepeccenxxxxxA}{\ensuremath{0.3716}}           
\newcommand{\hatcurLBiikepeccenxxxxxA}{\ensuremath{0.3464}}          
\newcommand{\hatcurLBiCeccenxxxxxA}{\ensuremath{0.3335}}             
\newcommand{\hatcurLBiiCeccenxxxxxA}{\ensuremath{0.3503}}            
\newcommand{\hatcurLBiMeccenxxxxxA}{\ensuremath{0.4561}}             
\newcommand{\hatcurLBiiMeccenxxxxxA}{\ensuremath{0.3301}}            
\newcommand{\hatcurLBiSoneeccenxxxxxA}{\ensuremath{0.0689}}            
\newcommand{\hatcurLBiiSoneeccenxxxxxA}{\ensuremath{0.1890}}           
\newcommand{\hatcurLBiStwoeccenxxxxxA}{\ensuremath{0.0582}}            
\newcommand{\hatcurLBiiStwoeccenxxxxxA}{\ensuremath{0.1489}}           
\newcommand{\hatcurLBiSthreeeccenxxxxxA}{\ensuremath{0.0560}}            
\newcommand{\hatcurLBiiSthreeeccenxxxxxA}{\ensuremath{0.1294}}           
\newcommand{\hatcurLBiSfoureccenxxxxxA}{\ensuremath{0.0640}}            
\newcommand{\hatcurLBiiSfoureccenxxxxxA}{\ensuremath{0.1031}}           
\newcommand{\hatcurISOmeccenxxxxxA}{\ensuremath{0.6034\pm0.0090}}    
\newcommand{\hatcurISOmshorteccenxxxxxA}{\ensuremath{0.60}}          
\newcommand{\hatcurISOmlongeccenxxxxxA}{\ensuremath{0.6034\pm0.0090}} 
\newcommand{\hatcurISOreccenxxxxxA}{\ensuremath{0.5739\pm0.0043}}    
\newcommand{\hatcurISOrshorteccenxxxxxA}{\ensuremath{0.57}}          
\newcommand{\hatcurISOrlongeccenxxxxxA}{\ensuremath{0.5739\pm0.0043}} 
\newcommand{\hatcurISOrhoeccenxxxxxA}{\ensuremath{4.48\pm0.13}}      
\newcommand{\hatcurISOrholongeccenxxxxxA}{\ensuremath{4.48\pm0.13}}  
\newcommand{\hatcurISOloggeccenxxxxxA}{\ensuremath{4.6998\pm0.0099}} 
\newcommand{\hatcurISOlumeccenxxxxxA}{\ensuremath{0.0604\pm0.0012}}  
\newcommand{\hatcurISOlumshorteccenxxxxxA}{\ensuremath{0.06}}        
\newcommand{\hatcurISOteffeccenxxxxxA}{\ensuremath{3778\pm10}}       
\newcommand{\hatcurISOzfeheccenxxxxxA}{\ensuremath{0.508_{-0.020}^{+0.033}}} 
\newcommand{\hatcurISOageeccenxxxxxA}{\ensuremath{9.3\pm5.0}}        
\newcommand{\hatcurISOspececcenxxxxxA}{M}                            
\newcommand{\hatcurRVKeccenxxxxxA}{\ensuremath{347\pm21}}            
\newcommand{\hatcurRVKtwosiglimeccenxxxxxA}{\ensuremath{<371.2}}     
\newcommand{\hatcurRVrkeccenxxxxxA}{\ensuremath{-0.006\pm0.072}}     
\newcommand{\hatcurRVrheccenxxxxxA}{\ensuremath{0.025\pm0.092}}      
\newcommand{\hatcurRVkeccenxxxxxA}{\ensuremath{-0.000\pm0.011}}      
\newcommand{\hatcurRVheccenxxxxxA}{\ensuremath{0.0016_{-0.0092}^{+0.0151}}} 
\newcommand{\hatcurRVtroneeccenxxxxxA}{\ensuremath{0\pm0}}           
\newcommand{\hatcurRVtrtwoeccenxxxxxA}{\ensuremath{0\pm0}}           
\newcommand{\hatcurRVgammaeccenxxxxxA}{\ensuremath{15852\pm20}}      
\newcommand{\hatcurRVjittereccenxxxxxA}{\ensuremath{0\pm39}}         
\newcommand{\hatcurRVjittertwosiglimeccenxxxxxA}{\ensuremath{<95.2}} 
\newcommand{\hatcurRVfitrmseccenxxxxxA}{\ensuremath{.1fym}}          %
\newcommand{\hatcurRVecceneccenxxxxxA}{\ensuremath{0.010\pm0.015}}   
\newcommand{\hatcurRVeccentwosiglimeccenxxxxxA}{\ensuremath{<0.044}} 
\newcommand{\hatcurRVomegaeccenxxxxxA}{\ensuremath{141\pm96}}        
\newcommand{\hatcurPPieccenxxxxxA}{\ensuremath{87.53\pm0.35}}        
\newcommand{\hatcurPPgeccenxxxxxA}{\ensuremath{34.1\pm2.7}}          
\newcommand{\hatcurPPloggeccenxxxxxA}{\ensuremath{3.533\pm0.035}}    
\newcommand{\hatcurPPareccenxxxxxA}{\ensuremath{8.932\pm0.086}}      
\newcommand{\hatcurPPareleccenxxxxxA}{\ensuremath{0.02387\pm0.00012}} 
\newcommand{\hatcurPPrhoeccenxxxxxA}{\ensuremath{1.66\pm0.16}}       
\newcommand{\hatcurPPmeccenxxxxxA}{\ensuremath{1.463\pm0.090}}       
\newcommand{\hatcurPPmtwosiglimeccenxxxxxA}{\ensuremath{<1.57}}      
\newcommand{\hatcurPPmshorteccenxxxxxA}{\ensuremath{1.46}}           
\newcommand{\hatcurPPmlongeccenxxxxxA}{\ensuremath{1.463\pm0.090}}   
\newcommand{\hatcurPPmeeccenxxxxxA}{\ensuremath{465\pm29}}           
\newcommand{\hatcurPPmeshorteccenxxxxxA}{\ensuremath{465.0}}         
\newcommand{\hatcurPPmelongeccenxxxxxA}{\ensuremath{465\pm29}}       
\newcommand{\hatcurPPreccenxxxxxA}{\ensuremath{1.031_{-0.023}^{+0.017}}} 
\newcommand{\hatcurPPrshorteccenxxxxxA}{\ensuremath{1.03}}           
\newcommand{\hatcurPPrlongeccenxxxxxA}{\ensuremath{1.031_{-0.023}^{+0.017}}} 
\newcommand{\hatcurPPreeccenxxxxxA}{\ensuremath{11.55_{-0.26}^{+0.19}}} 
\newcommand{\hatcurPPreshorteccenxxxxxA}{\ensuremath{11.6}}          
\newcommand{\hatcurPPrelongeccenxxxxxA}{\ensuremath{11.55_{-0.26}^{+0.19}}} 
\newcommand{\hatcurPPmrcorreccenxxxxxA}{\ensuremath{-0.16}}          
\newcommand{\hatcurPPteffeccenxxxxxA}{\ensuremath{893.5\pm4.3}}      
\newcommand{\hatcurPPthetaeccenxxxxxA}{\ensuremath{0.1120\pm0.0075}} 
\newcommand{\hatcurPPfluxperieccenxxxxxA}{\ensuremath{1.479\pm0.057}} 
\newcommand{\hatcurPPfluxperidimeccenxxxxxA}{\ensuremath{8}}         
\newcommand{\hatcurPPfluxapeccenxxxxxA}{\ensuremath{1.410\pm0.045}}  
\newcommand{\hatcurPPfluxapdimeccenxxxxxA}{\ensuremath{8}}           
\newcommand{\hatcurPPfluxavgeccenxxxxxA}{\ensuremath{1.445\pm0.028}} 
\newcommand{\hatcurPPfluxavgdimeccenxxxxxA}{\ensuremath{8}}          
\newcommand{\hatcurPPfluxavglogeccenxxxxxA}{\ensuremath{8.1599\pm0.0084}} 
\newcommand{\hatcurXsecphaseeccenxxxxxA}{\ensuremath{0.4998\pm0.0071}} 
\newcommand{\hatcurXsecondaryeccenxxxxxA}{\ensuremath{2458377.305\pm0.012}} 
\newcommand{\hatcurXsecdureccenxxxxxA}{\ensuremath{0.0691\pm0.0020}} 
\newcommand{\hatcurXsecingdureccenxxxxxA}{\ensuremath{0.01245\pm0.00045}} 
\newcommand{\hatcurPPphiconjeccenxxxxxA}{\ensuremath{-0.02\pm0.26}}  
\newcommand{\hatcurPPperieccenxxxxxA}{\ensuremath{2458376.48\pm0.45}} 
\newcommand{\hatcurPPaequiveccenxxxxxA}{\ensuremath{0.09710\pm0.00093}} 
\newcommand{\hatcurPPtcirceccenxxxxxA}{\ensuremath{29.5_{-2.9}^{+4.2}}} 
\newcommand{\hatcurPPtinfalleccenxxxxxA}{\ensuremath{387\pm34}}      
\newcommand{\hatcurXdisteccenxxxxxA}{\ensuremath{285.9\pm2.6}}       
\newcommand{\hatcurXAveccenxxxxxA}{\ensuremath{0.1970_{-0.0150}^{+0.0100}}} 
\newcommand{\hatcurXdistredeccenxxxxxA}{\ensuremath{285.9\pm2.6}}    
\newcommand{\hatcurXEBVeccenxxxxxA}{\ensuremath{0.0640_{-0.0050}^{+0.0030}}} 
\newcommand{\hatcurCCpmraeccenxxxxxA}{\ensuremath{-38.86\pm0.12}}    
\newcommand{\hatcurCCpmdececcenxxxxxA}{\ensuremath{39.679\pm0.077}}  
\newcommand{\hatcurCCpmeccenxxxxxA}{\ensuremath{55.54\pm0.14}}       

\newcommand{\hatcurhtreccenxxxxxB}{HATS548-007}                      
\newcommand{\hatcurfieldeccenxxxxxB}{\ensuremath{string}}            
\newcommand{\hatcurCCraeccenxxxxxB}{\ensuremath{04^{\mathrm h}03^{\mathrm m}47.8440{\mathrm s}}}                   
\newcommand{\hatcurCCdececcenxxxxxB}{\ensuremath{-25{\arcdeg}24{\arcmin}32.1170{\arcsec}}}                 
\newcommand{\hatcurCCmageccenxxxxxB}{15.759}                         
\newcommand{\hatcurCCtwomasseccenxxxxxB}{2MASS~04034783-2524320}     
\newcommand{\hatcurCCgsceccenxxxxxB}{GSC~}                           
\newcommand{\hatcurCCgaiaeccenxxxxxB}{GAIA~5082914333903138048}      
\newcommand{\hatcurCCgaiadrtwoeccenxxxxxB}{GAIA~DR2~5082914338199586560} 
\newcommand{\hatcurCCtassmveccenxxxxxB}{\ensuremath{15.759\pm0.036}} 
\newcommand{\hatcurCCtassmvshorteccenxxxxxB}{\ensuremath{15.8}}      
\newcommand{\hatcurCCtassmBeccenxxxxxB}{\ensuremath{17.120\pm0.019}} 
\newcommand{\hatcurCCtassmBshorteccenxxxxxB}{\ensuremath{17.1}}      
\newcommand{\hatcurCCtassmIeccenxxxxxB}{\ensuremath{nff\pmnff}}      
\newcommand{\hatcurCCtassmIshorteccenxxxxxB}{\ensuremath{0.0}}       
\newcommand{\hatcurCCtassmgeccenxxxxxB}{\ensuremath{16.503\pm0.070}} 
\newcommand{\hatcurCCtassmgshorteccenxxxxxB}{\ensuremath{16.5}}      
\newcommand{\hatcurCCtassmreccenxxxxxB}{\ensuremath{15.156\pm0.090}} 
\newcommand{\hatcurCCtassmrshorteccenxxxxxB}{\ensuremath{15.2}}      
\newcommand{\hatcurCCtassmieccenxxxxxB}{\ensuremath{14.230\pm0.070}} 
\newcommand{\hatcurCCtassmishorteccenxxxxxB}{\ensuremath{14.2}}      
\newcommand{\hatcurCCparallaxeccenxxxxxB}{\ensuremath{5.100\pm0.029}} 
\newcommand{\hatcurCCgaiamGeccenxxxxxB}{\ensuremath{14.90330\pm0.00040}} 
\newcommand{\hatcurCCgaiamBPeccenxxxxxB}{\ensuremath{16.0189\pm0.0026}} 
\newcommand{\hatcurCCgaiamRPeccenxxxxxB}{\ensuremath{13.8535\pm0.0013}} 
\newcommand{\hatcurCCtwomassJmageccenxxxxxB}{\ensuremath{12.481\pm0.023}} 
\newcommand{\hatcurCCtwomassHmageccenxxxxxB}{\ensuremath{11.756\pm0.026}} 
\newcommand{\hatcurCCtwomassKmageccenxxxxxB}{\ensuremath{11.584\pm0.021}} 
\newcommand{\hatcurCCcitJmageccenxxxxxB}{\ensuremath{12.470\pm0.025}} 
\newcommand{\hatcurCCcitHmageccenxxxxxB}{\ensuremath{11.748\pm0.027}} 
\newcommand{\hatcurCCcitKmageccenxxxxxB}{\ensuremath{11.608\pm0.022}} 
\newcommand{\hatcurCCbbJmageccenxxxxxB}{\ensuremath{12.562\pm0.027}} 
\newcommand{\hatcurCCbbHmageccenxxxxxB}{\ensuremath{11.772\pm0.028}} 
\newcommand{\hatcurCCbbKmageccenxxxxxB}{\ensuremath{11.628\pm0.022}} 
\newcommand{\hatcurCCesoJmageccenxxxxxB}{\ensuremath{12.571\pm0.032}} 
\newcommand{\hatcurCCesoHmageccenxxxxxB}{\ensuremath{11.770\pm0.037}} 
\newcommand{\hatcurCCesoKmageccenxxxxxB}{\ensuremath{11.624\pm0.024}} 
\newcommand{\hatcurCCesoJHmageccenxxxxxB}{\ensuremath{0.801\pm0.045}} 
\newcommand{\hatcurCCesoJKmageccenxxxxxB}{\ensuremath{0.947\pm0.037}} 
\newcommand{\hatcurCCesoHKmageccenxxxxxB}{\ensuremath{0.145\pm0.042}} 
\newcommand{\hatcurCCWonemageccenxxxxxB}{\ensuremath{11.486\pm0.024}} 
\newcommand{\hatcurCCWtwomageccenxxxxxB}{\ensuremath{11.502\pm0.022}} 
\newcommand{\hatcurCCWthreemageccenxxxxxB}{\ensuremath{11.61\pm0.19}} 
\newcommand{\hatcurCCWfourmageccenxxxxxB}{\ensuremath{nff\pmnff}}    
\newcommand{\hatcurLCdipeccenxxxxxB}{\ensuremath{0.0}}               
\newcommand{\hatcurLCrprstareccenxxxxxB}{\ensuremath{0.1550\pm0.0025}} 
\newcommand{\hatcurLCbsqeccenxxxxxB}{\ensuremath{0.153_{-0.053}^{+0.044}}} 
\newcommand{\hatcurLCimpeccenxxxxxB}{\ensuremath{0.391_{-0.075}^{+0.053}}} 
\newcommand{\hatcurLCzetaeccenxxxxxB}{\ensuremath{29.59\pm0.42}}     
\newcommand{\hatcurLCdureccenxxxxxB}{\ensuremath{0.07982\pm0.00084}} 
\newcommand{\hatcurLCdurshorteccenxxxxxB}{\ensuremath{0.0798}}       
\newcommand{\hatcurLCdurhreccenxxxxxB}{\ensuremath{1.916\pm0.020}}   
\newcommand{\hatcurLCdurhrshorteccenxxxxxB}{\ensuremath{1.916}}      
\newcommand{\hatcurLCqeccenxxxxxB}{\ensuremath{0.02860\pm0.00031}}   
\newcommand{\hatcurLCqshorteccenxxxxxB}{\ensuremath{0.029}}          
\newcommand{\hatcurLCingdureccenxxxxxB}{\ensuremath{0.01239\pm0.00075}} 
\newcommand{\hatcurLCPeccenxxxxxB}{\ensuremath{2.78865562\pm0.00000098}} 
\newcommand{\hatcurLCPprececcenxxxxxB}{\ensuremath{2.7886556}}       
\newcommand{\hatcurLCPshorteccenxxxxxB}{\ensuremath{2.7887}}         
\newcommand{\hatcurLCTeccenxxxxxB}{\ensuremath{2458354.49852\pm0.00026}} 
\newcommand{\hatcurLCTAeccenxxxxxB}{\ensuremath{2456840.25850\pm0.00064}} 
\newcommand{\hatcurLCTBeccenxxxxxB}{\ensuremath{2459110.22417\pm0.00034}} 
\newcommand{\hatcurLChatnetmAeccenxxxxxB}{\ensuremath{14.49375\pm0.00050}} 
\newcommand{\hatcurLCiblendAeccenxxxxxB}{\ensuremath{0.834\pm0.090}} 
\newcommand{\hatcurLChatnetmBeccenxxxxxB}{\ensuremath{14.502440\pm0.000045}} 
\newcommand{\hatcurLCiblendBeccenxxxxxB}{\ensuremath{0.953\pm0.030}} 
\newcommand{\hatcurLChatnetmCeccenxxxxxB}{\ensuremath{-0.000690\pm0.000062}} 
\newcommand{\hatcurLCiblendCeccenxxxxxB}{\ensuremath{0.942\pm0.033}} 
\newcommand{\hatcurLChatnetmDeccenxxxxxB}{\ensuremath{-0.000530\pm0.000067}} 
\newcommand{\hatcurLCiblendDeccenxxxxxB}{\ensuremath{0.938\pm0.028}} 
\newcommand{\hatcurLCrhoeccenxxxxxB}{\ensuremath{4.260_{-0.076}^{+0.115}}} 
\newcommand{\hatcurSMEiteffeccenxxxxxB}{\ensuremath{3812\pm79}}      
\newcommand{\hatcurSMEizfeheccenxxxxxB}{\ensuremath{0.28\pm0.12}}    
\newcommand{\hatcurSMEizfehshorteccenxxxxxB}{\ensuremath{0.28}}      
\newcommand{\hatcurSMEiloggeccenxxxxxB}{\ensuremath{4.50\pm0.50}}    
\newcommand{\hatcurSMEivsineccenxxxxxB}{\ensuremath{0\pm50}}         
\newcommand{\hatcurSMEivmaceccenxxxxxB}{\ensuremath{nff\pmnff}}      
\newcommand{\hatcurSMEivmiceccenxxxxxB}{\ensuremath{nff\pmnff}}      
\newcommand{\hatcurextraerrMJeccenxxxxxB}{\ensuremath{0\pm0}}        
\newcommand{\hatcurextraerrMJtwosiglimeccenxxxxxB}{\ensuremath{<0.0200}} 
\newcommand{\hatcurextraerrMHeccenxxxxxB}{\ensuremath{0\pm0}}        
\newcommand{\hatcurextraerrMHtwosiglimeccenxxxxxB}{\ensuremath{<0.0200}} 
\newcommand{\hatcurextraerrMKseccenxxxxxB}{\ensuremath{0\pm0}}       
\newcommand{\hatcurextraerrMKstwosiglimeccenxxxxxB}{\ensuremath{<0.0200}} 
\newcommand{\hatcurextraerrMGeccenxxxxxB}{\ensuremath{0\pm0}}        
\newcommand{\hatcurextraerrMGtwosiglimeccenxxxxxB}{\ensuremath{<0.0200}} 
\newcommand{\hatcurextraerrMBPtwoeccenxxxxxB}{\ensuremath{0\pm0}}    
\newcommand{\hatcurextraerrMBPtwotwosiglimeccenxxxxxB}{\ensuremath{<0.0200}} 
\newcommand{\hatcurextraerrMRPeccenxxxxxB}{\ensuremath{0\pm0}}       
\newcommand{\hatcurextraerrMRPtwosiglimeccenxxxxxB}{\ensuremath{<0.0200}} 
\newcommand{\hatcurextraerrMgeccenxxxxxB}{\ensuremath{0\pm0}}        
\newcommand{\hatcurextraerrMgtwosiglimeccenxxxxxB}{\ensuremath{<0.0200}} 
\newcommand{\hatcurextraerrMreccenxxxxxB}{\ensuremath{0\pm0}}        
\newcommand{\hatcurextraerrMrtwosiglimeccenxxxxxB}{\ensuremath{<0.0200}} 
\newcommand{\hatcurextraerrMieccenxxxxxB}{\ensuremath{0\pm0}}        
\newcommand{\hatcurextraerrMitwosiglimeccenxxxxxB}{\ensuremath{<0.0200}} 
\newcommand{\hatcurextraerrMWoneeccenxxxxxB}{\ensuremath{0\pm0}}     
\newcommand{\hatcurextraerrMWonetwosiglimeccenxxxxxB}{\ensuremath{<0.0200}} 
\newcommand{\hatcurextraerrMWtwoeccenxxxxxB}{\ensuremath{0\pm0}}     
\newcommand{\hatcurextraerrMWtwotwosiglimeccenxxxxxB}{\ensuremath{<0.0200}} 
\newcommand{\hatcurLBiBeccenxxxxxB}{\ensuremath{0.5844}}             
\newcommand{\hatcurLBiiBeccenxxxxxB}{\ensuremath{0.2352}}            
\newcommand{\hatcurLBiVeccenxxxxxB}{\ensuremath{0.4945}}             
\newcommand{\hatcurLBiiVeccenxxxxxB}{\ensuremath{0.2944}}            
\newcommand{\hatcurLBiReccenxxxxxB}{\ensuremath{0.4370}}             
\newcommand{\hatcurLBiiReccenxxxxxB}{\ensuremath{0.2876}}            
\newcommand{\hatcurLBiIeccenxxxxxB}{\ensuremath{0.2655}}             
\newcommand{\hatcurLBiiIeccenxxxxxB}{\ensuremath{0.3147}}            
\newcommand{\hatcurLBiueccenxxxxxB}{\ensuremath{0.6157}}             
\newcommand{\hatcurLBiiueccenxxxxxB}{\ensuremath{0.2014}}            
\newcommand{\hatcurLBigeccenxxxxxB}{\ensuremath{0.42\pm0.14}}        
\newcommand{\hatcurLBiigeccenxxxxxB}{\ensuremath{0.28\pm0.16}}       
\newcommand{\hatcurLBireccenxxxxxB}{\ensuremath{0.31\pm0.14}}        
\newcommand{\hatcurLBiireccenxxxxxB}{\ensuremath{0.24\pm0.17}}       
\newcommand{\hatcurLBiieccenxxxxxB}{\ensuremath{0.3087}}             
\newcommand{\hatcurLBiiieccenxxxxxB}{\ensuremath{0.3075}}            
\newcommand{\hatcurLBizeccenxxxxxB}{\ensuremath{0.18\pm0.11}}        
\newcommand{\hatcurLBiizeccenxxxxxB}{\ensuremath{0.11\pm0.15}}       
\newcommand{\hatcurLBiJeccenxxxxxB}{\ensuremath{0.1590}}             
\newcommand{\hatcurLBiiJeccenxxxxxB}{\ensuremath{0.2457}}            
\newcommand{\hatcurLBiHeccenxxxxxB}{\ensuremath{0.1275}}             
\newcommand{\hatcurLBiiHeccenxxxxxB}{\ensuremath{0.2607}}            
\newcommand{\hatcurLBiKeccenxxxxxB}{\ensuremath{0.0959}}             
\newcommand{\hatcurLBiiKeccenxxxxxB}{\ensuremath{0.2414}}            
\newcommand{\hatcurLBiTeccenxxxxxB}{\ensuremath{0.32\pm0.12}}        
\newcommand{\hatcurLBiiTeccenxxxxxB}{\ensuremath{0.34\pm0.16}}       
\newcommand{\hatcurLBikepeccenxxxxxB}{\ensuremath{0.3797}}           
\newcommand{\hatcurLBiikepeccenxxxxxB}{\ensuremath{0.3390}}          
\newcommand{\hatcurLBiCeccenxxxxxB}{\ensuremath{0.3436}}             
\newcommand{\hatcurLBiiCeccenxxxxxB}{\ensuremath{0.3422}}            
\newcommand{\hatcurLBiMeccenxxxxxB}{\ensuremath{0.4623}}             
\newcommand{\hatcurLBiiMeccenxxxxxB}{\ensuremath{0.3228}}            
\newcommand{\hatcurLBiSoneeccenxxxxxB}{\ensuremath{0.0708}}            
\newcommand{\hatcurLBiiSoneeccenxxxxxB}{\ensuremath{0.1866}}           
\newcommand{\hatcurLBiStwoeccenxxxxxB}{\ensuremath{0.0602}}            
\newcommand{\hatcurLBiiStwoeccenxxxxxB}{\ensuremath{0.1477}}           
\newcommand{\hatcurLBiSthreeeccenxxxxxB}{\ensuremath{0.0569}}            
\newcommand{\hatcurLBiiSthreeeccenxxxxxB}{\ensuremath{0.1283}}           
\newcommand{\hatcurLBiSfoureccenxxxxxB}{\ensuremath{0.0638}}            
\newcommand{\hatcurLBiiSfoureccenxxxxxB}{\ensuremath{0.1041}}           
\newcommand{\hatcurISOmeccenxxxxxB}{\ensuremath{0.6037\pm0.0072}}    
\newcommand{\hatcurISOmshorteccenxxxxxB}{\ensuremath{0.60}}          
\newcommand{\hatcurISOmlongeccenxxxxxB}{\ensuremath{0.6037\pm0.0072}} 
\newcommand{\hatcurISOreccenxxxxxB}{\ensuremath{0.5843_{-0.0032}^{+0.0024}}} 
\newcommand{\hatcurISOrshorteccenxxxxxB}{\ensuremath{0.58}}          
\newcommand{\hatcurISOrlongeccenxxxxxB}{\ensuremath{0.5843_{-0.0032}^{+0.0024}}} 
\newcommand{\hatcurISOrhoeccenxxxxxB}{\ensuremath{4.260_{-0.076}^{+0.115}}} 
\newcommand{\hatcurISOrholongeccenxxxxxB}{\ensuremath{4.260_{-0.076}^{+0.115}}} 
\newcommand{\hatcurISOloggeccenxxxxxB}{\ensuremath{4.6851\pm0.0086}} 
\newcommand{\hatcurISOlumeccenxxxxxB}{\ensuremath{0.06343\pm0.00074}} 
\newcommand{\hatcurISOlumshorteccenxxxxxB}{\ensuremath{0.06}}        
\newcommand{\hatcurISOteffeccenxxxxxB}{\ensuremath{3790.9\pm6.2}}    
\newcommand{\hatcurISOzfeheccenxxxxxB}{\ensuremath{0.527_{-0.032}^{+0.061}}} 
\newcommand{\hatcurISOageeccenxxxxxB}{\ensuremath{13.8_{-5.0}^{+3.5}}} 
\newcommand{\hatcurISOspececcenxxxxxB}{***TODO***}                   
\newcommand{\hatcurRVKeccenxxxxxB}{\ensuremath{98.7\pm7.1}}          
\newcommand{\hatcurRVKtwosiglimeccenxxxxxB}{\ensuremath{<110.6}}     
\newcommand{\hatcurRVrkeccenxxxxxB}{\ensuremath{-0.03\pm0.10}}       
\newcommand{\hatcurRVrheccenxxxxxB}{\ensuremath{0.02\pm0.11}}        
\newcommand{\hatcurRVkeccenxxxxxB}{\ensuremath{-0.002_{-0.024}^{+0.015}}} 
\newcommand{\hatcurRVheccenxxxxxB}{\ensuremath{0.001\pm0.023}}       
\newcommand{\hatcurRVtroneeccenxxxxxB}{\ensuremath{0\pm0}}           
\newcommand{\hatcurRVtrtwoeccenxxxxxB}{\ensuremath{0\pm0}}           
\newcommand{\hatcurRVgammaeccenxxxxxB}{\ensuremath{39997.0\pm5.5}}   
\newcommand{\hatcurRVjittereccenxxxxxB}{\ensuremath{0.0\pm1.3}}      
\newcommand{\hatcurRVjittertwosiglimeccenxxxxxB}{\ensuremath{<3.2}}  
\newcommand{\hatcurRVfitrmseccenxxxxxB}{\ensuremath{.1fym}}          %
\newcommand{\hatcurRVecceneccenxxxxxB}{\ensuremath{0.020\pm0.021}}   
\newcommand{\hatcurRVeccentwosiglimeccenxxxxxB}{\ensuremath{<0.064}} 
\newcommand{\hatcurRVomegaeccenxxxxxB}{\ensuremath{156\pm88}}        
\newcommand{\hatcurPPieccenxxxxxB}{\ensuremath{88.14_{-0.23}^{+0.34}}} 
\newcommand{\hatcurPPgeccenxxxxxB}{\ensuremath{15.6\pm1.3}}          
\newcommand{\hatcurPPloggeccenxxxxxB}{\ensuremath{3.193\pm0.035}}    
\newcommand{\hatcurPPareccenxxxxxB}{\ensuremath{12.057_{-0.073}^{+0.107}}} 
\newcommand{\hatcurPPareleccenxxxxxB}{\ensuremath{0.03278\pm0.00013}} 
\newcommand{\hatcurPPrhoeccenxxxxxB}{\ensuremath{0.886\pm0.080}}     
\newcommand{\hatcurPPmeccenxxxxxB}{\ensuremath{0.488\pm0.035}}       
\newcommand{\hatcurPPmtwosiglimeccenxxxxxB}{\ensuremath{<0.55}}      
\newcommand{\hatcurPPmshorteccenxxxxxB}{\ensuremath{0.49}}           
\newcommand{\hatcurPPmlongeccenxxxxxB}{\ensuremath{0.488\pm0.035}}   
\newcommand{\hatcurPPmeeccenxxxxxB}{\ensuremath{155\pm11}}           
\newcommand{\hatcurPPmeshorteccenxxxxxB}{\ensuremath{155.2}}         
\newcommand{\hatcurPPmelongeccenxxxxxB}{\ensuremath{155\pm11}}       
\newcommand{\hatcurPPreccenxxxxxB}{\ensuremath{0.881\pm0.015}}       
\newcommand{\hatcurPPrshorteccenxxxxxB}{\ensuremath{0.88}}           
\newcommand{\hatcurPPrlongeccenxxxxxB}{\ensuremath{0.881\pm0.015}}   
\newcommand{\hatcurPPreeccenxxxxxB}{\ensuremath{9.88\pm0.17}}        
\newcommand{\hatcurPPreshorteccenxxxxxB}{\ensuremath{9.9}}           
\newcommand{\hatcurPPrelongeccenxxxxxB}{\ensuremath{9.88\pm0.17}}    
\newcommand{\hatcurPPmrcorreccenxxxxxB}{\ensuremath{-0.03}}          
\newcommand{\hatcurPPteffeccenxxxxxB}{\ensuremath{771.8\pm2.7}}      
\newcommand{\hatcurPPthetaeccenxxxxxB}{\ensuremath{0.0600\pm0.0044}} 
\newcommand{\hatcurPPfluxperieccenxxxxxB}{\ensuremath{8.35_{-0.24}^{+0.50}}} 
\newcommand{\hatcurPPfluxperidimeccenxxxxxB}{\ensuremath{7}}         
\newcommand{\hatcurPPfluxapeccenxxxxxB}{\ensuremath{7.72_{-0.40}^{+0.27}}} 
\newcommand{\hatcurPPfluxapdimeccenxxxxxB}{\ensuremath{7}}           
\newcommand{\hatcurPPfluxavgeccenxxxxxB}{\ensuremath{8.05\pm0.11}}   
\newcommand{\hatcurPPfluxavgdimeccenxxxxxB}{\ensuremath{7}}          
\newcommand{\hatcurPPfluxavglogeccenxxxxxB}{\ensuremath{7.9056\pm0.0061}} 
\newcommand{\hatcurXsecphaseeccenxxxxxB}{\ensuremath{0.499\pm0.014}} 
\newcommand{\hatcurXsecondaryeccenxxxxxB}{\ensuremath{2458355.889\pm0.038}} 
\newcommand{\hatcurXsecdureccenxxxxxB}{\ensuremath{0.0800\pm0.0029}} 
\newcommand{\hatcurXsecingdureccenxxxxxB}{\ensuremath{0.01239\pm0.00042}} 
\newcommand{\hatcurPPphiconjeccenxxxxxB}{\ensuremath{-0.07\pm0.24}}  
\newcommand{\hatcurPPperieccenxxxxxB}{\ensuremath{2458354.71\pm0.68}} 
\newcommand{\hatcurPPaequiveccenxxxxxB}{\ensuremath{0.13010\pm0.00091}} 
\newcommand{\hatcurPPtcirceccenxxxxxB}{\ensuremath{169\pm20}}        
\newcommand{\hatcurPPtinfalleccenxxxxxB}{\ensuremath{8380_{-620}^{+810}}} 
\newcommand{\hatcurXdisteccenxxxxxB}{\ensuremath{195.1\pm1.0}}       
\newcommand{\hatcurXAveccenxxxxxB}{\ensuremath{0.0560\pm0.0092}}     
\newcommand{\hatcurXdistredeccenxxxxxB}{\ensuremath{195.1\pm1.0}}    
\newcommand{\hatcurXEBVeccenxxxxxB}{\ensuremath{0.0180\pm0.0030}}    
\newcommand{\hatcurCCpmraeccenxxxxxB}{\ensuremath{12.872\pm0.038}}   
\newcommand{\hatcurCCpmdececcenxxxxxB}{\ensuremath{-1.753\pm0.048}}  
\newcommand{\hatcurCCpmeccenxxxxxB}{\ensuremath{12.991\pm0.061}}     

\newcommand{\hatcurhtreccenxxxxxC}{HATS597-003}                      
\newcommand{\hatcurfieldeccenxxxxxC}{\ensuremath{string}}            
\newcommand{\hatcurCCraeccenxxxxxC}{\ensuremath{04^{\mathrm h}41^{\mathrm m}21.5520{\mathrm s}}}                   
\newcommand{\hatcurCCdececcenxxxxxC}{\ensuremath{-32{\arcdeg}19{\arcmin}13.5029{\arcsec}}}                 
\newcommand{\hatcurCCmageccenxxxxxC}{NULL}                           
\newcommand{\hatcurCCtwomasseccenxxxxxC}{2MASS~04412154-3219128}     
\newcommand{\hatcurCCgsceccenxxxxxC}{GSC~}                           
\newcommand{\hatcurCCgaiaeccenxxxxxC}{GAIA~4877426571427909248}      
\newcommand{\hatcurCCgaiadrtwoeccenxxxxxC}{GAIA~DR2~4877426575724467456} 
\newcommand{\hatcurCCtassmveccenxxxxxC}{\ensuremath{nff\pmnff}}      
\newcommand{\hatcurCCtassmvshorteccenxxxxxC}{\ensuremath{0.0}}       
\newcommand{\hatcurCCtassmBeccenxxxxxC}{\ensuremath{nff\pmnff}}      
\newcommand{\hatcurCCtassmBshorteccenxxxxxC}{\ensuremath{0.0}}       
\newcommand{\hatcurCCtassmIeccenxxxxxC}{\ensuremath{nff\pmnff}}      
\newcommand{\hatcurCCtassmIshorteccenxxxxxC}{\ensuremath{0.0}}       
\newcommand{\hatcurCCtassmgeccenxxxxxC}{\ensuremath{nff\pmnff}}      
\newcommand{\hatcurCCtassmgshorteccenxxxxxC}{\ensuremath{0.0}}       
\newcommand{\hatcurCCtassmreccenxxxxxC}{\ensuremath{nff\pmnff}}      
\newcommand{\hatcurCCtassmrshorteccenxxxxxC}{\ensuremath{0.0}}       
\newcommand{\hatcurCCtassmieccenxxxxxC}{\ensuremath{nff\pmnff}}      
\newcommand{\hatcurCCtassmishorteccenxxxxxC}{\ensuremath{0.0}}       
\newcommand{\hatcurCCparallaxeccenxxxxxC}{\ensuremath{2.564\pm0.038}} 
\newcommand{\hatcurCCgaiamGeccenxxxxxC}{\ensuremath{15.79420\pm0.00060}} 
\newcommand{\hatcurCCgaiamBPeccenxxxxxC}{\ensuremath{16.6962\pm0.0057}} 
\newcommand{\hatcurCCgaiamRPeccenxxxxxC}{\ensuremath{14.8610\pm0.0020}} 
\newcommand{\hatcurCCtwomassJmageccenxxxxxC}{\ensuremath{13.690\pm0.029}} 
\newcommand{\hatcurCCtwomassHmageccenxxxxxC}{\ensuremath{12.984\pm0.024}} 
\newcommand{\hatcurCCtwomassKmageccenxxxxxC}{\ensuremath{12.812\pm0.033}} 
\newcommand{\hatcurCCcitJmageccenxxxxxC}{\ensuremath{13.680\pm0.030}} 
\newcommand{\hatcurCCcitHmageccenxxxxxC}{\ensuremath{12.976\pm0.025}} 
\newcommand{\hatcurCCcitKmageccenxxxxxC}{\ensuremath{12.836\pm0.034}} 
\newcommand{\hatcurCCbbJmageccenxxxxxC}{\ensuremath{13.771\pm0.032}} 
\newcommand{\hatcurCCbbHmageccenxxxxxC}{\ensuremath{13.000\pm0.026}} 
\newcommand{\hatcurCCbbKmageccenxxxxxC}{\ensuremath{12.856\pm0.034}} 
\newcommand{\hatcurCCesoJmageccenxxxxxC}{\ensuremath{13.779\pm0.037}} 
\newcommand{\hatcurCCesoHmageccenxxxxxC}{\ensuremath{12.998\pm0.036}} 
\newcommand{\hatcurCCesoKmageccenxxxxxC}{\ensuremath{12.852\pm0.035}} 
\newcommand{\hatcurCCesoJHmageccenxxxxxC}{\ensuremath{0.781\pm0.047}} 
\newcommand{\hatcurCCesoJKmageccenxxxxxC}{\ensuremath{0.928\pm0.049}} 
\newcommand{\hatcurCCesoHKmageccenxxxxxC}{\ensuremath{0.146\pm0.049}} 
\newcommand{\hatcurCCWonemageccenxxxxxC}{\ensuremath{12.758\pm0.024}} 
\newcommand{\hatcurCCWtwomageccenxxxxxC}{\ensuremath{12.784\pm0.026}} 
\newcommand{\hatcurCCWthreemageccenxxxxxC}{\ensuremath{nff\pmnff}}   
\newcommand{\hatcurCCWfourmageccenxxxxxC}{\ensuremath{nff\pmnff}}    
\newcommand{\hatcurLCdipeccenxxxxxC}{\ensuremath{0.0}}               
\newcommand{\hatcurLCrprstareccenxxxxxC}{\ensuremath{0.1784\pm0.0046}} 
\newcommand{\hatcurLCbsqeccenxxxxxC}{\ensuremath{0.066_{-0.043}^{+0.052}}} 
\newcommand{\hatcurLCimpeccenxxxxxC}{\ensuremath{0.257_{-0.104}^{+0.087}}} 
\newcommand{\hatcurLCzetaeccenxxxxxC}{\ensuremath{30.97\pm0.66}}     
\newcommand{\hatcurLCdureccenxxxxxC}{\ensuremath{0.0769\pm0.0012}}   
\newcommand{\hatcurLCdurshorteccenxxxxxC}{\ensuremath{0.0769}}       
\newcommand{\hatcurLCdurhreccenxxxxxC}{\ensuremath{1.845\pm0.029}}   
\newcommand{\hatcurLCdurhrshorteccenxxxxxC}{\ensuremath{1.845}}      
\newcommand{\hatcurLCqeccenxxxxxC}{\ensuremath{0.03960\pm0.00062}}   
\newcommand{\hatcurLCqshorteccenxxxxxC}{\ensuremath{0.040}}          
\newcommand{\hatcurLCingdureccenxxxxxC}{\ensuremath{0.01237\pm0.00071}} 
\newcommand{\hatcurLCPeccenxxxxxC}{\ensuremath{1.9416423\pm0.0000014}} 
\newcommand{\hatcurLCPprececcenxxxxxC}{\ensuremath{1.9416423}}       
\newcommand{\hatcurLCPshorteccenxxxxxC}{\ensuremath{1.9416}}         
\newcommand{\hatcurLCTeccenxxxxxC}{\ensuremath{2458410.96400\pm0.00055}} 
\newcommand{\hatcurLCTAeccenxxxxxC}{\ensuremath{2456543.1042\pm0.0014}} 
\newcommand{\hatcurLCTBeccenxxxxxC}{\ensuremath{2458746.86811\pm0.00060}} 
\newcommand{\hatcurLChatnetmAeccenxxxxxC}{\ensuremath{15.64063\pm0.00038}} 
\newcommand{\hatcurLCiblendAeccenxxxxxC}{\ensuremath{0.862\pm0.068}} 
\newcommand{\hatcurLChatnetmBeccenxxxxxC}{\ensuremath{-0.00303\pm0.00028}} 
\newcommand{\hatcurLCiblendBeccenxxxxxC}{\ensuremath{0.968\pm0.034}} 
\newcommand{\hatcurLCrhoeccenxxxxxC}{\ensuremath{3.85\pm0.16}}       
\newcommand{\hatcurSMEiteffeccenxxxxxC}{\ensuremath{3990\pm120}}     
\newcommand{\hatcurSMEizfeheccenxxxxxC}{\ensuremath{0.29\pm0.13}}    
\newcommand{\hatcurSMEizfehshorteccenxxxxxC}{\ensuremath{0.29}}      
\newcommand{\hatcurSMEiloggeccenxxxxxC}{\ensuremath{4.50\pm0.50}}    
\newcommand{\hatcurSMEivsineccenxxxxxC}{\ensuremath{0\pm50}}         
\newcommand{\hatcurSMEivmaceccenxxxxxC}{\ensuremath{nff\pmnff}}      
\newcommand{\hatcurSMEivmiceccenxxxxxC}{\ensuremath{nff\pmnff}}      
\newcommand{\hatcurextraerrMJeccenxxxxxC}{\ensuremath{0\pm0}}        
\newcommand{\hatcurextraerrMJtwosiglimeccenxxxxxC}{\ensuremath{<0.0200}} 
\newcommand{\hatcurextraerrMHeccenxxxxxC}{\ensuremath{0\pm0}}        
\newcommand{\hatcurextraerrMHtwosiglimeccenxxxxxC}{\ensuremath{<0.0200}} 
\newcommand{\hatcurextraerrMKseccenxxxxxC}{\ensuremath{0\pm0}}       
\newcommand{\hatcurextraerrMKstwosiglimeccenxxxxxC}{\ensuremath{<0.0200}} 
\newcommand{\hatcurextraerrMGeccenxxxxxC}{\ensuremath{0\pm0}}        
\newcommand{\hatcurextraerrMGtwosiglimeccenxxxxxC}{\ensuremath{<0.0200}} 
\newcommand{\hatcurextraerrMBPtwoeccenxxxxxC}{\ensuremath{0\pm0}}    
\newcommand{\hatcurextraerrMBPtwotwosiglimeccenxxxxxC}{\ensuremath{<0.0200}} 
\newcommand{\hatcurextraerrMRPeccenxxxxxC}{\ensuremath{0\pm0}}       
\newcommand{\hatcurextraerrMRPtwosiglimeccenxxxxxC}{\ensuremath{<0.0200}} 
\newcommand{\hatcurextraerrMWoneeccenxxxxxC}{\ensuremath{0\pm0}}     
\newcommand{\hatcurextraerrMWonetwosiglimeccenxxxxxC}{\ensuremath{<0.0200}} 
\newcommand{\hatcurextraerrMWtwoeccenxxxxxC}{\ensuremath{0\pm0}}     
\newcommand{\hatcurextraerrMWtwotwosiglimeccenxxxxxC}{\ensuremath{<0.0200}} 
\newcommand{\hatcurLBiBeccenxxxxxC}{\ensuremath{0.7014}}             
\newcommand{\hatcurLBiiBeccenxxxxxC}{\ensuremath{0.1182}}            
\newcommand{\hatcurLBiVeccenxxxxxC}{\ensuremath{0.5707}}             
\newcommand{\hatcurLBiiVeccenxxxxxC}{\ensuremath{0.2046}}            
\newcommand{\hatcurLBiReccenxxxxxC}{\ensuremath{0.4830}}             
\newcommand{\hatcurLBiiReccenxxxxxC}{\ensuremath{0.2327}}            
\newcommand{\hatcurLBiIeccenxxxxxC}{\ensuremath{0.32\pm0.15}}        
\newcommand{\hatcurLBiiIeccenxxxxxC}{\ensuremath{0.24\pm0.18}}       
\newcommand{\hatcurLBiueccenxxxxxC}{\ensuremath{0.7492}}             
\newcommand{\hatcurLBiiueccenxxxxxC}{\ensuremath{0.0747}}            
\newcommand{\hatcurLBigeccenxxxxxC}{\ensuremath{0.48\pm0.14}}        
\newcommand{\hatcurLBiigeccenxxxxxC}{\ensuremath{0.31\pm0.15}}       
\newcommand{\hatcurLBireccenxxxxxC}{\ensuremath{0.34\pm0.16}}        
\newcommand{\hatcurLBiireccenxxxxxC}{\ensuremath{0.32\pm0.16}}       
\newcommand{\hatcurLBiieccenxxxxxC}{\ensuremath{0.3712}}             
\newcommand{\hatcurLBiiieccenxxxxxC}{\ensuremath{0.2527}}            
\newcommand{\hatcurLBizeccenxxxxxC}{\ensuremath{0.2882}}             
\newcommand{\hatcurLBiizeccenxxxxxC}{\ensuremath{0.2510}}            
\newcommand{\hatcurLBiJeccenxxxxxC}{\ensuremath{0.2146}}             
\newcommand{\hatcurLBiiJeccenxxxxxC}{\ensuremath{0.2279}}            
\newcommand{\hatcurLBiHeccenxxxxxC}{\ensuremath{0.1632}}             
\newcommand{\hatcurLBiiHeccenxxxxxC}{\ensuremath{0.2566}}            
\newcommand{\hatcurLBiKeccenxxxxxC}{\ensuremath{0.1218}}             
\newcommand{\hatcurLBiiKeccenxxxxxC}{\ensuremath{0.2332}}            
\newcommand{\hatcurLBiTeccenxxxxxC}{\ensuremath{0.31\pm0.15}}        
\newcommand{\hatcurLBiiTeccenxxxxxC}{\ensuremath{0.35\pm0.18}}       
\newcommand{\hatcurLBikepeccenxxxxxC}{\ensuremath{0.4512}}           
\newcommand{\hatcurLBiikepeccenxxxxxC}{\ensuremath{0.2754}}          
\newcommand{\hatcurLBiCeccenxxxxxC}{\ensuremath{0.4199}}             
\newcommand{\hatcurLBiiCeccenxxxxxC}{\ensuremath{0.2799}}            
\newcommand{\hatcurLBiMeccenxxxxxC}{\ensuremath{0.5363}}             
\newcommand{\hatcurLBiiMeccenxxxxxC}{\ensuremath{0.2478}}            
\newcommand{\hatcurLBiSoneeccenxxxxxC}{\ensuremath{0.0864}}            
\newcommand{\hatcurLBiiSoneeccenxxxxxC}{\ensuremath{0.1752}}           
\newcommand{\hatcurLBiStwoeccenxxxxxC}{\ensuremath{0.0736}}            
\newcommand{\hatcurLBiiStwoeccenxxxxxC}{\ensuremath{0.1430}}           
\newcommand{\hatcurLBiSthreeeccenxxxxxC}{\ensuremath{0.0640}}            
\newcommand{\hatcurLBiiSthreeeccenxxxxxC}{\ensuremath{0.1238}}           
\newcommand{\hatcurLBiSfoureccenxxxxxC}{\ensuremath{0.0634}}            
\newcommand{\hatcurLBiiSfoureccenxxxxxC}{\ensuremath{0.1063}}           
\newcommand{\hatcurISOmeccenxxxxxC}{\ensuremath{0.668\pm0.016}}      
\newcommand{\hatcurISOmshorteccenxxxxxC}{\ensuremath{0.67}}          
\newcommand{\hatcurISOmlongeccenxxxxxC}{\ensuremath{0.668\pm0.016}}  
\newcommand{\hatcurISOreccenxxxxxC}{\ensuremath{0.6256\pm0.0079}}    
\newcommand{\hatcurISOrshorteccenxxxxxC}{\ensuremath{0.63}}          
\newcommand{\hatcurISOrlongeccenxxxxxC}{\ensuremath{0.6256\pm0.0079}} 
\newcommand{\hatcurISOrhoeccenxxxxxC}{\ensuremath{3.85\pm0.16}}      
\newcommand{\hatcurISOrholongeccenxxxxxC}{\ensuremath{3.85\pm0.16}}  
\newcommand{\hatcurISOloggeccenxxxxxC}{\ensuremath{4.672\pm0.015}}   
\newcommand{\hatcurISOlumeccenxxxxxC}{\ensuremath{0.0922\pm0.0028}}  
\newcommand{\hatcurISOlumshorteccenxxxxxC}{\ensuremath{0.09}}        
\newcommand{\hatcurISOteffeccenxxxxxC}{\ensuremath{4021\pm16}}       
\newcommand{\hatcurISOzfeheccenxxxxxC}{\ensuremath{0.338_{-0.051}^{+0.070}}} 
\newcommand{\hatcurISOageeccenxxxxxC}{\ensuremath{2.3_{-2.0}^{+6.7}}} 
\newcommand{\hatcurISOspececcenxxxxxC}{***TODO***}                   
\newcommand{\hatcurRVKeccenxxxxxC}{\ensuremath{594\pm28}}            
\newcommand{\hatcurRVKtwosiglimeccenxxxxxC}{\ensuremath{<635.4}}     
\newcommand{\hatcurRVrkeccenxxxxxC}{\ensuremath{0.160_{-0.107}^{+0.054}}} 
\newcommand{\hatcurRVrheccenxxxxxC}{\ensuremath{0.035\pm0.087}}      
\newcommand{\hatcurRVkeccenxxxxxC}{\ensuremath{0.029\pm0.022}}       
\newcommand{\hatcurRVheccenxxxxxC}{\ensuremath{0.006\pm0.018}}       
\newcommand{\hatcurRVtroneeccenxxxxxC}{\ensuremath{0\pm0}}           
\newcommand{\hatcurRVtrtwoeccenxxxxxC}{\ensuremath{0\pm0}}           
\newcommand{\hatcurRVgammaeccenxxxxxC}{\ensuremath{8599\pm26}}       
\newcommand{\hatcurRVjittereccenxxxxxC}{\ensuremath{0\pm26}}         
\newcommand{\hatcurRVjittertwosiglimeccenxxxxxC}{\ensuremath{<47.4}} 
\newcommand{\hatcurRVfitrmseccenxxxxxC}{\ensuremath{.1fym}}          %
\newcommand{\hatcurRVecceneccenxxxxxC}{\ensuremath{0.036\pm0.018}}   
\newcommand{\hatcurRVeccentwosiglimeccenxxxxxC}{\ensuremath{<0.062}} 
\newcommand{\hatcurRVomegaeccenxxxxxC}{\ensuremath{50\pm170}}        
\newcommand{\hatcurPPieccenxxxxxC}{\ensuremath{88.38\pm0.58}}        
\newcommand{\hatcurPPgeccenxxxxxC}{\ensuremath{58.7\pm4.7}}          
\newcommand{\hatcurPPloggeccenxxxxxC}{\ensuremath{3.769\pm0.035}}    
\newcommand{\hatcurPPareccenxxxxxC}{\ensuremath{9.17\pm0.13}}        
\newcommand{\hatcurPPareleccenxxxxxC}{\ensuremath{0.02666\pm0.00021}} 
\newcommand{\hatcurPPrhoeccenxxxxxC}{\ensuremath{2.71\pm0.28}}       
\newcommand{\hatcurPPmeccenxxxxxC}{\ensuremath{2.79\pm0.14}}         
\newcommand{\hatcurPPmtwosiglimeccenxxxxxC}{\ensuremath{<3.01}}      
\newcommand{\hatcurPPmshorteccenxxxxxC}{\ensuremath{2.79}}           
\newcommand{\hatcurPPmlongeccenxxxxxC}{\ensuremath{2.79\pm0.14}}     
\newcommand{\hatcurPPmeeccenxxxxxC}{\ensuremath{887\pm46}}           
\newcommand{\hatcurPPmeshorteccenxxxxxC}{\ensuremath{886.6}}         
\newcommand{\hatcurPPmelongeccenxxxxxC}{\ensuremath{887\pm46}}       
\newcommand{\hatcurPPreccenxxxxxC}{\ensuremath{1.085\pm0.031}}       
\newcommand{\hatcurPPrshorteccenxxxxxC}{\ensuremath{1.08}}           
\newcommand{\hatcurPPrlongeccenxxxxxC}{\ensuremath{1.085\pm0.031}}   
\newcommand{\hatcurPPreeccenxxxxxC}{\ensuremath{12.16\pm0.35}}       
\newcommand{\hatcurPPreshorteccenxxxxxC}{\ensuremath{12.2}}          
\newcommand{\hatcurPPrelongeccenxxxxxC}{\ensuremath{12.16\pm0.35}}   
\newcommand{\hatcurPPmrcorreccenxxxxxC}{\ensuremath{-0.07}}          
\newcommand{\hatcurPPteffeccenxxxxxC}{\ensuremath{939.7\pm6.5}}      
\newcommand{\hatcurPPthetaeccenxxxxxC}{\ensuremath{0.204\pm0.012}}   
\newcommand{\hatcurPPfluxperieccenxxxxxC}{\ensuremath{1.899\pm0.083}} 
\newcommand{\hatcurPPfluxperidimeccenxxxxxC}{\ensuremath{8}}         
\newcommand{\hatcurPPfluxapeccenxxxxxC}{\ensuremath{1.645\pm0.077}}  
\newcommand{\hatcurPPfluxapdimeccenxxxxxC}{\ensuremath{8}}           
\newcommand{\hatcurPPfluxavgeccenxxxxxC}{\ensuremath{1.765\pm0.049}} 
\newcommand{\hatcurPPfluxavgdimeccenxxxxxC}{\ensuremath{8}}          
\newcommand{\hatcurPPfluxavglogeccenxxxxxC}{\ensuremath{8.247\pm0.012}} 
\newcommand{\hatcurXsecphaseeccenxxxxxC}{\ensuremath{0.519\pm0.014}} 
\newcommand{\hatcurXsecondaryeccenxxxxxC}{\ensuremath{2458411.971\pm0.027}} 
\newcommand{\hatcurXsecdureccenxxxxxC}{\ensuremath{0.0778\pm0.0025}} 
\newcommand{\hatcurXsecingdureccenxxxxxC}{\ensuremath{0.01260\pm0.00063}} 
\newcommand{\hatcurPPphiconjeccenxxxxxC}{\ensuremath{0.19\pm0.16}}   
\newcommand{\hatcurPPperieccenxxxxxC}{\ensuremath{2458410.59\pm0.32}} 
\newcommand{\hatcurPPaequiveccenxxxxxC}{\ensuremath{0.0879\pm0.0012}} 
\newcommand{\hatcurPPtcirceccenxxxxxC}{\ensuremath{77\pm12}}         
\newcommand{\hatcurPPtinfalleccenxxxxxC}{\ensuremath{286\pm22}}      
\newcommand{\hatcurXdisteccenxxxxxC}{\ensuremath{390.7\pm5.6}}       
\newcommand{\hatcurXAveccenxxxxxC}{\ensuremath{0.062_{-0.014}^{+0.010}}} 
\newcommand{\hatcurXdistredeccenxxxxxC}{\ensuremath{390.7\pm5.6}}    
\newcommand{\hatcurXEBVeccenxxxxxC}{\ensuremath{0.0200_{-0.0050}^{+0.0030}}} 
\newcommand{\hatcurCCpmraeccenxxxxxC}{\ensuremath{-5.429\pm0.057}}   
\newcommand{\hatcurCCpmdececcenxxxxxC}{\ensuremath{-38.609\pm0.090}} 
\newcommand{\hatcurCCpmeccenxxxxxC}{\ensuremath{38.99\pm0.11}}       

\newcommand{\hatcurhtreccenxxxxxD}{HATS607-010}                      
\newcommand{\hatcurfieldeccenxxxxxD}{\ensuremath{string}}            
\newcommand{\hatcurCCraeccenxxxxxD}{\ensuremath{09^{\mathrm h}59^{\mathrm m}17.6640{\mathrm s}}}                   
\newcommand{\hatcurCCdececcenxxxxxD}{\ensuremath{-27{\arcdeg}23{\arcmin}34.1427{\arcsec}}}                 
\newcommand{\hatcurCCmageccenxxxxxD}{16.354}                         
\newcommand{\hatcurCCtwomasseccenxxxxxD}{2MASS~09591770-2723339}     
\newcommand{\hatcurCCgsceccenxxxxxD}{GSC~}                           
\newcommand{\hatcurCCgaiaeccenxxxxxD}{GAIA~5466556137225499136}      
\newcommand{\hatcurCCgaiadrtwoeccenxxxxxD}{GAIA~DR2~5466556141521710592} 
\newcommand{\hatcurCCtassmveccenxxxxxD}{\ensuremath{16.354\pm0.010}} 
\newcommand{\hatcurCCtassmvshorteccenxxxxxD}{\ensuremath{16.4}}      
\newcommand{\hatcurCCtassmBeccenxxxxxD}{\ensuremath{17.716\pm0.010}} 
\newcommand{\hatcurCCtassmBshorteccenxxxxxD}{\ensuremath{17.7}}      
\newcommand{\hatcurCCtassmIeccenxxxxxD}{\ensuremath{nff\pmnff}}      
\newcommand{\hatcurCCtassmIshorteccenxxxxxD}{\ensuremath{0.0}}       
\newcommand{\hatcurCCtassmgeccenxxxxxD}{\ensuremath{17.10\pm0.25}}   
\newcommand{\hatcurCCtassmgshorteccenxxxxxD}{\ensuremath{17.1}}      
\newcommand{\hatcurCCtassmreccenxxxxxD}{\ensuremath{15.746\pm0.060}} 
\newcommand{\hatcurCCtassmrshorteccenxxxxxD}{\ensuremath{15.7}}      
\newcommand{\hatcurCCtassmieccenxxxxxD}{\ensuremath{15.212\pm0.010}} 
\newcommand{\hatcurCCtassmishorteccenxxxxxD}{\ensuremath{15.2}}      
\newcommand{\hatcurCCparallaxeccenxxxxxD}{\ensuremath{2.265\pm0.056}} 
\newcommand{\hatcurCCgaiamGeccenxxxxxD}{\ensuremath{15.7364\pm0.0011}} 
\newcommand{\hatcurCCgaiamBPeccenxxxxxD}{\ensuremath{16.5518\pm0.0054}} 
\newcommand{\hatcurCCgaiamRPeccenxxxxxD}{\ensuremath{14.8547\pm0.0038}} 
\newcommand{\hatcurCCtwomassJmageccenxxxxxD}{\ensuremath{13.779\pm0.029}} 
\newcommand{\hatcurCCtwomassHmageccenxxxxxD}{\ensuremath{13.047\pm0.029}} 
\newcommand{\hatcurCCtwomassKmageccenxxxxxD}{\ensuremath{12.934\pm0.032}} 
\newcommand{\hatcurCCcitJmageccenxxxxxD}{\ensuremath{13.770\pm0.029}} 
\newcommand{\hatcurCCcitHmageccenxxxxxD}{\ensuremath{13.041\pm0.030}} 
\newcommand{\hatcurCCcitKmageccenxxxxxD}{\ensuremath{12.958\pm0.032}} 
\newcommand{\hatcurCCbbJmageccenxxxxxD}{\ensuremath{13.859\pm0.032}} 
\newcommand{\hatcurCCbbHmageccenxxxxxD}{\ensuremath{13.063\pm0.031}} 
\newcommand{\hatcurCCbbKmageccenxxxxxD}{\ensuremath{12.978\pm0.032}} 
\newcommand{\hatcurCCesoJmageccenxxxxxD}{\ensuremath{13.866\pm0.036}} 
\newcommand{\hatcurCCesoHmageccenxxxxxD}{\ensuremath{13.057\pm0.036}} 
\newcommand{\hatcurCCesoKmageccenxxxxxD}{\ensuremath{12.975\pm0.033}} 
\newcommand{\hatcurCCesoJHmageccenxxxxxD}{\ensuremath{0.809\pm0.021}} 
\newcommand{\hatcurCCesoJKmageccenxxxxxD}{\ensuremath{0.892\pm0.048}} 
\newcommand{\hatcurCCesoHKmageccenxxxxxD}{\ensuremath{0.083\pm0.048}} 
\newcommand{\hatcurCCWonemageccenxxxxxD}{\ensuremath{12.847\pm0.024}} 
\newcommand{\hatcurCCWtwomageccenxxxxxD}{\ensuremath{12.870\pm0.027}} 
\newcommand{\hatcurCCWthreemageccenxxxxxD}{\ensuremath{nff\pmnff}}   
\newcommand{\hatcurCCWfourmageccenxxxxxD}{\ensuremath{nff\pmnff}}    
\newcommand{\hatcurLCdipeccenxxxxxD}{\ensuremath{41.4}}              
\newcommand{\hatcurLCrprstareccenxxxxxD}{\ensuremath{0.1862\pm0.0021}} 
\newcommand{\hatcurLCbsqeccenxxxxxD}{\ensuremath{0.095_{-0.041}^{+0.033}}} 
\newcommand{\hatcurLCimpeccenxxxxxD}{\ensuremath{0.308_{-0.076}^{+0.049}}} 
\newcommand{\hatcurLCzetaeccenxxxxxD}{\ensuremath{26.02\pm0.24}}     
\newcommand{\hatcurLCdureccenxxxxxD}{\ensuremath{0.09256\pm0.00072}} 
\newcommand{\hatcurLCdurshorteccenxxxxxD}{\ensuremath{0.0926}}       
\newcommand{\hatcurLCdurhreccenxxxxxD}{\ensuremath{2.221\pm0.017}}   
\newcommand{\hatcurLCdurhrshorteccenxxxxxD}{\ensuremath{2.221}}      
\newcommand{\hatcurLCqeccenxxxxxD}{\ensuremath{0.03000\pm0.00024}}   
\newcommand{\hatcurLCqshorteccenxxxxxD}{\ensuremath{0.030}}          
\newcommand{\hatcurLCingdureccenxxxxxD}{\ensuremath{0.01585\pm0.00068}} 
\newcommand{\hatcurLCPeccenxxxxxD}{\ensuremath{3.0876261\pm0.0000016}} 
\newcommand{\hatcurLCPprececcenxxxxxD}{\ensuremath{3.0876261}}       
\newcommand{\hatcurLCPshorteccenxxxxxD}{\ensuremath{3.0876}}         
\newcommand{\hatcurLCTeccenxxxxxD}{\ensuremath{2459164.48242\pm0.00020}} 
\newcommand{\hatcurLCTAeccenxxxxxD}{\ensuremath{2455573.5733\pm0.0019}} 
\newcommand{\hatcurLCTBeccenxxxxxD}{\ensuremath{2459220.05968\pm0.00021}} 
\newcommand{\hatcurLCrhoeccenxxxxxD}{\ensuremath{3.54\pm0.15}}       
\newcommand{\hatcurSMEiteffeccenxxxxxD}{\ensuremath{4082\pm69}}      
\newcommand{\hatcurSMEizfeheccenxxxxxD}{\ensuremath{-0.12\pm0.10}}   
\newcommand{\hatcurSMEizfehshorteccenxxxxxD}{\ensuremath{-0.12}}     
\newcommand{\hatcurSMEiloggeccenxxxxxD}{\ensuremath{4.50\pm0.50}}    
\newcommand{\hatcurSMEivsineccenxxxxxD}{\ensuremath{0\pm50}}         
\newcommand{\hatcurSMEivmaceccenxxxxxD}{\ensuremath{nff\pmnff}}      
\newcommand{\hatcurSMEivmiceccenxxxxxD}{\ensuremath{nff\pmnff}}      
\newcommand{\hatcurextraerrMJeccenxxxxxD}{\ensuremath{0\pm0}}        
\newcommand{\hatcurextraerrMJtwosiglimeccenxxxxxD}{\ensuremath{<0.0200}} 
\newcommand{\hatcurextraerrMHeccenxxxxxD}{\ensuremath{0\pm0}}        
\newcommand{\hatcurextraerrMHtwosiglimeccenxxxxxD}{\ensuremath{<0.0200}} 
\newcommand{\hatcurextraerrMKseccenxxxxxD}{\ensuremath{0\pm0}}       
\newcommand{\hatcurextraerrMKstwosiglimeccenxxxxxD}{\ensuremath{<0.0200}} 
\newcommand{\hatcurextraerrMGeccenxxxxxD}{\ensuremath{0\pm0}}        
\newcommand{\hatcurextraerrMGtwosiglimeccenxxxxxD}{\ensuremath{<0.0200}} 
\newcommand{\hatcurextraerrMBPtwoeccenxxxxxD}{\ensuremath{0\pm0}}    
\newcommand{\hatcurextraerrMBPtwotwosiglimeccenxxxxxD}{\ensuremath{<0.0200}} 
\newcommand{\hatcurextraerrMRPeccenxxxxxD}{\ensuremath{0\pm0}}       
\newcommand{\hatcurextraerrMRPtwosiglimeccenxxxxxD}{\ensuremath{<0.0200}} 
\newcommand{\hatcurextraerrMgeccenxxxxxD}{\ensuremath{0\pm0}}        
\newcommand{\hatcurextraerrMgtwosiglimeccenxxxxxD}{\ensuremath{<0.0200}} 
\newcommand{\hatcurextraerrMreccenxxxxxD}{\ensuremath{0\pm0}}        
\newcommand{\hatcurextraerrMrtwosiglimeccenxxxxxD}{\ensuremath{<0.0200}} 
\newcommand{\hatcurextraerrMieccenxxxxxD}{\ensuremath{0\pm0}}        
\newcommand{\hatcurextraerrMitwosiglimeccenxxxxxD}{\ensuremath{<0.0200}} 
\newcommand{\hatcurextraerrMWoneeccenxxxxxD}{\ensuremath{0\pm0}}     
\newcommand{\hatcurextraerrMWonetwosiglimeccenxxxxxD}{\ensuremath{<0.0200}} 
\newcommand{\hatcurextraerrMWtwoeccenxxxxxD}{\ensuremath{0\pm0}}     
\newcommand{\hatcurextraerrMWtwotwosiglimeccenxxxxxD}{\ensuremath{<0.0200}} 
\newcommand{\hatcurLBiBeccenxxxxxD}{\ensuremath{0.7848}}             
\newcommand{\hatcurLBiiBeccenxxxxxD}{\ensuremath{0.0472}}            
\newcommand{\hatcurLBiVeccenxxxxxD}{\ensuremath{0.6317}}             
\newcommand{\hatcurLBiiVeccenxxxxxD}{\ensuremath{0.1535}}            
\newcommand{\hatcurLBiReccenxxxxxD}{\ensuremath{0.5264}}             
\newcommand{\hatcurLBiiReccenxxxxxD}{\ensuremath{0.1979}}            
\newcommand{\hatcurLBiIeccenxxxxxD}{\ensuremath{0.3783}}             
\newcommand{\hatcurLBiiIeccenxxxxxD}{\ensuremath{0.2304}}            
\newcommand{\hatcurLBiueccenxxxxxD}{\ensuremath{0.8436}}             
\newcommand{\hatcurLBiiueccenxxxxxD}{\ensuremath{-0.0118}}           
\newcommand{\hatcurLBigeccenxxxxxD}{\ensuremath{0.50\pm0.12}}        
\newcommand{\hatcurLBiigeccenxxxxxD}{\ensuremath{0.37\pm0.15}}       
\newcommand{\hatcurLBireccenxxxxxD}{\ensuremath{0.39\pm0.12}}        
\newcommand{\hatcurLBiireccenxxxxxD}{\ensuremath{0.31\pm0.17}}       
\newcommand{\hatcurLBiieccenxxxxxD}{\ensuremath{0.33\pm0.11}}        
\newcommand{\hatcurLBiiieccenxxxxxD}{\ensuremath{0.23\pm0.18}}       
\newcommand{\hatcurLBizeccenxxxxxD}{\ensuremath{0.25\pm0.11}}        
\newcommand{\hatcurLBiizeccenxxxxxD}{\ensuremath{0.23\pm0.16}}       
\newcommand{\hatcurLBiJeccenxxxxxD}{\ensuremath{0.2428}}             
\newcommand{\hatcurLBiiJeccenxxxxxD}{\ensuremath{0.2237}}            
\newcommand{\hatcurLBiHeccenxxxxxD}{\ensuremath{0.1724}}             
\newcommand{\hatcurLBiiHeccenxxxxxD}{\ensuremath{0.2648}}            
\newcommand{\hatcurLBiKeccenxxxxxD}{\ensuremath{0.1300}}             
\newcommand{\hatcurLBiiKeccenxxxxxD}{\ensuremath{0.2371}}            
\newcommand{\hatcurLBiTeccenxxxxxD}{\ensuremath{0.53\pm0.15}}        
\newcommand{\hatcurLBiiTeccenxxxxxD}{\ensuremath{0.38\pm0.15}}       
\newcommand{\hatcurLBikepeccenxxxxxD}{\ensuremath{0.5067}}           
\newcommand{\hatcurLBiikepeccenxxxxxD}{\ensuremath{0.2326}}          
\newcommand{\hatcurLBiCeccenxxxxxD}{\ensuremath{0.4758}}             
\newcommand{\hatcurLBiiCeccenxxxxxD}{\ensuremath{0.2392}}            
\newcommand{\hatcurLBiMeccenxxxxxD}{\ensuremath{0.5991}}             
\newcommand{\hatcurLBiiMeccenxxxxxD}{\ensuremath{0.1948}}            
\newcommand{\hatcurLBiSoneeccenxxxxxD}{\ensuremath{0.0945}}            
\newcommand{\hatcurLBiiSoneeccenxxxxxD}{\ensuremath{0.1746}}           
\newcommand{\hatcurLBiStwoeccenxxxxxD}{\ensuremath{0.0810}}            
\newcommand{\hatcurLBiiStwoeccenxxxxxD}{\ensuremath{0.1431}}           
\newcommand{\hatcurLBiSthreeeccenxxxxxD}{\ensuremath{0.0688}}            
\newcommand{\hatcurLBiiSthreeeccenxxxxxD}{\ensuremath{0.1236}}           
\newcommand{\hatcurLBiSfoureccenxxxxxD}{\ensuremath{0.0659}}            
\newcommand{\hatcurLBiiSfoureccenxxxxxD}{\ensuremath{0.1065}}           
\newcommand{\hatcurISOmeccenxxxxxD}{\ensuremath{0.661\pm0.011}}      
\newcommand{\hatcurISOmshorteccenxxxxxD}{\ensuremath{0.66}}          
\newcommand{\hatcurISOmlongeccenxxxxxD}{\ensuremath{0.661\pm0.011}}  
\newcommand{\hatcurISOreccenxxxxxD}{\ensuremath{0.6412\pm0.0068}}    
\newcommand{\hatcurISOrshorteccenxxxxxD}{\ensuremath{0.64}}          
\newcommand{\hatcurISOrlongeccenxxxxxD}{\ensuremath{0.6412\pm0.0068}} 
\newcommand{\hatcurISOrhoeccenxxxxxD}{\ensuremath{3.54\pm0.15}}      
\newcommand{\hatcurISOrholongeccenxxxxxD}{\ensuremath{3.54\pm0.15}}  
\newcommand{\hatcurISOloggeccenxxxxxD}{\ensuremath{4.644\pm0.015}}   
\newcommand{\hatcurISOlumeccenxxxxxD}{\ensuremath{0.1020\pm0.0026}}  
\newcommand{\hatcurISOlumshorteccenxxxxxD}{\ensuremath{0.10}}        
\newcommand{\hatcurISOteffeccenxxxxxD}{\ensuremath{4075\pm11}}       
\newcommand{\hatcurISOzfeheccenxxxxxD}{\ensuremath{0.253_{-0.011}^{+0.019}}} 
\newcommand{\hatcurISOageeccenxxxxxD}{\ensuremath{9.6\pm5.2}}        
\newcommand{\hatcurISOspececcenxxxxxD}{K}                            
\newcommand{\hatcurRVKeccenxxxxxD}{\ensuremath{251\pm18}}            
\newcommand{\hatcurRVKtwosiglimeccenxxxxxD}{\ensuremath{<276.0}}     
\newcommand{\hatcurRVrkeccenxxxxxD}{\ensuremath{0.064_{-0.104}^{+0.078}}} 
\newcommand{\hatcurRVrheccenxxxxxD}{\ensuremath{0.029_{-0.115}^{+0.083}}} 
\newcommand{\hatcurRVkeccenxxxxxD}{\ensuremath{0.0062_{-0.0097}^{+0.0173}}} 
\newcommand{\hatcurRVheccenxxxxxD}{\ensuremath{0.003\pm0.015}}       
\newcommand{\hatcurRVtroneeccenxxxxxD}{\ensuremath{0\pm0}}           
\newcommand{\hatcurRVtrtwoeccenxxxxxD}{\ensuremath{0\pm0}}           
\newcommand{\hatcurRVgammaeccenxxxxxD}{\ensuremath{-7758.1\pm9.7}}   
\newcommand{\hatcurRVjittereccenxxxxxD}{\ensuremath{1\pm17}}         
\newcommand{\hatcurRVjittertwosiglimeccenxxxxxD}{\ensuremath{<40.5}} 
\newcommand{\hatcurRVfitrmseccenxxxxxD}{\ensuremath{.1fym}}          %
\newcommand{\hatcurRVecceneccenxxxxxD}{\ensuremath{0.016\pm0.014}}   
\newcommand{\hatcurRVeccentwosiglimeccenxxxxxD}{\ensuremath{<0.045}} 
\newcommand{\hatcurRVomegaeccenxxxxxD}{\ensuremath{100\pm130}}       
\newcommand{\hatcurPPieccenxxxxxD}{\ensuremath{88.53_{-0.23}^{+0.37}}} 
\newcommand{\hatcurPPgeccenxxxxxD}{\ensuremath{25.2\pm2.1}}          
\newcommand{\hatcurPPloggeccenxxxxxD}{\ensuremath{3.401\pm0.039}}    
\newcommand{\hatcurPPareccenxxxxxD}{\ensuremath{12.13\pm0.18}}       
\newcommand{\hatcurPPareleccenxxxxxD}{\ensuremath{0.03616\pm0.00021}} 
\newcommand{\hatcurPPrhoeccenxxxxxD}{\ensuremath{1.08\pm0.10}}       
\newcommand{\hatcurPPmeccenxxxxxD}{\ensuremath{1.37\pm0.10}}         
\newcommand{\hatcurPPmtwosiglimeccenxxxxxD}{\ensuremath{<1.50}}      
\newcommand{\hatcurPPmshorteccenxxxxxD}{\ensuremath{1.37}}           
\newcommand{\hatcurPPmlongeccenxxxxxD}{\ensuremath{1.37\pm0.10}}     
\newcommand{\hatcurPPmeeccenxxxxxD}{\ensuremath{435\pm32}}           
\newcommand{\hatcurPPmeshorteccenxxxxxD}{\ensuremath{434.8}}         
\newcommand{\hatcurPPmelongeccenxxxxxD}{\ensuremath{435\pm32}}       
\newcommand{\hatcurPPreccenxxxxxD}{\ensuremath{1.161\pm0.019}}       
\newcommand{\hatcurPPrshorteccenxxxxxD}{\ensuremath{1.16}}           
\newcommand{\hatcurPPrlongeccenxxxxxD}{\ensuremath{1.161\pm0.019}}   
\newcommand{\hatcurPPreeccenxxxxxD}{\ensuremath{13.01\pm0.22}}       
\newcommand{\hatcurPPreshorteccenxxxxxD}{\ensuremath{13.0}}          
\newcommand{\hatcurPPrelongeccenxxxxxD}{\ensuremath{13.01\pm0.22}}   
\newcommand{\hatcurPPmrcorreccenxxxxxD}{\ensuremath{-0.11}}          
\newcommand{\hatcurPPteffeccenxxxxxD}{\ensuremath{827.1\pm6.1}}      
\newcommand{\hatcurPPthetaeccenxxxxxD}{\ensuremath{0.1286\pm0.0096}} 
\newcommand{\hatcurPPfluxperieccenxxxxxD}{\ensuremath{1.097_{-0.035}^{+0.048}}} 
\newcommand{\hatcurPPfluxperidimeccenxxxxxD}{\ensuremath{8}}         
\newcommand{\hatcurPPfluxapeccenxxxxxD}{\ensuremath{1.024\pm0.042}}  
\newcommand{\hatcurPPfluxapdimeccenxxxxxD}{\ensuremath{8}}           
\newcommand{\hatcurPPfluxavgeccenxxxxxD}{\ensuremath{1.061\pm0.031}} 
\newcommand{\hatcurPPfluxavgdimeccenxxxxxD}{\ensuremath{8}}          
\newcommand{\hatcurPPfluxavglogeccenxxxxxD}{\ensuremath{8.026\pm0.013}} 
\newcommand{\hatcurXsecphaseeccenxxxxxD}{\ensuremath{0.504\pm0.010}} 
\newcommand{\hatcurXsecondaryeccenxxxxxD}{\ensuremath{2459166.039\pm0.031}} 
\newcommand{\hatcurXsecdureccenxxxxxD}{\ensuremath{0.0932\pm0.0023}} 
\newcommand{\hatcurXsecingdureccenxxxxxD}{\ensuremath{0.01592\pm0.00061}} 
\newcommand{\hatcurPPphiconjeccenxxxxxD}{\ensuremath{0.14_{-0.34}^{+0.20}}} 
\newcommand{\hatcurPPperieccenxxxxxD}{\ensuremath{2459164.04\pm0.78}} 
\newcommand{\hatcurPPaequiveccenxxxxxD}{\ensuremath{0.1133\pm0.0017}} 
\newcommand{\hatcurPPtcirceccenxxxxxD}{\ensuremath{197\pm25}}        
\newcommand{\hatcurPPtinfalleccenxxxxxD}{\ensuremath{3730\pm440}}    
\newcommand{\hatcurXdisteccenxxxxxD}{\ensuremath{414.5\pm5.7}}       
\newcommand{\hatcurXAveccenxxxxxD}{\ensuremath{0.1140_{-0.0150}^{+0.0090}}} 
\newcommand{\hatcurXdistredeccenxxxxxD}{\ensuremath{414.5\pm5.7}}    
\newcommand{\hatcurXEBVeccenxxxxxD}{\ensuremath{0.0370_{-0.0050}^{+0.0030}}} 
\newcommand{\hatcurCCpmraeccenxxxxxD}{\ensuremath{-24.64\pm0.10}}    
\newcommand{\hatcurCCpmdececcenxxxxxD}{\ensuremath{-8.286\pm0.098}}  
\newcommand{\hatcurCCpmeccenxxxxxD}{\ensuremath{25.99\pm0.14}}       

\newcommand{\hatcurCCbbHmageccen}[1]{\ifnum#1=74 %
\hatcurCCbbHmageccenxxxxxA
\else
\ifnum#1=75 %
\hatcurCCbbHmageccenxxxxxB
\else
\ifnum#1=76 %
\hatcurCCbbHmageccenxxxxxC
\else
\ifnum#1=77 %
\hatcurCCbbHmageccenxxxxxD
\else
??????\fi
\fi
\fi
\fi
}
\newcommand{\hatcurCCbbJmageccen}[1]{\ifnum#1=74 %
\hatcurCCbbJmageccenxxxxxA
\else
\ifnum#1=75 %
\hatcurCCbbJmageccenxxxxxB
\else
\ifnum#1=76 %
\hatcurCCbbJmageccenxxxxxC
\else
\ifnum#1=77 %
\hatcurCCbbJmageccenxxxxxD
\else
??????\fi
\fi
\fi
\fi
}
\newcommand{\hatcurCCbbKmageccen}[1]{\ifnum#1=74 %
\hatcurCCbbKmageccenxxxxxA
\else
\ifnum#1=75 %
\hatcurCCbbKmageccenxxxxxB
\else
\ifnum#1=76 %
\hatcurCCbbKmageccenxxxxxC
\else
\ifnum#1=77 %
\hatcurCCbbKmageccenxxxxxD
\else
??????\fi
\fi
\fi
\fi
}
\newcommand{\hatcurCCcitHmageccen}[1]{\ifnum#1=74 %
\hatcurCCcitHmageccenxxxxxA
\else
\ifnum#1=75 %
\hatcurCCcitHmageccenxxxxxB
\else
\ifnum#1=76 %
\hatcurCCcitHmageccenxxxxxC
\else
\ifnum#1=77 %
\hatcurCCcitHmageccenxxxxxD
\else
??????\fi
\fi
\fi
\fi
}
\newcommand{\hatcurCCcitJmageccen}[1]{\ifnum#1=74 %
\hatcurCCcitJmageccenxxxxxA
\else
\ifnum#1=75 %
\hatcurCCcitJmageccenxxxxxB
\else
\ifnum#1=76 %
\hatcurCCcitJmageccenxxxxxC
\else
\ifnum#1=77 %
\hatcurCCcitJmageccenxxxxxD
\else
??????\fi
\fi
\fi
\fi
}
\newcommand{\hatcurCCcitKmageccen}[1]{\ifnum#1=74 %
\hatcurCCcitKmageccenxxxxxA
\else
\ifnum#1=75 %
\hatcurCCcitKmageccenxxxxxB
\else
\ifnum#1=76 %
\hatcurCCcitKmageccenxxxxxC
\else
\ifnum#1=77 %
\hatcurCCcitKmageccenxxxxxD
\else
??????\fi
\fi
\fi
\fi
}
\newcommand{\hatcurCCdececcen}[1]{\ifnum#1=74 %
\hatcurCCdececcenxxxxxA
\else
\ifnum#1=75 %
\hatcurCCdececcenxxxxxB
\else
\ifnum#1=76 %
\hatcurCCdececcenxxxxxC
\else
\ifnum#1=77 %
\hatcurCCdececcenxxxxxD
\else
??????\fi
\fi
\fi
\fi
}
\newcommand{\hatcurCCesoHKmageccen}[1]{\ifnum#1=74 %
\hatcurCCesoHKmageccenxxxxxA
\else
\ifnum#1=75 %
\hatcurCCesoHKmageccenxxxxxB
\else
\ifnum#1=76 %
\hatcurCCesoHKmageccenxxxxxC
\else
\ifnum#1=77 %
\hatcurCCesoHKmageccenxxxxxD
\else
??????\fi
\fi
\fi
\fi
}
\newcommand{\hatcurCCesoHmageccen}[1]{\ifnum#1=74 %
\hatcurCCesoHmageccenxxxxxA
\else
\ifnum#1=75 %
\hatcurCCesoHmageccenxxxxxB
\else
\ifnum#1=76 %
\hatcurCCesoHmageccenxxxxxC
\else
\ifnum#1=77 %
\hatcurCCesoHmageccenxxxxxD
\else
??????\fi
\fi
\fi
\fi
}
\newcommand{\hatcurCCesoJHmageccen}[1]{\ifnum#1=74 %
\hatcurCCesoJHmageccenxxxxxA
\else
\ifnum#1=75 %
\hatcurCCesoJHmageccenxxxxxB
\else
\ifnum#1=76 %
\hatcurCCesoJHmageccenxxxxxC
\else
\ifnum#1=77 %
\hatcurCCesoJHmageccenxxxxxD
\else
??????\fi
\fi
\fi
\fi
}
\newcommand{\hatcurCCesoJKmageccen}[1]{\ifnum#1=74 %
\hatcurCCesoJKmageccenxxxxxA
\else
\ifnum#1=75 %
\hatcurCCesoJKmageccenxxxxxB
\else
\ifnum#1=76 %
\hatcurCCesoJKmageccenxxxxxC
\else
\ifnum#1=77 %
\hatcurCCesoJKmageccenxxxxxD
\else
??????\fi
\fi
\fi
\fi
}
\newcommand{\hatcurCCesoJmageccen}[1]{\ifnum#1=74 %
\hatcurCCesoJmageccenxxxxxA
\else
\ifnum#1=75 %
\hatcurCCesoJmageccenxxxxxB
\else
\ifnum#1=76 %
\hatcurCCesoJmageccenxxxxxC
\else
\ifnum#1=77 %
\hatcurCCesoJmageccenxxxxxD
\else
??????\fi
\fi
\fi
\fi
}
\newcommand{\hatcurCCesoKmageccen}[1]{\ifnum#1=74 %
\hatcurCCesoKmageccenxxxxxA
\else
\ifnum#1=75 %
\hatcurCCesoKmageccenxxxxxB
\else
\ifnum#1=76 %
\hatcurCCesoKmageccenxxxxxC
\else
\ifnum#1=77 %
\hatcurCCesoKmageccenxxxxxD
\else
??????\fi
\fi
\fi
\fi
}
\newcommand{\hatcurCCgaiadrtwoeccen}[1]{\ifnum#1=74 %
\hatcurCCgaiadrtwoeccenxxxxxA
\else
\ifnum#1=75 %
\hatcurCCgaiadrtwoeccenxxxxxB
\else
\ifnum#1=76 %
\hatcurCCgaiadrtwoeccenxxxxxC
\else
\ifnum#1=77 %
\hatcurCCgaiadrtwoeccenxxxxxD
\else
??????\fi
\fi
\fi
\fi
}
\newcommand{\hatcurCCgaiaeccen}[1]{\ifnum#1=74 %
\hatcurCCgaiaeccenxxxxxA
\else
\ifnum#1=75 %
\hatcurCCgaiaeccenxxxxxB
\else
\ifnum#1=76 %
\hatcurCCgaiaeccenxxxxxC
\else
\ifnum#1=77 %
\hatcurCCgaiaeccenxxxxxD
\else
??????\fi
\fi
\fi
\fi
}
\newcommand{\hatcurCCgaiamBPeccen}[1]{\ifnum#1=74 %
\hatcurCCgaiamBPeccenxxxxxA
\else
\ifnum#1=75 %
\hatcurCCgaiamBPeccenxxxxxB
\else
\ifnum#1=76 %
\hatcurCCgaiamBPeccenxxxxxC
\else
\ifnum#1=77 %
\hatcurCCgaiamBPeccenxxxxxD
\else
??????\fi
\fi
\fi
\fi
}
\newcommand{\hatcurCCgaiamGeccen}[1]{\ifnum#1=74 %
\hatcurCCgaiamGeccenxxxxxA
\else
\ifnum#1=75 %
\hatcurCCgaiamGeccenxxxxxB
\else
\ifnum#1=76 %
\hatcurCCgaiamGeccenxxxxxC
\else
\ifnum#1=77 %
\hatcurCCgaiamGeccenxxxxxD
\else
??????\fi
\fi
\fi
\fi
}
\newcommand{\hatcurCCgaiamRPeccen}[1]{\ifnum#1=74 %
\hatcurCCgaiamRPeccenxxxxxA
\else
\ifnum#1=75 %
\hatcurCCgaiamRPeccenxxxxxB
\else
\ifnum#1=76 %
\hatcurCCgaiamRPeccenxxxxxC
\else
\ifnum#1=77 %
\hatcurCCgaiamRPeccenxxxxxD
\else
??????\fi
\fi
\fi
\fi
}
\newcommand{\hatcurCCgsceccen}[1]{\ifnum#1=74 %
\hatcurCCgsceccenxxxxxA
\else
\ifnum#1=75 %
\hatcurCCgsceccenxxxxxB
\else
\ifnum#1=76 %
\hatcurCCgsceccenxxxxxC
\else
\ifnum#1=77 %
\hatcurCCgsceccenxxxxxD
\else
??????\fi
\fi
\fi
\fi
}
\newcommand{\hatcurCCmageccen}[1]{\ifnum#1=74 %
\hatcurCCmageccenxxxxxA
\else
\ifnum#1=75 %
\hatcurCCmageccenxxxxxB
\else
\ifnum#1=76 %
\hatcurCCmageccenxxxxxC
\else
\ifnum#1=77 %
\hatcurCCmageccenxxxxxD
\else
??????\fi
\fi
\fi
\fi
}
\newcommand{\hatcurCCparallaxeccen}[1]{\ifnum#1=74 %
\hatcurCCparallaxeccenxxxxxA
\else
\ifnum#1=75 %
\hatcurCCparallaxeccenxxxxxB
\else
\ifnum#1=76 %
\hatcurCCparallaxeccenxxxxxC
\else
\ifnum#1=77 %
\hatcurCCparallaxeccenxxxxxD
\else
??????\fi
\fi
\fi
\fi
}
\newcommand{\hatcurCCpmdececcen}[1]{\ifnum#1=74 %
\hatcurCCpmdececcenxxxxxA
\else
\ifnum#1=75 %
\hatcurCCpmdececcenxxxxxB
\else
\ifnum#1=76 %
\hatcurCCpmdececcenxxxxxC
\else
\ifnum#1=77 %
\hatcurCCpmdececcenxxxxxD
\else
??????\fi
\fi
\fi
\fi
}
\newcommand{\hatcurCCpmeccen}[1]{\ifnum#1=74 %
\hatcurCCpmeccenxxxxxA
\else
\ifnum#1=75 %
\hatcurCCpmeccenxxxxxB
\else
\ifnum#1=76 %
\hatcurCCpmeccenxxxxxC
\else
\ifnum#1=77 %
\hatcurCCpmeccenxxxxxD
\else
??????\fi
\fi
\fi
\fi
}
\newcommand{\hatcurCCpmraeccen}[1]{\ifnum#1=74 %
\hatcurCCpmraeccenxxxxxA
\else
\ifnum#1=75 %
\hatcurCCpmraeccenxxxxxB
\else
\ifnum#1=76 %
\hatcurCCpmraeccenxxxxxC
\else
\ifnum#1=77 %
\hatcurCCpmraeccenxxxxxD
\else
??????\fi
\fi
\fi
\fi
}
\newcommand{\hatcurCCraeccen}[1]{\ifnum#1=74 %
\hatcurCCraeccenxxxxxA
\else
\ifnum#1=75 %
\hatcurCCraeccenxxxxxB
\else
\ifnum#1=76 %
\hatcurCCraeccenxxxxxC
\else
\ifnum#1=77 %
\hatcurCCraeccenxxxxxD
\else
??????\fi
\fi
\fi
\fi
}
\newcommand{\hatcurCCtassmBeccen}[1]{\ifnum#1=74 %
\hatcurCCtassmBeccenxxxxxA
\else
\ifnum#1=75 %
\hatcurCCtassmBeccenxxxxxB
\else
\ifnum#1=76 %
\hatcurCCtassmBeccenxxxxxC
\else
\ifnum#1=77 %
\hatcurCCtassmBeccenxxxxxD
\else
??????\fi
\fi
\fi
\fi
}
\newcommand{\hatcurCCtassmBshorteccen}[1]{\ifnum#1=74 %
\hatcurCCtassmBshorteccenxxxxxA
\else
\ifnum#1=75 %
\hatcurCCtassmBshorteccenxxxxxB
\else
\ifnum#1=76 %
\hatcurCCtassmBshorteccenxxxxxC
\else
\ifnum#1=77 %
\hatcurCCtassmBshorteccenxxxxxD
\else
??????\fi
\fi
\fi
\fi
}
\newcommand{\hatcurCCtassmgeccen}[1]{\ifnum#1=74 %
\hatcurCCtassmgeccenxxxxxA
\else
\ifnum#1=75 %
\hatcurCCtassmgeccenxxxxxB
\else
\ifnum#1=76 %
\hatcurCCtassmgeccenxxxxxC
\else
\ifnum#1=77 %
\hatcurCCtassmgeccenxxxxxD
\else
??????\fi
\fi
\fi
\fi
}
\newcommand{\hatcurCCtassmgshorteccen}[1]{\ifnum#1=74 %
\hatcurCCtassmgshorteccenxxxxxA
\else
\ifnum#1=75 %
\hatcurCCtassmgshorteccenxxxxxB
\else
\ifnum#1=76 %
\hatcurCCtassmgshorteccenxxxxxC
\else
\ifnum#1=77 %
\hatcurCCtassmgshorteccenxxxxxD
\else
??????\fi
\fi
\fi
\fi
}
\newcommand{\hatcurCCtassmieccen}[1]{\ifnum#1=74 %
\hatcurCCtassmieccenxxxxxA
\else
\ifnum#1=75 %
\hatcurCCtassmieccenxxxxxB
\else
\ifnum#1=76 %
\hatcurCCtassmieccenxxxxxC
\else
\ifnum#1=77 %
\hatcurCCtassmieccenxxxxxD
\else
??????\fi
\fi
\fi
\fi
}
\newcommand{\hatcurCCtassmIeccen}[1]{\ifnum#1=74 %
\hatcurCCtassmIeccenxxxxxA
\else
\ifnum#1=75 %
\hatcurCCtassmIeccenxxxxxB
\else
\ifnum#1=76 %
\hatcurCCtassmIeccenxxxxxC
\else
\ifnum#1=77 %
\hatcurCCtassmIeccenxxxxxD
\else
??????\fi
\fi
\fi
\fi
}
\newcommand{\hatcurCCtassmishorteccen}[1]{\ifnum#1=74 %
\hatcurCCtassmishorteccenxxxxxA
\else
\ifnum#1=75 %
\hatcurCCtassmishorteccenxxxxxB
\else
\ifnum#1=76 %
\hatcurCCtassmishorteccenxxxxxC
\else
\ifnum#1=77 %
\hatcurCCtassmishorteccenxxxxxD
\else
??????\fi
\fi
\fi
\fi
}
\newcommand{\hatcurCCtassmIshorteccen}[1]{\ifnum#1=74 %
\hatcurCCtassmIshorteccenxxxxxA
\else
\ifnum#1=75 %
\hatcurCCtassmIshorteccenxxxxxB
\else
\ifnum#1=76 %
\hatcurCCtassmIshorteccenxxxxxC
\else
\ifnum#1=77 %
\hatcurCCtassmIshorteccenxxxxxD
\else
??????\fi
\fi
\fi
\fi
}
\newcommand{\hatcurCCtassmreccen}[1]{\ifnum#1=74 %
\hatcurCCtassmreccenxxxxxA
\else
\ifnum#1=75 %
\hatcurCCtassmreccenxxxxxB
\else
\ifnum#1=76 %
\hatcurCCtassmreccenxxxxxC
\else
\ifnum#1=77 %
\hatcurCCtassmreccenxxxxxD
\else
??????\fi
\fi
\fi
\fi
}
\newcommand{\hatcurCCtassmrshorteccen}[1]{\ifnum#1=74 %
\hatcurCCtassmrshorteccenxxxxxA
\else
\ifnum#1=75 %
\hatcurCCtassmrshorteccenxxxxxB
\else
\ifnum#1=76 %
\hatcurCCtassmrshorteccenxxxxxC
\else
\ifnum#1=77 %
\hatcurCCtassmrshorteccenxxxxxD
\else
??????\fi
\fi
\fi
\fi
}
\newcommand{\hatcurCCtassmveccen}[1]{\ifnum#1=74 %
\hatcurCCtassmveccenxxxxxA
\else
\ifnum#1=75 %
\hatcurCCtassmveccenxxxxxB
\else
\ifnum#1=76 %
\hatcurCCtassmveccenxxxxxC
\else
\ifnum#1=77 %
\hatcurCCtassmveccenxxxxxD
\else
??????\fi
\fi
\fi
\fi
}
\newcommand{\hatcurCCtassmvshorteccen}[1]{\ifnum#1=74 %
\hatcurCCtassmvshorteccenxxxxxA
\else
\ifnum#1=75 %
\hatcurCCtassmvshorteccenxxxxxB
\else
\ifnum#1=76 %
\hatcurCCtassmvshorteccenxxxxxC
\else
\ifnum#1=77 %
\hatcurCCtassmvshorteccenxxxxxD
\else
??????\fi
\fi
\fi
\fi
}
\newcommand{\hatcurCCtwomasseccen}[1]{\ifnum#1=74 %
\hatcurCCtwomasseccenxxxxxA
\else
\ifnum#1=75 %
\hatcurCCtwomasseccenxxxxxB
\else
\ifnum#1=76 %
\hatcurCCtwomasseccenxxxxxC
\else
\ifnum#1=77 %
\hatcurCCtwomasseccenxxxxxD
\else
??????\fi
\fi
\fi
\fi
}
\newcommand{\hatcurCCtwomassHmageccen}[1]{\ifnum#1=74 %
\hatcurCCtwomassHmageccenxxxxxA
\else
\ifnum#1=75 %
\hatcurCCtwomassHmageccenxxxxxB
\else
\ifnum#1=76 %
\hatcurCCtwomassHmageccenxxxxxC
\else
\ifnum#1=77 %
\hatcurCCtwomassHmageccenxxxxxD
\else
??????\fi
\fi
\fi
\fi
}
\newcommand{\hatcurCCtwomassJmageccen}[1]{\ifnum#1=74 %
\hatcurCCtwomassJmageccenxxxxxA
\else
\ifnum#1=75 %
\hatcurCCtwomassJmageccenxxxxxB
\else
\ifnum#1=76 %
\hatcurCCtwomassJmageccenxxxxxC
\else
\ifnum#1=77 %
\hatcurCCtwomassJmageccenxxxxxD
\else
??????\fi
\fi
\fi
\fi
}
\newcommand{\hatcurCCtwomassKmageccen}[1]{\ifnum#1=74 %
\hatcurCCtwomassKmageccenxxxxxA
\else
\ifnum#1=75 %
\hatcurCCtwomassKmageccenxxxxxB
\else
\ifnum#1=76 %
\hatcurCCtwomassKmageccenxxxxxC
\else
\ifnum#1=77 %
\hatcurCCtwomassKmageccenxxxxxD
\else
??????\fi
\fi
\fi
\fi
}
\newcommand{\hatcurCCWfourmageccen}[1]{\ifnum#1=74 %
\hatcurCCWfourmageccenxxxxxA
\else
\ifnum#1=75 %
\hatcurCCWfourmageccenxxxxxB
\else
\ifnum#1=76 %
\hatcurCCWfourmageccenxxxxxC
\else
\ifnum#1=77 %
\hatcurCCWfourmageccenxxxxxD
\else
??????\fi
\fi
\fi
\fi
}
\newcommand{\hatcurCCWonemageccen}[1]{\ifnum#1=74 %
\hatcurCCWonemageccenxxxxxA
\else
\ifnum#1=75 %
\hatcurCCWonemageccenxxxxxB
\else
\ifnum#1=76 %
\hatcurCCWonemageccenxxxxxC
\else
\ifnum#1=77 %
\hatcurCCWonemageccenxxxxxD
\else
??????\fi
\fi
\fi
\fi
}
\newcommand{\hatcurCCWthreemageccen}[1]{\ifnum#1=74 %
\hatcurCCWthreemageccenxxxxxA
\else
\ifnum#1=75 %
\hatcurCCWthreemageccenxxxxxB
\else
\ifnum#1=76 %
\hatcurCCWthreemageccenxxxxxC
\else
\ifnum#1=77 %
\hatcurCCWthreemageccenxxxxxD
\else
??????\fi
\fi
\fi
\fi
}
\newcommand{\hatcurCCWtwomageccen}[1]{\ifnum#1=74 %
\hatcurCCWtwomageccenxxxxxA
\else
\ifnum#1=75 %
\hatcurCCWtwomageccenxxxxxB
\else
\ifnum#1=76 %
\hatcurCCWtwomageccenxxxxxC
\else
\ifnum#1=77 %
\hatcurCCWtwomageccenxxxxxD
\else
??????\fi
\fi
\fi
\fi
}
\newcommand{\hatcurextraerrMBPtwoeccen}[1]{\ifnum#1=75 %
\hatcurextraerrMBPtwoeccenxxxxxB
\else
\ifnum#1=76 %
\hatcurextraerrMBPtwoeccenxxxxxC
\else
\ifnum#1=77 %
\hatcurextraerrMBPtwoeccenxxxxxD
\else
??????\fi
\fi
\fi
}
\newcommand{\hatcurextraerrMBPtwotwosiglimeccen}[1]{\ifnum#1=75 %
\hatcurextraerrMBPtwotwosiglimeccenxxxxxB
\else
\ifnum#1=76 %
\hatcurextraerrMBPtwotwosiglimeccenxxxxxC
\else
\ifnum#1=77 %
\hatcurextraerrMBPtwotwosiglimeccenxxxxxD
\else
??????\fi
\fi
\fi
}
\newcommand{\hatcurextraerrMgeccen}[1]{\ifnum#1=75 %
\hatcurextraerrMgeccenxxxxxB
\else
\ifnum#1=77 %
\hatcurextraerrMgeccenxxxxxD
\else
??????\fi
\fi
}
\newcommand{\hatcurextraerrMGeccen}[1]{\ifnum#1=74 %
\hatcurextraerrMGeccenxxxxxA
\else
\ifnum#1=75 %
\hatcurextraerrMGeccenxxxxxB
\else
\ifnum#1=76 %
\hatcurextraerrMGeccenxxxxxC
\else
\ifnum#1=77 %
\hatcurextraerrMGeccenxxxxxD
\else
??????\fi
\fi
\fi
\fi
}
\newcommand{\hatcurextraerrMgtwosiglimeccen}[1]{\ifnum#1=75 %
\hatcurextraerrMgtwosiglimeccenxxxxxB
\else
\ifnum#1=77 %
\hatcurextraerrMgtwosiglimeccenxxxxxD
\else
??????\fi
\fi
}
\newcommand{\hatcurextraerrMGtwosiglimeccen}[1]{\ifnum#1=74 %
\hatcurextraerrMGtwosiglimeccenxxxxxA
\else
\ifnum#1=75 %
\hatcurextraerrMGtwosiglimeccenxxxxxB
\else
\ifnum#1=76 %
\hatcurextraerrMGtwosiglimeccenxxxxxC
\else
\ifnum#1=77 %
\hatcurextraerrMGtwosiglimeccenxxxxxD
\else
??????\fi
\fi
\fi
\fi
}
\newcommand{\hatcurextraerrMHeccen}[1]{\ifnum#1=74 %
\hatcurextraerrMHeccenxxxxxA
\else
\ifnum#1=75 %
\hatcurextraerrMHeccenxxxxxB
\else
\ifnum#1=76 %
\hatcurextraerrMHeccenxxxxxC
\else
\ifnum#1=77 %
\hatcurextraerrMHeccenxxxxxD
\else
??????\fi
\fi
\fi
\fi
}
\newcommand{\hatcurextraerrMHtwosiglimeccen}[1]{\ifnum#1=74 %
\hatcurextraerrMHtwosiglimeccenxxxxxA
\else
\ifnum#1=75 %
\hatcurextraerrMHtwosiglimeccenxxxxxB
\else
\ifnum#1=76 %
\hatcurextraerrMHtwosiglimeccenxxxxxC
\else
\ifnum#1=77 %
\hatcurextraerrMHtwosiglimeccenxxxxxD
\else
??????\fi
\fi
\fi
\fi
}
\newcommand{\hatcurextraerrMieccen}[1]{\ifnum#1=75 %
\hatcurextraerrMieccenxxxxxB
\else
\ifnum#1=77 %
\hatcurextraerrMieccenxxxxxD
\else
??????\fi
\fi
}
\newcommand{\hatcurextraerrMitwosiglimeccen}[1]{\ifnum#1=75 %
\hatcurextraerrMitwosiglimeccenxxxxxB
\else
\ifnum#1=77 %
\hatcurextraerrMitwosiglimeccenxxxxxD
\else
??????\fi
\fi
}
\newcommand{\hatcurextraerrMJeccen}[1]{\ifnum#1=74 %
\hatcurextraerrMJeccenxxxxxA
\else
\ifnum#1=75 %
\hatcurextraerrMJeccenxxxxxB
\else
\ifnum#1=76 %
\hatcurextraerrMJeccenxxxxxC
\else
\ifnum#1=77 %
\hatcurextraerrMJeccenxxxxxD
\else
??????\fi
\fi
\fi
\fi
}
\newcommand{\hatcurextraerrMJtwosiglimeccen}[1]{\ifnum#1=74 %
\hatcurextraerrMJtwosiglimeccenxxxxxA
\else
\ifnum#1=75 %
\hatcurextraerrMJtwosiglimeccenxxxxxB
\else
\ifnum#1=76 %
\hatcurextraerrMJtwosiglimeccenxxxxxC
\else
\ifnum#1=77 %
\hatcurextraerrMJtwosiglimeccenxxxxxD
\else
??????\fi
\fi
\fi
\fi
}
\newcommand{\hatcurextraerrMKseccen}[1]{\ifnum#1=74 %
\hatcurextraerrMKseccenxxxxxA
\else
\ifnum#1=75 %
\hatcurextraerrMKseccenxxxxxB
\else
\ifnum#1=76 %
\hatcurextraerrMKseccenxxxxxC
\else
\ifnum#1=77 %
\hatcurextraerrMKseccenxxxxxD
\else
??????\fi
\fi
\fi
\fi
}
\newcommand{\hatcurextraerrMKstwosiglimeccen}[1]{\ifnum#1=74 %
\hatcurextraerrMKstwosiglimeccenxxxxxA
\else
\ifnum#1=75 %
\hatcurextraerrMKstwosiglimeccenxxxxxB
\else
\ifnum#1=76 %
\hatcurextraerrMKstwosiglimeccenxxxxxC
\else
\ifnum#1=77 %
\hatcurextraerrMKstwosiglimeccenxxxxxD
\else
??????\fi
\fi
\fi
\fi
}
\newcommand{\hatcurextraerrMreccen}[1]{\ifnum#1=75 %
\hatcurextraerrMreccenxxxxxB
\else
\ifnum#1=77 %
\hatcurextraerrMreccenxxxxxD
\else
??????\fi
\fi
}
\newcommand{\hatcurextraerrMRPeccen}[1]{\ifnum#1=75 %
\hatcurextraerrMRPeccenxxxxxB
\else
\ifnum#1=76 %
\hatcurextraerrMRPeccenxxxxxC
\else
\ifnum#1=77 %
\hatcurextraerrMRPeccenxxxxxD
\else
??????\fi
\fi
\fi
}
\newcommand{\hatcurextraerrMRPtwosiglimeccen}[1]{\ifnum#1=75 %
\hatcurextraerrMRPtwosiglimeccenxxxxxB
\else
\ifnum#1=76 %
\hatcurextraerrMRPtwosiglimeccenxxxxxC
\else
\ifnum#1=77 %
\hatcurextraerrMRPtwosiglimeccenxxxxxD
\else
??????\fi
\fi
\fi
}
\newcommand{\hatcurextraerrMrtwosiglimeccen}[1]{\ifnum#1=75 %
\hatcurextraerrMrtwosiglimeccenxxxxxB
\else
\ifnum#1=77 %
\hatcurextraerrMrtwosiglimeccenxxxxxD
\else
??????\fi
\fi
}
\newcommand{\hatcurextraerrMWoneeccen}[1]{\ifnum#1=74 %
\hatcurextraerrMWoneeccenxxxxxA
\else
\ifnum#1=75 %
\hatcurextraerrMWoneeccenxxxxxB
\else
\ifnum#1=76 %
\hatcurextraerrMWoneeccenxxxxxC
\else
\ifnum#1=77 %
\hatcurextraerrMWoneeccenxxxxxD
\else
??????\fi
\fi
\fi
\fi
}
\newcommand{\hatcurextraerrMWonetwosiglimeccen}[1]{\ifnum#1=74 %
\hatcurextraerrMWonetwosiglimeccenxxxxxA
\else
\ifnum#1=75 %
\hatcurextraerrMWonetwosiglimeccenxxxxxB
\else
\ifnum#1=76 %
\hatcurextraerrMWonetwosiglimeccenxxxxxC
\else
\ifnum#1=77 %
\hatcurextraerrMWonetwosiglimeccenxxxxxD
\else
??????\fi
\fi
\fi
\fi
}
\newcommand{\hatcurextraerrMWtwoeccen}[1]{\ifnum#1=74 %
\hatcurextraerrMWtwoeccenxxxxxA
\else
\ifnum#1=75 %
\hatcurextraerrMWtwoeccenxxxxxB
\else
\ifnum#1=76 %
\hatcurextraerrMWtwoeccenxxxxxC
\else
\ifnum#1=77 %
\hatcurextraerrMWtwoeccenxxxxxD
\else
??????\fi
\fi
\fi
\fi
}
\newcommand{\hatcurextraerrMWtwotwosiglimeccen}[1]{\ifnum#1=74 %
\hatcurextraerrMWtwotwosiglimeccenxxxxxA
\else
\ifnum#1=75 %
\hatcurextraerrMWtwotwosiglimeccenxxxxxB
\else
\ifnum#1=76 %
\hatcurextraerrMWtwotwosiglimeccenxxxxxC
\else
\ifnum#1=77 %
\hatcurextraerrMWtwotwosiglimeccenxxxxxD
\else
??????\fi
\fi
\fi
\fi
}
\newcommand{\hatcurfieldeccen}[1]{\ifnum#1=74 %
\hatcurfieldeccenxxxxxA
\else
\ifnum#1=75 %
\hatcurfieldeccenxxxxxB
\else
\ifnum#1=76 %
\hatcurfieldeccenxxxxxC
\else
\ifnum#1=77 %
\hatcurfieldeccenxxxxxD
\else
??????\fi
\fi
\fi
\fi
}
\newcommand{\hatcurhtreccen}[1]{\ifnum#1=74 %
\hatcurhtreccenxxxxxA
\else
\ifnum#1=75 %
\hatcurhtreccenxxxxxB
\else
\ifnum#1=76 %
\hatcurhtreccenxxxxxC
\else
\ifnum#1=77 %
\hatcurhtreccenxxxxxD
\else
??????\fi
\fi
\fi
\fi
}
\newcommand{\hatcurISOageeccen}[1]{\ifnum#1=74 %
\hatcurISOageeccenxxxxxA
\else
\ifnum#1=75 %
\hatcurISOageeccenxxxxxB
\else
\ifnum#1=76 %
\hatcurISOageeccenxxxxxC
\else
\ifnum#1=77 %
\hatcurISOageeccenxxxxxD
\else
??????\fi
\fi
\fi
\fi
}
\newcommand{\hatcurISOloggeccen}[1]{\ifnum#1=74 %
\hatcurISOloggeccenxxxxxA
\else
\ifnum#1=75 %
\hatcurISOloggeccenxxxxxB
\else
\ifnum#1=76 %
\hatcurISOloggeccenxxxxxC
\else
\ifnum#1=77 %
\hatcurISOloggeccenxxxxxD
\else
??????\fi
\fi
\fi
\fi
}
\newcommand{\hatcurISOlumeccen}[1]{\ifnum#1=74 %
\hatcurISOlumeccenxxxxxA
\else
\ifnum#1=75 %
\hatcurISOlumeccenxxxxxB
\else
\ifnum#1=76 %
\hatcurISOlumeccenxxxxxC
\else
\ifnum#1=77 %
\hatcurISOlumeccenxxxxxD
\else
??????\fi
\fi
\fi
\fi
}
\newcommand{\hatcurISOlumshorteccen}[1]{\ifnum#1=74 %
\hatcurISOlumshorteccenxxxxxA
\else
\ifnum#1=75 %
\hatcurISOlumshorteccenxxxxxB
\else
\ifnum#1=76 %
\hatcurISOlumshorteccenxxxxxC
\else
\ifnum#1=77 %
\hatcurISOlumshorteccenxxxxxD
\else
??????\fi
\fi
\fi
\fi
}
\newcommand{\hatcurISOmeccen}[1]{\ifnum#1=74 %
\hatcurISOmeccenxxxxxA
\else
\ifnum#1=75 %
\hatcurISOmeccenxxxxxB
\else
\ifnum#1=76 %
\hatcurISOmeccenxxxxxC
\else
\ifnum#1=77 %
\hatcurISOmeccenxxxxxD
\else
??????\fi
\fi
\fi
\fi
}
\newcommand{\hatcurISOmlongeccen}[1]{\ifnum#1=74 %
\hatcurISOmlongeccenxxxxxA
\else
\ifnum#1=75 %
\hatcurISOmlongeccenxxxxxB
\else
\ifnum#1=76 %
\hatcurISOmlongeccenxxxxxC
\else
\ifnum#1=77 %
\hatcurISOmlongeccenxxxxxD
\else
??????\fi
\fi
\fi
\fi
}
\newcommand{\hatcurISOmshorteccen}[1]{\ifnum#1=74 %
\hatcurISOmshorteccenxxxxxA
\else
\ifnum#1=75 %
\hatcurISOmshorteccenxxxxxB
\else
\ifnum#1=76 %
\hatcurISOmshorteccenxxxxxC
\else
\ifnum#1=77 %
\hatcurISOmshorteccenxxxxxD
\else
??????\fi
\fi
\fi
\fi
}
\newcommand{\hatcurISOreccen}[1]{\ifnum#1=74 %
\hatcurISOreccenxxxxxA
\else
\ifnum#1=75 %
\hatcurISOreccenxxxxxB
\else
\ifnum#1=76 %
\hatcurISOreccenxxxxxC
\else
\ifnum#1=77 %
\hatcurISOreccenxxxxxD
\else
??????\fi
\fi
\fi
\fi
}
\newcommand{\hatcurISOrhoeccen}[1]{\ifnum#1=74 %
\hatcurISOrhoeccenxxxxxA
\else
\ifnum#1=75 %
\hatcurISOrhoeccenxxxxxB
\else
\ifnum#1=76 %
\hatcurISOrhoeccenxxxxxC
\else
\ifnum#1=77 %
\hatcurISOrhoeccenxxxxxD
\else
??????\fi
\fi
\fi
\fi
}
\newcommand{\hatcurISOrholongeccen}[1]{\ifnum#1=74 %
\hatcurISOrholongeccenxxxxxA
\else
\ifnum#1=75 %
\hatcurISOrholongeccenxxxxxB
\else
\ifnum#1=76 %
\hatcurISOrholongeccenxxxxxC
\else
\ifnum#1=77 %
\hatcurISOrholongeccenxxxxxD
\else
??????\fi
\fi
\fi
\fi
}
\newcommand{\hatcurISOrlongeccen}[1]{\ifnum#1=74 %
\hatcurISOrlongeccenxxxxxA
\else
\ifnum#1=75 %
\hatcurISOrlongeccenxxxxxB
\else
\ifnum#1=76 %
\hatcurISOrlongeccenxxxxxC
\else
\ifnum#1=77 %
\hatcurISOrlongeccenxxxxxD
\else
??????\fi
\fi
\fi
\fi
}
\newcommand{\hatcurISOrshorteccen}[1]{\ifnum#1=74 %
\hatcurISOrshorteccenxxxxxA
\else
\ifnum#1=75 %
\hatcurISOrshorteccenxxxxxB
\else
\ifnum#1=76 %
\hatcurISOrshorteccenxxxxxC
\else
\ifnum#1=77 %
\hatcurISOrshorteccenxxxxxD
\else
??????\fi
\fi
\fi
\fi
}
\newcommand{\hatcurISOspececcen}[1]{\ifnum#1=74 %
\hatcurISOspececcenxxxxxA
\else
\ifnum#1=75 %
\hatcurISOspececcenxxxxxB
\else
\ifnum#1=76 %
\hatcurISOspececcenxxxxxC
\else
\ifnum#1=77 %
\hatcurISOspececcenxxxxxD
\else
??????\fi
\fi
\fi
\fi
}
\newcommand{\hatcurISOteffeccen}[1]{\ifnum#1=74 %
\hatcurISOteffeccenxxxxxA
\else
\ifnum#1=75 %
\hatcurISOteffeccenxxxxxB
\else
\ifnum#1=76 %
\hatcurISOteffeccenxxxxxC
\else
\ifnum#1=77 %
\hatcurISOteffeccenxxxxxD
\else
??????\fi
\fi
\fi
\fi
}
\newcommand{\hatcurISOzfeheccen}[1]{\ifnum#1=74 %
\hatcurISOzfeheccenxxxxxA
\else
\ifnum#1=75 %
\hatcurISOzfeheccenxxxxxB
\else
\ifnum#1=76 %
\hatcurISOzfeheccenxxxxxC
\else
\ifnum#1=77 %
\hatcurISOzfeheccenxxxxxD
\else
??????\fi
\fi
\fi
\fi
}
\newcommand{\hatcurLBiBeccen}[1]{\ifnum#1=74 %
\hatcurLBiBeccenxxxxxA
\else
\ifnum#1=75 %
\hatcurLBiBeccenxxxxxB
\else
\ifnum#1=76 %
\hatcurLBiBeccenxxxxxC
\else
\ifnum#1=77 %
\hatcurLBiBeccenxxxxxD
\else
??????\fi
\fi
\fi
\fi
}
\newcommand{\hatcurLBiCeccen}[1]{\ifnum#1=74 %
\hatcurLBiCeccenxxxxxA
\else
\ifnum#1=75 %
\hatcurLBiCeccenxxxxxB
\else
\ifnum#1=76 %
\hatcurLBiCeccenxxxxxC
\else
\ifnum#1=77 %
\hatcurLBiCeccenxxxxxD
\else
??????\fi
\fi
\fi
\fi
}
\newcommand{\hatcurLBigeccen}[1]{\ifnum#1=74 %
\hatcurLBigeccenxxxxxA
\else
\ifnum#1=75 %
\hatcurLBigeccenxxxxxB
\else
\ifnum#1=76 %
\hatcurLBigeccenxxxxxC
\else
\ifnum#1=77 %
\hatcurLBigeccenxxxxxD
\else
??????\fi
\fi
\fi
\fi
}
\newcommand{\hatcurLBiHeccen}[1]{\ifnum#1=74 %
\hatcurLBiHeccenxxxxxA
\else
\ifnum#1=75 %
\hatcurLBiHeccenxxxxxB
\else
\ifnum#1=76 %
\hatcurLBiHeccenxxxxxC
\else
\ifnum#1=77 %
\hatcurLBiHeccenxxxxxD
\else
??????\fi
\fi
\fi
\fi
}
\newcommand{\hatcurLBiiBeccen}[1]{\ifnum#1=74 %
\hatcurLBiiBeccenxxxxxA
\else
\ifnum#1=75 %
\hatcurLBiiBeccenxxxxxB
\else
\ifnum#1=76 %
\hatcurLBiiBeccenxxxxxC
\else
\ifnum#1=77 %
\hatcurLBiiBeccenxxxxxD
\else
??????\fi
\fi
\fi
\fi
}
\newcommand{\hatcurLBiiCeccen}[1]{\ifnum#1=74 %
\hatcurLBiiCeccenxxxxxA
\else
\ifnum#1=75 %
\hatcurLBiiCeccenxxxxxB
\else
\ifnum#1=76 %
\hatcurLBiiCeccenxxxxxC
\else
\ifnum#1=77 %
\hatcurLBiiCeccenxxxxxD
\else
??????\fi
\fi
\fi
\fi
}
\newcommand{\hatcurLBiieccen}[1]{\ifnum#1=74 %
\hatcurLBiieccenxxxxxA
\else
\ifnum#1=75 %
\hatcurLBiieccenxxxxxB
\else
\ifnum#1=76 %
\hatcurLBiieccenxxxxxC
\else
\ifnum#1=77 %
\hatcurLBiieccenxxxxxD
\else
??????\fi
\fi
\fi
\fi
}
\newcommand{\hatcurLBiIeccen}[1]{\ifnum#1=74 %
\hatcurLBiIeccenxxxxxA
\else
\ifnum#1=75 %
\hatcurLBiIeccenxxxxxB
\else
\ifnum#1=76 %
\hatcurLBiIeccenxxxxxC
\else
\ifnum#1=77 %
\hatcurLBiIeccenxxxxxD
\else
??????\fi
\fi
\fi
\fi
}
\newcommand{\hatcurLBiigeccen}[1]{\ifnum#1=74 %
\hatcurLBiigeccenxxxxxA
\else
\ifnum#1=75 %
\hatcurLBiigeccenxxxxxB
\else
\ifnum#1=76 %
\hatcurLBiigeccenxxxxxC
\else
\ifnum#1=77 %
\hatcurLBiigeccenxxxxxD
\else
??????\fi
\fi
\fi
\fi
}
\newcommand{\hatcurLBiiHeccen}[1]{\ifnum#1=74 %
\hatcurLBiiHeccenxxxxxA
\else
\ifnum#1=75 %
\hatcurLBiiHeccenxxxxxB
\else
\ifnum#1=76 %
\hatcurLBiiHeccenxxxxxC
\else
\ifnum#1=77 %
\hatcurLBiiHeccenxxxxxD
\else
??????\fi
\fi
\fi
\fi
}
\newcommand{\hatcurLBiiieccen}[1]{\ifnum#1=74 %
\hatcurLBiiieccenxxxxxA
\else
\ifnum#1=75 %
\hatcurLBiiieccenxxxxxB
\else
\ifnum#1=76 %
\hatcurLBiiieccenxxxxxC
\else
\ifnum#1=77 %
\hatcurLBiiieccenxxxxxD
\else
??????\fi
\fi
\fi
\fi
}
\newcommand{\hatcurLBiiIeccen}[1]{\ifnum#1=74 %
\hatcurLBiiIeccenxxxxxA
\else
\ifnum#1=75 %
\hatcurLBiiIeccenxxxxxB
\else
\ifnum#1=76 %
\hatcurLBiiIeccenxxxxxC
\else
\ifnum#1=77 %
\hatcurLBiiIeccenxxxxxD
\else
??????\fi
\fi
\fi
\fi
}
\newcommand{\hatcurLBiiJeccen}[1]{\ifnum#1=74 %
\hatcurLBiiJeccenxxxxxA
\else
\ifnum#1=75 %
\hatcurLBiiJeccenxxxxxB
\else
\ifnum#1=76 %
\hatcurLBiiJeccenxxxxxC
\else
\ifnum#1=77 %
\hatcurLBiiJeccenxxxxxD
\else
??????\fi
\fi
\fi
\fi
}
\newcommand{\hatcurLBiiKeccen}[1]{\ifnum#1=74 %
\hatcurLBiiKeccenxxxxxA
\else
\ifnum#1=75 %
\hatcurLBiiKeccenxxxxxB
\else
\ifnum#1=76 %
\hatcurLBiiKeccenxxxxxC
\else
\ifnum#1=77 %
\hatcurLBiiKeccenxxxxxD
\else
??????\fi
\fi
\fi
\fi
}
\newcommand{\hatcurLBiikepeccen}[1]{\ifnum#1=74 %
\hatcurLBiikepeccenxxxxxA
\else
\ifnum#1=75 %
\hatcurLBiikepeccenxxxxxB
\else
\ifnum#1=76 %
\hatcurLBiikepeccenxxxxxC
\else
\ifnum#1=77 %
\hatcurLBiikepeccenxxxxxD
\else
??????\fi
\fi
\fi
\fi
}
\newcommand{\hatcurLBiiMeccen}[1]{\ifnum#1=74 %
\hatcurLBiiMeccenxxxxxA
\else
\ifnum#1=75 %
\hatcurLBiiMeccenxxxxxB
\else
\ifnum#1=76 %
\hatcurLBiiMeccenxxxxxC
\else
\ifnum#1=77 %
\hatcurLBiiMeccenxxxxxD
\else
??????\fi
\fi
\fi
\fi
}
\newcommand{\hatcurLBiireccen}[1]{\ifnum#1=74 %
\hatcurLBiireccenxxxxxA
\else
\ifnum#1=75 %
\hatcurLBiireccenxxxxxB
\else
\ifnum#1=76 %
\hatcurLBiireccenxxxxxC
\else
\ifnum#1=77 %
\hatcurLBiireccenxxxxxD
\else
??????\fi
\fi
\fi
\fi
}
\newcommand{\hatcurLBiiReccen}[1]{\ifnum#1=74 %
\hatcurLBiiReccenxxxxxA
\else
\ifnum#1=75 %
\hatcurLBiiReccenxxxxxB
\else
\ifnum#1=76 %
\hatcurLBiiReccenxxxxxC
\else
\ifnum#1=77 %
\hatcurLBiiReccenxxxxxD
\else
??????\fi
\fi
\fi
\fi
}
\newcommand{\hatcurLBiiSfoureccen}[1]{\ifnum#1=74 %
\hatcurLBiiSfoureccenxxxxxA
\else
\ifnum#1=75 %
\hatcurLBiiSfoureccenxxxxxB
\else
\ifnum#1=76 %
\hatcurLBiiSfoureccenxxxxxC
\else
\ifnum#1=77 %
\hatcurLBiiSfoureccenxxxxxD
\else
??????\fi
\fi
\fi
\fi
}
\newcommand{\hatcurLBiiSoneeccen}[1]{\ifnum#1=74 %
\hatcurLBiiSoneeccenxxxxxA
\else
\ifnum#1=75 %
\hatcurLBiiSoneeccenxxxxxB
\else
\ifnum#1=76 %
\hatcurLBiiSoneeccenxxxxxC
\else
\ifnum#1=77 %
\hatcurLBiiSoneeccenxxxxxD
\else
??????\fi
\fi
\fi
\fi
}
\newcommand{\hatcurLBiiSthreeeccen}[1]{\ifnum#1=74 %
\hatcurLBiiSthreeeccenxxxxxA
\else
\ifnum#1=75 %
\hatcurLBiiSthreeeccenxxxxxB
\else
\ifnum#1=76 %
\hatcurLBiiSthreeeccenxxxxxC
\else
\ifnum#1=77 %
\hatcurLBiiSthreeeccenxxxxxD
\else
??????\fi
\fi
\fi
\fi
}
\newcommand{\hatcurLBiiStwoeccen}[1]{\ifnum#1=74 %
\hatcurLBiiStwoeccenxxxxxA
\else
\ifnum#1=75 %
\hatcurLBiiStwoeccenxxxxxB
\else
\ifnum#1=76 %
\hatcurLBiiStwoeccenxxxxxC
\else
\ifnum#1=77 %
\hatcurLBiiStwoeccenxxxxxD
\else
??????\fi
\fi
\fi
\fi
}
\newcommand{\hatcurLBiiTeccen}[1]{\ifnum#1=74 %
\hatcurLBiiTeccenxxxxxA
\else
\ifnum#1=75 %
\hatcurLBiiTeccenxxxxxB
\else
\ifnum#1=76 %
\hatcurLBiiTeccenxxxxxC
\else
\ifnum#1=77 %
\hatcurLBiiTeccenxxxxxD
\else
??????\fi
\fi
\fi
\fi
}
\newcommand{\hatcurLBiiueccen}[1]{\ifnum#1=74 %
\hatcurLBiiueccenxxxxxA
\else
\ifnum#1=75 %
\hatcurLBiiueccenxxxxxB
\else
\ifnum#1=76 %
\hatcurLBiiueccenxxxxxC
\else
\ifnum#1=77 %
\hatcurLBiiueccenxxxxxD
\else
??????\fi
\fi
\fi
\fi
}
\newcommand{\hatcurLBiiVeccen}[1]{\ifnum#1=74 %
\hatcurLBiiVeccenxxxxxA
\else
\ifnum#1=75 %
\hatcurLBiiVeccenxxxxxB
\else
\ifnum#1=76 %
\hatcurLBiiVeccenxxxxxC
\else
\ifnum#1=77 %
\hatcurLBiiVeccenxxxxxD
\else
??????\fi
\fi
\fi
\fi
}
\newcommand{\hatcurLBiizeccen}[1]{\ifnum#1=74 %
\hatcurLBiizeccenxxxxxA
\else
\ifnum#1=75 %
\hatcurLBiizeccenxxxxxB
\else
\ifnum#1=76 %
\hatcurLBiizeccenxxxxxC
\else
\ifnum#1=77 %
\hatcurLBiizeccenxxxxxD
\else
??????\fi
\fi
\fi
\fi
}
\newcommand{\hatcurLBiJeccen}[1]{\ifnum#1=74 %
\hatcurLBiJeccenxxxxxA
\else
\ifnum#1=75 %
\hatcurLBiJeccenxxxxxB
\else
\ifnum#1=76 %
\hatcurLBiJeccenxxxxxC
\else
\ifnum#1=77 %
\hatcurLBiJeccenxxxxxD
\else
??????\fi
\fi
\fi
\fi
}
\newcommand{\hatcurLBiKeccen}[1]{\ifnum#1=74 %
\hatcurLBiKeccenxxxxxA
\else
\ifnum#1=75 %
\hatcurLBiKeccenxxxxxB
\else
\ifnum#1=76 %
\hatcurLBiKeccenxxxxxC
\else
\ifnum#1=77 %
\hatcurLBiKeccenxxxxxD
\else
??????\fi
\fi
\fi
\fi
}
\newcommand{\hatcurLBikepeccen}[1]{\ifnum#1=74 %
\hatcurLBikepeccenxxxxxA
\else
\ifnum#1=75 %
\hatcurLBikepeccenxxxxxB
\else
\ifnum#1=76 %
\hatcurLBikepeccenxxxxxC
\else
\ifnum#1=77 %
\hatcurLBikepeccenxxxxxD
\else
??????\fi
\fi
\fi
\fi
}
\newcommand{\hatcurLBiMeccen}[1]{\ifnum#1=74 %
\hatcurLBiMeccenxxxxxA
\else
\ifnum#1=75 %
\hatcurLBiMeccenxxxxxB
\else
\ifnum#1=76 %
\hatcurLBiMeccenxxxxxC
\else
\ifnum#1=77 %
\hatcurLBiMeccenxxxxxD
\else
??????\fi
\fi
\fi
\fi
}
\newcommand{\hatcurLBireccen}[1]{\ifnum#1=74 %
\hatcurLBireccenxxxxxA
\else
\ifnum#1=75 %
\hatcurLBireccenxxxxxB
\else
\ifnum#1=76 %
\hatcurLBireccenxxxxxC
\else
\ifnum#1=77 %
\hatcurLBireccenxxxxxD
\else
??????\fi
\fi
\fi
\fi
}
\newcommand{\hatcurLBiReccen}[1]{\ifnum#1=74 %
\hatcurLBiReccenxxxxxA
\else
\ifnum#1=75 %
\hatcurLBiReccenxxxxxB
\else
\ifnum#1=76 %
\hatcurLBiReccenxxxxxC
\else
\ifnum#1=77 %
\hatcurLBiReccenxxxxxD
\else
??????\fi
\fi
\fi
\fi
}
\newcommand{\hatcurLBiSfoureccen}[1]{\ifnum#1=74 %
\hatcurLBiSfoureccenxxxxxA
\else
\ifnum#1=75 %
\hatcurLBiSfoureccenxxxxxB
\else
\ifnum#1=76 %
\hatcurLBiSfoureccenxxxxxC
\else
\ifnum#1=77 %
\hatcurLBiSfoureccenxxxxxD
\else
??????\fi
\fi
\fi
\fi
}
\newcommand{\hatcurLBiSoneeccen}[1]{\ifnum#1=74 %
\hatcurLBiSoneeccenxxxxxA
\else
\ifnum#1=75 %
\hatcurLBiSoneeccenxxxxxB
\else
\ifnum#1=76 %
\hatcurLBiSoneeccenxxxxxC
\else
\ifnum#1=77 %
\hatcurLBiSoneeccenxxxxxD
\else
??????\fi
\fi
\fi
\fi
}
\newcommand{\hatcurLBiSthreeeccen}[1]{\ifnum#1=74 %
\hatcurLBiSthreeeccenxxxxxA
\else
\ifnum#1=75 %
\hatcurLBiSthreeeccenxxxxxB
\else
\ifnum#1=76 %
\hatcurLBiSthreeeccenxxxxxC
\else
\ifnum#1=77 %
\hatcurLBiSthreeeccenxxxxxD
\else
??????\fi
\fi
\fi
\fi
}
\newcommand{\hatcurLBiStwoeccen}[1]{\ifnum#1=74 %
\hatcurLBiStwoeccenxxxxxA
\else
\ifnum#1=75 %
\hatcurLBiStwoeccenxxxxxB
\else
\ifnum#1=76 %
\hatcurLBiStwoeccenxxxxxC
\else
\ifnum#1=77 %
\hatcurLBiStwoeccenxxxxxD
\else
??????\fi
\fi
\fi
\fi
}
\newcommand{\hatcurLBiTeccen}[1]{\ifnum#1=74 %
\hatcurLBiTeccenxxxxxA
\else
\ifnum#1=75 %
\hatcurLBiTeccenxxxxxB
\else
\ifnum#1=76 %
\hatcurLBiTeccenxxxxxC
\else
\ifnum#1=77 %
\hatcurLBiTeccenxxxxxD
\else
??????\fi
\fi
\fi
\fi
}
\newcommand{\hatcurLBiueccen}[1]{\ifnum#1=74 %
\hatcurLBiueccenxxxxxA
\else
\ifnum#1=75 %
\hatcurLBiueccenxxxxxB
\else
\ifnum#1=76 %
\hatcurLBiueccenxxxxxC
\else
\ifnum#1=77 %
\hatcurLBiueccenxxxxxD
\else
??????\fi
\fi
\fi
\fi
}
\newcommand{\hatcurLBiVeccen}[1]{\ifnum#1=74 %
\hatcurLBiVeccenxxxxxA
\else
\ifnum#1=75 %
\hatcurLBiVeccenxxxxxB
\else
\ifnum#1=76 %
\hatcurLBiVeccenxxxxxC
\else
\ifnum#1=77 %
\hatcurLBiVeccenxxxxxD
\else
??????\fi
\fi
\fi
\fi
}
\newcommand{\hatcurLBizeccen}[1]{\ifnum#1=74 %
\hatcurLBizeccenxxxxxA
\else
\ifnum#1=75 %
\hatcurLBizeccenxxxxxB
\else
\ifnum#1=76 %
\hatcurLBizeccenxxxxxC
\else
\ifnum#1=77 %
\hatcurLBizeccenxxxxxD
\else
??????\fi
\fi
\fi
\fi
}
\newcommand{\hatcurLCbsqeccen}[1]{\ifnum#1=74 %
\hatcurLCbsqeccenxxxxxA
\else
\ifnum#1=75 %
\hatcurLCbsqeccenxxxxxB
\else
\ifnum#1=76 %
\hatcurLCbsqeccenxxxxxC
\else
\ifnum#1=77 %
\hatcurLCbsqeccenxxxxxD
\else
??????\fi
\fi
\fi
\fi
}
\newcommand{\hatcurLCdipeccen}[1]{\ifnum#1=74 %
\hatcurLCdipeccenxxxxxA
\else
\ifnum#1=75 %
\hatcurLCdipeccenxxxxxB
\else
\ifnum#1=76 %
\hatcurLCdipeccenxxxxxC
\else
\ifnum#1=77 %
\hatcurLCdipeccenxxxxxD
\else
??????\fi
\fi
\fi
\fi
}
\newcommand{\hatcurLCdureccen}[1]{\ifnum#1=74 %
\hatcurLCdureccenxxxxxA
\else
\ifnum#1=75 %
\hatcurLCdureccenxxxxxB
\else
\ifnum#1=76 %
\hatcurLCdureccenxxxxxC
\else
\ifnum#1=77 %
\hatcurLCdureccenxxxxxD
\else
??????\fi
\fi
\fi
\fi
}
\newcommand{\hatcurLCdurhreccen}[1]{\ifnum#1=74 %
\hatcurLCdurhreccenxxxxxA
\else
\ifnum#1=75 %
\hatcurLCdurhreccenxxxxxB
\else
\ifnum#1=76 %
\hatcurLCdurhreccenxxxxxC
\else
\ifnum#1=77 %
\hatcurLCdurhreccenxxxxxD
\else
??????\fi
\fi
\fi
\fi
}
\newcommand{\hatcurLCdurhrshorteccen}[1]{\ifnum#1=74 %
\hatcurLCdurhrshorteccenxxxxxA
\else
\ifnum#1=75 %
\hatcurLCdurhrshorteccenxxxxxB
\else
\ifnum#1=76 %
\hatcurLCdurhrshorteccenxxxxxC
\else
\ifnum#1=77 %
\hatcurLCdurhrshorteccenxxxxxD
\else
??????\fi
\fi
\fi
\fi
}
\newcommand{\hatcurLCdurshorteccen}[1]{\ifnum#1=74 %
\hatcurLCdurshorteccenxxxxxA
\else
\ifnum#1=75 %
\hatcurLCdurshorteccenxxxxxB
\else
\ifnum#1=76 %
\hatcurLCdurshorteccenxxxxxC
\else
\ifnum#1=77 %
\hatcurLCdurshorteccenxxxxxD
\else
??????\fi
\fi
\fi
\fi
}
\newcommand{\hatcurLChatnetmAeccen}[1]{\ifnum#1=74 %
\hatcurLChatnetmAeccenxxxxxA
\else
\ifnum#1=75 %
\hatcurLChatnetmAeccenxxxxxB
\else
\ifnum#1=76 %
\hatcurLChatnetmAeccenxxxxxC
\else
??????\fi
\fi
\fi
}
\newcommand{\hatcurLChatnetmBeccen}[1]{\ifnum#1=74 %
\hatcurLChatnetmBeccenxxxxxA
\else
\ifnum#1=75 %
\hatcurLChatnetmBeccenxxxxxB
\else
\ifnum#1=76 %
\hatcurLChatnetmBeccenxxxxxC
\else
??????\fi
\fi
\fi
}
\newcommand{\hatcurLChatnetmCeccen}[1]{\ifnum#1=75 %
\hatcurLChatnetmCeccenxxxxxB
\else
??????\fi
}
\newcommand{\hatcurLChatnetmDeccen}[1]{\ifnum#1=75 %
\hatcurLChatnetmDeccenxxxxxB
\else
??????\fi
}
\newcommand{\hatcurLChatnetmeccen}[1]{\ifnum#1=77 %
\hatcurLChatnetmeccenxxxxxD
\else
??????\fi
}
\newcommand{\hatcurLCiblendAeccen}[1]{\ifnum#1=74 %
\hatcurLCiblendAeccenxxxxxA
\else
\ifnum#1=75 %
\hatcurLCiblendAeccenxxxxxB
\else
\ifnum#1=76 %
\hatcurLCiblendAeccenxxxxxC
\else
??????\fi
\fi
\fi
}
\newcommand{\hatcurLCiblendBeccen}[1]{\ifnum#1=74 %
\hatcurLCiblendBeccenxxxxxA
\else
\ifnum#1=75 %
\hatcurLCiblendBeccenxxxxxB
\else
\ifnum#1=76 %
\hatcurLCiblendBeccenxxxxxC
\else
??????\fi
\fi
\fi
}
\newcommand{\hatcurLCiblendCeccen}[1]{\ifnum#1=75 %
\hatcurLCiblendCeccenxxxxxB
\else
??????\fi
}
\newcommand{\hatcurLCiblendDeccen}[1]{\ifnum#1=75 %
\hatcurLCiblendDeccenxxxxxB
\else
??????\fi
}
\newcommand{\hatcurLCiblendeccen}[1]{\ifnum#1=77 %
\hatcurLCiblendeccenxxxxxD
\else
??????\fi
}
\newcommand{\hatcurLCimpeccen}[1]{\ifnum#1=74 %
\hatcurLCimpeccenxxxxxA
\else
\ifnum#1=75 %
\hatcurLCimpeccenxxxxxB
\else
\ifnum#1=76 %
\hatcurLCimpeccenxxxxxC
\else
\ifnum#1=77 %
\hatcurLCimpeccenxxxxxD
\else
??????\fi
\fi
\fi
\fi
}
\newcommand{\hatcurLCingdureccen}[1]{\ifnum#1=74 %
\hatcurLCingdureccenxxxxxA
\else
\ifnum#1=75 %
\hatcurLCingdureccenxxxxxB
\else
\ifnum#1=76 %
\hatcurLCingdureccenxxxxxC
\else
\ifnum#1=77 %
\hatcurLCingdureccenxxxxxD
\else
??????\fi
\fi
\fi
\fi
}
\newcommand{\hatcurLCPeccen}[1]{\ifnum#1=74 %
\hatcurLCPeccenxxxxxA
\else
\ifnum#1=75 %
\hatcurLCPeccenxxxxxB
\else
\ifnum#1=76 %
\hatcurLCPeccenxxxxxC
\else
\ifnum#1=77 %
\hatcurLCPeccenxxxxxD
\else
??????\fi
\fi
\fi
\fi
}
\newcommand{\hatcurLCPprececcen}[1]{\ifnum#1=74 %
\hatcurLCPprececcenxxxxxA
\else
\ifnum#1=75 %
\hatcurLCPprececcenxxxxxB
\else
\ifnum#1=76 %
\hatcurLCPprececcenxxxxxC
\else
\ifnum#1=77 %
\hatcurLCPprececcenxxxxxD
\else
??????\fi
\fi
\fi
\fi
}
\newcommand{\hatcurLCPshorteccen}[1]{\ifnum#1=74 %
\hatcurLCPshorteccenxxxxxA
\else
\ifnum#1=75 %
\hatcurLCPshorteccenxxxxxB
\else
\ifnum#1=76 %
\hatcurLCPshorteccenxxxxxC
\else
\ifnum#1=77 %
\hatcurLCPshorteccenxxxxxD
\else
??????\fi
\fi
\fi
\fi
}
\newcommand{\hatcurLCqeccen}[1]{\ifnum#1=74 %
\hatcurLCqeccenxxxxxA
\else
\ifnum#1=75 %
\hatcurLCqeccenxxxxxB
\else
\ifnum#1=76 %
\hatcurLCqeccenxxxxxC
\else
\ifnum#1=77 %
\hatcurLCqeccenxxxxxD
\else
??????\fi
\fi
\fi
\fi
}
\newcommand{\hatcurLCqshorteccen}[1]{\ifnum#1=74 %
\hatcurLCqshorteccenxxxxxA
\else
\ifnum#1=75 %
\hatcurLCqshorteccenxxxxxB
\else
\ifnum#1=76 %
\hatcurLCqshorteccenxxxxxC
\else
\ifnum#1=77 %
\hatcurLCqshorteccenxxxxxD
\else
??????\fi
\fi
\fi
\fi
}
\newcommand{\hatcurLCrhoeccen}[1]{\ifnum#1=74 %
\hatcurLCrhoeccenxxxxxA
\else
\ifnum#1=75 %
\hatcurLCrhoeccenxxxxxB
\else
\ifnum#1=76 %
\hatcurLCrhoeccenxxxxxC
\else
\ifnum#1=77 %
\hatcurLCrhoeccenxxxxxD
\else
??????\fi
\fi
\fi
\fi
}
\newcommand{\hatcurLCrprstareccen}[1]{\ifnum#1=74 %
\hatcurLCrprstareccenxxxxxA
\else
\ifnum#1=75 %
\hatcurLCrprstareccenxxxxxB
\else
\ifnum#1=76 %
\hatcurLCrprstareccenxxxxxC
\else
\ifnum#1=77 %
\hatcurLCrprstareccenxxxxxD
\else
??????\fi
\fi
\fi
\fi
}
\newcommand{\hatcurLCTAeccen}[1]{\ifnum#1=74 %
\hatcurLCTAeccenxxxxxA
\else
\ifnum#1=75 %
\hatcurLCTAeccenxxxxxB
\else
\ifnum#1=76 %
\hatcurLCTAeccenxxxxxC
\else
\ifnum#1=77 %
\hatcurLCTAeccenxxxxxD
\else
??????\fi
\fi
\fi
\fi
}
\newcommand{\hatcurLCTBeccen}[1]{\ifnum#1=74 %
\hatcurLCTBeccenxxxxxA
\else
\ifnum#1=75 %
\hatcurLCTBeccenxxxxxB
\else
\ifnum#1=76 %
\hatcurLCTBeccenxxxxxC
\else
\ifnum#1=77 %
\hatcurLCTBeccenxxxxxD
\else
??????\fi
\fi
\fi
\fi
}
\newcommand{\hatcurLCTeccen}[1]{\ifnum#1=74 %
\hatcurLCTeccenxxxxxA
\else
\ifnum#1=75 %
\hatcurLCTeccenxxxxxB
\else
\ifnum#1=76 %
\hatcurLCTeccenxxxxxC
\else
\ifnum#1=77 %
\hatcurLCTeccenxxxxxD
\else
??????\fi
\fi
\fi
\fi
}
\newcommand{\hatcurLCzetaeccen}[1]{\ifnum#1=74 %
\hatcurLCzetaeccenxxxxxA
\else
\ifnum#1=75 %
\hatcurLCzetaeccenxxxxxB
\else
\ifnum#1=76 %
\hatcurLCzetaeccenxxxxxC
\else
\ifnum#1=77 %
\hatcurLCzetaeccenxxxxxD
\else
??????\fi
\fi
\fi
\fi
}
\newcommand{\hatcurPPaequiveccen}[1]{\ifnum#1=74 %
\hatcurPPaequiveccenxxxxxA
\else
\ifnum#1=75 %
\hatcurPPaequiveccenxxxxxB
\else
\ifnum#1=76 %
\hatcurPPaequiveccenxxxxxC
\else
\ifnum#1=77 %
\hatcurPPaequiveccenxxxxxD
\else
??????\fi
\fi
\fi
\fi
}
\newcommand{\hatcurPPareccen}[1]{\ifnum#1=74 %
\hatcurPPareccenxxxxxA
\else
\ifnum#1=75 %
\hatcurPPareccenxxxxxB
\else
\ifnum#1=76 %
\hatcurPPareccenxxxxxC
\else
\ifnum#1=77 %
\hatcurPPareccenxxxxxD
\else
??????\fi
\fi
\fi
\fi
}
\newcommand{\hatcurPPareleccen}[1]{\ifnum#1=74 %
\hatcurPPareleccenxxxxxA
\else
\ifnum#1=75 %
\hatcurPPareleccenxxxxxB
\else
\ifnum#1=76 %
\hatcurPPareleccenxxxxxC
\else
\ifnum#1=77 %
\hatcurPPareleccenxxxxxD
\else
??????\fi
\fi
\fi
\fi
}
\newcommand{\hatcurPPfluxapdimeccen}[1]{\ifnum#1=74 %
\hatcurPPfluxapdimeccenxxxxxA
\else
\ifnum#1=75 %
\hatcurPPfluxapdimeccenxxxxxB
\else
\ifnum#1=76 %
\hatcurPPfluxapdimeccenxxxxxC
\else
\ifnum#1=77 %
\hatcurPPfluxapdimeccenxxxxxD
\else
??????\fi
\fi
\fi
\fi
}
\newcommand{\hatcurPPfluxapeccen}[1]{\ifnum#1=74 %
\hatcurPPfluxapeccenxxxxxA
\else
\ifnum#1=75 %
\hatcurPPfluxapeccenxxxxxB
\else
\ifnum#1=76 %
\hatcurPPfluxapeccenxxxxxC
\else
\ifnum#1=77 %
\hatcurPPfluxapeccenxxxxxD
\else
??????\fi
\fi
\fi
\fi
}
\newcommand{\hatcurPPfluxavgdimeccen}[1]{\ifnum#1=74 %
\hatcurPPfluxavgdimeccenxxxxxA
\else
\ifnum#1=75 %
\hatcurPPfluxavgdimeccenxxxxxB
\else
\ifnum#1=76 %
\hatcurPPfluxavgdimeccenxxxxxC
\else
\ifnum#1=77 %
\hatcurPPfluxavgdimeccenxxxxxD
\else
??????\fi
\fi
\fi
\fi
}
\newcommand{\hatcurPPfluxavgeccen}[1]{\ifnum#1=74 %
\hatcurPPfluxavgeccenxxxxxA
\else
\ifnum#1=75 %
\hatcurPPfluxavgeccenxxxxxB
\else
\ifnum#1=76 %
\hatcurPPfluxavgeccenxxxxxC
\else
\ifnum#1=77 %
\hatcurPPfluxavgeccenxxxxxD
\else
??????\fi
\fi
\fi
\fi
}
\newcommand{\hatcurPPfluxavglogeccen}[1]{\ifnum#1=74 %
\hatcurPPfluxavglogeccenxxxxxA
\else
\ifnum#1=75 %
\hatcurPPfluxavglogeccenxxxxxB
\else
\ifnum#1=76 %
\hatcurPPfluxavglogeccenxxxxxC
\else
\ifnum#1=77 %
\hatcurPPfluxavglogeccenxxxxxD
\else
??????\fi
\fi
\fi
\fi
}
\newcommand{\hatcurPPfluxperidimeccen}[1]{\ifnum#1=74 %
\hatcurPPfluxperidimeccenxxxxxA
\else
\ifnum#1=75 %
\hatcurPPfluxperidimeccenxxxxxB
\else
\ifnum#1=76 %
\hatcurPPfluxperidimeccenxxxxxC
\else
\ifnum#1=77 %
\hatcurPPfluxperidimeccenxxxxxD
\else
??????\fi
\fi
\fi
\fi
}
\newcommand{\hatcurPPfluxperieccen}[1]{\ifnum#1=74 %
\hatcurPPfluxperieccenxxxxxA
\else
\ifnum#1=75 %
\hatcurPPfluxperieccenxxxxxB
\else
\ifnum#1=76 %
\hatcurPPfluxperieccenxxxxxC
\else
\ifnum#1=77 %
\hatcurPPfluxperieccenxxxxxD
\else
??????\fi
\fi
\fi
\fi
}
\newcommand{\hatcurPPgeccen}[1]{\ifnum#1=74 %
\hatcurPPgeccenxxxxxA
\else
\ifnum#1=75 %
\hatcurPPgeccenxxxxxB
\else
\ifnum#1=76 %
\hatcurPPgeccenxxxxxC
\else
\ifnum#1=77 %
\hatcurPPgeccenxxxxxD
\else
??????\fi
\fi
\fi
\fi
}
\newcommand{\hatcurPPieccen}[1]{\ifnum#1=74 %
\hatcurPPieccenxxxxxA
\else
\ifnum#1=75 %
\hatcurPPieccenxxxxxB
\else
\ifnum#1=76 %
\hatcurPPieccenxxxxxC
\else
\ifnum#1=77 %
\hatcurPPieccenxxxxxD
\else
??????\fi
\fi
\fi
\fi
}
\newcommand{\hatcurPPloggeccen}[1]{\ifnum#1=74 %
\hatcurPPloggeccenxxxxxA
\else
\ifnum#1=75 %
\hatcurPPloggeccenxxxxxB
\else
\ifnum#1=76 %
\hatcurPPloggeccenxxxxxC
\else
\ifnum#1=77 %
\hatcurPPloggeccenxxxxxD
\else
??????\fi
\fi
\fi
\fi
}
\newcommand{\hatcurPPmeccen}[1]{\ifnum#1=74 %
\hatcurPPmeccenxxxxxA
\else
\ifnum#1=75 %
\hatcurPPmeccenxxxxxB
\else
\ifnum#1=76 %
\hatcurPPmeccenxxxxxC
\else
\ifnum#1=77 %
\hatcurPPmeccenxxxxxD
\else
??????\fi
\fi
\fi
\fi
}
\newcommand{\hatcurPPmeeccen}[1]{\ifnum#1=74 %
\hatcurPPmeeccenxxxxxA
\else
\ifnum#1=75 %
\hatcurPPmeeccenxxxxxB
\else
\ifnum#1=76 %
\hatcurPPmeeccenxxxxxC
\else
\ifnum#1=77 %
\hatcurPPmeeccenxxxxxD
\else
??????\fi
\fi
\fi
\fi
}
\newcommand{\hatcurPPmelongeccen}[1]{\ifnum#1=74 %
\hatcurPPmelongeccenxxxxxA
\else
\ifnum#1=75 %
\hatcurPPmelongeccenxxxxxB
\else
\ifnum#1=76 %
\hatcurPPmelongeccenxxxxxC
\else
\ifnum#1=77 %
\hatcurPPmelongeccenxxxxxD
\else
??????\fi
\fi
\fi
\fi
}
\newcommand{\hatcurPPmeshorteccen}[1]{\ifnum#1=74 %
\hatcurPPmeshorteccenxxxxxA
\else
\ifnum#1=75 %
\hatcurPPmeshorteccenxxxxxB
\else
\ifnum#1=76 %
\hatcurPPmeshorteccenxxxxxC
\else
\ifnum#1=77 %
\hatcurPPmeshorteccenxxxxxD
\else
??????\fi
\fi
\fi
\fi
}
\newcommand{\hatcurPPmlongeccen}[1]{\ifnum#1=74 %
\hatcurPPmlongeccenxxxxxA
\else
\ifnum#1=75 %
\hatcurPPmlongeccenxxxxxB
\else
\ifnum#1=76 %
\hatcurPPmlongeccenxxxxxC
\else
\ifnum#1=77 %
\hatcurPPmlongeccenxxxxxD
\else
??????\fi
\fi
\fi
\fi
}
\newcommand{\hatcurPPmrcorreccen}[1]{\ifnum#1=74 %
\hatcurPPmrcorreccenxxxxxA
\else
\ifnum#1=75 %
\hatcurPPmrcorreccenxxxxxB
\else
\ifnum#1=76 %
\hatcurPPmrcorreccenxxxxxC
\else
\ifnum#1=77 %
\hatcurPPmrcorreccenxxxxxD
\else
??????\fi
\fi
\fi
\fi
}
\newcommand{\hatcurPPmshorteccen}[1]{\ifnum#1=74 %
\hatcurPPmshorteccenxxxxxA
\else
\ifnum#1=75 %
\hatcurPPmshorteccenxxxxxB
\else
\ifnum#1=76 %
\hatcurPPmshorteccenxxxxxC
\else
\ifnum#1=77 %
\hatcurPPmshorteccenxxxxxD
\else
??????\fi
\fi
\fi
\fi
}
\newcommand{\hatcurPPmtwosiglimeccen}[1]{\ifnum#1=74 %
\hatcurPPmtwosiglimeccenxxxxxA
\else
\ifnum#1=75 %
\hatcurPPmtwosiglimeccenxxxxxB
\else
\ifnum#1=76 %
\hatcurPPmtwosiglimeccenxxxxxC
\else
\ifnum#1=77 %
\hatcurPPmtwosiglimeccenxxxxxD
\else
??????\fi
\fi
\fi
\fi
}
\newcommand{\hatcurPPperieccen}[1]{\ifnum#1=74 %
\hatcurPPperieccenxxxxxA
\else
\ifnum#1=75 %
\hatcurPPperieccenxxxxxB
\else
\ifnum#1=76 %
\hatcurPPperieccenxxxxxC
\else
\ifnum#1=77 %
\hatcurPPperieccenxxxxxD
\else
??????\fi
\fi
\fi
\fi
}
\newcommand{\hatcurPPphiconjeccen}[1]{\ifnum#1=74 %
\hatcurPPphiconjeccenxxxxxA
\else
\ifnum#1=75 %
\hatcurPPphiconjeccenxxxxxB
\else
\ifnum#1=76 %
\hatcurPPphiconjeccenxxxxxC
\else
\ifnum#1=77 %
\hatcurPPphiconjeccenxxxxxD
\else
??????\fi
\fi
\fi
\fi
}
\newcommand{\hatcurPPreccen}[1]{\ifnum#1=74 %
\hatcurPPreccenxxxxxA
\else
\ifnum#1=75 %
\hatcurPPreccenxxxxxB
\else
\ifnum#1=76 %
\hatcurPPreccenxxxxxC
\else
\ifnum#1=77 %
\hatcurPPreccenxxxxxD
\else
??????\fi
\fi
\fi
\fi
}
\newcommand{\hatcurPPreeccen}[1]{\ifnum#1=74 %
\hatcurPPreeccenxxxxxA
\else
\ifnum#1=75 %
\hatcurPPreeccenxxxxxB
\else
\ifnum#1=76 %
\hatcurPPreeccenxxxxxC
\else
\ifnum#1=77 %
\hatcurPPreeccenxxxxxD
\else
??????\fi
\fi
\fi
\fi
}
\newcommand{\hatcurPPrelongeccen}[1]{\ifnum#1=74 %
\hatcurPPrelongeccenxxxxxA
\else
\ifnum#1=75 %
\hatcurPPrelongeccenxxxxxB
\else
\ifnum#1=76 %
\hatcurPPrelongeccenxxxxxC
\else
\ifnum#1=77 %
\hatcurPPrelongeccenxxxxxD
\else
??????\fi
\fi
\fi
\fi
}
\newcommand{\hatcurPPreshorteccen}[1]{\ifnum#1=74 %
\hatcurPPreshorteccenxxxxxA
\else
\ifnum#1=75 %
\hatcurPPreshorteccenxxxxxB
\else
\ifnum#1=76 %
\hatcurPPreshorteccenxxxxxC
\else
\ifnum#1=77 %
\hatcurPPreshorteccenxxxxxD
\else
??????\fi
\fi
\fi
\fi
}
\newcommand{\hatcurPPrhoeccen}[1]{\ifnum#1=74 %
\hatcurPPrhoeccenxxxxxA
\else
\ifnum#1=75 %
\hatcurPPrhoeccenxxxxxB
\else
\ifnum#1=76 %
\hatcurPPrhoeccenxxxxxC
\else
\ifnum#1=77 %
\hatcurPPrhoeccenxxxxxD
\else
??????\fi
\fi
\fi
\fi
}
\newcommand{\hatcurPPrlongeccen}[1]{\ifnum#1=74 %
\hatcurPPrlongeccenxxxxxA
\else
\ifnum#1=75 %
\hatcurPPrlongeccenxxxxxB
\else
\ifnum#1=76 %
\hatcurPPrlongeccenxxxxxC
\else
\ifnum#1=77 %
\hatcurPPrlongeccenxxxxxD
\else
??????\fi
\fi
\fi
\fi
}
\newcommand{\hatcurPPrshorteccen}[1]{\ifnum#1=74 %
\hatcurPPrshorteccenxxxxxA
\else
\ifnum#1=75 %
\hatcurPPrshorteccenxxxxxB
\else
\ifnum#1=76 %
\hatcurPPrshorteccenxxxxxC
\else
\ifnum#1=77 %
\hatcurPPrshorteccenxxxxxD
\else
??????\fi
\fi
\fi
\fi
}
\newcommand{\hatcurPPtcirceccen}[1]{\ifnum#1=74 %
\hatcurPPtcirceccenxxxxxA
\else
\ifnum#1=75 %
\hatcurPPtcirceccenxxxxxB
\else
\ifnum#1=76 %
\hatcurPPtcirceccenxxxxxC
\else
\ifnum#1=77 %
\hatcurPPtcirceccenxxxxxD
\else
??????\fi
\fi
\fi
\fi
}
\newcommand{\hatcurPPteffeccen}[1]{\ifnum#1=74 %
\hatcurPPteffeccenxxxxxA
\else
\ifnum#1=75 %
\hatcurPPteffeccenxxxxxB
\else
\ifnum#1=76 %
\hatcurPPteffeccenxxxxxC
\else
\ifnum#1=77 %
\hatcurPPteffeccenxxxxxD
\else
??????\fi
\fi
\fi
\fi
}
\newcommand{\hatcurPPthetaeccen}[1]{\ifnum#1=74 %
\hatcurPPthetaeccenxxxxxA
\else
\ifnum#1=75 %
\hatcurPPthetaeccenxxxxxB
\else
\ifnum#1=76 %
\hatcurPPthetaeccenxxxxxC
\else
\ifnum#1=77 %
\hatcurPPthetaeccenxxxxxD
\else
??????\fi
\fi
\fi
\fi
}
\newcommand{\hatcurPPtinfalleccen}[1]{\ifnum#1=74 %
\hatcurPPtinfalleccenxxxxxA
\else
\ifnum#1=75 %
\hatcurPPtinfalleccenxxxxxB
\else
\ifnum#1=76 %
\hatcurPPtinfalleccenxxxxxC
\else
\ifnum#1=77 %
\hatcurPPtinfalleccenxxxxxD
\else
??????\fi
\fi
\fi
\fi
}
\newcommand{\hatcurRVecceneccen}[1]{\ifnum#1=74 %
\hatcurRVecceneccenxxxxxA
\else
\ifnum#1=75 %
\hatcurRVecceneccenxxxxxB
\else
\ifnum#1=76 %
\hatcurRVecceneccenxxxxxC
\else
\ifnum#1=77 %
\hatcurRVecceneccenxxxxxD
\else
??????\fi
\fi
\fi
\fi
}
\newcommand{\hatcurRVeccentwosiglimeccen}[1]{\ifnum#1=74 %
\hatcurRVeccentwosiglimeccenxxxxxA
\else
\ifnum#1=75 %
\hatcurRVeccentwosiglimeccenxxxxxB
\else
\ifnum#1=76 %
\hatcurRVeccentwosiglimeccenxxxxxC
\else
\ifnum#1=77 %
\hatcurRVeccentwosiglimeccenxxxxxD
\else
??????\fi
\fi
\fi
\fi
}
\newcommand{\hatcurRVfitrmseccen}[1]{\ifnum#1=74 %
\hatcurRVfitrmseccenxxxxxA
\else
\ifnum#1=75 %
\hatcurRVfitrmseccenxxxxxB
\else
\ifnum#1=76 %
\hatcurRVfitrmseccenxxxxxC
\else
\ifnum#1=77 %
\hatcurRVfitrmseccenxxxxxD
\else
??????\fi
\fi
\fi
\fi
}
\newcommand{\hatcurRVgammaeccen}[1]{\ifnum#1=74 %
\hatcurRVgammaeccenxxxxxA
\else
\ifnum#1=75 %
\hatcurRVgammaeccenxxxxxB
\else
\ifnum#1=76 %
\hatcurRVgammaeccenxxxxxC
\else
\ifnum#1=77 %
\hatcurRVgammaeccenxxxxxD
\else
??????\fi
\fi
\fi
\fi
}
\newcommand{\hatcurRVheccen}[1]{\ifnum#1=74 %
\hatcurRVheccenxxxxxA
\else
\ifnum#1=75 %
\hatcurRVheccenxxxxxB
\else
\ifnum#1=76 %
\hatcurRVheccenxxxxxC
\else
\ifnum#1=77 %
\hatcurRVheccenxxxxxD
\else
??????\fi
\fi
\fi
\fi
}
\newcommand{\hatcurRVjittereccen}[1]{\ifnum#1=74 %
\hatcurRVjittereccenxxxxxA
\else
\ifnum#1=75 %
\hatcurRVjittereccenxxxxxB
\else
\ifnum#1=76 %
\hatcurRVjittereccenxxxxxC
\else
\ifnum#1=77 %
\hatcurRVjittereccenxxxxxD
\else
??????\fi
\fi
\fi
\fi
}
\newcommand{\hatcurRVjittertwosiglimeccen}[1]{\ifnum#1=74 %
\hatcurRVjittertwosiglimeccenxxxxxA
\else
\ifnum#1=75 %
\hatcurRVjittertwosiglimeccenxxxxxB
\else
\ifnum#1=76 %
\hatcurRVjittertwosiglimeccenxxxxxC
\else
\ifnum#1=77 %
\hatcurRVjittertwosiglimeccenxxxxxD
\else
??????\fi
\fi
\fi
\fi
}
\newcommand{\hatcurRVkeccen}[1]{\ifnum#1=74 %
\hatcurRVkeccenxxxxxA
\else
\ifnum#1=75 %
\hatcurRVkeccenxxxxxB
\else
\ifnum#1=76 %
\hatcurRVkeccenxxxxxC
\else
\ifnum#1=77 %
\hatcurRVkeccenxxxxxD
\else
??????\fi
\fi
\fi
\fi
}
\newcommand{\hatcurRVKeccen}[1]{\ifnum#1=74 %
\hatcurRVKeccenxxxxxA
\else
\ifnum#1=75 %
\hatcurRVKeccenxxxxxB
\else
\ifnum#1=76 %
\hatcurRVKeccenxxxxxC
\else
\ifnum#1=77 %
\hatcurRVKeccenxxxxxD
\else
??????\fi
\fi
\fi
\fi
}
\newcommand{\hatcurRVKtwosiglimeccen}[1]{\ifnum#1=74 %
\hatcurRVKtwosiglimeccenxxxxxA
\else
\ifnum#1=75 %
\hatcurRVKtwosiglimeccenxxxxxB
\else
\ifnum#1=76 %
\hatcurRVKtwosiglimeccenxxxxxC
\else
\ifnum#1=77 %
\hatcurRVKtwosiglimeccenxxxxxD
\else
??????\fi
\fi
\fi
\fi
}
\newcommand{\hatcurRVomegaeccen}[1]{\ifnum#1=74 %
\hatcurRVomegaeccenxxxxxA
\else
\ifnum#1=75 %
\hatcurRVomegaeccenxxxxxB
\else
\ifnum#1=76 %
\hatcurRVomegaeccenxxxxxC
\else
\ifnum#1=77 %
\hatcurRVomegaeccenxxxxxD
\else
??????\fi
\fi
\fi
\fi
}
\newcommand{\hatcurRVrheccen}[1]{\ifnum#1=74 %
\hatcurRVrheccenxxxxxA
\else
\ifnum#1=75 %
\hatcurRVrheccenxxxxxB
\else
\ifnum#1=76 %
\hatcurRVrheccenxxxxxC
\else
\ifnum#1=77 %
\hatcurRVrheccenxxxxxD
\else
??????\fi
\fi
\fi
\fi
}
\newcommand{\hatcurRVrkeccen}[1]{\ifnum#1=74 %
\hatcurRVrkeccenxxxxxA
\else
\ifnum#1=75 %
\hatcurRVrkeccenxxxxxB
\else
\ifnum#1=76 %
\hatcurRVrkeccenxxxxxC
\else
\ifnum#1=77 %
\hatcurRVrkeccenxxxxxD
\else
??????\fi
\fi
\fi
\fi
}
\newcommand{\hatcurRVtroneeccen}[1]{\ifnum#1=74 %
\hatcurRVtroneeccenxxxxxA
\else
\ifnum#1=75 %
\hatcurRVtroneeccenxxxxxB
\else
\ifnum#1=76 %
\hatcurRVtroneeccenxxxxxC
\else
\ifnum#1=77 %
\hatcurRVtroneeccenxxxxxD
\else
??????\fi
\fi
\fi
\fi
}
\newcommand{\hatcurRVtrtwoeccen}[1]{\ifnum#1=74 %
\hatcurRVtrtwoeccenxxxxxA
\else
\ifnum#1=75 %
\hatcurRVtrtwoeccenxxxxxB
\else
\ifnum#1=76 %
\hatcurRVtrtwoeccenxxxxxC
\else
\ifnum#1=77 %
\hatcurRVtrtwoeccenxxxxxD
\else
??????\fi
\fi
\fi
\fi
}
\newcommand{\hatcurSMEiloggeccen}[1]{\ifnum#1=74 %
\hatcurSMEiloggeccenxxxxxA
\else
\ifnum#1=75 %
\hatcurSMEiloggeccenxxxxxB
\else
\ifnum#1=76 %
\hatcurSMEiloggeccenxxxxxC
\else
\ifnum#1=77 %
\hatcurSMEiloggeccenxxxxxD
\else
??????\fi
\fi
\fi
\fi
}
\newcommand{\hatcurSMEiteffeccen}[1]{\ifnum#1=74 %
\hatcurSMEiteffeccenxxxxxA
\else
\ifnum#1=75 %
\hatcurSMEiteffeccenxxxxxB
\else
\ifnum#1=76 %
\hatcurSMEiteffeccenxxxxxC
\else
\ifnum#1=77 %
\hatcurSMEiteffeccenxxxxxD
\else
??????\fi
\fi
\fi
\fi
}
\newcommand{\hatcurSMEivmaceccen}[1]{\ifnum#1=74 %
\hatcurSMEivmaceccenxxxxxA
\else
\ifnum#1=75 %
\hatcurSMEivmaceccenxxxxxB
\else
\ifnum#1=76 %
\hatcurSMEivmaceccenxxxxxC
\else
\ifnum#1=77 %
\hatcurSMEivmaceccenxxxxxD
\else
??????\fi
\fi
\fi
\fi
}
\newcommand{\hatcurSMEivmiceccen}[1]{\ifnum#1=74 %
\hatcurSMEivmiceccenxxxxxA
\else
\ifnum#1=75 %
\hatcurSMEivmiceccenxxxxxB
\else
\ifnum#1=76 %
\hatcurSMEivmiceccenxxxxxC
\else
\ifnum#1=77 %
\hatcurSMEivmiceccenxxxxxD
\else
??????\fi
\fi
\fi
\fi
}
\newcommand{\hatcurSMEivsineccen}[1]{\ifnum#1=74 %
\hatcurSMEivsineccenxxxxxA
\else
\ifnum#1=75 %
\hatcurSMEivsineccenxxxxxB
\else
\ifnum#1=76 %
\hatcurSMEivsineccenxxxxxC
\else
\ifnum#1=77 %
\hatcurSMEivsineccenxxxxxD
\else
??????\fi
\fi
\fi
\fi
}
\newcommand{\hatcurSMEizfeheccen}[1]{\ifnum#1=74 %
\hatcurSMEizfeheccenxxxxxA
\else
\ifnum#1=75 %
\hatcurSMEizfeheccenxxxxxB
\else
\ifnum#1=76 %
\hatcurSMEizfeheccenxxxxxC
\else
\ifnum#1=77 %
\hatcurSMEizfeheccenxxxxxD
\else
??????\fi
\fi
\fi
\fi
}
\newcommand{\hatcurSMEizfehshorteccen}[1]{\ifnum#1=74 %
\hatcurSMEizfehshorteccenxxxxxA
\else
\ifnum#1=75 %
\hatcurSMEizfehshorteccenxxxxxB
\else
\ifnum#1=76 %
\hatcurSMEizfehshorteccenxxxxxC
\else
\ifnum#1=77 %
\hatcurSMEizfehshorteccenxxxxxD
\else
??????\fi
\fi
\fi
\fi
}
\newcommand{\hatcurXAveccen}[1]{\ifnum#1=74 %
\hatcurXAveccenxxxxxA
\else
\ifnum#1=75 %
\hatcurXAveccenxxxxxB
\else
\ifnum#1=76 %
\hatcurXAveccenxxxxxC
\else
\ifnum#1=77 %
\hatcurXAveccenxxxxxD
\else
??????\fi
\fi
\fi
\fi
}
\newcommand{\hatcurXdisteccen}[1]{\ifnum#1=74 %
\hatcurXdisteccenxxxxxA
\else
\ifnum#1=75 %
\hatcurXdisteccenxxxxxB
\else
\ifnum#1=76 %
\hatcurXdisteccenxxxxxC
\else
\ifnum#1=77 %
\hatcurXdisteccenxxxxxD
\else
??????\fi
\fi
\fi
\fi
}
\newcommand{\hatcurXdistredeccen}[1]{\ifnum#1=74 %
\hatcurXdistredeccenxxxxxA
\else
\ifnum#1=75 %
\hatcurXdistredeccenxxxxxB
\else
\ifnum#1=76 %
\hatcurXdistredeccenxxxxxC
\else
\ifnum#1=77 %
\hatcurXdistredeccenxxxxxD
\else
??????\fi
\fi
\fi
\fi
}
\newcommand{\hatcurXEBVeccen}[1]{\ifnum#1=74 %
\hatcurXEBVeccenxxxxxA
\else
\ifnum#1=75 %
\hatcurXEBVeccenxxxxxB
\else
\ifnum#1=76 %
\hatcurXEBVeccenxxxxxC
\else
\ifnum#1=77 %
\hatcurXEBVeccenxxxxxD
\else
??????\fi
\fi
\fi
\fi
}
\newcommand{\hatcurXsecdureccen}[1]{\ifnum#1=74 %
\hatcurXsecdureccenxxxxxA
\else
\ifnum#1=75 %
\hatcurXsecdureccenxxxxxB
\else
\ifnum#1=76 %
\hatcurXsecdureccenxxxxxC
\else
\ifnum#1=77 %
\hatcurXsecdureccenxxxxxD
\else
??????\fi
\fi
\fi
\fi
}
\newcommand{\hatcurXsecingdureccen}[1]{\ifnum#1=74 %
\hatcurXsecingdureccenxxxxxA
\else
\ifnum#1=75 %
\hatcurXsecingdureccenxxxxxB
\else
\ifnum#1=76 %
\hatcurXsecingdureccenxxxxxC
\else
\ifnum#1=77 %
\hatcurXsecingdureccenxxxxxD
\else
??????\fi
\fi
\fi
\fi
}
\newcommand{\hatcurXsecondaryeccen}[1]{\ifnum#1=74 %
\hatcurXsecondaryeccenxxxxxA
\else
\ifnum#1=75 %
\hatcurXsecondaryeccenxxxxxB
\else
\ifnum#1=76 %
\hatcurXsecondaryeccenxxxxxC
\else
\ifnum#1=77 %
\hatcurXsecondaryeccenxxxxxD
\else
??????\fi
\fi
\fi
\fi
}
\newcommand{\hatcurXsecphaseeccen}[1]{\ifnum#1=74 %
\hatcurXsecphaseeccenxxxxxA
\else
\ifnum#1=75 %
\hatcurXsecphaseeccenxxxxxB
\else
\ifnum#1=76 %
\hatcurXsecphaseeccenxxxxxC
\else
\ifnum#1=77 %
\hatcurXsecphaseeccenxxxxxD
\else
??????\fi
\fi
\fi
\fi
}

\newcommand{\hatcurxxxxxA}{HATS-74A}
\newcommand{\hatcurbxxxxxA}{HATS-74Ab}
\newcommand{\hatcurcxxxxxA}{HATS-74Ac}

\newcommand{\hatcurplanetnumxxxxxA}{74}

\newcommand{\hatcurCCtwomassshortxxxxxA}{11240360-1933257}
\newcommand{\hatcurCCtoixxxxxA}{737.01}
\newcommand{\hatcurCCtoibarexxxxxA}{737}
\newcommand{\hatcurCCticxxxxxA}{219189765}

\newcommand{\hatcurRotPerxxxxxA}{\ensuremath{4.745422 \pm 0.000040}}

\newcommand{\hatcurISOmBxxxxxA}{\ensuremath{0.2284\pm0.0078}}

\newcommand{\hatcurRVgammaabsxxxxxA}{\hatcurRVgamma{\hatcurplanetnumxxxxxA}}                           

\newcommand{\hatcurRVgammarelxxxxxA}{\hatcurRVgamma{\hatcurplanetnumxxxxxA}}                           

\newcommand{\hatcurCCtassvixxxxxA}{\ensuremath{NULL\pm NULL}}                  

\newcommand{\hatcurSMEversionxxxxxA}{i}                                       

\newcommand{\hatcurisoshortxxxxxA}{YY}
\newcommand{\hatcurisofullxxxxxA}{Yonsei-Yale (YY)}
\newcommand{\hatcurisocitexxxxxA}{yi:2001}

\newcommand{\hatcurlumindxxxxxA}{\arstar}

\newcommand{\hatcurjhkfilsetxxxxxA}{ESO}

\newcommand{\hatcurSMEteffxxxxxA}{\ifthenelse{\equal{\hatcurSMEversionxxxxxA}{i}}{\hatcurSMEiteff{\hatcurplanetnumxxxxxA}}{\hatcurSMEiiteff{\hatcurplanetnumxxxxxA}}}
\newcommand{\hatcurSMEzfehxxxxxA}{\ifthenelse{\equal{\hatcurSMEversionxxxxxA}{i}}{\hatcurSMEizfeh{\hatcurplanetnumxxxxxA}}{\hatcurSMEiizfeh{\hatcurplanetnumxxxxxA}}}
\newcommand{\hatcurSMEzfehshortxxxxxA}{\ifthenelse{\equal{\hatcurSMEversionxxxxxA}{i}}{\hatcurSMEizfehshort{\hatcurplanetnumxxxxxA}}{\hatcurSMEiizfehshort{\hatcurplanetnumxxxxxA}}}
\newcommand{\hatcurSMEloggxxxxxA}{\ifthenelse{\equal{\hatcurSMEversionxxxxxA}{i}}{\hatcurSMEilogg{\hatcurplanetnumxxxxxA}}{\hatcurSMEiilogg{\hatcurplanetnumxxxxxA}}}
\newcommand{\hatcurSMEvsinxxxxxA}{\ifthenelse{\equal{\hatcurSMEversionxxxxxA}{i}}{\hatcurSMEivsin{\hatcurplanetnumxxxxxA}}{\hatcurSMEiivsin{\hatcurplanetnumxxxxxA}}}
\newcommand{\hatcurSMEvmacxxxxxA}{\ifthenelse{\equal{\hatcurSMEversionxxxxxA}{i}}{\hatcurSMEivmac{\hatcurplanetnumxxxxxA}}{\hatcurSMEiivmac{\hatcurplanetnumxxxxxA}}}
\newcommand{\hatcurSMEvmicxxxxxA}{\ifthenelse{\equal{\hatcurSMEversionxxxxxA}{i}}{\hatcurSMEivmic{\hatcurplanetnumxxxxxA}}{\hatcurSMEiivmic{\hatcurplanetnumxxxxxA}}}

\newcommand{\hatcurxxxxxB}{HATS-75}
\newcommand{\hatcurbxxxxxB}{HATS-75b}
\newcommand{\hatcurcxxxxxB}{HATS-75c}

\newcommand{\hatcurplanetnumxxxxxB}{75}

\newcommand{\hatcurCCtwomassshortxxxxxB}{04034783-2524320}
\newcommand{\hatcurCCtoixxxxxB}{552.01}
\newcommand{\hatcurCCtoibarexxxxxB}{552}
\newcommand{\hatcurCCticxxxxxB}{44737596}

\newcommand{\hatcurRotPerxxxxxB}{\ensuremath{35.0435 \pm 0.0017}}

\newcommand{\hatcurISOmBxxxxxB}{\ensuremath{<0.38}}

\newcommand{\hatcurRVgammaabsxxxxxB}{\hatcurRVgamma{\hatcurplanetnumxxxxxB}}                           

\newcommand{\hatcurRVgammarelxxxxxB}{\hatcurRVgamma{\hatcurplanetnumxxxxxB}}                           

\newcommand{\hatcurCCtassvixxxxxB}{\ensuremath{NULL\pm NULL}}                  

\newcommand{\hatcurSMEversionxxxxxB}{i}                                       

\newcommand{\hatcurisoshortxxxxxB}{YY}
\newcommand{\hatcurisofullxxxxxB}{Yonsei-Yale (YY)}
\newcommand{\hatcurisocitexxxxxB}{yi:2001}

\newcommand{\hatcurlumindxxxxxB}{\arstar}

\newcommand{\hatcurjhkfilsetxxxxxB}{ESO}

\newcommand{\hatcurSMEteffxxxxxB}{\ifthenelse{\equal{\hatcurSMEversionxxxxxB}{i}}{\hatcurSMEiteff{\hatcurplanetnumxxxxxB}}{\hatcurSMEiiteff{\hatcurplanetnumxxxxxB}}}
\newcommand{\hatcurSMEzfehxxxxxB}{\ifthenelse{\equal{\hatcurSMEversionxxxxxB}{i}}{\hatcurSMEizfeh{\hatcurplanetnumxxxxxB}}{\hatcurSMEiizfeh{\hatcurplanetnumxxxxxB}}}
\newcommand{\hatcurSMEzfehshortxxxxxB}{\ifthenelse{\equal{\hatcurSMEversionxxxxxB}{i}}{\hatcurSMEizfehshort{\hatcurplanetnumxxxxxB}}{\hatcurSMEiizfehshort{\hatcurplanetnumxxxxxB}}}
\newcommand{\hatcurSMEloggxxxxxB}{\ifthenelse{\equal{\hatcurSMEversionxxxxxB}{i}}{\hatcurSMEilogg{\hatcurplanetnumxxxxxB}}{\hatcurSMEiilogg{\hatcurplanetnumxxxxxB}}}
\newcommand{\hatcurSMEvsinxxxxxB}{\ifthenelse{\equal{\hatcurSMEversionxxxxxB}{i}}{\hatcurSMEivsin{\hatcurplanetnumxxxxxB}}{\hatcurSMEiivsin{\hatcurplanetnumxxxxxB}}}
\newcommand{\hatcurSMEvmacxxxxxB}{\ifthenelse{\equal{\hatcurSMEversionxxxxxB}{i}}{\hatcurSMEivmac{\hatcurplanetnumxxxxxB}}{\hatcurSMEiivmac{\hatcurplanetnumxxxxxB}}}
\newcommand{\hatcurSMEvmicxxxxxB}{\ifthenelse{\equal{\hatcurSMEversionxxxxxB}{i}}{\hatcurSMEivmic{\hatcurplanetnumxxxxxB}}{\hatcurSMEiivmic{\hatcurplanetnumxxxxxB}}}

\newcommand{\hatcurxxxxxC}{HATS-76}
\newcommand{\hatcurbxxxxxC}{HATS-76b}
\newcommand{\hatcurcxxxxxC}{HATS-76c}

\newcommand{\hatcurplanetnumxxxxxC}{76}

\newcommand{\hatcurCCtwomassshortxxxxxC}{04412154-3219128}
\newcommand{\hatcurCCtoixxxxxC}{555.01}
\newcommand{\hatcurCCtoibarexxxxxC}{555}
\newcommand{\hatcurCCticxxxxxC}{170849515}

\newcommand{\hatcurRotPerxxxxxC}{\ensuremath{15.16063 \pm 0.00048}}

\newcommand{\hatcurISOmBxxxxxC}{\ensuremath{<0.24}}

\newcommand{\hatcurRVgammaabsxxxxxC}{\hatcurRVgamma{\hatcurplanetnumxxxxxC}}                           

\newcommand{\hatcurRVgammarelxxxxxC}{\hatcurRVgamma{\hatcurplanetnumxxxxxC}}                           

\newcommand{\hatcurCCtassvixxxxxC}{\ensuremath{NULL\pm NULL}}                  

\newcommand{\hatcurSMEversionxxxxxC}{i}                                       

\newcommand{\hatcurisoshortxxxxxC}{YY}
\newcommand{\hatcurisofullxxxxxC}{Yonsei-Yale (YY)}
\newcommand{\hatcurisocitexxxxxC}{yi:2001}

\newcommand{\hatcurlumindxxxxxC}{\arstar}

\newcommand{\hatcurjhkfilsetxxxxxC}{ESO}

\newcommand{\hatcurSMEteffxxxxxC}{\ifthenelse{\equal{\hatcurSMEversionxxxxxC}{i}}{\hatcurSMEiteff{\hatcurplanetnumxxxxxC}}{\hatcurSMEiiteff{\hatcurplanetnumxxxxxC}}}
\newcommand{\hatcurSMEzfehxxxxxC}{\ifthenelse{\equal{\hatcurSMEversionxxxxxC}{i}}{\hatcurSMEizfeh{\hatcurplanetnumxxxxxC}}{\hatcurSMEiizfeh{\hatcurplanetnumxxxxxC}}}
\newcommand{\hatcurSMEzfehshortxxxxxC}{\ifthenelse{\equal{\hatcurSMEversionxxxxxC}{i}}{\hatcurSMEizfehshort{\hatcurplanetnumxxxxxC}}{\hatcurSMEiizfehshort{\hatcurplanetnumxxxxxC}}}
\newcommand{\hatcurSMEloggxxxxxC}{\ifthenelse{\equal{\hatcurSMEversionxxxxxC}{i}}{\hatcurSMEilogg{\hatcurplanetnumxxxxxC}}{\hatcurSMEiilogg{\hatcurplanetnumxxxxxC}}}
\newcommand{\hatcurSMEvsinxxxxxC}{\ifthenelse{\equal{\hatcurSMEversionxxxxxC}{i}}{\hatcurSMEivsin{\hatcurplanetnumxxxxxC}}{\hatcurSMEiivsin{\hatcurplanetnumxxxxxC}}}
\newcommand{\hatcurSMEvmacxxxxxC}{\ifthenelse{\equal{\hatcurSMEversionxxxxxC}{i}}{\hatcurSMEivmac{\hatcurplanetnumxxxxxC}}{\hatcurSMEiivmac{\hatcurplanetnumxxxxxC}}}
\newcommand{\hatcurSMEvmicxxxxxC}{\ifthenelse{\equal{\hatcurSMEversionxxxxxC}{i}}{\hatcurSMEivmic{\hatcurplanetnumxxxxxC}}{\hatcurSMEiivmic{\hatcurplanetnumxxxxxC}}}

\newcommand{\hatcurxxxxxD}{HATS-77}
\newcommand{\hatcurbxxxxxD}{HATS-77b}
\newcommand{\hatcurcxxxxxD}{HATS-77c}

\newcommand{\hatcurplanetnumxxxxxD}{77}

\newcommand{\hatcurCCtwomassshortxxxxxD}{09591770-2723339}
\newcommand{\hatcurCCtoixxxxxD}{730.01}
\newcommand{\hatcurCCtoibarexxxxxD}{730}
\newcommand{\hatcurCCticxxxxxD}{11561667}

\newcommand{\hatcurRotPerxxxxxD}{\ensuremath{\cdots}}

\newcommand{\hatcurISOmBxxxxxD}{\ensuremath{<0.53}}

\newcommand{\hatcurRVgammaabsxxxxxD}{\hatcurRVgamma{\hatcurplanetnumxxxxxD}}                           

\newcommand{\hatcurRVgammarelxxxxxD}{\hatcurRVgamma{\hatcurplanetnumxxxxxD}}                           

\newcommand{\hatcurCCtassvixxxxxD}{\ensuremath{NULL\pm NULL}}                  

\newcommand{\hatcurSMEversionxxxxxD}{i}                                       

\newcommand{\hatcurisoshortxxxxxD}{YY}
\newcommand{\hatcurisofullxxxxxD}{Yonsei-Yale (YY)}
\newcommand{\hatcurisocitexxxxxD}{yi:2001}

\newcommand{\hatcurlumindxxxxxD}{\arstar}

\newcommand{\hatcurjhkfilsetxxxxxD}{ESO}

\newcommand{\hatcurSMEteffxxxxxD}{\ifthenelse{\equal{\hatcurSMEversionxxxxxD}{i}}{\hatcurSMEiteff{\hatcurplanetnumxxxxxD}}{\hatcurSMEiiteff{\hatcurplanetnumxxxxxD}}}
\newcommand{\hatcurSMEzfehxxxxxD}{\ifthenelse{\equal{\hatcurSMEversionxxxxxD}{i}}{\hatcurSMEizfeh{\hatcurplanetnumxxxxxD}}{\hatcurSMEiizfeh{\hatcurplanetnumxxxxxD}}}
\newcommand{\hatcurSMEzfehshortxxxxxD}{\ifthenelse{\equal{\hatcurSMEversionxxxxxD}{i}}{\hatcurSMEizfehshort{\hatcurplanetnumxxxxxD}}{\hatcurSMEiizfehshort{\hatcurplanetnumxxxxxD}}}
\newcommand{\hatcurSMEloggxxxxxD}{\ifthenelse{\equal{\hatcurSMEversionxxxxxD}{i}}{\hatcurSMEilogg{\hatcurplanetnumxxxxxD}}{\hatcurSMEiilogg{\hatcurplanetnumxxxxxD}}}
\newcommand{\hatcurSMEvsinxxxxxD}{\ifthenelse{\equal{\hatcurSMEversionxxxxxD}{i}}{\hatcurSMEivsin{\hatcurplanetnumxxxxxD}}{\hatcurSMEiivsin{\hatcurplanetnumxxxxxD}}}
\newcommand{\hatcurSMEvmacxxxxxD}{\ifthenelse{\equal{\hatcurSMEversionxxxxxD}{i}}{\hatcurSMEivmac{\hatcurplanetnumxxxxxD}}{\hatcurSMEiivmac{\hatcurplanetnumxxxxxD}}}
\newcommand{\hatcurSMEvmicxxxxxD}{\ifthenelse{\equal{\hatcurSMEversionxxxxxD}{i}}{\hatcurSMEivmic{\hatcurplanetnumxxxxxD}}{\hatcurSMEiivmic{\hatcurplanetnumxxxxxD}}}

\newcommand{\hatcur}[1]{\ifnum#1=74 %
\hatcurxxxxxA
\else
\ifnum#1=75 %
\hatcurxxxxxB
\else
\ifnum#1=76 %
\hatcurxxxxxC
\else
\ifnum#1=77 %
\hatcurxxxxxD
\else
??????\fi
\fi
\fi
\fi
}
\newcommand{\hatcurb}[1]{\ifnum#1=74 %
\hatcurbxxxxxA
\else
\ifnum#1=75 %
\hatcurbxxxxxB
\else
\ifnum#1=76 %
\hatcurbxxxxxC
\else
\ifnum#1=77 %
\hatcurbxxxxxD
\else
??????\fi
\fi
\fi
\fi
}
\newcommand{\hatcurc}[1]{\ifnum#1=74 %
\hatcurcxxxxxA
\else
\ifnum#1=75 %
\hatcurcxxxxxB
\else
\ifnum#1=76 %
\hatcurcxxxxxC
\else
\ifnum#1=77 %
\hatcurcxxxxxD
\else
??????\fi
\fi
\fi
\fi
}
\newcommand{\hatcurCCtassvi}[1]{\ifnum#1=74 %
\hatcurCCtassvixxxxxA
\else
\ifnum#1=75 %
\hatcurCCtassvixxxxxB
\else
\ifnum#1=76 %
\hatcurCCtassvixxxxxC
\else
\ifnum#1=77 %
\hatcurCCtassvixxxxxD
\else
??????\fi
\fi
\fi
\fi
}
\newcommand{\hatcurCCtic}[1]{\ifnum#1=74 %
\hatcurCCticxxxxxA
\else
\ifnum#1=75 %
\hatcurCCticxxxxxB
\else
\ifnum#1=76 %
\hatcurCCticxxxxxC
\else
\ifnum#1=77 %
\hatcurCCticxxxxxD
\else
??????\fi
\fi
\fi
\fi
}
\newcommand{\hatcurCCtoi}[1]{\ifnum#1=74 %
\hatcurCCtoixxxxxA
\else
\ifnum#1=75 %
\hatcurCCtoixxxxxB
\else
\ifnum#1=76 %
\hatcurCCtoixxxxxC
\else
\ifnum#1=77 %
\hatcurCCtoixxxxxD
\else
??????\fi
\fi
\fi
\fi
}

\newcommand{\hatcurCCtoibare}[1]{\ifnum#1=74 %
\hatcurCCtoibarexxxxxA
\else
\ifnum#1=75 %
\hatcurCCtoibarexxxxxB
\else
\ifnum#1=76 %
\hatcurCCtoibarexxxxxC
\else
\ifnum#1=77 %
\hatcurCCtoibarexxxxxD
\else
??????\fi
\fi
\fi
\fi
}

\newcommand{\hatcurCCtwomassshort}[1]{\ifnum#1=74 %
\hatcurCCtwomassshortxxxxxA
\else
\ifnum#1=75 %
\hatcurCCtwomassshortxxxxxB
\else
\ifnum#1=76 %
\hatcurCCtwomassshortxxxxxC
\else
\ifnum#1=77 %
\hatcurCCtwomassshortxxxxxD
\else
??????\fi
\fi
\fi
\fi
}
\newcommand{\hatcurisocite}[1]{\ifnum#1=74 %
\hatcurisocitexxxxxA
\else
\ifnum#1=75 %
\hatcurisocitexxxxxB
\else
\ifnum#1=76 %
\hatcurisocitexxxxxC
\else
\ifnum#1=77 %
\hatcurisocitexxxxxD
\else
??????\fi
\fi
\fi
\fi
}
\newcommand{\hatcurisofull}[1]{\ifnum#1=74 %
\hatcurisofullxxxxxA
\else
\ifnum#1=75 %
\hatcurisofullxxxxxB
\else
\ifnum#1=76 %
\hatcurisofullxxxxxC
\else
\ifnum#1=77 %
\hatcurisofullxxxxxD
\else
??????\fi
\fi
\fi
\fi
}
\newcommand{\hatcurisoshort}[1]{\ifnum#1=74 %
\hatcurisoshortxxxxxA
\else
\ifnum#1=75 %
\hatcurisoshortxxxxxB
\else
\ifnum#1=76 %
\hatcurisoshortxxxxxC
\else
\ifnum#1=77 %
\hatcurisoshortxxxxxD
\else
??????\fi
\fi
\fi
\fi
}
\newcommand{\hatcurjhkfilset}[1]{\ifnum#1=74 %
\hatcurjhkfilsetxxxxxA
\else
\ifnum#1=75 %
\hatcurjhkfilsetxxxxxB
\else
\ifnum#1=76 %
\hatcurjhkfilsetxxxxxC
\else
\ifnum#1=77 %
\hatcurjhkfilsetxxxxxD
\else
??????\fi
\fi
\fi
\fi
}
\newcommand{\hatcurlumind}[1]{\ifnum#1=74 %
\hatcurlumindxxxxxA
\else
\ifnum#1=75 %
\hatcurlumindxxxxxB
\else
\ifnum#1=76 %
\hatcurlumindxxxxxC
\else
\ifnum#1=77 %
\hatcurlumindxxxxxD
\else
??????\fi
\fi
\fi
\fi
}
\newcommand{\hatcurplanetnum}[1]{\ifnum#1=74 %
\hatcurplanetnumxxxxxA
\else
\ifnum#1=75 %
\hatcurplanetnumxxxxxB
\else
\ifnum#1=76 %
\hatcurplanetnumxxxxxC
\else
\ifnum#1=77 %
\hatcurplanetnumxxxxxD
\else
??????\fi
\fi
\fi
\fi
}
\newcommand{\hatcurRotPer}[1]{\ifnum#1=74 %
\hatcurRotPerxxxxxA
\else
\ifnum#1=75 %
\hatcurRotPerxxxxxB
\else
\ifnum#1=76 %
\hatcurRotPerxxxxxC
\else
\ifnum#1=77 %
\hatcurRotPerxxxxxD
\else
??????\fi
\fi
\fi
\fi
}
\newcommand{\hatcurISOmB}[1]{\ifnum#1=74 %
\hatcurISOmBxxxxxA
\else
\ifnum#1=75 %
\hatcurISOmBxxxxxB
\else
\ifnum#1=76 %
\hatcurISOmBxxxxxC
\else
\ifnum#1=77 %
\hatcurISOmBxxxxxD
\else
??????\fi
\fi
\fi
\fi
}
\newcommand{\hatcurRVgammaabs}[1]{\ifnum#1=74 %
\hatcurRVgammaabsxxxxxA
\else
\ifnum#1=75 %
\hatcurRVgammaabsxxxxxB
\else
\ifnum#1=76 %
\hatcurRVgammaabsxxxxxC
\else
\ifnum#1=77 %
\hatcurRVgammaabsxxxxxD
\else
??????\fi
\fi
\fi
\fi
}
\newcommand{\hatcurRVgammarel}[1]{\ifnum#1=74 %
\hatcurRVgammarelxxxxxA
\else
\ifnum#1=75 %
\hatcurRVgammarelxxxxxB
\else
\ifnum#1=76 %
\hatcurRVgammarelxxxxxC
\else
\ifnum#1=77 %
\hatcurRVgammarelxxxxxD
\else
??????\fi
\fi
\fi
\fi
}
\newcommand{\hatcurSMElogg}[1]{\ifnum#1=74 %
\hatcurSMEloggxxxxxA
\else
\ifnum#1=75 %
\hatcurSMEloggxxxxxB
\else
\ifnum#1=76 %
\hatcurSMEloggxxxxxC
\else
\ifnum#1=77 %
\hatcurSMEloggxxxxxD
\else
??????\fi
\fi
\fi
\fi
}
\newcommand{\hatcurSMEteff}[1]{\ifnum#1=74 %
\hatcurSMEteffxxxxxA
\else
\ifnum#1=75 %
\hatcurSMEteffxxxxxB
\else
\ifnum#1=76 %
\hatcurSMEteffxxxxxC
\else
\ifnum#1=77 %
\hatcurSMEteffxxxxxD
\else
??????\fi
\fi
\fi
\fi
}
\newcommand{\hatcurSMEversion}[1]{\ifnum#1=74 %
\hatcurSMEversionxxxxxA
\else
\ifnum#1=75 %
\hatcurSMEversionxxxxxB
\else
\ifnum#1=76 %
\hatcurSMEversionxxxxxC
\else
\ifnum#1=77 %
\hatcurSMEversionxxxxxD
\else
??????\fi
\fi
\fi
\fi
}
\newcommand{\hatcurSMEvmac}[1]{\ifnum#1=74 %
\hatcurSMEvmacxxxxxA
\else
\ifnum#1=75 %
\hatcurSMEvmacxxxxxB
\else
\ifnum#1=76 %
\hatcurSMEvmacxxxxxC
\else
\ifnum#1=77 %
\hatcurSMEvmacxxxxxD
\else
??????\fi
\fi
\fi
\fi
}
\newcommand{\hatcurSMEvmic}[1]{\ifnum#1=74 %
\hatcurSMEvmicxxxxxA
\else
\ifnum#1=75 %
\hatcurSMEvmicxxxxxB
\else
\ifnum#1=76 %
\hatcurSMEvmicxxxxxC
\else
\ifnum#1=77 %
\hatcurSMEvmicxxxxxD
\else
??????\fi
\fi
\fi
\fi
}
\newcommand{\hatcurSMEvsin}[1]{\ifnum#1=74 %
\hatcurSMEvsinxxxxxA
\else
\ifnum#1=75 %
\hatcurSMEvsinxxxxxB
\else
\ifnum#1=76 %
\hatcurSMEvsinxxxxxC
\else
\ifnum#1=77 %
\hatcurSMEvsinxxxxxD
\else
??????\fi
\fi
\fi
\fi
}
\newcommand{\hatcurSMEzfeh}[1]{\ifnum#1=74 %
\hatcurSMEzfehxxxxxA
\else
\ifnum#1=75 %
\hatcurSMEzfehxxxxxB
\else
\ifnum#1=76 %
\hatcurSMEzfehxxxxxC
\else
\ifnum#1=77 %
\hatcurSMEzfehxxxxxD
\else
??????\fi
\fi
\fi
\fi
}
\newcommand{\hatcurSMEzfehshort}[1]{\ifnum#1=74 %
\hatcurSMEzfehshortxxxxxA
\else
\ifnum#1=75 %
\hatcurSMEzfehshortxxxxxB
\else
\ifnum#1=76 %
\hatcurSMEzfehshortxxxxxC
\else
\ifnum#1=77 %
\hatcurSMEzfehshortxxxxxD
\else
??????\fi
\fi
\fi
\fi
}

\newcounter{planetcounter}

\newboolean{emulateapj}

\setboolean{emulateapj}{true}

\newboolean{rvtablelong}
\setboolean{rvtablelong}{true}

\newboolean{astroph}

\setboolean{astroph}{true}

\shortauthors{Jord\'an et al.}
\shorttitle{\hatcur{74}\lowercase{b}, \hatcur{75}\lowercase{b}, \hatcur{76}\lowercase{b} and \hatcur{77}\lowercase{b}}
\ifthenelse{\boolean{emulateapj}}{
    
}{
    
}

\begin{document}

\title{
\hatcur{74}\lowercase{b}, \hatcur{75}\lowercase{b}, \hatcur{76}\lowercase{b} and \hatcur{77}\lowercase{b}: Four Transiting Giant Planets Around K and M Dwarfs\footnote{The HATSouth network is operated by a collaboration consisting of
Princeton University (PU), the Max Planck Institute f\"ur Astronomie
(MPIA), the Australian National University (ANU), and the Universidad Adolfo Ib\'a\~nez (UAI).
The station at Las Campanas
Observatory (LCO) of the Carnegie Institute is operated by PU in
conjunction with UAI, the station at the High Energy Spectroscopic
Survey (H.E.S.S.) site is operated in conjunction with MPIA, and the
station at Siding Spring Observatory (SSO) is operated jointly with
ANU.
 Based in
 part on observations made with the MPG~2.2\,m Telescope at the ESO Observatory in La Silla.
Based on observations collected at the European Southern Observatory.
}
}

\correspondingauthor{Andr\'es Jord\'an}
\email{andres.jordan@uai.cl}

\author[0000-0002-5389-3944]{Andr\'es Jord\'an}
\affiliation{Facultad de Ingenier\'ia y Ciencias, Universidad Adolfo Ib\'a\~nez, Av.\ Diagonal las Torres 2640, Pe\~nalol\'en, Santiago, Chile}
\affiliation{Millennium Institute for Astrophysics, Chile}

\author[0000-0001-8732-6166]{J.~D.\ Hartman}
\affil{Department of Astrophysical Sciences, Princeton University, NJ 08544, USA}

\author[0000-0001-6023-1335]{D.\ Bayliss}
\affil{Department of Physics, University of Warwick, Gibbet Hill Road, Coventry CV4 7AL, UK}

\author[0000-0001-7204-6727]{G.\ \'A. Bakos}
\affil{Department of Astrophysical Sciences, Princeton University, NJ 08544, USA}

\author[0000-0002-9158-7315]{R.\ Brahm}
\affiliation{Facultad de Ingenier\'ia y Ciencias, Universidad Adolfo Ib\'a\~nez, Av.\ Diagonal las Torres 2640, Pe\~nalol\'en, Santiago, Chile}
\affiliation{Millennium Institute for Astrophysics, Chile}

\author[0000-0001-7904-4441]{E.~M.\ Bryant}
\affiliation{Department of Physics, University of Warwick, Gibbet Hill Road, Coventry CV4 7AL, UK}
\affiliation{Centre for Exoplanets and Habitability, University of Warwick, Gibbet Hill Road, Coventry CV4 7AL, UK}

\author{Z.\ Csubry}
\affil{Department of Astrophysical Sciences, Princeton University, NJ 08544, USA}

\author{Th.\ Henning}
\affil{Max Planck Institute for Astronomy, K{\"{o}}nigstuhl 17, 69117 - Heidelberg, Germany}

\author{M.\ Hobson}
\affiliation{Millennium Institute for Astrophysics, Chile}
\affil{Max Planck Institute for Astronomy, K{\"{o}}nigstuhl 17, 69117 - Heidelberg, Germany}

\author[0000-0002-9428-8732]{L.\ Mancini}
\affil{Department of Physics, University of Rome Tor Vergata, Via della
Ricerca Scientifica 1, I-00133 - Roma, Italy}
\affil{INAF - Astrophysical Observatory of Turin, Via Osservatorio 20, I-10025 - Pino Torinese, Italy}
\affil{Max Planck Institute for Astronomy, K{\"{o}}nigstuhl 17, 69117 - Heidelberg, Germany}

\author[0000-0003-4464-1371]{K.\ Penev}
\affil{Department of Physics, University of Texas at Dallas, Richardson, TX 75080, USA}

\author[0000-0003-2935-7196]{M.\ Rabus}
\affil{Departamento de Matem\'atica y F\'isica Aplicadas, Facultad de Ingenier\'ia, Universidad Cat\'olica de la Sant\'isima Concepci\'on, Alonso de Rivera 2850, Concepci\'on, Chile}

\author[0000-0001-7070-3842]{V.\ Suc}
\affiliation{Facultad de Ingenier\'ia y Ciencias, Universidad Adolfo Ib\'a\~nez, Av.\ Diagonal las Torres 2640, Pe\~nalol\'en, Santiago, Chile}
\affiliation{Millennium Institute for Astrophysics, Chile}

\author[0000-0002-0455-9384]{M.\ de~Val-Borro}
\affil{Astrochemistry Laboratory, Goddard Space Flight Center, NASA, 8800 Greenbelt Rd, Greenbelt, MD 20771, USA}

\author{J.\ Wallace}
\affiliation{Department of Astrophysical Sciences, Princeton University, NJ 08544, USA}

\author[0000-0003-1464-9276]{K.\ Barkaoui}
\affiliation{Astrobiology Research Unit, Universit\'e de Li\`ege, 19C All\'ee du 6 Ao\^ut, 4000 Li\`ege, Belgium}
\affiliation{Oukaimeden Observatory, High Energy Physics and Astrophysics Laboratory, Cadi Ayyad University, Marrakech, Morocco}

\author[0000-0002-5741-3047]{David R. Ciardi}
\affiliation{NASA Exoplanet Science Institute - Caltech/IPAC, Pasadena, CA 91125 USA}

\author[0000-0001-6588-9574]{K.\ A.\ Collins}
\affiliation{Center for Astrophysics \textbar \ Harvard \& Smithsonian, 60 Garden Street, Cambridge, MA 02138, USA}

\author{E.\ Esparza-Borges}
\affiliation{Departamento de Astrof\'\i sica, Universidad de La Laguna (ULL), 38206, La Laguna, Tenerife, Spain}

\author[0000-0001-9800-6248]{E. Furlan}
\affiliation{NASA Exoplanet Science Institute, Caltech/IPAC, Mail Code 100-22, 1200 E. California Blvd., Pasadena, CA 91125, USA}

\author[0000-0002-4503-9705]{T.\ Gan}
\affiliation{Department of Astronomy and Tsinghua Centre for Astrophysics, Tsinghua University, Beijing 100084, China}

\author{M.\ Ghachoui}
\affiliation{Astrobiology Research Unit, Universit\'e de Li\`ege, 19C All\'ee du 6 Ao\^ut, 4000 Li\`ege, Belgium}
\affiliation{Oukaimeden Observatory, High Energy Physics and Astrophysics Laboratory, Cadi Ayyad University, Marrakech, Morocco}

\author{M.\ Gillon}
\affiliation{Astrobiology Research Unit, Universit\'e de Li\`ege, 19C All\'ee du 6 Ao\^ut, 4000 Li\`ege, Belgium}

\author{S.\ Howell}
\affiliation{NASA Ames Research Center, Moffett Field, CA 94035, USA}

\author{E.~Jehin}
\affiliation{Space sciences, Technologies and Astrophysics Research (STAR) Institute, Universit\'e de Li\`ege, Belgium}

\author[0000-0002-4909-5763]{A.~Fukui}
\affiliation{Komaba Institute for Science, The University of Tokyo, 3-8-1 Komaba, Meguro, Tokyo 153-8902, Japan}
\affiliation{Instituto de Astrof\'\i sica de Canarias (IAC), 38205 La Laguna, Tenerife, Spain}

\author{K.~Kawauchi}
\affiliation{Instituto de Astrof\'\i sica de Canarias (IAC), 38205 La Laguna, Tenerife, Spain}

\author[0000-0002-4881-3620]{J.~H.\ Livingston}
\affiliation{Department of Astronomy, The University of Tokyo, Hongo 7-3-1, Bunkyo-ku, Tokyo, 113-0033, Japan}

\author{R.\ Luque}
\affiliation{Instituto de Astrof\'isica de Andaluc\'ia (IAA-CSIC), Glorieta de la Astronom\'ia s/n, 18008 Granada, Spain}

\author[0000-0001-7233-7508]{R.~Matson}
\affiliation{U.S. Naval Observatory, Washington, D.C. 20392, USA}

\author[0000-0003-0593-1560]{E.\ C.\ Matthews}
\affil{Geneva Observatory, University of Geneva, Chemin Pegasi 51, 1290 Versoix, Switzerland}

\author[0000-0002-4047-4724]{H.P.~Osborn}
\affiliation{Kavli Institute for Space Sciences, Massacussets Institute of Technology, Cambridge, MA 02138, USA}
\affiliation{NCCR/Planet S, Centre for Space and Habitability, University of Bern, Switzerland}

\author[0000-0001-9087-1245]{F.\ Murgas}
\affiliation{Instituto de Astrof\'\i sica de Canarias (IAC), 38205 La Laguna, Tenerife, Spain}
\affiliation{Departamento de Astrof\'\i sica, Universidad de La Laguna (ULL), 38206, La Laguna, Tenerife, Spain}

\author{E.\ Palle}
\affiliation{Instituto de Astrof\'\i sica de Canarias (IAC), 38205 La Laguna, Tenerife, Spain}
\affiliation{Departamento de Astrof\'\i sica, Universidad de La Laguna (ULL), 38206, La Laguna, Tenerife, Spain}

\author[0000-0002-8961-0352]{W.\ C. Waalkes}
\affiliation{Department of Astrophysical and Planetary Sciences, University of Colorado, Boulder, CO 80309, USA}


\begin{abstract}

\setcounter{footnote}{10}
The relative rarity of giant planets around low mass stars compared with solar-type stars is a key prediction from core accretion planet formation theory.  In this paper we report on the discovery of four gas giant planets that transit low mass late K and early M dwarfs. The planets \hatcurb{74} (TOI~\hatcurCCtoibare{74}b), \hatcurb{75} (TOI~\hatcurCCtoibare{75}b), \hatcurb{76} (TOI~\hatcurCCtoibare{76}b), and \hatcurb{77} (TOI~\hatcurCCtoibare{77}b), were all discovered from the HATSouth photometric survey and followed-up using TESS and other photometric facilities.  We use the new ESPRESSO facility at the VLT to confirm and systems and measure their masses.
We find that that planets have masses of \hatcurPPm{74}\,\mjup, \hatcurPPm{75}\,\mjup, \hatcurPPm{76}\,\mjup\ and \hatcurPPm{77}\,\mjup, respectively, and radii of \hatcurPPr{74}\,\rjup, \hatcurPPr{75}\,\rjup, \hatcurPPr{76}\,\rjup, and \hatcurPPr{77}\,\rjup, respectively.  The planets all orbit close to their host stars with orbital periods ranging from \hatcurLCPshort{74}\,d to \hatcurLCPshort{77}\,d.  With further work we aim to test core accretion theory by using these and further discoveries to quantify the occurrence rate of giant planets around low mass host stars.
\setcounter{footnote}{0}
\end{abstract}

\keywords{
    planetary systems ---
    stars: individual (
\setcounter{planetcounter}{1}
\hatcur{74},
TOI~\hatcurCCtoibare{74},
TIC~\hatcurCCtic{74}\loopcommanoperiod
\setcounter{planetcounter}{2}
\hatcur{75},
TOI~\hatcurCCtoibare{75},
TIC~\hatcurCCtic{75}\loopcommanoperiod
\setcounter{planetcounter}{3}
\hatcur{76},
TOI~\hatcurCCtoibare{76},
TIC~\hatcurCCtic{76}\loopcommanoperiod
\setcounter{planetcounter}{4}
\hatcur{77},
TOI~\hatcurCCtoibare{77},
TIC~\hatcurCCtic{77}\loopcommanoperiod
\setcounter{planetcounter}{5}
)
    techniques: spectroscopic, photometric
}


\section{Introduction}
\label{sec:introduction}

One of the basic quantities of interest in exoplanetary science is the planet occurrence rate
expressed as a function of both the properties of the planets and the stars that host them.
A significant early result was the realization that occurrence of gas giants scaled with stellar metallicity, in the sense that more metal-rich stars were more likely to host gas giants \citep[e.g.][]{gonzalez:1998,santos:2004,fischer:2005}.  This  provided strong support to the core-accretion scenario for the formation of short period gas giants, illustrating how occurrence rates can provide stringent tests for the processes that drive the formation and evolution of planetary systems.

Various surveys using the whole spectrum of exoplanet detection techniques continued advancing towards a better determination of occurrence rates \citep[for recent reviews, see][]{mulders:2018, Zhu:2021}, and in particular to determining the joint dependence of occurrence rate with metallicity and mass. Despite significant progress, there remain large gaps in our understanding of many classes of planetary system.  One of them is the occurrence of giant planets around low-mass stars with masses $M \lesssim 0.6$ $M_\odot$, which corresponds to stars of type M and later. While some studies suggested that the occurrence rate increased with stellar mass and was significantly higher for FGK hosts as compared to M dwarf hosts \citep[][]{johnson:2010,clanton:2014,montet:2014}, others have shown that this result was not statistically significant \citep{mortier:2013,gaidos:2014,obermeier:2016} and conclude the data were consistent with no dependence on stellar mass.  The recent radial velocity study of M dwarfs by \citet{sabotta:2021} cannot rule out the giant planet occurrence rate being the same for M and G dwarf hosts. The Kepler mission allowed great progress in the determination of occurrence rates down to Earth-size planets \citep[e.g.][]{hsu:2019}, but it did not improve significantly the situation for giant planets around M dwarfs.
Giant planets are very rare in comparison to sub-Neptunes, and M dwarfs are intrinsically faint. As a result, very few giant planets around M dwarfs were uncovered by Kepler. Indeed, of the  137 giant planets ($R_p > 0.6$ $R_J$) validated by Kepler, only two are orbiting M dwarfs  \citep{doyle:2011,johnson:2012}.

Formation models based on the core-accretion paradigm predict that M dwarf systems should form very few, if any, giant planets. This is a consequence of the lack of sufficient mass surface density and the increased orbital timescales around low-mass stars \citep[e.g.,][]{laughlin:2004,ida:2005}. The occurrence rate of giant planets is predicted by recent models to decrease from their value for FGK dwarfs down to zero in the stellar mass range 0.7 $M_\odot$--0.3 $M_\odot$ \citep{Burn:2021}. This prediction is currently not well tested observationally due to the very low number of M stars monitored in exoplanet surveys, although the recent discovery of a giant planet with a minimum mass 0.46 $M_J$ around a 0.123 $M_\odot$ star \citep{morales:2019} is already providing some tension for this prediction. Therefore, systematically uncovering these systems is of importance as it allows us to map the planet formation efficiency in a region of parameter space where dramatic changes are expected.

In order to discover significant numbers of giants around low mass stars it is necessary to scan larger regions of the sky and go deeper, often to magnitudes $V \gtrsim 16$, which in turn makes the confirmation via radial velocities significantly more challenging.
The TESS mission is surveying the whole sky, providing new candidate giants around low-mass stars. There is a synergy in this search with ground-based surveys, particularly those such as HATSouth \citep{bakos:2013:hatsouth} that have a larger aperture than TESS and can therefore provide competitive photometric accuracy at the faint magnitude of the typical target of interest. In this work we present the discovery of four giant planets around early M and late K dwarfs with stellar masses in the range $0.6-0.65 \, M_\odot$, a result of a systematic effort to discover giant planets around low mass stars exploiting the synergies between TESS and ground-based surveys.  The paper is structured as follows: in \S2 we describe the
data which were used to perform the global modeling of the system as described in \S3. The results are discussed in \S4.

\section{Observations}
\label{sec:obs}

Figures~\ref{fig:hats74}, \ref{fig:hats75}, \ref{fig:hats76} and
\ref{fig:hats77} show the observations collected for
\hatcur{74}, \hatcur{75}, \hatcur{76} and \hatcur{77},
respectively. Each figure shows the HATSouth light curve used to
detect the transits, the ground-based follow-up transit light curves,
the high-precision RVs, and the
catalog broad-band photometry, including parallax corrections from
Gaia~DR2 \citep{gaiadr2}, used in characterizing the host stars. We also show the TESS
light curves for each system in
Figures~\ref{fig:hats74tess}, \ref{fig:hats75tess},
\ref{fig:hats76tess} and \ref{fig:hats77tess}. Below we
describe the observations of these objects that were collected and analyzed here.

\subsection{Photometric detection}
\label{sec:detection}

All four of the systems presented here were discovered as transiting
planet candidates by the HATSouth ground-based transiting planet
survey \citep{bakos:2013:hatsouth} as we discuss in
Section~\ref{sec:hatsouth}.  Following the detection of transits for these four systems by HATSouth, we proposed for short-cadence NASA {\em TESS} observations for all of these systems through the NASA {\em TESS} Guest Investigator Program (G011214). All four objects showed clear transits in the {\em TESS} data (Section~\ref{sec:tess}), and were independently selected as transit candidates, based on these observations, by the {\em TESS} team.

\subsubsection{HATSouth}
\label{sec:hatsouth}

\begin{figure*}[!ht]
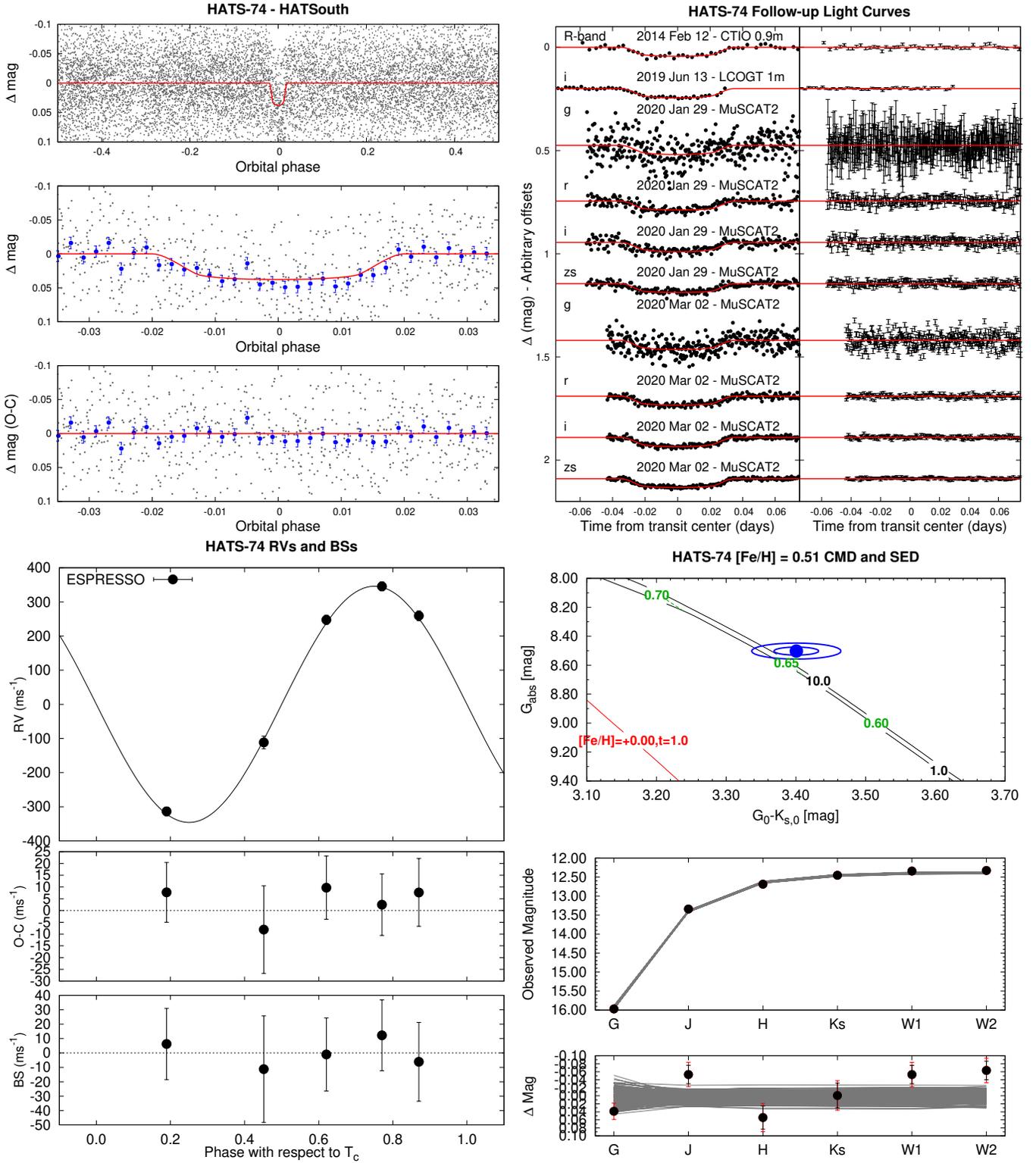

 {
 \centering
 \leavevmode
 \includegraphics[width={1.0\linewidth}]{\hatcurhtr{74}-banner}
}
 {
 \centering
 \leavevmode
 \includegraphics[width={0.5\linewidth}]{\hatcurhtr{74}-hs}%
 \hfil
 \includegraphics[width={0.5\linewidth}]{\hatcurhtr{74}-lc}%
 }
 {
 \centering
 \leavevmode
 \includegraphics[width={0.5\linewidth}]{\hatcurhtr{74}-rv}%
 \hfil
 \includegraphics[width={0.5\linewidth}]{\hatcurhtr{74}-iso-gk-gabs-isofeh-SED}%
 }
\caption{
    Observations used to confirm the transiting planet system \hatcur{74}, excluding data from the NASA {\em TESS} mission which are shown in Figure~\ref{fig:hats74tess}. {\em Top Left:} Phase-folded unbinned HATSouth light curve. The
    top panel shows the full light curve, the middle panel shows
    the light curve zoomed-in on the transit, and the bottom panel shows the residuals from the best-fit model zoomed-in on the transit. The solid lines show the
    model fits to the light curves. The dark filled circles show the light curves binned in phase with a bin
    size of 0.002. (Caption continued on next page.)
\label{fig:hats74}
}
    \end{figure*}

HATSouth uses a network of 24 telescopes, each 0.18\,m in aperture,
and 4K$\times$4K front-side-illuminated CCD cameras. These are
attached to a total of six fully-automated mounts, each with an
associated enclosure, which are in turn located at three sites around
the Southern hemisphere. The three sites are Las Campanas Observatory (LCO) in
Chile, the site of the H.E.S.S.\ gamma-ray observatory in Namibia, and
Siding Spring Observatory (SSO) in Australia. The operations and
observing procedures of the network were described by
\citet{bakos:2013:hatsouth}, while the method for reducing the images
to trend-filtered light curves and searching for candidate transiting
planets were described by \citet{penev:2013:hats1}. We note that the
trend-filtering makes use of the Trend-Filtering Algorithm (TFA) of
\citet{kovacs:2005:TFA}, while transit signals are detected using the
Box-fitting Least Squares (BLS) method of \citet{kovacs:2002:BLS}. The
HATSouth observations of each system are summarized in
\reftabl{photobs}, and displayed in Figures~\ref{fig:hats74},
\ref{fig:hats75}, \ref{fig:hats76}, and \ref{fig:hats77}, while the
light curve data are made available in \reftabl{phfu}.

%
%

%
%
\addtocounter{figure}{-1}
    \begin{figure*}[!ht]
\caption{
    (Caption continued from previous page.)
{\em Top Right:} Unbinned follow-up transit light curves
    corrected for instrumental trends fitted
    simultaneously with the transit model, which is overplotted.
    The dates, filters and instruments used are indicated.  The residuals are shown on the right-hand-side in the same order as the original light curves.  The error bars represent the photon
    and background shot noise, plus the readout noise. Note that these
    uncertainties are scaled up in the fitting procedure to achieve a
    reduced $\chi^2$ of unity, but the uncertainties shown in the plot
    have not been scaled.
{\em Bottom Left:}
Radial velocities and bisector span measurements from ESPRESSO/VLT phased with respect to the mid-transit time.
The top panel shows the phased radial velocity measurements together with the best-fit model.
The center-of-mass velocity has been subtracted. The middle panel shows the velocity $O\!-\!C$ residuals.
The error bars include the estimated jitter, which is varied as a free parameter in the fitting.
The bottom panel shows the phased bisector span measurements.
{\em Bottom Right:} Color-magnitude diagram (CMD) and spectral energy distribution (SED). The top panel shows the absolute $G$ magnitude vs.\ the de-reddened $G - K_{S}$ color compared to
  theoretical isochrones (black lines) and stellar evolution tracks
  (green lines) from the MIST models interpolated at
  the best-estimate value for the metallicity of the host. The age
  of each isochrone is listed in black in Gyr, while the mass of each
  evolution track is listed in green in solar masses. The solid red lines show isochrones at higher and lower metallicities than the best-estimate value, with the metallicity and age in Gyr of each isochrone labelled on the plot. The filled
  blue circles show the measured reddening- and distance-corrected
  values from Gaia DR2 and 2MASS, while the blue lines indicate
  the $1\sigma$ and $2\sigma$ confidence regions, including the
  estimated systematic errors in the photometry. The middle panel shows the SED as measured via broadband photometry through the listed filters. Here we plot the observed magnitudes without correcting for distance or extinction. Overplotted are 200 model SEDs randomly selected from the MCMC posterior distribution produced through the global analysis (gray lines).
The model makes use of the predicted absolute magnitudes in each bandpass from the MIST isochrones, the distance to the system (constrained largely via Gaia DR2) and extinction (constrained from the SED with a prior coming from the {\sc mwdust} 3D Galactic extinction model).
The bottom panel shows the $O\!-\!C$ residuals from the best-fit model SED. The errors listed in the catalogs for the broad-band photometry measurements are shown with black lines, while the errors including an assumed 0.02\,mag systematic uncertainty, which is added in quadrature to the listed uncertainties, are shown with red lines. These latter uncertainties are what we use in the fit.
\label{fig:hats74:labcontinue}}
    \end{figure*}

%
%
    \begin{figure*}[!ht]
 {
 \centering
 \leavevmode
 \includegraphics[width={1.0\linewidth}]{\hatcurhtr{75}-banner}
}
 {
 \centering
 \leavevmode
 \includegraphics[width={0.5\linewidth}]{\hatcurhtr{75}-hs}%
 \hfil
 \includegraphics[width={0.5\linewidth}]{\hatcurhtr{75}-lc}%
 }
 {
 \centering
 \leavevmode
 \includegraphics[width={0.5\linewidth}]{\hatcurhtr{75}-rv}%
 \hfil
 \includegraphics[width={0.5\linewidth}]{\hatcurhtr{75}-iso-gk-gabs-isofeh-SED}%
 }
\caption{
    Same as Figure~\ref{fig:hats74}, here we show the observations of \hatcur{75} together with our best-fit model. The {\em TESS} light curve for this system is shown in Figure~\ref{fig:hats75tess}.
\label{fig:hats75}
}
    \end{figure*}

%
%
    \begin{figure*}[!ht]
 {
 \centering
 \leavevmode
 \includegraphics[width={1.0\linewidth}]{\hatcurhtr{76}-banner}
}
 {
 \centering
 \leavevmode
 \includegraphics[width={0.5\linewidth}]{\hatcurhtr{76}-hs}%
 \hfil
 \includegraphics[width={0.5\linewidth}]{\hatcurhtr{76}-lc}%
 }
 {
 \centering
 \leavevmode
 \includegraphics[width={0.5\linewidth}]{\hatcurhtr{76}-rv}%
 \hfil
 \includegraphics[width={0.5\linewidth}]{\hatcurhtr{76}-iso-gk-gabs-isofeh-SED}%
 }
\caption{
    Same as Figure~\ref{fig:hats74}, here we show the observations of \hatcur{76} together with our best-fit model. The {\em TESS} light curve for this system is shown in Figure~\ref{fig:hats76tess}.
\label{fig:hats76}
}
    \end{figure*}

%
%
    \begin{figure*}[!ht]
 {
 \centering
 \leavevmode
 \includegraphics[width={1.0\linewidth}]{\hatcurhtr{77}-banner}
}
 {
 \centering
 \leavevmode
 \includegraphics[width={0.5\linewidth}]{\hatcurhtr{77}-hs}%
 \hfil
 \includegraphics[width={0.5\linewidth}]{\hatcurhtr{77}-lc}%
 }
 {
 \centering
 \leavevmode
 \includegraphics[width={0.5\linewidth}]{\hatcurhtr{77}-rv}%
 \hfil
 \includegraphics[width={0.5\linewidth}]{\hatcurhtr{77}-iso-gk-gabs-isofeh-SED}%
 }
\caption{
    Same as Figure~\ref{fig:hats74}, here we show the observations of \hatcur{77} together with our best-fit model. The {\em TESS} light curve for this system is shown in Figure~\ref{fig:hats77tess}.
\label{fig:hats77}
}
    \end{figure*}

    \begin{figure*}[!ht]
 {
 \centering
 \leavevmode
 \includegraphics[width={1.0\linewidth}]{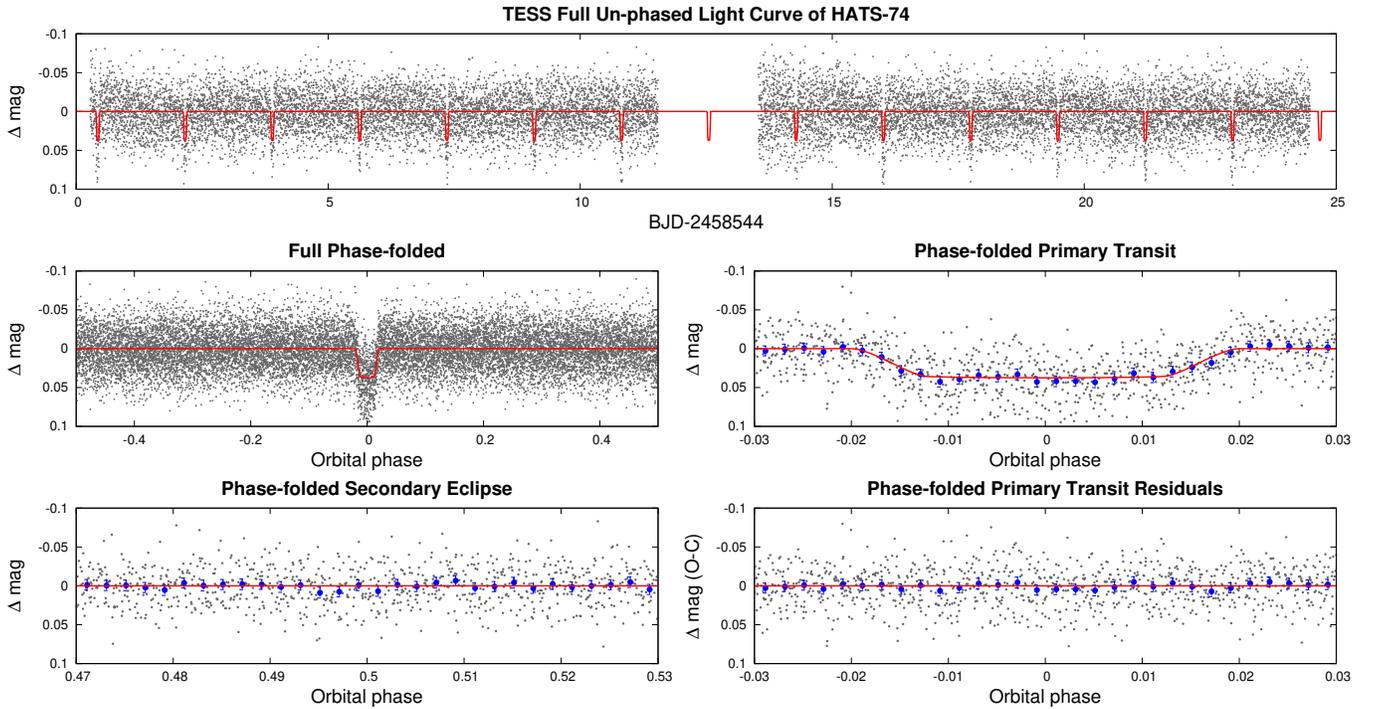}
}
\caption{
    {\em TESS} short-cadence light curve for \hatcur{74}. We show the full un-phased light curve as a function of time ({\em top}), the full phase-folded light curve ({\em middle left}), the phase-folded light curve zoomed-in on the planetary transit ({\em middle right}), the phase-folded light curve zoomed-in on the secondary eclipse ({\em bottom left}), and the residuals from the best-fit model, phase-folded and zoomed-in on the planetary transit ({\em bottom right}). The solid line in each panel shows the model fit to the light curve. The dark filled circles show the light curve binned in phase with a bin size of 0.002. Other observations included in our analysis of this system are shown in Figure~\ref{fig:hats74}.
\label{fig:hats74tess}
}
    \end{figure*}

    \begin{figure*}[!ht]
 {
 \centering
 \leavevmode
 \includegraphics[width={1.0\linewidth}]{\hatcurhtr{75}-TESS}
}
\caption{
    Similar to Figure~\ref{fig:hats74tess}, here we show the {\em TESS} short-cadence light curve for \hatcur{75}.  Other observations included in our analysis of this system are shown in Figure~\ref{fig:hats75}.
\label{fig:hats75tess}
}
    \end{figure*}

    \begin{figure*}[!ht]
 {
 \centering
 \leavevmode
 \includegraphics[width={1.0\linewidth}]{\hatcurhtr{76}-TESS}
}
\caption{
    Similar to Figure~\ref{fig:hats74tess}, here we show the {\em TESS} short-cadence light curve for \hatcur{76}. Other observations included in our analysis of this system are shown in Figure~\ref{fig:hats76}.
\label{fig:hats76tess}
}
    \end{figure*}

    \begin{figure*}[!ht]
 {
 \centering
 \leavevmode
 \includegraphics[width={1.0\linewidth}]{\hatcurhtr{77}-TESS}
}
\caption{
    Similar to Figure~\ref{fig:hats77tess}, here we show the {\em TESS} short-cadence light curve for \hatcur{77}. Other observations included in our analysis of this system are shown in Figures~\ref{fig:hats77}.
\label{fig:hats77tess}
}
    \end{figure*}

\clearpage

\subsubsection{TESS}
\label{sec:tess}

All four systems were observed by the NASA {\em TESS} mission as
summarized in Table~\ref{tab:photobs}. Observations were carried out
in short-cadence mode through the {\em TESS} Guest-Investigator
program (G011214; PI Bakos) to observe HATSouth transiting planet
candidates with {\em TESS}. The short-cadence observations were
reduced to light curves by the NASA Science Processing Operations
Center (SPOC) Pipeline at NASA Ames Research Center
\citep{jenkinsSPOC2016,jenkins:2010}. Multiple threshold crossing
events were identified for each target, and all four objects were
selected as transiting planet candidates and assigned {\em TESS}
Object of Interest (TOI) identifiers (TOI~\hatcurCCtoi{74},
TOI~\hatcurCCtoi{75}, TOI~\hatcurCCtoi{76}, and TOI~\hatcurCCtoi{77},
respectively). Each target passed all of the data validation tests
conducted by the pipeline, including no discernable difference between
odd and even transits, no evidence for a weak secondary event, no
evidence for stronger transits in a halo aperture compared to the
optimal aperture used to extract the light curve, strong evidence that
the target is not a false alarm due to correlated noise, and no
evidence for variations in the difference image centroid.

We obtained the SPOC PDC light curves
\citep{2014PASP..126..100S,2012PASP..124.1000S} for all four objects
from the Barbara A.\ Mikulski Archive for Space Telescopes
(MAST). These light curves have been corrected for dilution from any
other sources in the {\em TESS} Input Catalog \citep[TIC;][]{stassun:2019} that
are blended with the targets in the {\em TESS} observations. The {\em
  TESS} light curves show clear transit signals for all four systems
that are fully consistent with the transit signals detected with
HATSouth as shown in Fig.~\ref{fig:hats74tess},
Fig.~\ref{fig:hats75tess}, Fig.~\ref{fig:hats76tess}, and
Fig.~\ref{fig:hats77tess}. The {\em TESS} light curve data that we
included in the analyses are listed in Table~\ref{tab:phfu}.

\hatcur{74} is blended in the {\em TESS} images with a $0\farcs84$
neighbor with $\Delta G = 3.183$\,mag, which we denote HATS-74B in this work. The neighbor is resolved in the
Gaia DR2 catalog, and was also detected in high-spatial-resolution
imaging (Section~\ref{sec:highresimaging}). The neighbor is blended
with the target in all of the time-series photometric observations
which we carried out, and in all of the catalog photometry except the
Gaia~DR2 $G$-band measurement, and we discuss our methods for
correcting these data in Section~\ref{sec:transitmodel}.

There are no known sources blended with either \hatcur{75} or
\hatcur{76} in the {\em TESS} images down to $G \la 20$\,mag.

There are two sources that are within 2 pixels of \hatcur{77} in the
{\em TESS} images, including one object with $\Delta G = 1.04$\,mag at
a separation of $24\farcs1$, and an object with $\Delta G = 3.52$\,mag
at a separation of $28\farcs8$. These two objects are fully resolved
from \hatcur{77} in all of the other observations included in the
analysis of this system.

\subsubsection{Photometric Rotation Periods}
\label{sec:photper}

    \begin{figure*}[!ht]
 {
 \centering
 \leavevmode
 \includegraphics[width={0.5\linewidth}]{\hatcurhtr{74}-GLS}
 \hfil
 \includegraphics[width={0.5\linewidth}]{\hatcurhtr{75}-GLS}
 }
\caption{
Detection of a $P = \hatcurRotPer{74}$\,days photometric rotation period signal in the {\em TESS} light curve of \hatcur{74} ({\em left}) and $P = \hatcurRotPer{75}$\,days signal in the HATSouth light curve of \hatcur{75} ({\em right}). In each case we show the following panels. {\em Top:} the Generalized Lomb-Scargle (GLS) periodogram of the light curve after subtracting the best-fit transit model. The horizontal blue line shows the bootstrap-calibrated $10^{-3}$ false alarm probability level for \hatcur{74} and the $10^{-5}$ false alarm probability level for \hatcur{75}. {\em Second from top:} The Box-fitting Least Squares (BLS) periodogram of the same light curve. For \hatcur{74} there is a peak in the BLS periodogram at twice the period of the strongest peak in the GLS periodogram. For \hatcur{75} no significant peak is present in the BLS periodogram. {\em Second from bottom:} The {\em TESS} ({\em left}) and HATSouth ({\em right}) light curve phase-folded at the peak GLS period. The gray points show the individual photometric measurements, while the dark red filled squares show the observations binned in phase with a bin size of 0.02. {\em Bottom:} Same as the second from bottom, here we restrict the vertical range of the plot to better show the variation seen in the phase-binned measurements.\label{fig:gls:hats7475}
}

    \end{figure*}

    \begin{figure*}[!ht]
 {
 \centering
 \leavevmode
 \includegraphics[width={0.5\linewidth}]{\hatcurhtr{76}-GLS}
 }
\caption{
Similar to Fig.~\ref{fig:gls:hats7475}, here we show the detection of a $P = \hatcurRotPer{76}$\,days photometric rotation period in the HATSouth light curve of \hatcur{76}.\label{fig:gls:hats76}
}
    \end{figure*}

We also searched the HATSouth and {\em TESS} light curves for other
periodic signals using the Generalized Lomb-Scargle method
\citep[GLS;][]{zechmeister:2009}, and for additional transit signals
by applying a second iteration of BLS. Both of these searches were
performed on the residual light curves after subtracting the best-fit
transit models. We analyzed the HATSouth and {\em TESS} light curves
separately for each object.

We detect no evidence for additional variability in the HATSouth light
curve of \hatcur{74}. The highest peak in the GLS periodogram of the
HATSouth residual light curve of \hatcur{74} has a 95\% confidence
upper-limit on the semi-amplitude of 4.9\,mmag, and the highest peak in
the BLS periodogram has a transit depth of 16.7\,mmag. The {\em TESS}
light curve, however, does show evidence for a periodic signal, with a
period of \hatcurRotPer{74}\,days and a semi-amplitude of $1.86
\pm 0.41$\,mmag (Fig.~\ref{fig:gls:hats7475}, left) that we interpret as the
photometric rotation period of the star. The GLS false alarm
probability, calibrated using a bootstrap procedure, is $10^{-4}$. No
significant additional transit signals are revealed by BLS, with the
highest peak in the BLS spectrum having a transit depth of 4.3\,mmag.

The GLS analysis of the HATSouth light curve of \hatcur{75} reveals a
significant periodic signal, with a period of \hatcurRotPer{75}\,days, semi-amplitude of $1.71 \pm 0.21$\,mmag, and false-alarm
probability of $10^{-11}$ (Fig.~\ref{fig:gls:hats7475}, right).  The BLS
analysis identifies this same signal in the HATSouth light curve, but
the morphology of the phase-folded light curve clearly indicates that
this is rotational variability due to starspots, rather than a transit
signal. No other transit signals are seen in the HATSouth light
curve. We do not detect any variability in the residual {\em TESS}
light curve of \hatcur{75}. Note that the rotation period identified
by HATSouth is too long to be detectable in the {\em TESS} data given
the 27\,day window for each sector, and the detrending procedures that
remove low-frequency variability from the {\em TESS} light curves.

We detect a $P = \hatcurRotPer{76}$\,day periodic signal in the
HATSouth light curve of \hatcur{76} using GLS
(Fig.~\ref{fig:gls:hats76}). The signal has a semi-amplitude of $7.46
\pm 0.66$\,mmag, and a false alarm probability of $10^{-32}$. BLS also
identifies this same signal, but it is clearly starspot-induced
rotational variability. No other signals are identified by BLS. As for
\hatcur{75}, we do not detect the rotational variability in the {\em
  TESS} light curve of \hatcur{76} due to the long period relative to
the 27\,day observing window. No additional signals are detected by
BLS in the {\em TESS} light curve either.

For \hatcur{77}, no signals are identified in either the HATSouth or
{\em TESS} light curves. The highest peak in the GLS periodogram of
the HATSouth light curve has a 95\% confidence upper-limit on its
semi-amplitude of 4.1\,mmag. The corresponding upper-limit for the
{\em TESS} light curve is 3.1\,mmag. The highest peak in the BLS
spectrum of the HATSouth residuals has a depth of 13\,mmag, while for
the {\em TESS} residuals it is 6.0\,mmag.

\subsection{Spectroscopic Observations}
\label{sec:obsspec}

The spectroscopic observations carried out to confirm and characterize
the four transiting planet systems presented here are summarized in
\reftabl{specobs}. The facilities used include: the \'Echelle
spectrograph on the du~Pont~2.54\,m\footnote{\url{http://www.lco.cl/?epkb_post_type_1=echelle-spectrograph-users-manual}}, WiFeS on the
ANU~2.3\,m \citep{dopita:2007}, ARCES on the ARC~3.5\,m \citep{wang:2003},
FEROS on the MPG~2.2\,m \citep{kaufer:1998}, ARCoIRIS on the
Blanco~4\,m \citep{arcoiris:2016}, and ESPRESSO on the VLT~8.2\,m
\citep{pepe:2021}. The du~Pont, WiFeS, ARCES and FEROS
observations were obtained only for \hatcur{74}, while all four
systems were observed with ARCoIRIS and ESPRESSO.

A 1200\,s exposure of \hatcur{74} was obtained with the \'Echelle
spectrograph on the du~Pont~2.54\,m telescope at Las Campanas
Observatory in Chile on 2011 May 18. The spectrum had a resolution of
$R \equiv \lambda/\Delta\lambda = 40000$, and covered the wavelength
range of $3700$--$7000$\,\AA. Th-Ar\ lamp spectra were obtained before
and after the observation to calibrate the wavelength scale of the
science spectrum. The spectrum was extracted from the observation and
analyzed following the procedure used by \citet{jordan:2014:hats4} to
reduce observations from the Coralie and FEROS spectrographs.  The
spectrum had S/N$ = 7$, and thus only a low precision RV measurement
was possible, and estimates of the stellar spectroscopic parameters
based on this observation are unreliable.

\hatcur{74} was subsequently observed with the Wide Field Spectrograph
\citep[WiFeS;][]{dopita:2007} on the ANU~2.3\,m telescope at SSO.  The
WiFeS data were reduced and analyzed following
\citet{bayliss:2013:hats3}.  We obtained two spectra at a resolution
of $R = 3000$ to determine the effective temperature, surface gravity,
and metallicity of the star, while four spectra were obtained at $R =
7000$ to search for large amplitude RV variations that would indicate
the presence of a stellar-mass companion.
  The RV measurements extracted from the $R = 7000$ spectra
had very large uncertainties (median value of $30$\,\kms), and were
not useful for ruling out large amplitude RV variations.

Three optical spectra of \hatcur{74} were obtained with the
Astrophysics Research Consortium \'Echelle Spectrograph
\citep[ARCES;][]{wang:2003} on the Astrophysics Research Consortium
(ARC)~3.5\,m telescope at Apache Point Observatory in New Mexico. The
observations were performed and reduced to wavelength-calibrated
spectra in the manner discussed by \citet{brahm:2015:hats910}. The
observations were then analyzed using the Spectral Parameter
Classification program \citep[SPC;][]{buchhave:2012:SPC}. We found
that two of the spectra had S/N that were too low to yield useful
measurements, while a third spectrum obtained with $S/N \sim 9$
yielded an RV measurement of $17.9 \pm 0.6$\,\kms.  The atmospheric
parameters estimated from the spectra hinted at a cool surface
temperature of $\teffstar \sim 4000$\,K, but due to the low S/N,
reliable surface gravity, metallicity and \vsini\ measurements could
not be derived from these observations.

A single FEROS observation of \hatcur{74} was obtained and reduced to
a wavelength-calibrated spectrum, and RV and BS measurements using the
CERES software package \citep{brahm:2017:ceres}. The CERES package
produced a high effective temperature of $7000$\,K and low metallicity
of $\feh = -2.0$\,dex, which we find to often be the case for M dwarfs
with temperatures below the $\teffstar = 4000$\,K lower limit of the
model spectra used for cross-correlation by the package. The
measurements RV of $15.87 \pm 0.11$\,\kms\ is consistent with the
higher-precision RV measurements of the system obtained with ESPRESSO.

\begin{figure*}[ht!]
\includegraphics[width={1.0\linewidth}]{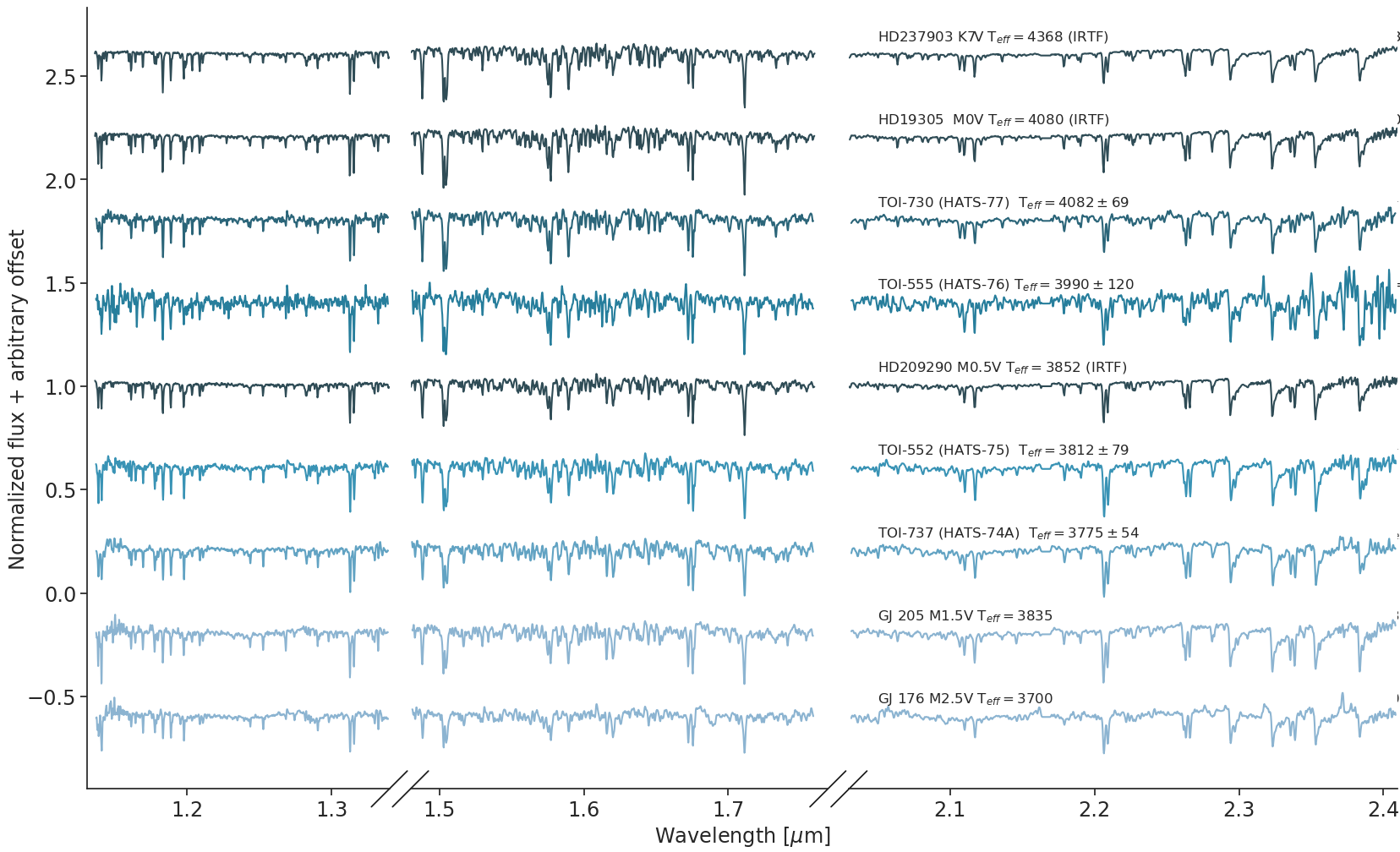}
\caption{ARCoIRIS spectra for our targets, along with IRTF/SpeX and ArcoIRIS spectra for M dwarfs with known stellar atmospheric parameters. The ARCoIRIS spectra have been degraded to the resolution of IRTF/SpeX for this figure. The effective temperatures derived using the procedure of \cite{Newton:2014,Newton:2015} are indicated over the K band spectra.}
\label{fig:arcoiris}
\end{figure*}

Because all four objects have surface temperatures that are too low to
apply the ZASPE package of \citep{brahm:2017:zaspe}, we obtained
near-infrared spectra of all four systems using the ``Astronomy
Research using the Cornell Infra Red Imaging Spectrograph'' (ARCoIRIS)
instrument on the Blanco~4\,m at CTIO \citep{arcoiris:2016}.
ARCoIRIS is a fixed slit spectrograph. It reaches a resolution of R$\sim$3500 over a large wavelength range from 0.80 to 2.47 microns by cross-dispersing the reflected grating light. For each science frame, we used the Fowler readout mode and took subsequent CuHeAr lamp spectra. We interleaved telluric standard star observations in the course of the night, matching the spectral type A0V and close in air mass. All stars were observed in an ABBA pattern. We analyzed the raw ARCoIRIS frames using SpexTool \citep{Spextool,Spextool:telluric}, obtaining wavelength calibrated and telluric corrected spectra. Given the mixed cloud conditions, we did not attempt to flux calibrate the spectra. Stellar atmospheric parameters were obtained by downgrading our spectra to match the IRTF/SpeX resolution and applying the procedure described in \cite{Newton:2014,Newton:2015}. These are the atmospheric parameters that we adopt for the joint analysis  discussion in Section~\ref{sec:transitmodel}. We also obtained ARCoIRIS spectra of two known M-dwarfs (namely GJ176 and GJ205) for which we applied the same procedure. The ARCoIRIS spectra for our targets, along with IRTF/SpeX and ARCoIRIS spectra for standards, are shown in Figure~\ref{fig:arcoiris}.

In order to detect the radial velocity variation of each star due to the transiting
companions, and thereby determine the mass of each transiting
companion, and confirm these as transiting planet systems, we obtained
VLT~8.2\,m/ESPRESSO observations of all four objects. A large
telescope was needed to perform these observations due to the
faintness, in optical bandpasses, of the host stars.  The observations
were carried out through the queue service mode between 2019 September
and 2019 December. We used an exposure time of 1800\,s for all
observations, and obtained five exposures each for \hatcur{74},
\hatcur{75} and \hatcur{77}, and four exposures for \hatcur{76}. The
observing and reduction procedures were the same as those discussed in
\citep{bakos:2020:hats71}, making use of version 2.0.0 of the ESPRESSO
pipeline in the ESO Reflex environment
\citep{freudling:2013} to produce high-precision RVs and bisector span measurements via
cross-correlation with an M2 spectra mask.  During the observations, the target
was placed on fiber A, while fiber B pointed to the sky for a simultaneous monitoring
of the sky.
For all four objects the resulting
RVs showed clear variations in phase with the transit ephemerides, and consistent
with Keplerian orbital variations due to transiting planets. The
phase-folded observations are shown in Fig.~\ref{fig:hats74},
Fig.~\ref{fig:hats75}, Fig.~\ref{fig:hats76}, and
Fig.~\ref{fig:hats77}, while the radial velocity and bisector span data are made available in
Table~\ref{tab:rvs}.

\startlongtable
    \begin{deluxetable*}{llrrrr}
\tablewidth{0pc}
\tabletypesize{\scriptsize}
\tablecaption{
    Summary of photometric observations
    \label{tab:photobs}
}
\tablehead{
    \multicolumn{1}{c}{Instrument/Field\tablenotemark{a}} &
    \multicolumn{1}{c}{Date(s)} &
    \multicolumn{1}{c}{\# Images\tablenotemark{b}} &
    \multicolumn{1}{c}{Cadence\tablenotemark{c}} &
    \multicolumn{1}{c}{Filter} &
    \multicolumn{1}{c}{Precision\tablenotemark{d}} \\
    \multicolumn{1}{c}{} &
    \multicolumn{1}{c}{} &
    \multicolumn{1}{c}{} &
    \multicolumn{1}{c}{(sec)} &
    \multicolumn{1}{c}{} &
    \multicolumn{1}{c}{(mmag)}
}
\startdata
\sidehead{\textbf{\hatcur{74}}}
~~~~HS-1/G563 & 2010 Jan--2010 Aug & 3875 & 277 & $r$ & 44.4 \\
~~~~HS-3/G563 & 2010 Jan--2010 Aug & 5197 & 281 & $r$ & 46.0 \\
~~~~HS-5/G563 & 2010 Jan--2010 Aug & 636 & 271 & $r$ & 41.9 \\
~~~~{\em TESS}/Sector~9 & 2019 Mar 1--25 & 15918 & 120 & $T$ & 24.8 \\
~~~~CTIO~0.9\,m & 2014 Feb 12 & 74 & 240 & $R$ & 7.5 \\
~~~~LCO~1\,m/Sinistro & 2019 Jun 13 & 50 & 206 & $i$ & 4.6 \\
~~~~TCS~1.5\,m/MuSCAT2 & 2020 Jan 29 & 371 & 30 & $g$ & 59.0 \\
~~~~TCS~1.5\,m/MuSCAT2 & 2020 Jan 29 & 189 & 60 & $r$ & 16.6 \\
~~~~TCS~1.5\,m/MuSCAT2 & 2020 Jan 29 & 189 & 60 & $i$ & 16.5 \\
~~~~TCS~1.5\,m/MuSCAT2 & 2020 Jan 29 & 189 & 60 & $z_{S}$ & 10.7 \\
~~~~TSC~1.5\,m/MuSCAT2 & 2020 Mar 2 & 384 & 30 & $g$ & 33.8 \\
~~~~TCS~1.5\,m/MuSCAT2 & 2020 Mar 2 & 196 & 60 & $r$ & 9.7 \\
~~~~TCS~1.5\,m/MuSCAT2 & 2020 Mar 2 & 195 & 60 & $i$ & 7.0 \\
~~~~TCS~1.5\,m/MuSCAT2 & 2020 Mar 2 & 195 & 60 & $z_{S}$ & 5.5 \\
\sidehead{\textbf{\hatcur{75}}}
~~~~HS-1/G548.focus & 2014 Sep--2015 Apr & 1677 & 1071 & $r$ & 53.2 \\
~~~~HS-2/G548.focus & 2014 Jun--2015 Mar & 2011 & 1209 & $r$ & 52.6 \\
~~~~HS-3/G548.focus & 2014 Sep--2015 Mar & 1486 & 1217 & $r$ & 52.4 \\
~~~~HS-4/G548.focus & 2014 Jun--2015 Mar & 1702 & 1222 & $r$ & 51.0 \\
~~~~HS-5/G548.focus & 2014 Sep--2015 Mar & 1381 & 1232 & $r$ & 52.1 \\
~~~~HS-6/G548.focus & 2014 Jul--2015 Mar & 1672 & 1200 & $r$ & 51.5 \\
~~~~HS-1/G548 & 2014 Sep--2015 Apr & 6547 & 287 & $r$ & 28.5 \\
~~~~HS-2/G548 & 2014 Jun--2015 Apr & 7590 & 348 & $r$ & 28.2 \\
~~~~HS-3/G548 & 2014 Sep--2015 Mar & 5284 & 352 & $r$ & 27.4 \\
~~~~HS-4/G548 & 2014 Jun--2015 Mar & 5976 & 352 & $r$ & 26.9 \\
~~~~HS-5/G548 & 2014 Sep--2015 Mar & 4945 & 359 & $r$ & 31.6 \\
~~~~HS-6/G548 & 2014 Jul--2015 Mar & 5956 & 351 & $r$ & 30.0 \\
~~~~{\em TESS}/Sector~4 & 2018 Oct--Nov & 14368 & 120 & $T$ & 13.9 \\
~~~~{\em TESS}/Sector~5 & 2018 Nov--Dec & 16376 & 120 & $T$ & 13.1 \\
~~~~LCO~1\,m/Sinistro & 2019 Sep 23 & 44 & 276 & $g^{\prime}$ & 2.8 \\
~~~~LCO~0.4\,m & 2020 Jan 2 & 38 & 314 & $z_{S}$ & 11.6 \\
~~~~LCO~1\,m/Sinistro & 2020 Sep 17 & 83 & 181 & $z_{S}$ & 2.6 \\
\sidehead{\textbf{\hatcur{76}}}
~~~~HS-1/G597 & 2014 Jan--2014 Mar & 1228 & 286 & $r$ & 38.6 \\
~~~~HS-3/G597 & 2013 Sep--2014 Feb & 4540 & 285 & $r$ & 44.4 \\
~~~~HS-5/G597 & 2013 Sep--2014 Mar & 4915 & 278 & $r$ & 44.7 \\
~~~~{\em TESS}/Sector~5 & 2018 Nov--Dec & 16362 & 120 & $T$ & 36.1 \\
~~~~LCO~1\,m/Sinistro & 2019 Sep 20 & 44 & 327 & $g^{\prime}$ & 4.9 \\
~~~~TRAPPIST-South & 2020 Oct 20 & 48 & 130 & $I+z$ & 9.1 \\
~~~~TRAPPIST-South & 2020 Oct 29 & 138 & 130 & $I+z$ & 10.2 \\
\sidehead{\textbf{\hatcur{77}}}
~~~~HS-1/G607 & 2011 Jan--2012 Jun & 6703 & 289 & $r$ & 43.3 \\
~~~~HS-3/G607 & 2011 Jan--2012 Jun & 3179 & 294 & $r$ & 48.4 \\
~~~~HS-5/G607 & 2011 Jan--2012 Jun & 2544 & 288 & $r$ & 44.6 \\
~~~~{\em TESS}/Sector~9 & 2019 Mar 1--25 & 15726 & 120 & $T$ & 39.4 \\
~~~~LCO~1\,m/Sinistro & 2019 Jun 14 & 55 & 147 & $i^{\prime}$ & 5.9 \\
~~~~Mt.\ Stuart~0.3\,m & 2020 May 31 & 66 & 191 & $g^{\prime}$ & 32.8 \\
~~~~LCO~2\,m/MuSCAT3 & 2021 Jan 5 & 44 & 404 & $g$ & 2.2 \\
~~~~LCO~2\,m/MuSCAT3 & 2021 Jan 5 & 73 & 244 & $r$ & 1.1 \\
~~~~LCO~2\,m/MuSCAT3 & 2021 Jan 5 & 92 & 194 & $i$ & 1.2 \\
~~~~LCO~2\,m/MuSCAT3 & 2021 Jan 5 & 44 & 404 & $z_{S}$ & 1.5 \\
\enddata \tablenotetext{a}{ For HATSouth data we list the HATSouth
  unit, CCD and field name from which the observations are taken. HS-1
  and -2 are located at Las Campanas Observatory in Chile, HS-3 and -4
  are located at the H.E.S.S. site in Namibia, and HS-5 and -6 are
  located at Siding Spring Observatory in Australia. Each unit has 4
  CCDs. Each field corresponds to one of 838 fixed pointings used to
  cover the full 4$\pi$ celestial sphere. All data from a given
  HATSouth field and CCD number are reduced together, while detrending
  through External Parameter Decorrelation (EPD) is done independently
  for each unique unit+CCD+field combination. Observations with ``.focus'' included in the name are from light curves derived from focusing frames, which are shorter 30\,s exposures that are taken every 20--30 minutes to refine the focus of the cameras.
}
\tablenotetext{b}{ Excluding any outliers or other data not included in the modelling. }
\tablenotetext{c}{ The median time between consecutive images rounded
  to the nearest second. Due to factors such as weather, the
  day--night cycle, guiding and focus corrections the cadence is only
  approximately uniform over short timescales.  }
\tablenotetext{d}{
  The RMS of the residuals from the best-fit model. Note that in the case of HATSouth and {\em TESS} observations the transit may appear artificially shallower due to over-filtering and/or blending from unresolved neighbors. As a result the S/N of the transit may be less than what would be calculated from $\rpl/\rstar$ and the RMS estimates given here. }
    \end{deluxetable*}

%
%
    \begin{deluxetable*}{llrrrrl}
\tablewidth{0pc}
\tablecaption{
    Light curve data for \hatcur{74}, \hatcur{75}, \hatcur{76}, and \hatcur{77}\label{tab:phfu}.
}
\tablehead{
    \colhead{Object\tablenotemark{a}} &
    \colhead{BJD\tablenotemark{b}} &
    \colhead{Mag\tablenotemark{c}} &
    \colhead{\ensuremath{\sigma_{\rm Mag}}} &
    \colhead{Mag(orig)\tablenotemark{d}} &
    \colhead{Filter} &
    \colhead{Instrument} 
}
\startdata
HATS-74 & $ 2455254.76774 $ & $  -0.08602 $ & $   0.05337 $ & $ \cdots $ & $ r$ &         HS\\
HATS-74 & $ 2455322.31035 $ & $   0.04862 $ & $   0.03805 $ & $ \cdots $ & $ r$ &         HS\\
HATS-74 & $ 2455280.74579 $ & $  -0.01962 $ & $   0.02762 $ & $ \cdots $ & $ r$ &         HS\\
HATS-74 & $ 2455379.46209 $ & $   0.05838 $ & $   0.04105 $ & $ \cdots $ & $ r$ &         HS\\
HATS-74 & $ 2455263.42771 $ & $   0.01115 $ & $   0.03608 $ & $ \cdots $ & $ r$ &         HS\\
HATS-74 & $ 2455294.60127 $ & $   0.03434 $ & $   0.03152 $ & $ \cdots $ & $ r$ &         HS\\
HATS-74 & $ 2455247.84123 $ & $  -0.02098 $ & $   0.02971 $ & $ \cdots $ & $ r$ &         HS\\
HATS-74 & $ 2455289.40588 $ & $  -0.00697 $ & $   0.08795 $ & $ \cdots $ & $ r$ &         HS\\
HATS-74 & $ 2455374.26713 $ & $   0.06199 $ & $   0.04867 $ & $ \cdots $ & $ r$ &         HS\\
HATS-74 & $ 2455327.50711 $ & $   0.01042 $ & $   0.06350 $ & $ \cdots $ & $ r$ &         HS\\
\enddata
\tablenotetext{a}{
    Either \hatcur{74}, \hatcur{75}, \hatcur{76}, or \hatcur{77}.
}
\tablenotetext{b}{
    Barycentric Julian Dates in this paper are reported on the
    Barycentric Dynamical Time (TDB) system.
} \tablenotetext{c}{
    The out-of-transit level has been subtracted. For observations
    made with the HATSouth instruments (identified by ``HS'' in the
    ``Instrument'' column) these magnitudes have been corrected for
    trends using the EPD and TFA procedures applied {\em prior} to
    fitting the transit model. This procedure may lead to an
    artificial dilution in the transit depths. For several of these
    systems neighboring stars are blended into the TESS observations
    as well. The blend factors for the HATSouth and TESS light curves
    are listed in Table~\ref{tab:planetparam}. For observations made
    with follow-up instruments (anything other than ``HS'', or ``TESS''
    in the ``Instrument'' column), the magnitudes have
    been corrected for a quadratic trend in time, and for variations
    correlated with up to three PSF shape parameters, fit
    simultaneously with the transit.
}
\tablenotetext{d}{
    Raw magnitude values without correction for the quadratic trend in
    time, or for trends correlated with the seeing. These are only
    reported for the follow-up observations.
}
\tablecomments{
    This table is available in a machine-readable form in the online
    journal.  A portion is shown here for guidance regarding its form
    and content.
}
    \end{deluxetable*}

    \begin{deluxetable*}{llrrrrr}
\tablewidth{0pc}
\tabletypesize{\scriptsize}
\tablecaption{
    Summary of spectroscopy observations.
    \label{tab:specobs}
}
\tablehead{
    \multicolumn{1}{c}{Instrument}          &
    \multicolumn{1}{c}{UT Date(s)}             &
    \multicolumn{1}{c}{\# Spec.}   &
    \multicolumn{1}{c}{Res.}          &
    \multicolumn{1}{c}{S/N Range\tablenotemark{a}}           &
    \multicolumn{1}{c}{$\gamma_{\rm RV}$\tablenotemark{b}} &
    \multicolumn{1}{c}{RV Precision\tablenotemark{c}} \\
    &
    &
    &
    \multicolumn{1}{c}{$\Delta \lambda$/$\lambda$/1000} &
    &
    \multicolumn{1}{c}{(\kms)}              &
    \multicolumn{1}{c}{(\ms)}
}
\startdata
\noalign{\vskip -3pt}
\sidehead{\textbf{\hatcur{74}}}\\
\noalign{\vskip -13pt}
du~Pont~2.54\,m/Echelle & 2011 May 18 & 1 & 30 & 7 & $5.4$ & $\sim 10000$ \\
ANU~2.3\,m/WiFeS & 2011 Jun 6 & 2 & 3 & 43--46 & $\cdots$ & $\cdots$ \\
ARC~3.5\,m/ARCES & 2012 Apr--2013 Feb & 3 & 31.5 & 7--9 & $17.9$ & $\sim 1000$ \\
ANU~2.3\,m/WiFeS & 2013 Mar--Apr & 4 & 7 & $\cdots$ & $8.8$ & $30000$ \\
MPG~2.2\,m/FEROS & 2013 May 12 & 1 & 48 & 14 & 15.868 & 100 \\
Blanco~4\,m/ARCoIRIS & 2017 Jun 8--9 & 2 & 3.5 & 120 & $\cdots$ & $\cdots$ \\
VLT~8.2\,m/ESPRESSO & 2019 Dec 26--31 & 5 & 140 & $\cdots$ & 15.853 & 11.0 \\
\noalign{\vskip -3pt}
\sidehead{\textbf{\hatcur{75}}}\\
\noalign{\vskip -13pt}
Blanco~4\,m/ARCoIRIS & 2017 Dec 2 & 1 & 3.5 & 90 & $\cdots$ & $\cdots$ \\
VLT~8.2\,m/ESPRESSO & 2019 Sep--Oct & 5 & 140 & $\cdots$ & 39.995 & 2.9 \\
\noalign{\vskip -3pt}
\sidehead{\textbf{\hatcur{76}}}\\
\noalign{\vskip -13pt}
Blanco~4\,m/ARCoIRIS & 2017 Dec 2 & 1 & 3.5 & 40 & $\cdots$ & $\cdots$ \\
VLT~8.2\,m/ESPRESSO & 2019 Sep--Oct & 4 & 140 & $\cdots$ & 8.601 & 35.7 \\
\noalign{\vskip -3pt}
\sidehead{\textbf{\hatcur{77}}}\\
\noalign{\vskip -13pt}
Blanco~4\,m/ARCoIRIS & 2017 Jun 8--9 & 2 & 3.5 & 70--100 & $\cdots$ & $\cdots$ \\
VLT~8.2\,m/ESPRESSO & 2019 Dec 1--27 & 5 & 140 & $\cdots$ & -7.759 & 25.0 \\
\enddata
\tablenotetext{a}{
    S/N per resolution element near 5180\,\AA. This was not measured for all of the instruments. For the ARCoIRIS NIR spectra, we list the S/N in the $H$-band.
}
\tablenotetext{b}{
    For high-precision RV observations included in the orbit determination this is the zero-point RV from the best-fit orbit. For other instruments it is the mean value. We only provide this quantity when applicable.
}
\tablenotetext{c}{
    For high-precision RV observations included in the orbit
    determination this is the scatter in the RV residuals from the
    best-fit orbit (which may include astrophysical jitter), for other
    instruments this is either an estimate of the precision (not
    including jitter), or the measured standard deviation.  We only provide this quantity when applicable.
}
    \end{deluxetable*}


\subsection{Photometric follow-up observations}
\label{sec:phot}

We obtained additional follow-up time-series photometry for each
system using larger 0.3\,m--2\,m ground-based telescopes to obtain
higher photometric-precision light curves from
higher-spatial-resolution images than those available from HATSouth or
{\em TESS}. As summarized in Table~\ref{tab:photobs}, the facilities
that we made use of for this purpose include: the imager on the
CTIO~0.9\,m \citep{subasavage:2010}, the imagers on the Las Cumbres Observatory \citep{brown:2013:lcogt}
0.4\,m network (LCO~0.4\,m), the Sinistro imagers on the
Las Cumbres Observatory 1\,m network (LCO~1\,m), the
MuSCAT2 imager \citep{narita:2019} on the 1.5\,m Telescopio Carlos Sanchez (TCS) at Teide Observatory, the MuSCAT3 imager \citep{narita:2020} at LCO's 2\,m telescope at Haleakala Observatory, and the imager on the Mt.~Stuart 0.3\,m
telescope near Dunedin, New Zealand. The CTIO~0.9\,m observations were carried out
by the HATSouth team, while the other observations were carried out by
members of the {\em TESS} Follow-up Observing Program \citep[TFOP;][]{collins:2018}, and were made available to the community through the ExoFOP-TESS portal\footnote{\url{https://exofop.ipac.caltech.edu/tess/index.php}}. For the TFOP observations, we used the {\tt TESS Transit Finder}, which is a customized version of the {\tt Tapir} software package \citep{Jensen:2013}, to schedule the observations, and the photometric data were extracted using {\tt AstroImageJ} \citep{Collins:2017}.

\subsection{Search for Resolved Stellar Companions}
\label{sec:highresimaging}

High-spatial-resolution images were obtained for all four objects by
members of TFOP, and made available on ExoFOP-TESS as part of the standard process
for vetting transit candidates and properly accounting for transit dilution that may be caused
by the presence of stellar companions \citep{ciardi:2015, schlieder:2021}. Optical speckle imaging
was carried out at 562\,nm and 832\,nm with the twin Zorro and 'Alopeke imagers\footnote {https://www.gemini.edu/sciops/instruments/alopeke-zorro/} (Scott et al.\ 2021, in press) mounted on the Gemini~8\,m South and North telescopes, respectively.
Near-infrared (NIR) adaptive optics (AO) imaging in Br$\gamma$ and the $J$-band was performed with the NIRC2 instrument on the Keck~2\,m for \hatcur{74}, and in the $K_{s}$-band with the NaCo instrument on
the VLT~8\,m for \hatcur{75}. Finally, $J$ and $K_{s}$-band imaging of
\hatcur{77} was obtained with WHIRC on the ARC~3.5\,m. The
observations with this latter instrument were gathered by the HATSouth
team before the {\em TESS} mission.  The optical and near-infrared techniques complement each other in terms of resolution and sensitivity to yield a more complete picture of the presence of near-by and (possibly) bound companions, with the optical speckle typically have better resolution and the NIR AO having better sensitivity.  The various observations are described below.

\hatcur{76} and \hatcur{77} were observed with Zorro, and \hatcur{74} and \hatcur{75} were observed with ’Alopeke.
The two instruments provide simultaneous speckle imaging in two bands (562nm and 832 nm) with output data products including a reconstructed image with robust contrast limits on companion detections \citep[e.g.,][]{howell:2016}. Images were collected and subjected to Fourier analysis in our standard reduction pipeline \citep[see][]{howell:2011}.  We find that all four targets are single stars to within the contrast achieved by the observations (4-5 magnitudes) from the diffraction limit (20 mas) out to 1.2”. At the distances of these HATS stars ($d=195$ to 414 pc) these angular limits correspond to spatial limits of 4-8 AU out to 230--500 AU. We note that AO imaging reveals a companion to \hatcur{74} at  $0\farcs844$ which was not immediately apparent in the 'Alopeke data.

\begin{figure*}[!ht]
 {
 \centering
 \leavevmode
 \includegraphics[width={0.5\linewidth}]{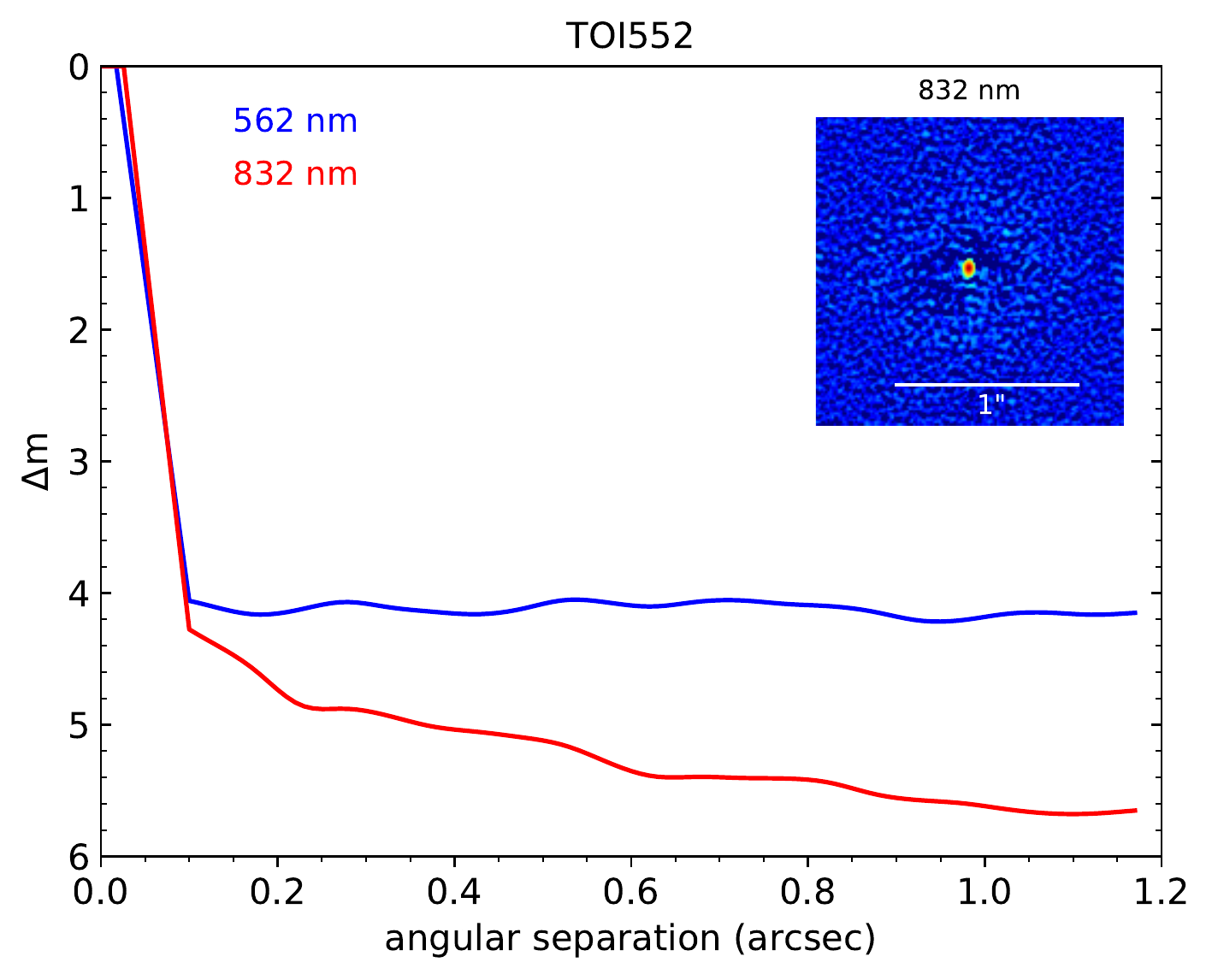}
\hfil
 \includegraphics[width={0.5\linewidth}]{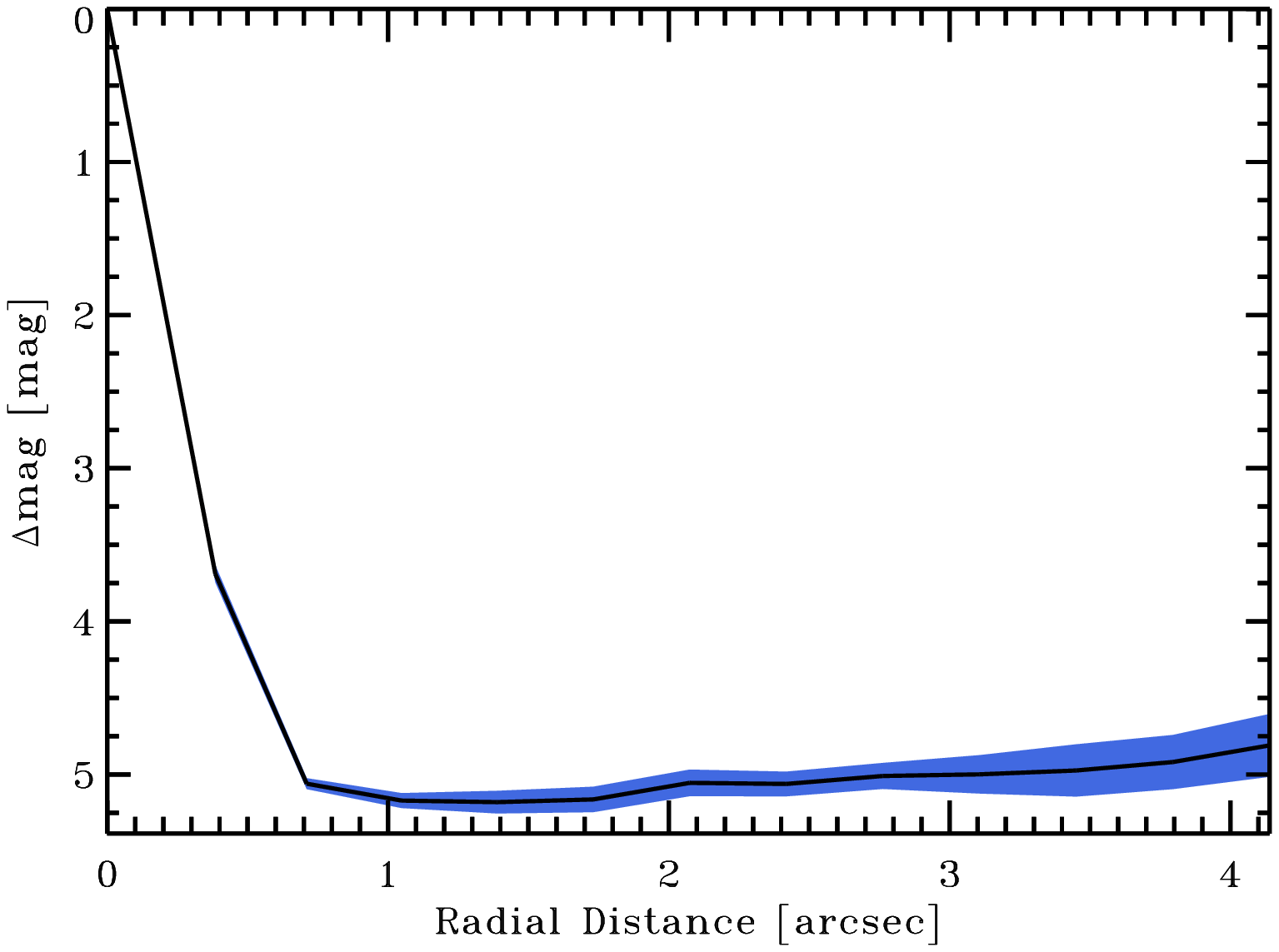}
 }
\caption{
    ({\em Left}) Contrast curves for \hatcur{75} (TOI~\hatcurCCtoibare{75}) derived from high-resolution images at 562\,nm and 832\,nm obtained with 'Alopeke/Gemini~8\,m. The 832\,nm image is also shown in the inset. ({\em right}) Contrast curve for \hatcur{75} derived from high-resolution $K_{s}$-band imaging with NaCo/VLT.
\label{fig:hats75highresimage}
}
    \end{figure*}

 The NaCo \citep{lenzen:2003,rousset:2003} data were collected in a nine-point grid dither pattern, with the star position moved 2$\arcsec$ for each exposure. We ensured the star was within the upper left quadrant of the detector for all images, since other quadrants of the detector suffer from light- and dark- column striping in the images. We collected 9 individual frames, each with exposure time 75\,s, using the Ks filter. The dither pattern allows for a sky background frame to be constructed from a median-combination of the science frames themselves. We reduced the raw data using a custom code, which performs badpixel and flatfield correction, subtracts the sky background, aligns the stellar position between images and finally co-adds the nine individual frames. We visually inspected images to search for companions, and did not find companions anywhere in the field of view, which extends to at least 4.9$\arcsec$ from the star in all directions. To quantify the sensitivity of the final image, we injected fake companions into the data cube at a range of separations and position angles. We retrieved these fake companions, and measured the S/N of each fake companion. We then scaled the flux of the fake companions, such that they could be retrieved at 5$\sigma$. Finally, the sensitivity was averaged over position angle. We are sensitive to companions 3.7\,mag fainter than the host beyond 400mas, and to companions 5\,mag fainter than the host in the background-limited regime beyond ~700\,mas. Our NaCo detection limits as a function of radius for \hatcur{75} are shown in Figure~\ref{fig:hats75highresimage}.

\hatcur{74} was observed with the NIRC2 instrument on Keck-II behind the natural guide star AO system \citep{wizinowich:2000}.  The observations were made on 2019~Jun~10 UT in the standard 3-point dither pattern that is used with NIRC2 to avoid the left lower quadrant of the detector which is typically noisier than the other three quadrants. The dither pattern step size was $3\arcsec$ and was repeated twice, with each dither offset from the previous dither by $0.5\arcsec$. The camera was in the narrow-angle mode with a full field of view of $\sim10\arcsec$ and a pixel scale of approximately $0.0099442\arcsec$ per pixel. The observations were made in the narrow-band $Br\gamma$ filter $(\lambda_o = 2.1686; \Delta\lambda = 0.0326\,\mu$m) and the narrow-band $J_{cont}$ filter ($\lambda_o = 1.2132; \Delta\lambda = 0.0198\,\mu$m), each with an integration time of 60 seconds with one coadd per frame for a total of 540 seconds on target per filter.

The AO data were processed and analyzed with a custom set of IDL tools.  The science frames were flat-fielded and sky-subtracted.  The flat fields were generated from a median average of dark subtracted flats taken on-sky, and the flats were normalized such that the median value of the flats is unity.  Sky frames were generated from the median average of the 9 dithered science frames; each science image was then sky-subtracted and flat-fielded.  The reduced science frames were combined into a single combined image using a intra-pixel interpolation that conserves flux, shifts the individual dithered frames by the appropriate fractional pixels, and median-coadds the frames.  The final resolution of the combined dithers was determined from the full-width half-maximum of the point spread function; 0.064\arcsec\ and 0.121\arcsec\ for $Br\gamma$ and $J_{cont}$ observations, respectively.  The sensitivities of the final combined AO image were determined by injecting simulated sources azimuthally around the primary target every $20^\circ $ at separations of integer multiples of the central source's FWHM \citep{furlan:2017}. The brightness of each injected source was scaled until standard aperture photometry detected it with $5\sigma $ significance. The resulting brightness of the injected sources relative to the target set the contrast limits at that injection location. The final $5\sigma $ limit at each separation was determined from the average of all of the determined limits at that separation and the uncertainty on the limit was set by the rms dispersion of the azimuthal slices at a given radial distance.

Additional imaging results are available for all four
objects from Gaia~DR2 \citep{gaiadr2}, which is sensitive to neighbors
with $G \la 20$\,mag down to a limiting resolution of $\sim 1\arcsec$
\citep[e.g.,][]{ziegler:2018}.

\begin{figure*}[!ht]
 {
 \centering
 \leavevmode
 \includegraphics[width={0.5\linewidth}]{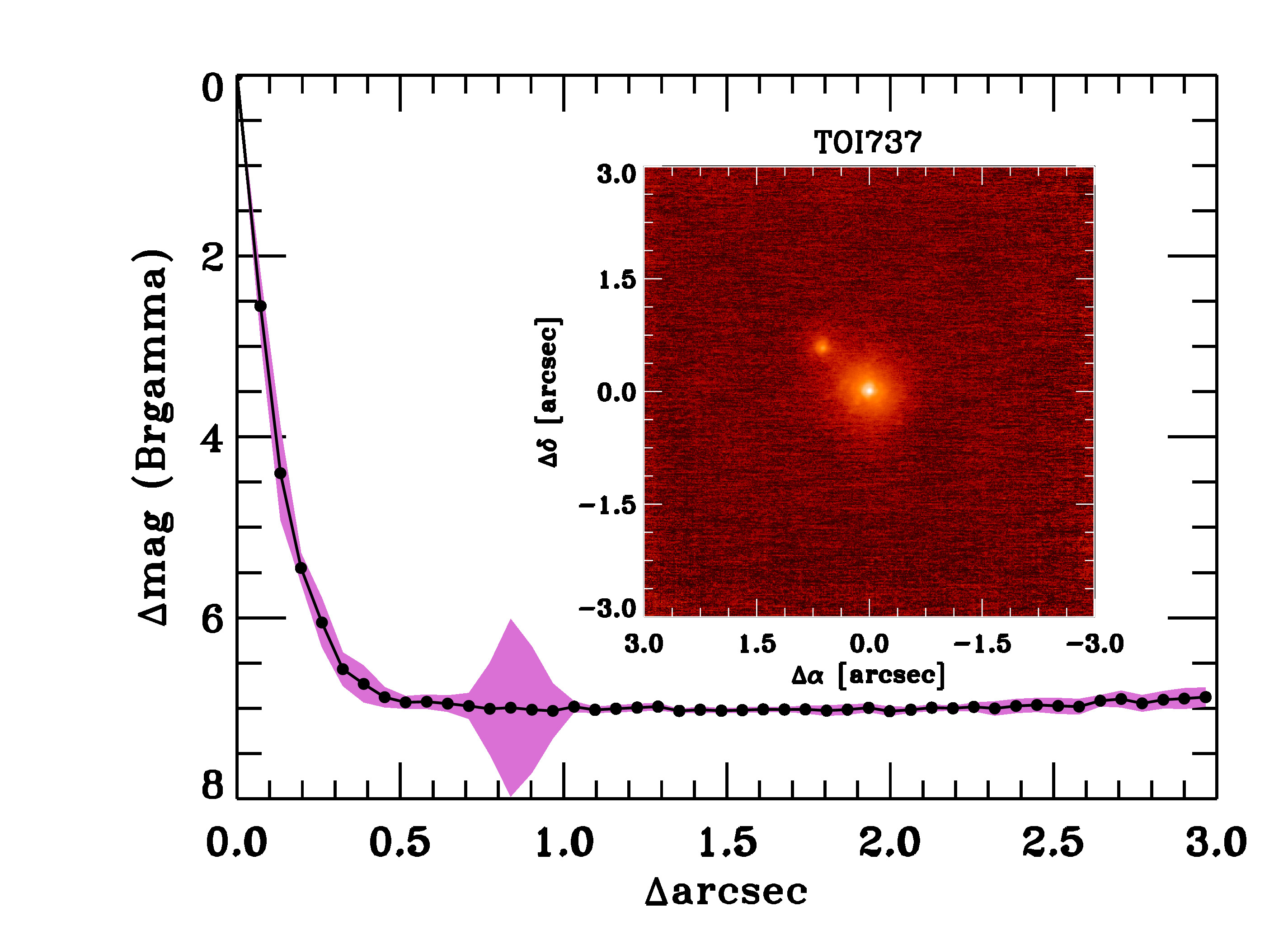}
\hfil
 \includegraphics[width={0.5\linewidth}]{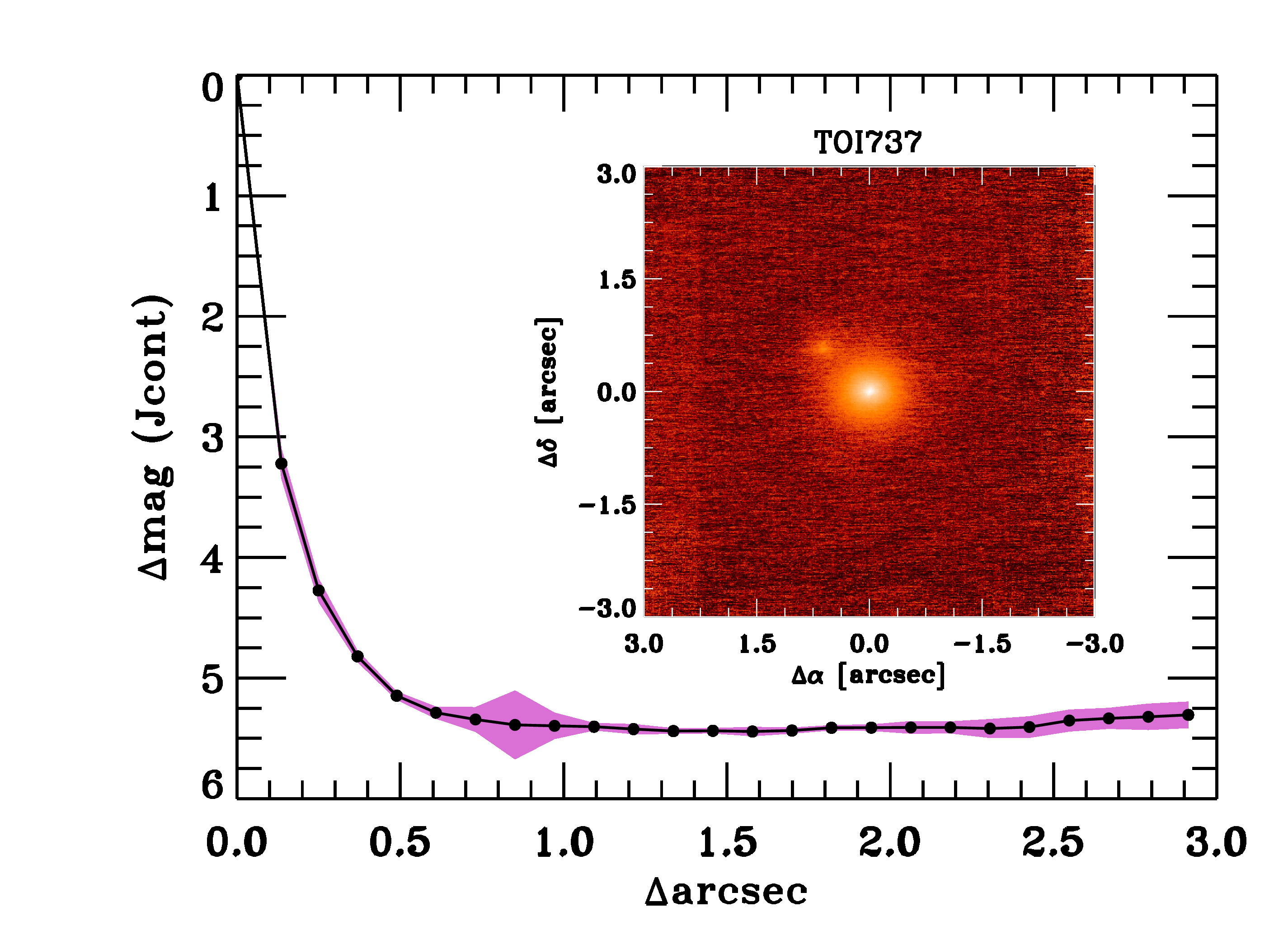}
 \includegraphics[width={1.0\linewidth}]{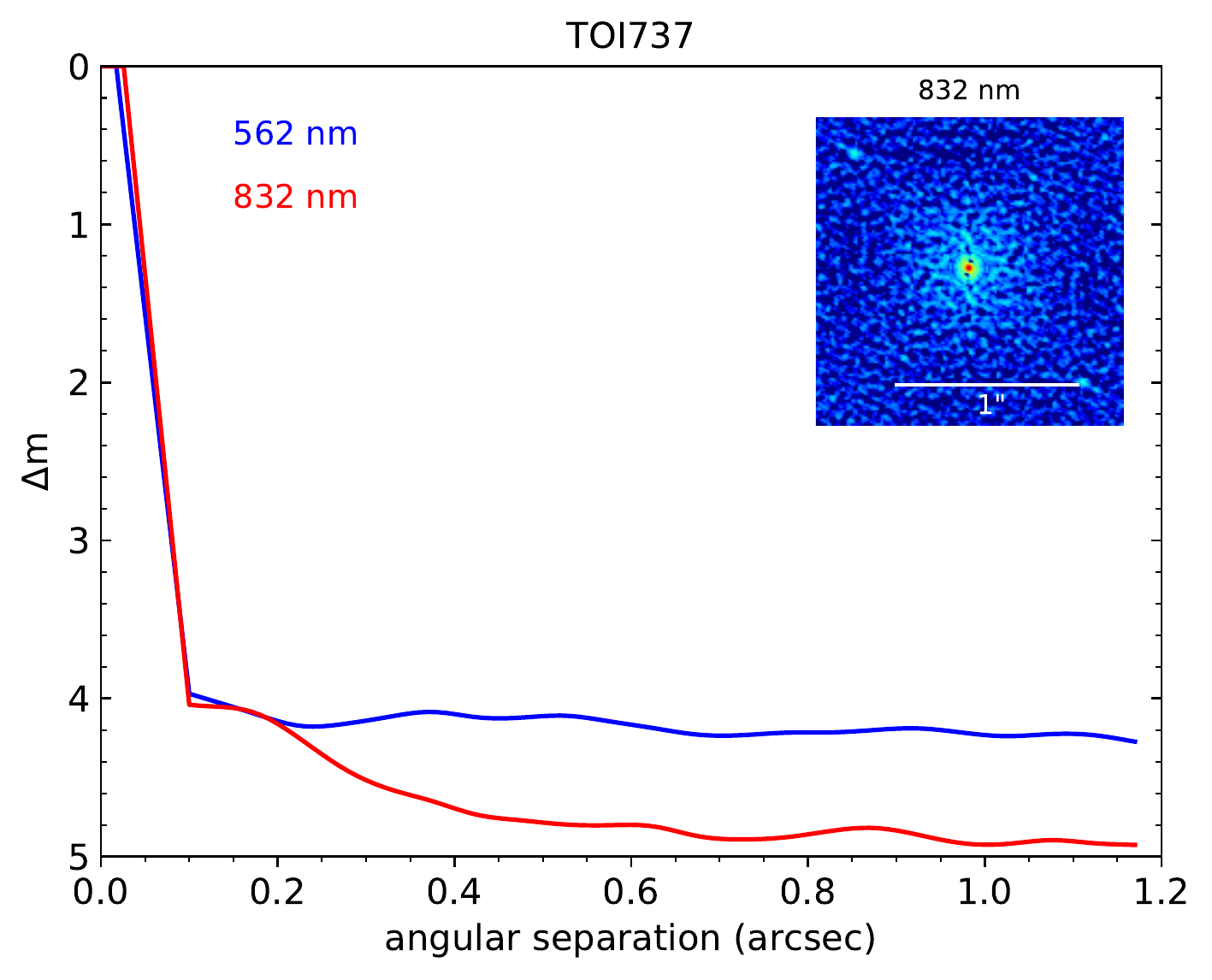}
 }
\caption{
    ({\em Top}) High-resolution images in Br$\gamma$ ({\em left}) and $J$-continuum ({\em middle}) and associated contrast curve for \hatcur{74} (TOI~\hatcurCCtoibare{74}) from Keck-2/NIRC2. The $0\farcs844$ neighbor HATS-74B is apparent in these images. ({\em Bottom}) Contrast curves at 562\,nm and 832\,nm from the 'Alopeke/Gemini~8\,m observations of \hatcur{74}. The 832\,nm image is also shown in the inset.
\label{fig:hats74highresimage}
}
    \end{figure*}

Based on the NIRC2 observations we find that \hatcur{74} has a $0\farcs844 \pm 0\farcs0014$ neighbor at a position angle of 46$^{\circ}$ east of north. The neighbor has magnitudes, relative to \hatcur{74}, of $\Delta Br\gamma = 2.615 \pm 0.013$\,mag and $\Delta J = 2.642 \pm 0.030$\,mag (Fig.~\ref{fig:hats74highresimage}). The neighbor is not obviously apparent in the 'Alopeke images, however (Fig.~\ref{fig:hats74highresimage}). Finally, the neighbor is also listed in the Gaia~DR2 catalog with a projected separation of $0\farcs84$ and a relative magnitude of $\Delta G = 3.1830 \pm 0.0049$\,mag. The neighbor has a parallax of $3.80 \pm 0.63$\,mas and a proper motion of $\mu_{\rm R.A.} = -41.4 \pm 1.4$\,\masy, and $\mu_{\rm Dec.} = 42.6 \pm 1.5$\,\masy, which are consistent with the values listed for \hatcur{74} ($\pi = \hatcurCCparallax{74}$\,mas, $\mu_{\rm R.A.} = \hatcurCCpmra{74}$\,\masy, and $\mu_{\rm Dec.} = \hatcurCCpmdec{74}$\,\masy), indicating that the neighbor is very likely a bound companion to \hatcur{74}, and we henceforth refer to it has HATS-74B. Given the distance to \hatcur{74} (Table~\ref{tab:stellarderived}), the measured angular separation between \hatcur{74} and HATS-74B corresponds to a projected physical separation of $238.4 \pm 3.9$\,AU.  Assuming HATS-74B is a main sequence companion to \hatcur{74} with the same age, metallicity, distance and extinction, then from the blend analysis that we discuss in Section~\ref{sec:blendmodel} we find that HATS-74B has a stellar mass of \hatcurISOmB{74}\,\msun.

\begin{figure*}[!ht]
 {
 \centering
 \leavevmode
 \includegraphics[width={1.0\linewidth}]{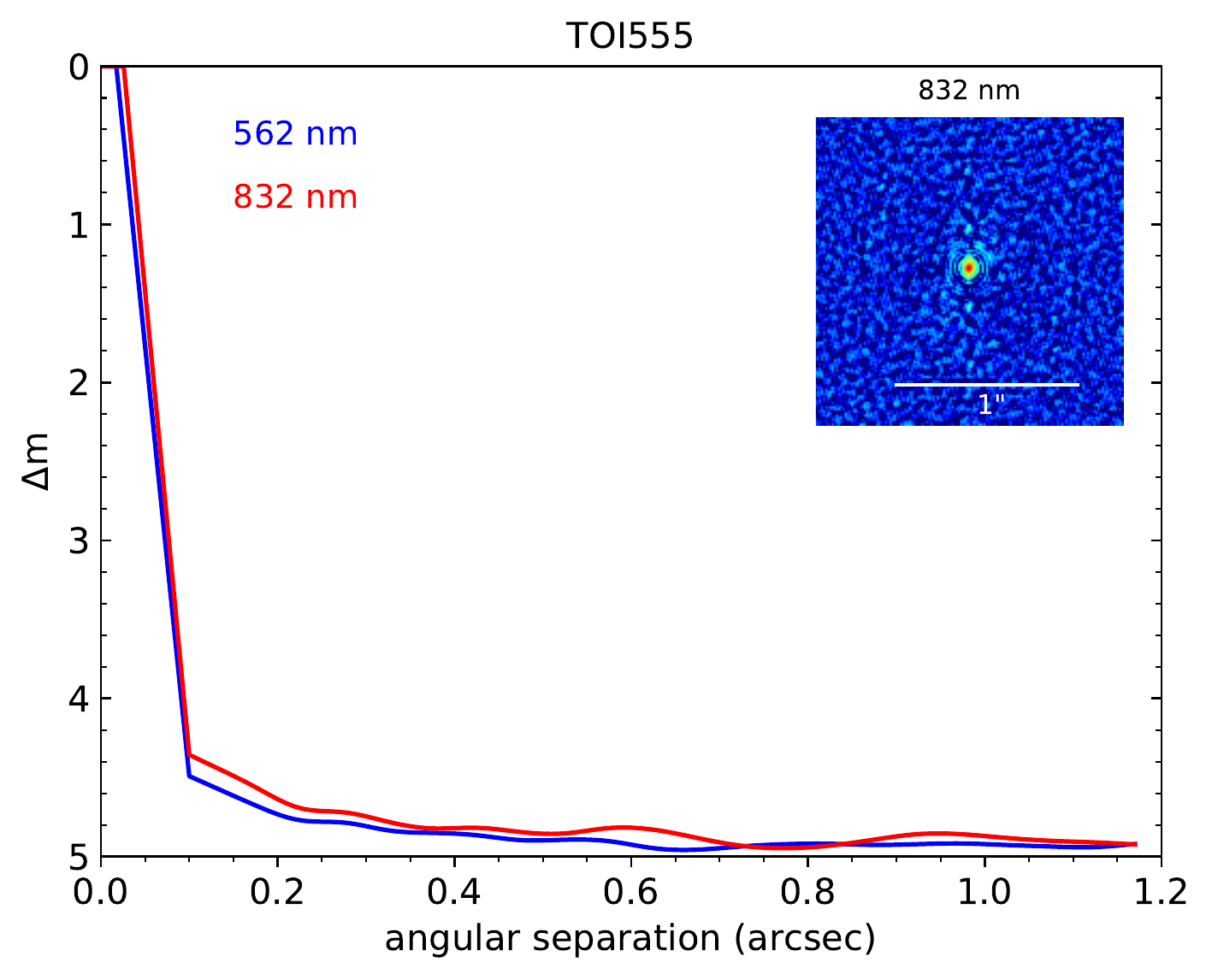}
 }
\caption{
    Contrast curves for \hatcur{76} (TOI~\hatcurCCtoibare{76}) derived from high-resolution images at 562\,nm and 832\,nm obtained with Zorro/Gemini~8\,m. The 832\,nm image is also shown in the inset.
\label{fig:hats76highresimage}
}
    \end{figure*}

\begin{figure*}[!ht]
 {
 \centering
 \leavevmode
 \includegraphics[width={1.0\linewidth}]{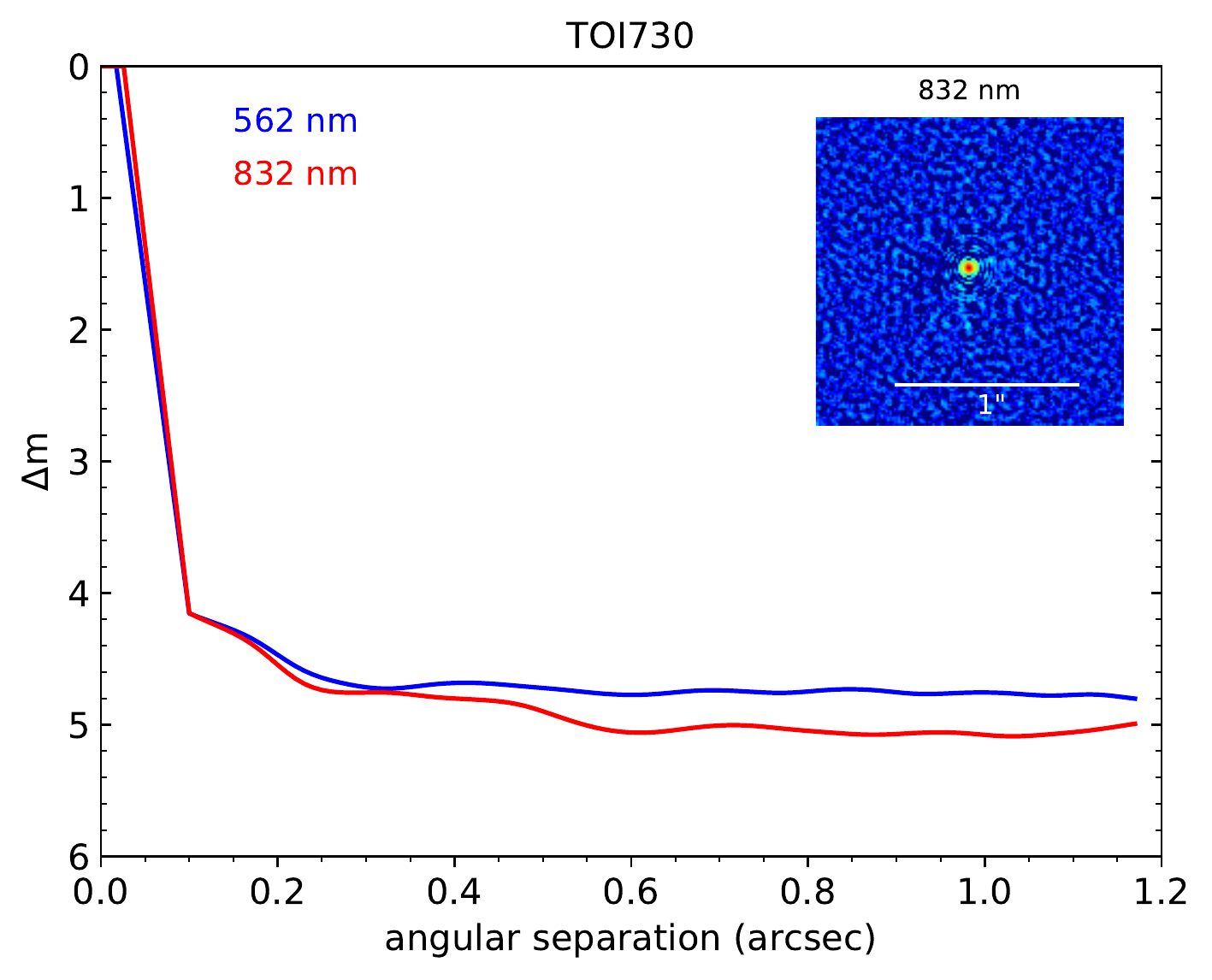}
 }
\caption{
    Contrast curves for \hatcur{77} (TOI~\hatcurCCtoibare{77}) derived from high-resolution images at 562\,nm and 832\,nm obtained with Zorro/Gemini~8\,m. The 832\,nm image is also shown in the inset.
\label{fig:hats77highresimage}
}
    \end{figure*}

No neighbors are detected for the other three systems, \hatcur{75},
\hatcur{76}, or
\hatcur{77}. Figures-\ref{fig:hats75highresimage}--\ref{fig:hats77highresimage}
show contrast limits on any resolved neighbors that are derived based
on the high-resolution imaging that we have reported for these three
objects.

\section{Analysis}
\label{sec:analysis}

\subsection{Transiting Planet Modelling}
\label{sec:transitmodel}

We perform a global fit to the light curves, radial velocities,
spectroscopically measured stellar atmospheric parameters,
broad-band photometry, and parallax from Gaia~DR2, using the methods described in \citet{hartman:2019:hats6069}, with modifications as summarized most
recently by \citet{bakos:2018:hats71}.
The fit is carried out using a modified version of the {\sc lfit} program which is included in the {\sc fitsh} software package \citep{pal:2012}. The light curves are modelled using the \citet{mandel:2002} transit model with quadratic limb-darkening. The limb darkening coefficients are allowed to vary in the fit. We place Gaussian prior constraints on the limb darkening coefficients using the tables of \citet{claret:2012,claret:2013} and \citet{claret:2018} and assume a prior uncertainty of $0.2$ for each coefficient.

We include in the model several parameters for the physical and observed properties of the host star, including the effective temperature, the metallicity, the distance modulus, and the $V$-band extinction $A_{V}$. These parameters are, in turn, constrained by the observed spectroscopic stellar atmospheric parameters (as measured in Section~\ref{sec:obsspec}), the photometry, and the parallax. Together with the parameters used to describe the transit and radial velocity observations, these parameters are sufficient to determine the bulk physical properties of the stars and their transiting planets. We fit the data using two different methods to relate the stellar mass to the stellar radius, metallicity and luminosity: (1) an empirical method which uses the stellar mean density measured from the transit and radial velocity observations to determine the stellar mass from the stellar radius, which is itself inferred from the effective temperature and luminosity \citep[this method is similar to that of, e.g.,][]{stassun:2017}, and (2) using version 1.2 of the MIST stellar evolution models \citep{dotter:2016,choi:2016,paxton:2011,paxton:2013,paxton:2015} to impose an additional constraint on the stellar relations that is typically tighter than the observed constraint on the stellar mean density. Note that here we take a different approach from prior HATSouth discovery papers which generally made use of the PARSEC stellar evolution models \citep{marigo:2017} instead. In each case, we assume  both the orbital eccentricity is zero and allow the eccentricity to be a free parameter.

A Differential Evolution Markov Chain Monte Carlo (DEMCMC) procedure is used to sample the
posterior parameter distribution.
See \citet{hartman:2019:hats6069} for a full list of the parameters that we vary, and their assumed priors. The fit includes the
optical broad-band photometry from Gaia~DR2 and APASS, NIR photometry
from 2MASS, and IR photometry from WISE. For WISE we exclude the W4
band for all systems as none of the objects were detected in that
bandpass, while for \hatcur{74}, \hatcur{76}, and \hatcur{77} we
also exclude the W3 bands. These observations, together with the stellar atmospheric parameters, the parallax, and the reddening, constrain the luminosity of the star. To model the reddening, we assume a $R_{V} = 3.1$ \citet{cardelli:1989} dust law parameterized by $A_{V}$, and use the {\sc mwdust} 3D Galactic extinction model \citep{bovy:2016} to place a prior constraint on its value.

For \hatcur{74} we excluded the Gaia~DR2 $BP$ and $RP$ measurements as we expect these to be contaminated in a non-trivial way from blending with the $0\farcs8$ neighbor HATS-74B (Section~\ref{sec:highresimaging}). For the $J$, $H$, $K_{S}$, $W1$ and $W2$ bandpasses HATS-74A and HATS-74B are completely blended. In these cases we estimated $\Delta {\rm mag}$ values in each bandpass from the MIST isochrones assuming a $0.23$\,\msun\ stellar mass for the companion, and that its age, metallicity, distance and redenning are the same as those for \hatcur{74}, determined in an initial iteration of the analysis. These $\Delta {\rm mag}$ values, which are given in the footnotes to Table~\ref{tab:stellarobserved}, were then used to subtract the flux from HATS-74B from each bandpass measurement. The corrected magnitudes are then included in the fit, and are what we list for \hatcur{74} in Table~\ref{tab:stellarobserved}.

We find that for all four transiting planet systems the orbits are
consistent with being circular when the eccentricities are varied, and
that the stellar parameters are more robustly constrained when
imposing the stellar evolution model constraints. We
therefore choose to adopt the parameters that stem from fixing the
orbit to be circular, and imposing the stellar evolution models as a
constraint on the stellar physical parameters.

The best-fit models are compared to the various observational data for
the four transiting planet systems in
Figures~\ref{fig:hats74}--\ref{fig:hats77tess}. The adopted stellar
parameters derived from the analysis are listed in
Table~\ref{tab:stellarderived}, while the adopted planetary parameters
are listed in Table~\ref{tab:planetparam}. We also list in
Table~\ref{tab:planetparam} the 95\% confidence upper limit on the
eccentricity that comes from allowing the eccentricity to vary in the
fit.

\subsection{Stellar Blend Modelling}
\label{sec:blendmodel}

We performed a blend modelling of each system following the procedure described in
\citet{hartman:2019:hats6069}. In summary, our blend modelling attempts to fit all of the
observations excepting the radial velocity data using various combinations of stars
with parameters constrained by the MIST models. We find
that for \hatcur{76} a model consisting of a single star with a
transiting planet provides a better fit (a greater likelihood or equivalently
lower $\chi^2$) to the light curves, spectroscopic stellar atmospheric
parameters, broad-band catalog photometry, and astrometric parallax
measurements than the best-fit blended stellar eclipsing binary
models. The blended stellar eclipsing binary models involve more free
parameters than the transiting planet model, and thus can be rejected
on the grounds that they are both poorer-fitting and higher complexity
models.  However, for \hatcur{75}, and \hatcur{77}, we
find that blends between a foreground star and a background eclipsing
binary provide somewhat better fits to these data than do
models consisting of a single star with a transiting planet. In these cases, comparably good fits to the data can be found for models consisting of a star with both a transiting planet and an unresolved stellar companion. For all three systems, the improvement in $\chi^2$ for the blends can be attributed to the increased number of free parameters that are included in these more complicated models. For these two systems we simulated radial velocities for the model blend scenarios by simulating composite cross-correlation functions. We find that for \hatcur{75} the blend scenarios that we considered would produce radial velocity variations in excess of $600$\,\ms\ that do not vary sinusoidally in phase with the transit ephemeris. This is in contrast to the observed  radial velocity variation that have $K = \hatcurRVK{75}$\,\ms\ and that are in phase with the transit ephemeris. Similarly, for \hatcur{77} we find that the simulated blended eclipsing binary radial velocities do not vary sinusoidally in phase with the ephemeris, and that the scatter is in excess of 1\,\kms, compared to the observed radial velocity variation with $K = \hatcurRVK{76}$\,\ms\ in phase with the transits. We conclude therefore that the blended eclipsing binary scenarios that might reproduce the photometric data for \hatcur{75} and \hatcur{77} can be ruled out on the grounds that they do not reproduce the radial velocity observations. We therefore consider \hatcur{75} and \hatcur{77}, like \hatcur{76}, to be confirmed transiting planet systems.

For \hatcur{74}, with its known resolved neighbor
(Section~\ref{sec:highresimaging}), we considered four scenarios: (1)
a transiting planet around the brighter source, with the fainter
source being a bound companion; (2) a transiting planet around the
fainter source, with the brighter source being a bound stellar
companion; (3) the brighter source being a blend between a bright
foreground star and a background stellar eclipsing binary, and the
fainter source being unrelated to either the foreground star or the
eclipsing binary; (4) the brighter source being a foreground star, and
the fainter source being a background stellar eclipsing binary. In all
cases we assume the 2MASS $J$, $H$, $K_{S}$, and the $W1$ and $W2$
photometry is blended between the two known sources,
while $\Delta J$ from the NIRC2 observations, and $G$ and the parallax
values from {\em Gaia} DR2 are unblended. The mass of both resolved
stars are varied in the fits. They are assumed either to have the same
age, distance, metallicity, and extinction, or independent values for
these, depending on the scenario
considered.

We find that scenarios (2) and (4) for \hatcur{74} do not fit the data
included in the modelling, and can be therefore easily ruled out. Scenario (3)
provides a somewhat better fit to the data than scenario (1), however,
it uses seven extra parameters and the improvement in $\chi^2$
can be fully attributed to the additional model complexity.
As an additional check on scenario (3) we
simulated radial velocity observations for 1000 blend scenarios drawn
randomly from the posterior chain. For each draw from the
chain we simulated the radial velocity observations in two ways:
(a) assuming the resolved star is also resolved in the ESPRESSO
observations; and (b) assuming it is not resolved in the ESPRESSO
observations. We find that in all cases the simulated radial velocities show much
larger variations (well in excess of 1\,\kms) compared to the observed
ones ($K = \hatcurRVK{74}$\,\ms), and, moreover, they do not exhibit a clean, in-phase
Keplerian orbital variation. We conclude therefore that scenario (3) is not
consistent with all of the observations, and confirm that \hatcur{74}
is a transiting planet system, with a resolved binary star
companion. The modelling carried out for scenario (1) yields a mass
for the binary star companion of \hatcurISOmB{74}\,\msun, which we
adopt in Table~\ref{tab:stellarderived}.

For \hatcur{75}, \hatcur{76} and \hatcur{77} we place limits on the presence of any unresolved binary star companions based on this modelling which we also list in Table~\ref{tab:stellarderived}. Additionally, we note that any such companion would need to satisfy the contrast limits based on the null detection of companions in the high resolution imaging discussed in Section~\ref{sec:highresimaging}.
Finally, we note that there is no evidence of correlation of the bisector span measurements with orbital phase for any of the systems.



\section{Discussion}
\label{sec:discussion}

\begin{figure*}[!ht]
 \includegraphics[width={1.0\linewidth}]{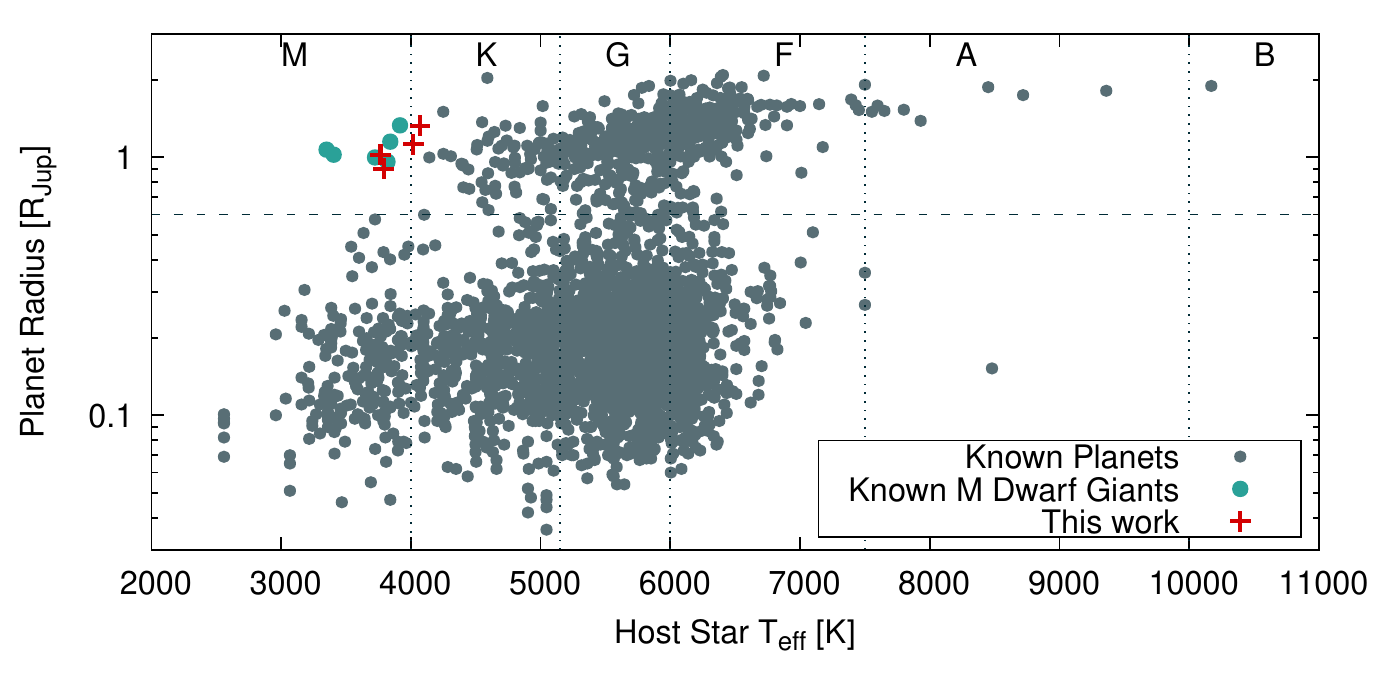}
 \caption{Planetary radius versus stellar effective temperature $T_{\rm eff}$ for confirmed transiting exoplanets. The dotted vertical lines indicate the $T_{\rm eff}$ values that roughly separate the host star spectral types as indicated on top, while the horizontal line at $R=0.6 R_J$ indicates the separation between giant and smaller planets. Different symbols are used to plot known giants around M dwarfs and the discoveries presented in this work as indicated in the legend. Data were downloaded from the NASA Exoplanet Archive on May 13 2021.}
 \label{fig:TeffRp}
\end{figure*}

We have presented in this work the discovery of four giant planets hosted by early M and late K dwarf stars. We place these discoveries in the context of known planets in Figure~\ref{fig:TeffRp}, where we plot radius versus effective temperature of the host star for all exoplanets that have these quantities measured. It is apparent in this figure that the discoveries presented in this work add significantly to the number of known transiting giant planets hosted by stars with effective temperatures $T_{\rm eff} \lesssim 4000$ K. As stated in the introduction, these kinds of systems are intrinsically rare and observationally challenging to confirm due to the faintness of the host stars.  To confirm these exoplanet, we need high
resolution stable spectrographs mounted on large aperture telescopes. In this work we used ESPRESSO mounted on the VLT in order to confirm and measure the masses for our discoveries. Also noteworthy in terms of the required discovery resources is the fact that these systems were first uncovered as candidates by the HATSouth survey, and observed by the TESS mission in high cadence by virtue of their nature as candidates from a ground-based survey.

\begin{figure*}[!ht]
 \includegraphics[width={1.0\linewidth}]{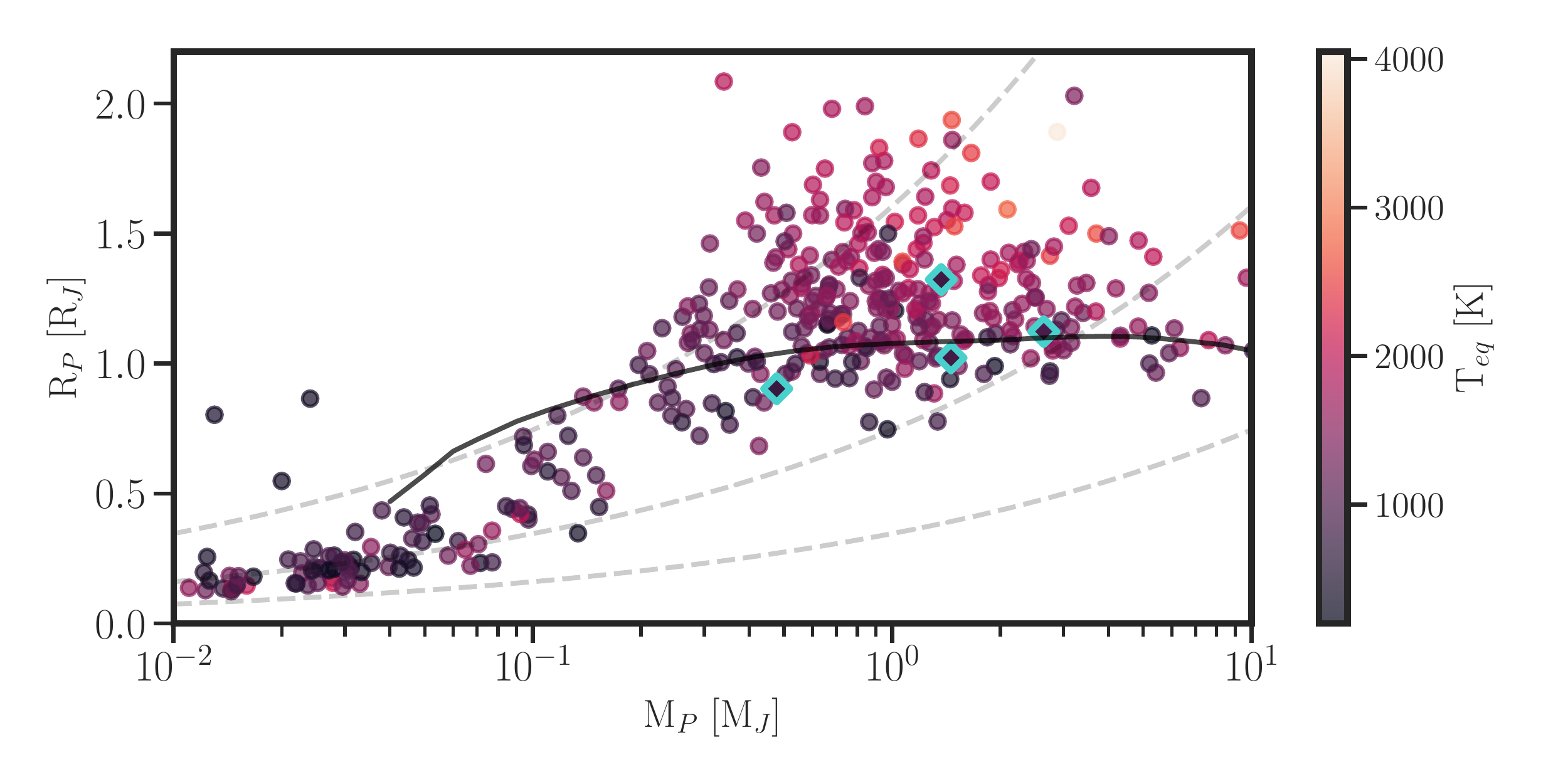}
 \caption{Mass--Radius diagram for the population of exoplanets with measured mass and radii. The points corresponding the discoveries presented in this work are indicated with diamonds. The color represents the equilibrium temperature of the planet. The dashed gray lines correspond to isodensity curves for 0.3, 3 and 30 g cm$^{-3}$ , respectively. The solid line corresponds to the predicted radius using the models of \citet{fortney:2007} for a planet with a 10 $M_\oplus$ central core.}
 \label{fig:MR}
\end{figure*}

In Figure~\ref{fig:MR} we plot out discoveries in the mass-radius plane along with other confirmed planets with measured masses and radii. We color-code in the figure the equilibrium temperature of each discovery. Despite the very short periods of our discoveries (in the range $P =1.7$--3.1 d), the average flux received by none of them exceeds $2\times 10^8$ erg cm sec$^{-2}$, the stellar irradiation value below which it has been shown
that the effects of irradiation on the planetary radius are negligible \citep[eg][]{demory:2011}. Therefore, we don't expect any of them to show anomalously large radii. This is borne out by our measurements for \hatcurb{74}, \hatcurb{75} and \hatcurb{76}, but \hatcurb{77} has an unexpectedly high radius of $\hatcurPPrlong{77}$, formally $>3\sigma$ higher than the expected radius for its mass. It is not possible to draw any conclusions from a single object which although formally receiving an irradiation that is below the value where radius inflation starts appearing it is still receiving a sizable irradiation of $\approx 10^8$ erg cm sec$^{-2}$. It will be interesting to see if, as we discover further giants planets around low mass stars, and especially systems with periods larger than those typical of hot Jupiters, more planets show larger radii than expected.

\begin{figure*}[!ht]
 \includegraphics[width={1.0\linewidth}]{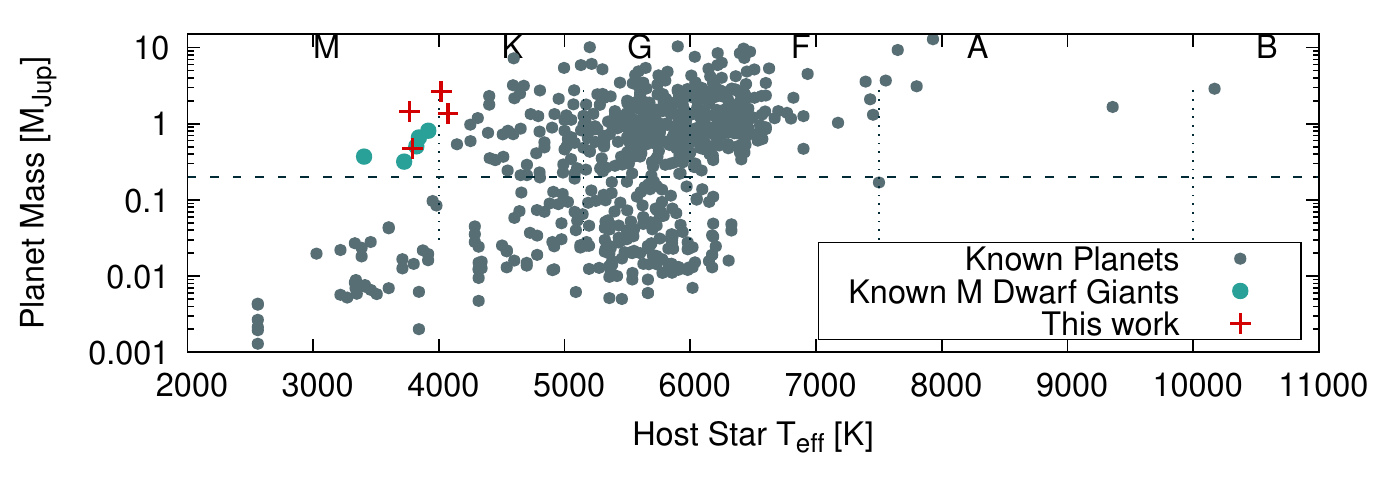}
 \caption{Mass versus effective temperature $T_{\rm eff}$ for confirmed transiting exoplanets. The dotted vertical lines indicate the $T_{\rm eff}$ values that roughly separate the spectral types as indicated on top, while the horizontal line at $M=0.2 M_\odot$ indicates the adopted boundary between giant and smaller planets. Different symbols are used to plot known giants around M dwarfs and the discoveries presented in this work as indicated in the legend. Data were downloaded from the NASA Exoplanet Archive on May 13 2021.}
 \label{fig:TeffMp}
\end{figure*}

 We plot in Figure~\ref{fig:TeffMp} the masses versus effective temperature for all planets that have both quantities measured in addition to their radii. We can see that three of the planets presented in this work have the higher masses known for transiting planets hosted by stars with $T_{\rm eff} \lesssim 4000$ K, with \hatcurb{76} having a mass of $\hatcurPPmlong{76}$, \hatcurb{74} a mass of $\hatcurPPmlong{74}$ and \hatcurb{77} a mass of $\hatcurPPmlong{77}$. If we include radial velocity planets these masses are not particularly remarkable, and confirm the fact that despite their lower average protoplanetary disk masses these systems can assemble very massive giant planets. This result is in accord with formation models that find that, while the occurrence rate of giant planets is expected to decrease for M dwarfs, the maximum planetary mass is not  \citep[e.g.,][]{Burn:2021}.

 The TESS mission, with its all-sky, high photometric precision survey, is allowing us to further the frontiers of exoplanet discoveries. While focused to a large degree on discovering small planets around nearby stars, the combination of its coverage with larger aperture facilities for photometric and spectroscopic follow-up, is allowing to efficiently target the rare class of giant planets around late K and later dwarfs with effective temperatures $\lesssim 4000$ K. In particular, the arrival of ESPRESSO to the VLT allows an efficiency in this quest heretofore unavailable in the southern hemisphere. We are undertaking a systematic search for these systems. With an increased sample of known giants around M dwarfs we expect to provide stronger constraints on the occurrence rates for these systems.  Such occurance rates have remained to date too uncertain to effectively constrain models despite their unique potential.  The clear prediction of core-accretion theory is that the efficiency of formation should decrease dramatically for stars with masses $M\lesssim 0.7 M_\odot$. A higher efficiency of formation, above what models based on core-accretion predict, could be traced to the inadequacy some basic assumptions. For example, a basic tenet is that disk mass is proportional to stellar mass. While this is observationally well established, the dispersion at a given mass is significant and there are instances of young M stars with  massive disks where gravitational instabilities may be a viable formation pathway for giant planets \citep[e.g., the young M0 star Elias 2-27, see][and references therein]{paneque:2021}. A well determined occurrence rate for giant planets around M dwarfs will be invaluable in providing stringent tests for the current favoured models of planetary formation and evolution.

\acknowledgements
AJ thanks Gijs Mulders for informative discussions.
Development of the HATSouth
project was funded by NSF MRI grant NSF/AST-0723074, operations have
been supported by NASA grants NNX09AB29G, NNX12AH91H, and NNX17AB61G, and follow-up
observations have received partial support from grant NSF/AST-1108686.
A.J.\ acknowledges support from FONDECYT project 1210718, and ANID - Millennium Science Initiative - ICN12\_009.
J.H.\ acknowledges partial support from the TESS GI Program, programs G011103 and G022117, through NASA grants 80NSSC19K0386 and 80NSSC19K1728.
L.M.\ acknowledges support from the Italian Minister of Instruction, University and Research (MIUR) through FFABR 2017 fund. L.M.\ acknowledges support from the University of Rome Tor Vergata through ``Mission: Sustainability 2016'' fund.
K.P.\ acknowledges support from NASA ATP grant 80NSSC18K1009.
M.R.\ acknowledges support from the Universidad Cat\'olica de lo Sant\'isima Concepci\'on grant DI-FIAI 03/2021.
R.L.\ acknowledges financial support from the Spanish Ministerio de Ciencia e Innovación, through project PID2019-109522GB-C52/AEI/10.13039/501100011033, and the Centre of Excellence "Severo Ochoa" award to the Instituto de Astrofísica de Andalucía (SEV-2017-0709).
Based in part on observations collected at the European Organisation for Astronomical Research in the Southern Hemisphere under ESO programme 0103.C-0449(A).
Part of this work has been carried out within the framework of the National Centre of Competence in Research PlanetS supported by the Swiss National Science Foundation. ECM acknowledges the financial support of the SNSF.
This work makes use of observations from the LCOGT network. Part of the LCOGT telescope time was granted by NOIRLab through the Mid-Scale Innovations Program (MSIP). MSIP is funded by NSF. This paper is partly based on observations made with the MuSCAT3 instrument, developed by the Astrobiology Center and under financial supports by JSPS KAKENHI (JP18H05439) and JST PRESTO (JPMJPR1775), at Faulkes Telescope North on Maui, HI, operated by the Las Cumbres Observatory. This work is partly supported by JSPS KAKENHI Grant Number JP20K14518.
Some of the Observations in the paper made use of the High-Resolution Imaging instrument(s) ‘Alopeke (and/or Zorro). ‘Alopeke (and/or Zorro) was funded by the NASA Exoplanet Exploration Program and built at the NASA Ames Research Center by Steve B. Howell, Nic Scott, Elliott P. Horch, and Emmett Quigley. Data were reduced using a software pipeline originally written by Elliott Horch and Mark Everett. ‘Alopeke (and/or Zorro) was mounted on the Gemini North (and/or South) telescope of the international Gemini Observatory, a program of NSF’s OIR Lab, which is managed by the Association of Universities for Research in Astronomy (AURA) under a cooperative agreement with the National Science Foundation. on behalf of the Gemini partnership: the National Science Foundation (United States), National Research Council (Canada), Agencia Nacional de Investigación y Desarrollo (Chile), Ministerio de Ciencia, Tecnología e Innovación (Argentina), Ministério da Ciência, Tecnologia, Inovações e Comunicações (Brazil), and Korea Astronomy and Space Science Institute (Republic of Korea).
We acknowledge the use of the AAVSO Photometric All-Sky Survey (APASS),
funded by the Robert Martin Ayers Sciences Fund, and the SIMBAD
database, operated at CDS, Strasbourg, France.
TRAPPIST-South is a project funded by the Belgian F.R.S.-FNRS under
grant PDR T.0120.21, with the participation of the Swiss FNS.  The
research leading to these results has received funding from the ARC
grant for Concerted Research Actions, financed by the
Wallonia-Brussels Federation. EJ and MG are F.R.S.-FNRS Senior
Research Associates.
This work has made use of data from the European Space Agency (ESA)
mission {\it Gaia} (\url{https://www.cosmos.esa.int/gaia}), processed by
the {\it Gaia} Data Processing and Analysis Consortium (DPAC,
\url{https://www.cosmos.esa.int/web/gaia/dpac/consortium}). Funding
for the DPAC has been provided by national institutions, in particular
the institutions participating in the {\it Gaia} Multilateral Agreement.
This research has made use of the NASA Exoplanet Archive, which is
operated by the California Institute of Technology, under contract
with the National Aeronautics and Space Administration under the
Exoplanet Exploration Program.
This research has made use NASA's Astrophysics Data System Bibliographic Services.

\facilities{HATSouth, TESS, LCOGT (0.4m, 1m, 2m MuSCAT3), CTIO 0.9m, MuSCAT2, TRAPPIST-South, Mt Stuart 0.3m,  VLT (ESPRESSO), Gemini ('Alopeke), Keck (NIRC2), VLT (NaCo), ARC 3.5m (WHIRC), Gaia, Exoplanet Archive}

\software{FITSH \citep{pal:2012}, BLS \citep{kovacs:2002:BLS},
  VARTOOLS \citep{hartman:2016:vartools}, CERES
  \citep{brahm:2017:ceres}, ZASPE \citep{brahm:2017:zaspe}, SPEX-tool
  \citep{cushing:2004,vacca:2004}, SExtractor \citep{bertin:1996},
  Astrometry.net \citep{lang:2010}, MWDUST \citep{bovy:2016}, TESSCut
  \citep{brasseur:2019}, Lightkurve \citep{lightkurve:2018}, Astropy \citep{astropy:2013,astropy:2018}, AstroImageJ \citep{Collins:2017}, TAPIR \citep{Jensen:2013}}


\bibliographystyle{aasjournal}
\bibliography{hats74-77}

\clearpage

%
%
    \begin{deluxetable*}{lccccl}
%
\tablewidth{0pc}
\tabletypesize{\tiny}
\tablecaption{
    Astrometric, Spectroscopic and Photometric parameters for \hatcur{74}, \hatcur{75}, \hatcur{76} and \hatcur{77}
    \label{tab:stellarobserved}
}
\tablehead{
    \multicolumn{1}{c}{} &
    \multicolumn{1}{c}{\bf HATS-74A} &
    \multicolumn{1}{c}{\bf HATS-75} &
    \multicolumn{1}{c}{\bf HATS-76} &
    \multicolumn{1}{c}{\bf HATS-77} &
    \multicolumn{1}{c}{} \\
    \multicolumn{1}{c}{~~~~~~~~Parameter~~~~~~~~} &
    \multicolumn{1}{c}{Value}                     &
    \multicolumn{1}{c}{Value}                     &
    \multicolumn{1}{c}{Value}                     &
    \multicolumn{1}{c}{Value}                     &
    \multicolumn{1}{c}{Source}
}
\startdata
\noalign{\vskip -3pt}
\sidehead{Astrometric properties and cross-identifications}
~~~~2MASS-ID\dotfill               & \hatcurCCtwomassshort{74}  & \hatcurCCtwomassshort{75} & \hatcurCCtwomassshort{76} & \hatcurCCtwomassshort{77} & \\
~~~~TIC-ID\dotfill                 & \hatcurCCtic{74} & \hatcurCCtic{75} & \hatcurCCtic{76} & \hatcurCCtic{77} & \\
~~~~TOI-ID\dotfill                 & \hatcurCCtoi{74} & \hatcurCCtoi{75} & \hatcurCCtoi{76} & \hatcurCCtoi{77} & \\
~~~~GAIA~DR2-ID\dotfill                 & \hatcurCCgaiadrtwoshort{74}      & \hatcurCCgaiadrtwoshort{75} & \hatcurCCgaiadrtwoshort{76} & \hatcurCCgaiadrtwoshort{77} & \\
~~~~R.A. (J2000)\dotfill            & \hatcurCCra{74}       & \hatcurCCra{75}    & \hatcurCCra{76}    & \hatcurCCra{77}    & GAIA DR2\\
~~~~Dec. (J2000)\dotfill            & \hatcurCCdec{74}      & \hatcurCCdec{75}   & \hatcurCCdec{76}   & \hatcurCCdec{77}   & GAIA DR2\\
~~~~$\mu_{\rm R.A.}$ (\masy)              & \hatcurCCpmra{74}     & \hatcurCCpmra{75} & \hatcurCCpmra{76} & \hatcurCCpmra{77} & GAIA DR2\\
~~~~$\mu_{\rm Dec.}$ (\masy)              & \hatcurCCpmdec{74}    & \hatcurCCpmdec{75} & \hatcurCCpmdec{76} & \hatcurCCpmdec{77} & GAIA DR2\\
~~~~parallax (mas)              & \hatcurCCparallax{74}    & \hatcurCCparallax{75} & \hatcurCCparallax{76} & \hatcurCCparallax{77} & GAIA DR2\\
\sidehead{Spectroscopic properties}
~~~~$\teffstar$ (K)\dotfill         &  \hatcurSMEteff{74}   & \hatcurSMEteff{75} & \hatcurSMEteff{76} & \hatcurSMEteff{77} & ARCoIRIS\tablenotemark{a}\\
~~~~$\feh$\dotfill                  &  \hatcurSMEzfeh{74}   & \hatcurSMEzfeh{75} & \hatcurSMEzfeh{76} & \hatcurSMEzfeh{77} & ARCoIRIS               \\
~~~~$\gamma_{\rm RV}$ (\ms)\dotfill&  $\hatcurRVgammaabs{74}$  & $\hatcurRVgammaabs{75}$ & $\hatcurRVgammaabs{76}$ & $\hatcurRVgammaabs{77}$ & ESPRESSO\tablenotemark{b}  \\
\sidehead{Photometric properties\tablenotemark{c}}
~~~~$P_{\rm rot}$ (d)\tablenotemark{d}   & $\hatcurRotPer{74}$ & $\hatcurRotPer{75}$ & $\hatcurRotPer{76}$ & $\hatcurRotPer{77}$ & HATSouth \\
~~~~$G$ (mag)\tablenotemark{e}\dotfill               &  \hatcurCCgaiamG{74}  & \hatcurCCgaiamG{75} & \hatcurCCgaiamG{76} & \hatcurCCgaiamG{77} & GAIA DR2 \\
~~~~$BP$ (mag)\tablenotemark{e}\dotfill               &  $\cdots$  & \hatcurCCgaiamBP{75} & \hatcurCCgaiamBP{76} & \hatcurCCgaiamBP{77} & GAIA DR2 \\
~~~~$RP$ (mag)\tablenotemark{e}\dotfill               &  $\cdots$  & \hatcurCCgaiamRP{75} & \hatcurCCgaiamRP{76} & \hatcurCCgaiamRP{77} & GAIA DR2 \\
~~~~$B$ (mag)\dotfill               &  $\cdots$  & \hatcurCCtassmB{75} & $\cdots$ & \hatcurCCtassmB{77} & APASS\tablenotemark{g} \\
~~~~$V$ (mag)\dotfill               &  $\cdots$  & \hatcurCCtassmv{75} & $\cdots$ & \hatcurCCtassmv{77} & APASS\tablenotemark{g} \\
~~~~$g$ (mag)\dotfill               &  $\cdots$  & \hatcurCCtassmg{75} & $\cdots$ & \hatcurCCtassmg{77} & APASS\tablenotemark{g} \\
~~~~$r$ (mag)\dotfill               &  $\cdots$  & \hatcurCCtassmr{75} & $\cdots$ & \hatcurCCtassmr{77} & APASS\tablenotemark{g} \\
~~~~$i$ (mag)\dotfill               &  $\cdots$  & \hatcurCCtassmi{75} & $\cdots$ & \hatcurCCtassmi{77} & APASS\tablenotemark{g} \\
~~~~$J$ (mag)\tablenotemark{f}\dotfill               &  \hatcurCCtwomassJmag{74} & \hatcurCCtwomassJmag{75} & \hatcurCCtwomassJmag{76} & \hatcurCCtwomassJmag{77} & 2MASS           \\
~~~~$H$ (mag)\tablenotemark{f}\dotfill               &  \hatcurCCtwomassHmag{74} & \hatcurCCtwomassHmag{75} & \hatcurCCtwomassHmag{76} & \hatcurCCtwomassHmag{77} & 2MASS           \\
~~~~$K_s$ (mag)\tablenotemark{f}\dotfill             &  \hatcurCCtwomassKmag{74} & \hatcurCCtwomassKmag{75} & \hatcurCCtwomassKmag{76} & \hatcurCCtwomassKmag{77} & 2MASS           \\
~~~~$W1$ (mag)\tablenotemark{f}\dotfill             &  \hatcurCCWonemag{74} & \hatcurCCWonemag{75} & \hatcurCCWonemag{76} & \hatcurCCWonemag{77} & WISE           \\
~~~~$W2$ (mag)\tablenotemark{f}\dotfill             &  \hatcurCCWtwomag{74} & \hatcurCCWtwomag{75} & \hatcurCCWtwomag{76} & \hatcurCCWtwomag{77} & WISE           \\
~~~~$W3$ (mag)\dotfill             &  $\cdots$ & \hatcurCCWthreemag{75} & $\cdots$ & $\cdots$ & WISE           \\
\enddata
\tablenotetext{a}{
    The parameters are estimated from the ARCoIRIS NIR spectra.
}
\tablenotetext{b}{
    The error on $\gamma_{\rm RV}$ is determined from the
    orbital fit to the RV measurements, and does not include the
    systematic uncertainty in transforming the velocities to the IAU
    standard system. The velocities have not been corrected for
    gravitational redshifts.
}
\tablenotetext{c}{
    We only include in the table catalog magnitudes that were included in our analysis of each system. In some cases magnitudes we list as $\cdots$ the bandpass magnitude for a source when this magnitude is available in the indicated source catalog. These magnitudes are excluded from the analysis for reasons discussed in Section~\ref{sec:transitmodel}.
}
\tablenotetext{d}{
    Photometric rotation period.
}
\tablenotetext{e}{
    The listed uncertainties for the Gaia DR2 photometry are taken from the catalog. For the analysis we assume additional systematic uncertainties of 0.002\,mag, 0.005\,mag and 0.003\,mag for the G, BP and RP bands, respectively.
}
\tablenotetext{f}{
    The listed $J$, $H$, $K_{S}$, $W1$, and $W2$ magnitudes for \hatcur{74} have been corrected for blending with HATS-74B, assuming this latter object is a $0.23$\,\msun\ main sequence star with the same age and metallicity as \hatcur{74} (Tab.~\ref{tab:stellarderived}). Following these assumptions, we adopt $\Delta J = 2.6418$, $\Delta H = 2.7294$, $\Delta K_{S} = 2.6473$, $\Delta W1 = 2.5259$ and $\Delta W2 = 2.3352$ between HATS-74B and \hatcur{74} in removing the contribution of HATS-74B from the catalog photometry.
}
\tablenotetext{g}{
    From APASS DR6 as
    listed in the UCAC 4 catalog \citep{zacharias:2013:ucac4}.
}
    \end{deluxetable*}

\tabletypesize{\scriptsize}
    \begin{deluxetable*}{lrrrrrrl}
%
\tablewidth{0pc}
\tablecaption{
    ESPRESSO relative radial velocities and Bisector Spans for \hatcur{74}, \hatcur{75}, \hatcur{76} and \hatcur{77}.
    \label{tab:rvs}
}
\tablehead{
    \colhead{System} &
    \colhead{BJD} &
    \colhead{RV\tablenotemark{a}} &
    \colhead{\ensuremath{\sigma_{\rm RV}}\tablenotemark{b}} &
    \colhead{BS\tablenotemark{c}} &
    \colhead{\ensuremath{\sigma_{\rm BS}}} &
    \colhead{Phase}\\
    \colhead{} &
    \colhead{\hbox{(2,450,000$+$)}} &
    \colhead{(\ms)} &
    \colhead{(\ms)} &
    \colhead{(\ms)} &
    \colhead{(\ms)} &
    \colhead{}
}
\startdata
HATS-74 & $ 8843.81635 $ & $   254.27 $ & $    14.40 $ & $   10.4 $ & $   27.4 $ &$   0.870 $ \\
HATS-74 & $ 8844.82333 $ & $  -116.73 $ & $    18.60 $ & $    5.3 $ & $   37.0 $ &$   0.451 $ \\
HATS-74 & $ 8846.84777 $ & $   242.27 $ & $    13.40 $ &$   15.5 $ & $   25.4 $ & $   0.620 $ \\
HATS-74 & $ 8847.83295 $ & $  -318.73 $ & $    12.70 $ &$   22.8 $ & $   24.7 $ & $   0.189 $ \\
HATS-74 & $ 8848.83946 $ & $   340.27 $ & $    13.10 $ &$   28.8 $ & $   24.6 $ & $   0.770 $ \\
HATS-75 & $ 8756.72534 $ & $   -93.43 $ & $    10.80 $ &$  -20.4 $ & $   21.6 $ & $   0.237 $ \\
HATS-75 & $ 8759.70279 $ & $   -90.43 $ & $     9.40 $ &$   36.1 $ & $   18.8 $ & $   0.304 $ \\
HATS-75 & $ 8760.72353 $ & $    91.57 $ & $    12.80 $ &$   17.9 $ & $   25.6 $ & $   0.670 $ \\
HATS-75 & $ 8761.73812 $ & $   -18.43 $ & $     8.60 $ &$   55.3 $ & $   17.1 $ & $   0.034 $ \\
HATS-75 & $ 8762.77554 $ & $   -49.43 $ & $     8.20 $ &$   45.8 $ & $   16.4 $ & $   0.406 $ \\
HATS-76 & $ 8756.74359 $ & $  -265.56 $ & $    24.90 $ &$   75.2 $ & $   49.8 $ & $   0.086 $ \\
HATS-76 & $ 8759.72348 $ & $   361.44 $ & $    19.70 $ &$   67.1 $ & $   39.4 $ & $   0.620 $ \\
HATS-76 & $ 8762.81602 $ & $  -565.56 $ & $    16.10 $ &$   62.7 $ & $   32.3 $ & $   0.213 $ \\
HATS-76 & $ 8779.83962 $ & $   101.44 $ & $    19.20 $ &$   34.9 $ & $   38.3 $ & $   0.981 $ \\
HATS-77 & $ 8818.79874 $ & $   -96.07 $ & $    27.00 $ &$   49.8 $ & $   54.0 $ & $   0.042 $ \\
HATS-77 & $ 8820.78227 $ & $   227.93 $ & $    12.90 $ &$   64.1 $ & $   25.8 $ & $   0.685 $ \\
HATS-77 & $ 8836.81364 $ & $   177.93 $ & $    17.20 $ &$   91.2 $ & $   34.4 $ & $   0.877 $ \\
HATS-77 & $ 8843.73764 $ & $  -143.07 $ & $    17.50 $ &$  108.3 $ & $   35.1 $ & $   0.119 $ \\
HATS-77 & $ 8844.74086 $ & $  -114.07 $ & $    13.10 $ &$  107.2 $ & $   26.1 $ & $   0.444 $ \\
\enddata
\tablenotetext{a}{
    The zero-point of these velocities is arbitrary. An overall offset
    $\gamma_{\rm rel}$ fitted to the orbit (and listed in Tab.~\ref{tab:stellarobserved}) has been subtracted for each system.
}
\tablenotetext{b}{
    Internal errors excluding the component of astrophysical jitter
    allowed to vary in the fit.
}
\tablenotetext{c}{
    Bisector span of the cross-correlation function profile.
}
    \end{deluxetable*}

%
%
    \begin{deluxetable*}{lcccc}
\tablewidth{0pc}
\tabletypesize{\footnotesize}
\tablecaption{
    Adopted derived stellar parameters for \hatcur{74}, \hatcur{75}, \hatcur{76} and \hatcur{77}.
    \label{tab:stellarderived}
}
\tablehead{
    \multicolumn{1}{c}{} &
    \multicolumn{1}{c}{\bf HATS-74A} &
    \multicolumn{1}{c}{\bf HATS-75} &
    \multicolumn{1}{c}{\bf HATS-76} &
    \multicolumn{1}{c}{\bf HATS-77} \\
    \multicolumn{1}{c}{~~~~~~~~Parameter~~~~~~~~} &
    \multicolumn{1}{c}{Value}                     &
    \multicolumn{1}{c}{Value}                     &
    \multicolumn{1}{c}{Value}                     &
    \multicolumn{1}{c}{Value}
}
\startdata
~~~~$\mstar$ ($\msun$)\dotfill      &  \hatcurISOmlong{74}   & \hatcurISOmlong{75} & \hatcurISOmlong{76} & \hatcurISOmlong{77} \\
~~~~$\rstar$ ($\rsun$)\dotfill      &  \hatcurISOrlong{74}   & \hatcurISOrlong{75} & \hatcurISOrlong{76} & \hatcurISOrlong{77} \\
~~~~$\loggstar$ (cgs)\dotfill       &  \hatcurISOlogg{74}    & \hatcurISOlogg{75} & \hatcurISOlogg{76} & \hatcurISOlogg{77} \\
~~~~$\rhostar$ (\gcmc)\dotfill       &  \hatcurLCrho{74}    & \hatcurLCrho{75} & \hatcurLCrho{76} & \hatcurLCrho{77} \\
~~~~$\lstar$ ($\lsun$)\dotfill      &  \hatcurISOlum{74}     & \hatcurISOlum{75} & \hatcurISOlum{76} & \hatcurISOlum{77} \\
~~~~$\teffstar$ (K)\dotfill      &  \hatcurISOteff{74} &  \hatcurISOteff{75} &  \hatcurISOteff{76} &  \hatcurISOteff{77} \\
~~~~$\feh$\dotfill      &  \hatcurISOzfeh{74} &  \hatcurISOzfeh{75} &  \hatcurISOzfeh{76} &  \hatcurISOzfeh{77} \\
~~~~Age (Gyr)\dotfill               &  \hatcurISOage{74}     & \hatcurISOage{75} & \hatcurISOage{76} & \hatcurISOage{77} \\
~~~~$A_{V}$ (mag)\dotfill               &  \hatcurXAv{74}     & \hatcurXAv{75} & \hatcurXAv{76} & \hatcurXAv{77} \\
~~~~Distance (pc)\dotfill           &  \hatcurXdistred{74}\phn  & \hatcurXdistred{75} & \hatcurXdistred{76} & \hatcurXdistred{77} \\
~~~~M$_{\rm B}$ (\msun)\tablenotemark{a} & $\hatcurISOmB{74}$ & $\hatcurISOmB{75}$ & $\hatcurISOmB{76}$ & $\hatcurISOmB{77}$ \\
\enddata
\tablenotetext{a}{
    For \hatcur{75}, \hatcur{76} and \hatcur{77} we list the 95\% confidence upper limit on the mass of any unresolved stellar companion based on modelling the system as a blend between a transiting planet system and an unresolved wide stellar binary companion (Section~\ref{sec:blendmodel}). For \hatcur{74} we list the estimated mass for the 0\farcs844 neighbor in Gaia~DR2 which we determined to be a common-proper-motion and common-parallax companion to \hatcur{74} (Section~\ref{sec:highresimaging}).
}
\tablecomments{
The listed parameters are those determined through the joint differential evolution Markov Chain analysis described in Section~\ref{sec:transitmodel}. For all four systems the RV observations are consistent with a circular orbit, and we assume a fixed circular orbit in generating the parameters listed here. Systematic errors in the bolometric correction tables or stellar evolution models are not included, and may dominate the error budget for some of these parameters.
}
    \end{deluxetable*}

%
    \begin{deluxetable*}{lcccc}
\tabletypesize{\tiny}
\tablecaption{Adopted orbital and planetary parameters for \hatcurb{74}, \hatcurb{75}, \hatcurb{76} and \hatcurb{77}\label{tab:planetparam}}
\tablehead{
    \multicolumn{1}{c}{} &
    \multicolumn{1}{c}{\bf HATS-74Ab} &
    \multicolumn{1}{c}{\bf HATS-75b} &
    \multicolumn{1}{c}{\bf HATS-76b} &
    \multicolumn{1}{c}{\bf HATS-77b} \\
    \multicolumn{1}{c}{~~~~~~~~~~~~~~~Parameter~~~~~~~~~~~~~~~} &
    \multicolumn{1}{c}{Value} &
    \multicolumn{1}{c}{Value} &
    \multicolumn{1}{c}{Value} &
    \multicolumn{1}{c}{Value}
}
\startdata
\noalign{\vskip -3pt}
\sidehead{\Lc{} parameters}
~~~$P$ (days)             \dotfill    & $\hatcurLCP{74}$ & $\hatcurLCP{75}$ & $\hatcurLCP{76}$ & $\hatcurLCP{77}$ \\
~~~$T_c$ (${\rm BJD\_{}TDB}$)
      \tablenotemark{a}   \dotfill    & $\hatcurLCT{74}$ & $\hatcurLCT{75}$ & $\hatcurLCT{76}$ & $\hatcurLCT{77}$ \\
~~~$T_{14}$ (days)
      \tablenotemark{a}   \dotfill    & $\hatcurLCdur{74}$ & $\hatcurLCdur{75}$ & $\hatcurLCdur{76}$ & $\hatcurLCdur{77}$ \\
~~~$T_{12} = T_{34}$ (days)
      \tablenotemark{a}   \dotfill    & $\hatcurLCingdur{74}$ & $\hatcurLCingdur{75}$ & $\hatcurLCingdur{76}$ & $\hatcurLCingdur{77}$ \\
~~~$\arstar$              \dotfill    & $\hatcurPPar{74}$ & $\hatcurPPar{75}$ & $\hatcurPPar{76}$ & $\hatcurPPar{77}$ \\
~~~$\zrstar$ \tablenotemark{b}             \dotfill    & $\hatcurLCzeta{74}$\phn & $\hatcurLCzeta{75}$\phn& $\hatcurLCzeta{76}$\phn& $\hatcurLCzeta{77}$\phn\\
~~~$\rpl/\rstar$          \dotfill    & $\hatcurLCrprstar{74}$ & $\hatcurLCrprstar{75}$& $\hatcurLCrprstar{76}$& $\hatcurLCrprstar{77}$\\
~~~$b^2$                  \dotfill    & $\hatcurLCbsq{74}$ & $\hatcurLCbsq{75}$& $\hatcurLCbsq{76}$& $\hatcurLCbsq{77}$\\
~~~$b \equiv a \cos i/\rstar$
                          \dotfill    & $\hatcurLCimp{74}$ & $\hatcurLCimp{75}$& $\hatcurLCimp{76}$& $\hatcurLCimp{77}$\\
~~~$i$ (deg)              \dotfill    & $\hatcurPPi{74}$\phn & $\hatcurPPi{75}$\phn& $\hatcurPPi{76}$\phn& $\hatcurPPi{77}$\phn\\
\sidehead{Dilution factors \tablenotemark{c}}
~~~HATSouth 1\dotfill & \hatcurLCiblendA{74} & \hatcurLCiblendA{75}& \hatcurLCiblendA{76}& \hatcurLCiblendA{77}\\
~~~HATSouth 2\dotfill & $\cdots$ & $\hatcurLCiblendB{75}$ & $\cdots$ & $\cdots$ \\
~~~{\em TESS} 1\dotfill & \hatcurLCiblendB{74} & \hatcurLCiblendC{75}& \hatcurLCiblendB{76}& \hatcurLCiblendB{77}\\
~~~{\em TESS} 2\dotfill & $\cdots$ & \hatcurLCiblendD{75}& $\cdots$ & $\cdots$ \\
\sidehead{Limb-darkening coefficients \tablenotemark{d}}
~~~$c_1,g$                  \dotfill    & $\hatcurLBig{74}$ & $\hatcurLBig{75}$& $\hatcurLBig{76}$& $\hatcurLBig{77}$\\
~~~$c_2,g$                  \dotfill    & $\hatcurLBiig{74}$ & $\hatcurLBiig{75}$& $\hatcurLBiig{76}$& $\hatcurLBiig{77}$\\
~~~$c_1,r$                  \dotfill    & $\hatcurLBir{74}$ & $\hatcurLBir{75}$& $\hatcurLBir{76}$& $\hatcurLBir{77}$\\
~~~$c_2,r$                  \dotfill    & $\hatcurLBiir{74}$ & $\hatcurLBiir{75}$& $\hatcurLBiir{76}$& $\hatcurLBiir{77}$\\
~~~$c_1,R$                  \dotfill    & $\hatcurLBiR{74}$ & $\cdots$ & $\cdots$& $\cdots$\\
~~~$c_2,R$                  \dotfill    & $\hatcurLBiiR{74}$ & $\cdots$ & $\cdots$& $\cdots$\\
~~~$c_1,i$                  \dotfill    & $\hatcurLBii{74}$ & $\cdots$ & $\cdots$& $\hatcurLBii{77}$\\
~~~$c_2,i$                  \dotfill    & $\hatcurLBiii{74}$ & $\cdots$ & $\cdots$& $\hatcurLBiii{77}$\\
~~~$c_1,zs$                  \dotfill    & $\hatcurLBiz{74}$ & $\hatcurLBiz{75}$& $\cdots$& $\cdots$\\
~~~$c_2,zs$                  \dotfill    & $\hatcurLBiiz{74}$ & $\hatcurLBiiz{75}$& $\cdots$& $\cdots$\\
~~~$c_1,T$                  \dotfill    & $\hatcurLBiT{74}$ & $\hatcurLBiT{75}$& $\hatcurLBiT{76}$& $\hatcurLBiT{77}$\\
~~~$c_2,T$                  \dotfill    & $\hatcurLBiiT{74}$ & $\hatcurLBiiT{75}$& $\hatcurLBiiT{76}$& $\hatcurLBiiT{77}$\\
\sidehead{RV parameters}
~~~$K$ (\ms)              \dotfill    & $\hatcurRVK{74}$\phn\phn & $\hatcurRVK{75}$\phn\phn& $\hatcurRVK{76}$\phn\phn& $\hatcurRVK{77}$\phn\phn\\
~~~$e$ \tablenotemark{e}               \dotfill    & $\hatcurRVeccentwosiglimeccen{74}$ & $\hatcurRVeccentwosiglimeccen{75}$ & $\hatcurRVeccentwosiglimeccen{76}$ & $\hatcurRVeccentwosiglimeccen{77}$ \\
~~~RV jitter ESPRESSO (\ms)        \dotfill    & $\hatcurRVjittertwosiglim{74}$ & $\hatcurRVjittertwosiglim{75}$& $\hatcurRVjittertwosiglim{76}$& $\hatcurRVjittertwosiglim{77}$\\
\sidehead{Planetary parameters}
~~~$\mpl$ ($\mjup$)       \dotfill    & $\hatcurPPmlong{74}$ & $\hatcurPPmlong{75}$& $\hatcurPPmlong{76}$& $\hatcurPPmlong{77}$\\
~~~$\rpl$ ($\rjup$)       \dotfill    & $\hatcurPPrlong{74}$ & $\hatcurPPrlong{75}$& $\hatcurPPrlong{76}$& $\hatcurPPrlong{77}$\\
~~~$C(\mpl,\rpl)$
    \tablenotemark{g}     \dotfill    & $\hatcurPPmrcorr{74}$ & $\hatcurPPmrcorr{75}$& $\hatcurPPmrcorr{76}$& $\hatcurPPmrcorr{77}$\\
~~~$\rhopl$ (\gcmc)       \dotfill    & $\hatcurPPrho{74}$ & $\hatcurPPrho{75}$& $\hatcurPPrho{76}$& $\hatcurPPrho{77}$\\
~~~$\log g_p$ (cgs)       \dotfill    & $\hatcurPPlogg{74}$ & $\hatcurPPlogg{75}$& $\hatcurPPlogg{76}$& $\hatcurPPlogg{77}$\\
~~~$a$ (AU)               \dotfill    & $\hatcurPParel{74}$ & $\hatcurPParel{75}$& $\hatcurPParel{76}$& $\hatcurPParel{77}$\\
~~~$T_{\rm eq}$ (K)        \dotfill   & $\hatcurPPteff{74}$ & $\hatcurPPteff{75}$& $\hatcurPPteff{76}$& $\hatcurPPteff{77}$\\
~~~$\Theta$ \tablenotemark{h} \dotfill & $\hatcurPPtheta{74}$ & $\hatcurPPtheta{75}$& $\hatcurPPtheta{76}$& $\hatcurPPtheta{77}$\\
~~~$\log_{10}\langle F \rangle$ (cgs) \tablenotemark{i}
                          \dotfill    & $\hatcurPPfluxavglog{74}$ & $\hatcurPPfluxavglog{75}$& $\hatcurPPfluxavglog{76}$& $\hatcurPPfluxavglog{77}$\\
\enddata
\tablecomments{
For all systems we adopt a model in which the orbit is assumed to be circular. See the discussion in Section~\ref{sec:transitmodel}.
}
\tablenotetext{a}{
    Times are in Barycentric Julian Date calculated on the Barycentric Dynamical Time (TDB) system.
    \ensuremath{T_c}: Reference epoch of
    mid transit that minimizes the correlation with the orbital
    period.
    \ensuremath{T_{12}}: total transit duration, time
    between first to last contact;
    \ensuremath{T_{12}=T_{34}}: ingress/egress time, time between first
    and second, or third and fourth contact.
}
\tablenotetext{b}{
   Reciprocal of the half duration of the transit used as a jump parameter in our MCMC analysis in place of $\arstar$. It is related to $\arstar$ by the expression $\zrstar = \arstar(2\pi(1+e\sin\omega))/(P\sqrt{1-b^2}\sqrt{1-e^2})$ \citep{bakos:2010:hat11}.
}
\tablenotetext{c}{
    Scaling factor applied to the model transit that is fit to the HATSouth and {\em TESS} light curves. This factor accounts for dilution of the transit due to blending from neighboring stars and/or over-filtering of the light curve.  These factors are varied in the fit, with independent values adopted for each light curve. For \hatcur{75} we list separately the independent dilution factors determined for the focus frame, and science frame HATSouth images, and for the {\em TESS} Sectors four and five light curves.
}
\tablenotetext{d}{
    Values for a quadratic law. The limb darkening parameters were
    directly varied in the fit, using the tabulations from
    \cite{claret:2012,claret:2013,claret:2018} to place Gaussian prior
    constraints on their values, assuming a prior uncertainty of $0.2$
    for each coefficient.
}
\tablenotetext{e}{
    The 95\% confidence upper limit on the eccentricity determined
    when $\sqrt{e}\cos\omega$ and $\sqrt{e}\sin\omega$ are allowed to
    vary in the fit.
}
\tablenotetext{f}{
    Term added in quadrature to the formal RV uncertainties for each
    instrument. This is treated as a free parameter in the fitting
    routine.
}
\tablenotetext{g}{
    Correlation coefficient between the planetary mass \mpl\ and radius
    \rpl\ estimated from the posterior parameter distribution.
}
\tablenotetext{h}{
    The Safronov number is given by $\Theta = \frac{1}{2}(V_{\rm
    esc}/V_{\rm orb})^2 = (a/\rpl)(\mpl / \mstar )$
    \citep[see][]{hansen:2007}.
}
\tablenotetext{i}{
    Incoming flux per unit surface area, averaged over the orbit.
}
    \end{deluxetable*}

\end{document}